\documentclass[smallextended]{svjour3}       
\usepackage{pgfplots}
\usepackage{tikz-3dplot}

\tdplotsetmaincoords{60}{135}
\pgfplotsset{compat=newest}
\usepackage{graphicx}
\usepackage{tikz}
\usetikzlibrary{calc,patterns,decorations.pathmorphing,decorations.markings, quotes,angles}
\usepackage{tikz-3dplot}

\makeatletter
\tikzset{
    dot diameter/.store in=\dot@diameter,
    dot diameter=3pt,
    dot spacing/.store in=\dot@spacing,
    dot spacing=10pt,
    dots/.style={
        line width=\dot@diameter,
        line cap=round,
        dash pattern=on 0pt off \dot@spacing
    }
}
\makeatother

\smartqed  
\usepackage[latin1]{inputenc}
\usepackage{graphicx}
\usepackage{tikz}
\usetikzlibrary{calc,patterns,decorations.pathmorphing,decorations.markings, quotes,angles}
\usepackage{tikz-3dplot}
\usepackage{amsmath,amsfonts,amssymb,amscd}
\usepackage{wasysym}
\usepackage{wrapfig}
\usepackage[colorlinks,linkcolor=blue,citecolor=blue,urlcolor=blue]{hyperref}
\usepackage{natbib}
\setcitestyle{round}

\usepackage{tikz}
\usetikzlibrary{calc,patterns,decorations.pathmorphing,decorations.markings}

\newcommand{\erm}{{\rm e} }

\usepackage[graphicx]{realboxes}

\newcommand{\Tr}{{\rm \!\ Tr\!\ } }

\newcommand{\Id}{\mathbb {I}  }

\newcommand{\K}{\mathrm{K}}

\newcommand{\R}{\mathbb{R}}
\newcommand{\C}{\mathbb{C}}
\newcommand{\Oc}{\mathcal{O}}

\newcommand{\gm}{\gamma}

\newcommand{\dt}{\delta}
\newcommand{\ep}{\varepsilon}
\newcommand{\om}{\omega}

\newcommand{\mat}[1]{\mathbf{#1}}
\newcommand{\dmat}[1]{\dot{\mat{#1}}}
\newcommand{\tmat}[1]{\tilde{\mat{#1}}}

\newcommand{\dd}{\mathrm{d}}

\DeclareMathOperator{\tr}{Tr}

\newcommand{\GMEMB}{\ensuremath{\mathcal{G}\mathcal{M}_\mathrm{EMB}}}

\newcommand{\ov}{\overline}
\newcommand{\Irm}{ {\,\rm I}}
\newcommand{\Io}{ {\,\rm I}_\circ}

\newcommand{\0} {\underline 0}

\def\R{{\mathbb R}}

\def\C{{\mathbb C}}

\newcommand{\nd} {\noindent}
\begin{document}

\title{Librations  of a body composed of a deformable mantle and a  fluid core.}


\author{    Clodoaldo Ragazzo    \and
  Gwena\"{e}l Bou\'{e}     \and
        Yeva Gevorgyan       \and
          Lucas S. Ruiz
}


\institute{C. Ragazzo  (ORCID 0000-0002-4277-4173)\at
 Instituto de Matem\'{a}tica e Estat\'{i}stica, Universidade de S\~{a}o Paulo, 05508-090 S\~{a}o Paulo, SP, Brazil\\
 \email{ragazzo@usp.br}
 \and
G. Bou\'{e} \at
              ASD/IMCCE,  CNRS-UMR8028,  Observatoire  de  Paris,  PSL  University,  Sorbonne  Universit{\'e},  77  Avenue  Denfert-Rochereau, 75014 Paris, France \\
              \email{gwenael.boue@obspm.fr}           
           \and
           Y. Gevorgyan \at
              Instituto de Matem\'{a}tica e Estat\'{i}stica, Universidade de S\~{a}o Paulo, 05508-090 S\~{a}o Paulo, SP, Brazil\\
\email{yeva@ime.usp.br}
\and
L.S. Ruiz \at
Instituto de Matem\'{a}tica e Computa\c{c}\~{a}o, Universidade Federal de Itajub\'{a}, 37500-903 Itajub\'{a}, MG, Brazil and\\
CFisUC, Department of Physics, University of Coimbra, Portugal\\
\email{lucasruiz@unifei.edu.br}
}

\date{Received: date / Accepted: date}

\maketitle

\begin{abstract}
  We present fully three-dimensional equations to  describe the rotations of a body made of a deformable mantle and a 
  fluid core.
 The model in its  essence is similar to that 
 used by  INPOP19a (Integration Plan\'{e}taire de l'Observatoire de Paris) \cite{INPOPa}, and  by
  JPL (Jet Propulsion Laboratory) \cite{JPL}, to represent the Moon. The intended advantages of our model are:
 straightforward use  of any  linear-viscoelastic model for the
 rheology of the  mantle; easy numerical implementation in time-domain (no time lags are necessary);
 all  parameters, including those
 related to the ``permanent deformation'', have a physical interpretation.

 The  paper also contains: 1) A physical model to explain
 the  usual lack of hydrostaticity of the mantle (permanent deformation).
2) Formulas for
 free librations of  bodies in and out-of  spin-orbit resonance
 that are valid for any  linear viscoelastic rheology of the mantle. 3) Formulas for the  offset
 between the mantle and the idealized rigid-body motion (Peale's Cassini states). 
 4) Applications to the librations of  Moon, Earth, and Mercury that are
 used for model validation.

\keywords{Tide \and Rheology  \and Libration \and Cassini states \and spin-orbit resonance}
\end{abstract}

\section{Introduction}
\label{intro}

 The interior structure of planets and satellites in the solar system may be very complex.
 Models with several fluid and solid layers
 have been proposed to explain a great variety of dynamical behaviour of
 planets and satellites, see, for instance, \cite{2010JGRE..115.9011B,beuthe2015tidal,2016JGRE..121.1378N,
  HugoSylvio2017, boue2017rotation, 2014Matsuyama, 2018Matsuyama,correia2019spin}.
The understanding  of the physics behind rotation and deformation of bodies  in the
solar system is a requirement for the research on  extrasolar systems.

A generalisation of the physical mechanisms at
play in solar system bodies to exoplanets is only achievable if the model parameters can be related to measurable physical quantities. As expected, more complex internal structure models require more parameters to be fit from the observational data. For the Moon and the Earth, with abundance of measurements, accurate ephemerides such as
 INPOP19a  \cite{INPOPa}, and 
  JPL DE440 and DE441 \cite{JPL}
  are produced and eventually used to fit the parameters. For the objects in the outer solar
  system most of our knowledge on the interior comes from the analysis of their rotational dynamics.
The amplitudes of tiny oscillations, called librations, about a perfect synchronous rotation
are  important  in inferring  the internal structure of these bodies.  
For the bodies beyond our solar system we will hardly have any measurements
beyond the rotational dynamics.

An effective rheological model with fewer parameters is easier to fit
and it can be a better model for the unreachable worlds. Hence it is important to establish correlation between the parameters of the effective and
extended rheological models and to establish the limits of
the reasonable performance of the effective rheologies. 

In this paper we present a class of models to describe the rotational motion of planets and
satellites with a solid  mantle and a fluid core. The models in their essence are similar to those
used by  INPOP and JPL to represent the Moon. 
 The intended advantages of our models are:
 straightforward use  of any  linear-viscoelastic model for the
 rheology of the  mantle; easy numerical implementation in time-domain (no time lags are necessary);
 all  parameters, including those
 related to the ``permanent deformation'', have a physical interpretation; and facility to introduce other effects as
 core-mantle magnetic coupling, deformation of the
 core-mantle boundary (CMB), and solid inner core, by means of a Lagrangian formulation.
 From the model we  obtain formulas for the eigenmodes of  free libration and for
 deviations  from  the usual  Cassini states.

 The paper is organised as follows.

 In Section \ref{preliminaries} we revisit the problem of libration and
 introduce  notation and  expressions used throughout. We introduce a concept of ``guiding frame''
 that is 
 an auxiliary reference frame equivalent  to the
 ``Terrestrial Intermediate Reference System (TIRS)'' defined in  \cite{iers2010} (Section
 [5.4.1]). The guiding frame   will be important in the theoretical analysis of the librations of bodies
 in and out-of spin-orbit resonance.

 Section \ref{spherical} contains a review of our previous results 
 about spherically symmetric bodies
 in the absence of external forces  \cite{Ragazzo17} and about bodies with a fluid core
 \cite{boue2017rotation} and \cite{boue2020cassini}. We pay special attention to the physical
 interpretation of a parameter ``$k_c$'' commonly used to describe the  hydrodynamic friction
 at the core-mantle boundary.

Section \ref{prestress} contains 
the concept of prestress frame.  This concept is used to describe the ``permanent triaxiality''
of bodies out of hydrostatic equilibrium.
We revive and old idea that the lack of hydrostaticity is a long transient state due to 
the viscosity of the mantle  \cite{mckenzie1966}. This idea combined  with our theory
of spherically symmetric bodies allow for the definition of the prestress frame.
Our prestress frame is not different from  the frame of the ``undistorted Moon'' used for instance
in  \cite{eckhardt1981theory}, \cite{Viswanathan19}, and  \cite{Folkner14}. With our approach
we can better understand the fact, implicitly assumed in all these references, that the undistorted frame
is a ``Tisserand frame'' for the mantle. 

Section \ref{equations} contains our main contribution:
an explicit system of ordinary differential equations for the rotational  motion of a body
made of a mantle and a fluid core. The rheology of the mantle can be given by any linear viscoelastic
model, including accurate approximations  \cite{gev2020} \cite{gev2021}
to models with infinite memory as the Andrade model.
The influence of   oceans and other fluids bounded to the mantle
can be incorporated into the rheology,  as far as these effects  can be considered in
a   spherically average sense. The equations can be easily combined with equations for the coupled 
motion of many-body systems (see section 8 of \cite{rr2017}).

In Section \ref{free2}  we present formulas for the eigenvalues and eigenvectors
of the rotational eigenmodes of free-libration  of bodies in and out-of  spin-orbit resonance.
The formulas are presented using Love numbers and are not bounded
to any particular rheological model. 
These formulas are new in  the
sense of their generality but in particular situations
they coincide with several other formulas in the literature.

In Section \ref{offsetsec} we investigate the effect of inertial forces due to  precession that
act differently upon the mantle and the core. As a result we obtain formulas for the angular
displacement of the mantle from the usual Cassini state 
of a rigid-body \cite{peale1969generalized}. These formulas
are new. They can be considered as  generalisations of those obtained in
\cite{baland2017obliquity} for Mercury modelled as a solid body with no fluid core.

In Section \ref{chandler} we use one of our libration formulas to show
how to calibrate the parameters of some rheological models (Kelvin-Voigt, Generalised Maxwell, and Andrade)
using the  Chandler's wobble period and its
quality factor both estimated from observations.  

In Section \ref{forcedlib} we present general comments about the usual linear approach
to forced libration, in particular its limitations in the description of parametric resonances.

In Section \ref{INPOP} we compare numerical integrations obtained with our model
and those obtained with   INPOP19a  \cite{INPOPa}.
The parameters of our model were calibrated according to those of INPOP.
The agreement between the results are excellent, since the small differences can be explained
by physical effects (high degree gravitational moments) not taken into account in our model.

 Section \ref{conclusions} is a conclusion where 
 we summarise and discuss the main results  in the paper.

 There are five Appendices. In Appendix
 \ref{hansen} we present formulas for the mean gravitational coefficients  of  bodies
 in a Cassini state as defined in \cite{peale1969generalized}.
 These coefficients appear in the linearized equations for librations. Appendix \ref{simple}
 contains an example to illustrate the cancellation of inertial forces that appear in
 the guiding-frame. Appendix \ref{class} contains an algorithm to separate 
  the Nearly Diurnal Free Wobble
  (NDFW) from the  Free Libration in Latitude (FLL), since the two modes have the same
  essential characteristics.
  Appendices \ref{Poinc} and \ref{offset} are related  to the dynamics
 of the fluid inside the  core.   In Appendix \ref{Poinc}
 we show that the well known  Poincar\'e-Hough flow for fluids with variable density
  leads to the same  equations for the angular
  momentum we have used. We remark that our equations are not bounded
  by the Poincar\'e-Hough model.  In Appendix \ref{offset} we study the offset 
  of the mean angular velocity of the core in the same way we did for  the mantle in Section
\ref{offsetsec}.   This Appendix has two parts. The first one is related to bodies
that are out of any spin-orbit resonance, e.g. the Earth,
and the second to bodies in spin-orbit resonance,
e.g.  Moon and Mercury. In the first we show that the mean angular velocity of the
fluid in the core coincides with the mean vorticity of the
Roberts-Stewartson (viscous) flow \cite{stewartson1963motion}, \cite{roberts1965motion}.
 In the second part
we show that if the mantle is rigid and the core-mantle friction is neglected, then our
formulas for the core and mantle offsets  coincide with those obtained in \cite{boue2020cassini}
for the Cassini states of bodies with a fluid core.

 \section{The problem of libration, preliminaries, and notation.}
\label{preliminaries}

\subsection{The problem and frames of reference.}

Most  large celestial bodies rotate almost steadily and with no deformation
about the  axis
of  largest moment of inertia. Rotational librations, or just librations,
are small deviations from this dominant motion. 

We aim to describe the libration of an extended body  of mass $m$
under the gravitational field of  $N$  point  masses $m_\beta$, $\beta=1,\ldots,N$.
The positions  of the  centres of mass of all
bodies  as a function of time  are supposed to be known.
Let $\kappa$ be  an orthonormal reference frame
with origin at the centre of mass of the extended body and 
with  axes that are parallel to the axes of a given inertial frame.
 We  assume that the extended body is small enough, compared
 to its distance to the point masses, such that the   Taylor  expansion
 of the gravitational force field of the point masses about 
 the origin of $\kappa$ can be truncated at first order with a negligible error.
 The zero order term of the Taylor
expansion  cancels out the  inertial   force that appears  in the
accelerated frame  $\kappa$, so that  the torque upon the body is determined by the
first order term of the Taylor expansion. The rotational dynamics of the extended body in
$\kappa$ happens as if $\kappa$ were an inertial frame and for this reason we will
refer to $\kappa$ as ``the inertial frame''.

If 
$\mathbf{I}_T:\kappa\to\kappa$ is the moment of inertia  operator and
$\boldsymbol{\pi}_T\in\kappa$ is  the angular momentum vector  
of the extended body  and 
 $\mathbf{r}_\beta(t)\in\kappa$ is the position of the point mass
$\beta$, then  
\begin{equation}
  \label{eqrot1}
  \dot{\boldsymbol{\pi}}_T =\sum_\beta
  \frac{3{\cal G}m_\beta}{\|\mat r_\beta\|^5}\mathbf{r}_\beta\times\mat I_T\,\mathbf{r}_\beta \, 
\end{equation}
is the Euler's equation for the motion of the extended body.

If the extended body is rigid, then there is  a  frame, the body frame,
in which the body  remains at rest. In particular, the
angular momentum of the body with respect to the body  frame
is null. If the body is deformable, then there still exists a frame with respect to
which the body angular momentum is null: the Tisserand frame. The angular velocity
$\boldsymbol{\omega}_T(t)\in\kappa$ of the Tisserand frame (the index $T$ stands simultaneously for total and
Tisserand)  is uniquely defined by both $\mat I_T(t)$
and  $\boldsymbol{\pi}_T(t)$ by means of $\boldsymbol{\pi}_T=\mat I_T \boldsymbol{\omega}_T$.
Integration of  $\boldsymbol{\omega}_T(t)$ defines an orthogonal transformation
 $\mat R_T:\K_T\to\kappa$
 that determines the motion of the Tisserand frame $\K_T$ within $\kappa$. The angular velocity
of   $\K_T$ can be interpreted as a  mean angular velocity of the body \citep{MunkMac}.

In this paper  the extended body is assumed to be  made of a deformable mantle along with  a fluid core.
The  core mantle boundary (CMB) is assumed to be rigid. 
The body  is supposed to satisfy the following  hypotheses:
\begin{equation}\begin{split}
    &\text{(a) the layers of constant density are almost spherical and ellipsoidal,}\\ &\text{(b)  deformations
are small, and}\\ &\text{(c) the  material of the  body
is incompressible.}
\end{split}\label{hypotheses}
\end{equation}

We will use different Tisserand frames to describe the average rotation of the mantle and of the core.
In table \ref{frames}
we list these frames and the respective
orthogonal transformation and angular velocities associated with them. 
\begin{table}
\begin{center}
  \caption{\label{frames}
    List of  ``body frames'' and  angular velocity vectors used  throughout the paper.
    If a vector is not represented in the inertial frame, then a second
    index is used to  show  in which frame  the vector is being represented, e.g, $
    \boldsymbol {\omega}_{c,m}\in\K_m$ means the representation of the Tisserand angular velocity
    of the core in the Tisserand frame of the mantle.
  }
\renewcommand{\arraystretch}{1.5}
\begin{tabular}{|l |  l |}
  \hline
  $\kappa$ & Inertial frame\\
  $\mat R_T:\K_T\to \kappa$& $\K_T$ Tisserand frame of the whole body \\
   $\mat R_m:\K_m\to \kappa$ & $\K_m$ Tisserand frame of the mantle \\
  $\mat R_c:\K_c\to \kappa$ & $\K_c$ Tisserand frame of the core\\
  $\mat R_p:\K_p\to \kappa$ & $\K_p$ Frame of the principal axes of inertia\\
  $\boldsymbol \omega_T\in\kappa$ & Tisserand angular velocity of the whole body\\
  $\boldsymbol \omega_m\in\kappa$ & Tisserand angular velocity of the mantle\\
      $\boldsymbol \omega_c\in\kappa$ & Tisserand angular velocity of the core\\
  $\boldsymbol \omega_{T,m}\in\K_m$ & Representation of $\boldsymbol \omega_T$ in the frame of the mantle\\
  $\boldsymbol \omega_{m,m}\in\K_m$ & Representation of $\boldsymbol \omega_m$ in the frame of the mantle\\
  $\boldsymbol \omega_{c,m}\in\K_m$ & Representation of $\boldsymbol \omega_c$ in the frame of the mantle\\
                                                                            \hline
\end{tabular}
\end{center}
\end{table}

\subsection{The ``hat map''
  and operators on different frames.}
\label{sechat}
Many times we will represent the angular velocity and the torque as matrices and not as vectors.
The reason for this unusual choice  is  our deformation theory that  uses the traceless part of the inertia
tensor as deformation variable.  In this paragraph we introduce some of the notation
to be used throughout the paper.

{\bf Vectors}  will be represented in bold face
and  small letters. {\bf Matrices} will be represented in bold face, except for the
the identity that will be represented as $\Id$.
Following \cite{holm}, to every vector $\mathbf{x}\in \mathbb{R}^3$
we associate an anti-symmetric operator  by means of the so-called ``hat map''
defined as
\begin{equation}
  \begin{split}
& \mathbf{x} = \begin{pmatrix} x_1 \\ x_2 \\ x_3 \end{pmatrix}  \in \mathbb{R}^3
\longmapsto 
\mathbf{\widehat x} = \begin{bmatrix}
0 & -x_3 & x_2 \\
x_3 & 0 & -x_1 \\
-x_2 & x_1 & 0
\end{bmatrix} \\
& \text{or}\\
&  \mathbf{\widehat x}_{ij} = -\sum_k\ep_{ijk} x_k\, ,
\end{split}\label{hatmap}
\end{equation}
where $\ep_{ijk}$ is the Levi-Civita  anti-symmetric tensor.
The inverse of the hat map is the  ``check map'' $\mat A\to \check{\mat A}$ that maps an antisymmetric
matrix to a vector: $\check{\mat A}_i=-\frac{1}{2} \sum_{jk}\epsilon_{ijk} A_{jk}$.
In table \ref{list} we present a list of formulas that are useful when
dealing with vectors, matrices, and the hat operator.
\begin{table}
\begin{center}
  \caption{\label{list} List of formulas involving  vectors ($\mathbf{x}$ and
    $\mathbf{y}$) and  matrices ($\Id$,
    $\mathbf{A}$, $\mathbf{B}$, $\mathbf{S}$, $\mathbf{R}$),
    where: $\Id$ is the identity,
    $\mathbf{S}$ is symmetric,
    $\mathbf{R}$ is a generic rotation matrix.}
\renewcommand{\arraystretch}{1.5}
\begin{tabular}{rclc}
  \hline
  $[\mathbf{A},\mathbf{B}]$&=&$\mathbf{A}\mathbf{B}-\mathbf{B}\mathbf{A}$& Commutator\\
  $\langle \mathbf{x},\mathbf{y}\rangle$&=&$\sum_i x_i y_i$& Inn. prod. vectors\\
   $\langle \mathbf{A},\mathbf{B}\rangle$ &=&$\frac{1}{2}
  \Tr\big(\mathbf{A}\mathbf{B}^{\rm T}\big)=
                                              \frac{1}{2}\sum_{ij}A_{ij}B_{ij}$& Inn. prod matrices\\
   $\big(\mathbf{x}\otimes\mathbf{y})_{ij}$&=&$x_iy_j$& Tensor product\\
  $ \mathbf{\widehat x}\,\mathbf{y}$&=&$\mathbf{x}\times \mathbf{y}$&(a)\\
$  \left[ \mathbf{\widehat x}\, , \mathbf{\widehat y}\right]$&=&$
    \widehat{\mathbf{x}\times \mathbf{y}}$&(b)\\
  $\langle \mathbf{x} ,\mathbf{y}\rangle$&=&$\langle \mathbf{\widehat x} ,
    \mathbf{\widehat y}\rangle=\frac{1}{2}
  \Tr\big(\mathbf{\widehat x}\mathbf{\widehat y}^{\rm T}\big)$&(c)\\
  $\left[\mathbf{\widehat x},\mathbf{A}\right]\mathbf{y}$&=&
   $                  \mathbf{x}\times \mathbf{A}\mathbf{y}+
    \mathbf{A}\big(\mathbf{y}\times\mathbf{x}\big)$&(d)\\
$ \langle\mathbf{x},\mathbf{A}\mathbf{x}\rangle$&=&$
   \Tr\big(\mathbf{A}\big)|\mathbf{x}|^2+2\langle\mathbf{A}, \mathbf{\widehat x}
 \mathbf{\widehat x}\rangle$&(e)\\
$ \left[ \mathbf{S},\mathbf{x}\otimes\mathbf{x}\right]$&=&$
  \widehat {\mathbf{x}\times\mathbf{S}\mathbf{x}}$&(f)\\
  $\widehat{\mathbf{R}\mathbf{x}}$&=&$\mathbf{R}\,\widehat{\mathbf{x}}\,\mathbf{R}^{-1}$&
                                                                                          (g)\\

  $\widehat{\mathbf{A}\mathbf{x}}$&=&$\tr\big(\mathbf{A}\big)\widehat{\mathbf{x}}-
                                                                                         \mathbf{A}^{\rm {T}}\widehat{\mathbf{x}}-\widehat{\mathbf{x}}\mathbf{A}$&(h)\\
$ \mathbf{x}\otimes\mathbf{x}-\frac{1}{3}|\mathbf{x}|^2\Id $&=& $\widehat{\mathbf x}
\widehat{\mathbf x}-
\frac{1}{3}\tr\big( \widehat{\mathbf x}\widehat{\mathbf x}\big)\Id$&(i)\\
\hline
\end{tabular}
\end{center}
\end{table}

 The angular velocity operator 
associated with the rotations of the mantle  $\mat R_m:\K_m\to\kappa$ is denoted as  
$\boldsymbol {\widehat \omega}_m:=\dot{\mat R}_m\mat R_m^{-1}:\kappa\to\kappa$ and the same applies to
all rotations that appear in table \ref{frames}.

The angular velocity operator
$\boldsymbol {\widehat \omega}_m$, the inertia operator, and other operators on $\kappa$
can be transformed to  other frames:
\begin{equation}\renewcommand\arraystretch{1.5}\begin{array}{rll}
\boldsymbol {\widehat \omega}_{m,m}&:=\mat R_m^{-1}\boldsymbol {\widehat \omega}_{m}\mat R_m=
\mat R_m^{-1}\dot{\mat R}_m \quad &\text{or}\quad \boldsymbol {\omega}_{m,m}:=
\mat R_m^{-1}\boldsymbol {\omega}_{m}\\
                  \boldsymbol {\widehat \omega}_{c,m}&:=\mat R_m^{-1}\boldsymbol {\widehat \omega}_{c}\mat R_m
&\text{or}\quad \boldsymbol {\omega}_{c,m}:=
                                                                                                                 \mat R_m^{-1}\boldsymbol {\omega}_{c}\,,\\
                                                                                                                 \mat I_{T,m}&:=\mat R_m^{-1}\mat I_T\mat R_m\,,\quad \text{etc}\,,
\end{array}
\label{notationomega}
\end{equation}
where the second index defines the space in which  the operator acts, e.g, $
\boldsymbol {\widehat \omega}_{c,m}:\K_m\to\K_m$.

We denote the rotation matrices about the coordinate axes as
 \begin{equation}\label{rotmat}\begin{array}{lrl}
   \mathbf{R_1}(\theta)&=\begin{bmatrix}
       1&0&0\\
0&\cos(\theta) & -\sin(\theta)  \\
0&\sin(\theta) & \ \  \cos(\theta)  
\end{bmatrix}\quad 
\mathbf{R_2}(\theta)&=\begin{bmatrix}
\ \  \cos(\theta) &0&  \sin(\theta)\\
0&1&0\\
-\sin(\theta) & 0& \cos(\theta)
\end{bmatrix}
                                 \\ & & \\
    \mathbf{R_3}(\theta)&=\begin{bmatrix}
\cos(\theta) & -\sin(\theta) & 0 \\
\sin(\theta) & \ \ \cos(\theta) & 0 \\
0 & 0 & 1
\end{bmatrix}\quad\text{with}& \quad
                               \Big(\frac{\dd}{\dd \theta}\mathbf{R_i}(\theta)\Big)
                               \,  \mathbf{R_i}^{-1}(\theta)|_{\theta=0}=\mathbf{\widehat e_i}\,.
                               \end{array}
\end{equation}
These matrices will be used to represent transformations between arbitrary frames.
\subsection{Inertia operators and deformation operators.}  
\label{subinertia}

The three hypotheses in (\ref{hypotheses}) imply that   \begin{equation}
  {\rm I}_{\circ}:=\frac{1}{3}\Tr\big(\mathbf{I}_T\big)
  \label{Io}
\end{equation}
is constant in time
\citep[this observation is due to G. Darwin]{rochester1974changes}. If the inertia operator
is split
 into  isotropic and   traceless parts
\begin{equation}
\mathbf{I}_T=\frac{\tr\mathbf{I}_T}{3}\Id+\bigg(\mathbf{I}_T-\frac{\tr\mathbf{I}_T}{3}\Id\bigg)
:=  {\rm I}_{\circ}\Id- {\rm I}_{\circ}\mathbf{B}_T
\label{iso}
\end{equation}
then $\dot {\mathbf{I}}_T= - {\rm I}_{\circ}\dot{\mathbf{B}}_T$. We call $\mat B_T:\kappa\to\kappa$
the deformation operator of the whole body (deformation with respect to the spherical configuration).
The same decomposition can be applied
to the moment of inertia operator of the mantle and of the core as summarised in Table \ref{inertial}.
Note that the trace does not depend on the frame, so ${\rm I}_{\circ}=\tr\mathbf{I}_{T,m}/3=
\tr\mathbf{I}_{T,c}/3=\ldots$. 
\begin{table}
\begin{center}
  \caption{\label{inertial} List of coefficients of inertia and relations among them used in the paper.
    If an operator is not represented in the inertial frame, then a second
    index is used to  show  in which frame  the operator is being represented, e.g, $
    \mat I_{m,m}:\K_m\to \K_m$ means the representation of the moment of inertia of the mantle
    in the Tisserand frame of the mantle.}
\renewcommand{\arraystretch}{1.5}
\begin{tabular}{|l |  l |}
  \hline
  $\mat I_m:\kappa\to\kappa$& Moment of inertia of the mantle\\
  $\mat I_c:\kappa\to\kappa$& Moment of inertia of the core\\
   $\mathbf{I}_T=\mathbf{I}_m+\mathbf{I}_c$& Total Moment of inertia                                                                       in $\kappa$\\
  $\mathbf{B}_m:\kappa\to\kappa$& Deformation operator of the mantle,
                                  $\mathbf{I}_m= {\rm I}_{\circ m}\Id- {\rm I}_{\circ m}\mathbf{B}_m $\\
  $\mathbf{B}_c:\kappa\to\kappa$& Deformation operator of the core,
                                  $\mathbf{I}_c= {\rm I}_{\circ c}\Id- {\rm I}_{\circ c}\mathbf{B}_c $\\
  $\mathbf{B}_T:\kappa\to\kappa$& Deformation operator of the whole body,
                                   $\mathbf{I}_T= {\rm I}_{\circ}\Id- {\rm I}_{\circ}\mathbf{B}_T $\\
  ${\rm I}_{\circ}=\frac{1}{3}\tr\big(\mat I_T\big)$& Mean total moment of inertia\\
  ${\rm I}_{\circ m}=\frac{1}{3}\tr\big(\mat I_m\big)$& Mean  moment of inertia of the mantle\\
  ${\rm I}_{\circ c}=\frac{1}{3}\tr\big(\mat I_c\big)$& Mean  moment of inertia of the core\\
   $f_\circ=\frac{{\rm I}_{\circ c}}{{\rm I}_{\circ m}}$& Parameter of significance of the core\\
  ${\rm I}_{\circ}={\rm I}_{\circ m}+{\rm I}_{\circ c}$& Mean moment of inertia identity\\
  $\mat I_T \boldsymbol{\om}_T=\mat I_m\boldsymbol{\omega}_m+\mat I_c\boldsymbol{\omega}_c$ &
                                                                                              Tisserand frames identity  (consequence of  $\boldsymbol{\pi}_T=\boldsymbol{\pi}_m+\boldsymbol{\pi}_c$)
  \\                                                                                      ${\rm I}_{\circ}\mathbf{B}_T={\rm I}_{\circ m}\mathbf{B}_m+{\rm I}_{\circ c}\mathbf{B}_c$&
Deformation operators identity\\
    $\mat I_{m,m}=\mat R_m^{-1}\mat I_m\mat R_m$& Representation of $\mat I_m$ in the mantle frame\\
  $\mat I_{c,m}=\mat R_m^{-1}\mat I_m\mat R_m$& Representation of $\mat I_c$ in the mantle frame\\
  \hline
\end{tabular}
\end{center}
\end{table}

Since the body remains almost
spherical for all time,  $|\mathbf{B}_T|\ll 1$,    
\begin{equation}
  \mathbf{I}_T^{-1}=\frac{1}{ {\rm I}_{\circ}}\big(\Id+\mathbf{B}_T\big)\,, 
\label{Iinverse}
\end{equation}
where terms of order  $|\mathbf{B}_T|^2$ were neglected. The same applies to  $ \mathbf{I}_m^{-1}$ and
 $ \mathbf{I}_c^{-1}$.

Using the hat map and the identities (f) and (h) in Table \ref{list} 
we can rewrite Euler's equation (\ref{eqrot1}) 
in matricial form\footnote{ Some expressions become simpler if we use the density tensor 
  $\mat M_T$
\cite{chandrasekhar1969ellipsoidal} defined as
\begin{equation}\begin{split}
    &M_{Tij}= \int \rho(\mat x)x_ix_jd\mat x^3\,,\quad \text{with:}\quad
    {\rm I}_{\circ}=\frac{2}{3}\Tr\big(\mathbf{M}_T\big)\,,\quad
    \mathbf{M}_T=\frac{1}{2} {\rm I}_{\circ}\Id+ {\rm I}_{\circ}\mathbf{B}_T\,,\\ & 
    \mat M_T=\frac{\Tr\mat I_T}{2}\Id-\mat I_T\,
    \quad \text{and}\quad \mat I_T=\Tr \big(\mat M_T\big)\Id-\mat M_T\,.
  \end{split}
  \label{M}
\end{equation}
For instance, $ \boldsymbol{\widehat \pi}_T=
  \mat M_T \, \boldsymbol{\widehat \omega}_T +
  \boldsymbol{\widehat \omega}_T\,\mat M_T$.
}
:
\begin{equation}
  \label{eqrot2}\begin{split}
  \dot{\boldsymbol{\widehat \pi}}_T &
  =[\mat I_T\, ,\mathbf{J}]=-\Io[\mat B_T\,,\mathbf{J}]\\
  \mathbf{J}&:=\sum_\beta
  \frac{3{\cal G}m_\beta}{\|\mat r_\beta\|^5}\mathbf{r}_\beta\otimes\mathbf{r}_\beta\quad\text{
    (tidal force matrix),}\\
  \boldsymbol{\widehat \pi}_T&=
  \Tr \big(\mat I_T\big) \boldsymbol{\widehat \omega}_T-
  \mat I_T \boldsymbol{\widehat \omega}_T-
  \boldsymbol{\widehat \omega}_T\mat I_T\,.
  \end{split}
\end{equation}
The tidal force  matrix $\mathbf{J}$ is symmetric and represents the whole external force field acting
upon the body while the anti-symmetric matrix $[\mat I_T,\mathbf{J}]=-\Io[\mat B_T\,,\mathbf{J}]$ is the
total torque matrix.

Using the check map $^\vee$ (the inverse of the hat map)  we can rewrite equation (\ref{eqrot1}) as
\begin{equation}
  \label{eqrot1.5}
  \dot{\boldsymbol{\pi}}_T =[\mat I_T\, ,\mathbf{J}]^\vee
\end{equation}
that is probably the most convenient form of Euler's equation, since it keeps the simplicity
of the vectorial form of the angular momentum, $\boldsymbol{\pi}_T=\mat I_T\boldsymbol\omega_T$,
  while  separates the force operator $\mat J$
  from the inertia operator $\mat I_T$ in the torque.

 \subsection{The guiding frame and  nominal (average) moment of inertia.}
\label{aerageI}
 
The operation of averaging, e. g.  
\begin{equation}
 \mat I_{T,m}\to
 \lim_{\tau\to\infty}\frac{1}{\tau}\int_0^\tau
 \underbrace{\mat R_m^{-1}(t){\mat I}_{T}(t)\mat R_m(t)}_{=\mat I_{T,m}(t)}dt
     \label{ovT}
\end{equation}
depends on the frame in which $\mat I_T$ is represented. The Tisserand frame of the mantle
(or of the whole body), which  would be a natural  frame for the averaging, is a priori unknown.
So, in order to give a meaning to  average (or nominal)
moments of inertia and to average forces we need to
first  define an operational reference frame: the  {\bf Guiding Frame}.

The definition of the guiding frame is based on the  ``libration hypothesis'', namely
\begin{equation}\begin{split}&\text{The extended body rotates almost steadily and with no}\\
    &\text{deformation about the  axis
of largest moment of inertia.}\end{split}\label{libhyp}\end{equation}
The ideal rigid-body motion
of the extended body will be  called 
the `` guiding motion''. The guiding motion can
be described as follows: $-$ the sidereal  angular speed $\omega$
 remains constant for all time, $-$ in
any time window containing many revolutions the spin axis stays nearly fixed
with respect to the inertial frame  $\kappa$, and $-$
eventually a slow motion  of the spin axis may occur. The guiding motion is realised by the
transformation  $\mat R_g:\K_g\to\kappa$, where  $\K_g$ is the ``guiding frame''.

The guiding motion can be factorised using an intermediate frame $\K_s$ that we call the ``slow frame''.
The factorisation is 
\begin{equation}
  \mathbf{R}_g(t)=\mathbf{R}_s(t)\mathbf{R_3}(\omega t):\K_g\to \kappa\, ,
 \label{Rg}
  \end{equation}
  where  $\mathbf{R_3}(\omega t):\K_g\to \K_s$ is the dominant-rotational motion  and
  $\mat R_s(t):\K_s\to\kappa$ is the  motion of the slow frame $\K_s$ within $\kappa$.
   In order to define the slow frame $\K_s$ 
   we use the notion of ``non-rotating origin''
   \cite{guinot1979basic} (see also \cite{capitaine1986non}) and impose that the projection of the
   angular velocity of the slow frame $\boldsymbol \omega_{s,s}\in\K_s$
   ($\boldsymbol {\widehat \omega}_{s,s}=\mat R_s^{-1}\dot{\mat R}_s$) 
   on the polar axis $\mat e_3\in\K_s$
   is null. With this definition the angular velocity of the guiding frame is given by
  \begin{equation}\begin{split}
&    \boldsymbol{\omega}_{g,g}=
    \omega \mathbf{e_3}+ \boldsymbol {\omega}_{s,g}\in \K_g\quad
    \big( \boldsymbol{\widehat \omega}_{g,g}=
    \mat R_g^{-1}\dot{\mat R}_g\big) 
    \, ,\\
    &\text{where:}\quad\boldsymbol{\omega}_{s,g}=\mat R_g^{-1}\boldsymbol {\omega}_s\,,\quad
    |\boldsymbol{\omega}_s|\ll \omega\,,\quad\langle\boldsymbol{\omega}_{s,g}\,,\mat e_3\rangle=0\,,
    \end{split}
     \label{omegag}
    \end{equation}
 and $\omega$ is the nominal 
 sidereal angular speed of the extended body, namely  the projection of the angular velocity
of the guiding motion  $\boldsymbol \omega_{g,g}$  on
the  polar axis $\mat e_3\in\K_g$. 
 If the spin axis does not move,
  then a basis  $\{\mathbf{e_1},\mathbf{e_2},\mathbf{e_3}\}$ of $\kappa$ is 
  chosen such that $\mathbf{R}_s=\Id$ and
  $\boldsymbol{\omega}_g=\omega \mathbf{e_3}$
  \footnote{Most  frames we have defined  are similar to  those
    used to describe the rotational motion of the Earth in the IERS2010, chapters 2 to 5 \cite{iers2010}. The
    correspondence is  the following (the number in brackets refers to a section in the IERS 2010):
    ``International Terrestrial Reference System (ITRS)[4.1.1]''$\to\K_m$,
    ``Terrestrial Intermediate Reference System (TIRS)[5.4.1]''$\to\K_g$, ``Celestial Intermediate Reference
    System (CIRS) [5.4.2 and 5.4.4]''$\to \K_s$, and
    ``Geocentric Celestial Reference System (GCRS) [5.4.4]''$\to \kappa$. Our  definition of
    $\K_m$  is different but related to that of the  ITRS after the identification of $\K_m$ with the
    prestress frame. Our definitions of  guiding frame and  slow frame are conventional as well as
    those of  TIRS and  CIRS in the  IERS2010 [5.3.2]. We decided to  give different names to
    reference systems already defined in the IERS2010 because those in the later have precise definitions,
    which applies   to the Earth,  while  ours $\K_g$ and $\K_s$ do not, since they are to be applied to any
    libration problem.}.

  In the guiding motion
the average, or  nominal, moment of inertia tensor  of the extended  body
is given by a diagonal matrix $\ov{\mat I}$ with
constant entries $\ov I_{1}\le \ov I_{2}< \ov I_{3}$. The nominal deformation matrix $\ov{\mat B}$ in $\K_g$
is defined by $\ov{\mat I}={\rm I}_{\circ}(\Id-\ov{ \mat B})$ and several nominal ellipticity coefficients
$\ov {\boldsymbol \alpha},\ov {\boldsymbol \beta},\ldots$ 
are defined in Table \ref{frames2}.

The libration hypothesis (\ref{libhyp}) implies that the   Tisserand frame 
of the mantle $\K_m$ (and also of the whole body $\K_T$) remains close to the guiding
frame $\K_g$. Therefore there exists a small  angular  (antisymmetric) matrix 
$\boldsymbol{\widehat \alpha}_m$ such that
\begin{equation}
  \mat R_g^{-1}\mat R_m=\exp\boldsymbol{\widehat \alpha}_m
  \approx\Id+\boldsymbol{\widehat \alpha}_m:\K_m\to\K_g\,.\label{apr1m}
  \end{equation}
  The component $\alpha_{mj}$ of the angular vector $\boldsymbol\alpha_m$ represents the angle of rotation
  about the axis $\mat e_j\in\K_g$, with the usual orientation,
  induced by $\mat R_g^{-1}\mat R_m:\K_m\to\K_g$\footnote{    
 In \cite{eckhardt1981theory}, for instance, three small
 angles $(\sigma,\rho,\tau)$ describe the deviation of the lunar orientation from the ideal Cassini state, which
 is our guiding motion.
 A computation using Eckhardt's parameterization 
 of the Moon's body frame and the approximation
 $\cos(I)=\cos(\iota_p)\approx 1$
 shows that $\tau$ is equal to our angle
$\alpha_{m3}$. The relation between $(\sigma,\rho)$ to $(\alpha_{m1},\alpha_{m2})$ is not so simple
and instead of these angles it is more convenient to use, as Eckhardt did,
 ``the selenographic unit vector to the pole of the
 ecliptic''  $\mat p=(p_1,p_2,p_3)$. In our notation, $\mat p= \mat R_m^{-1} \mat e_3=
\mat R_m^{-1}\mat R_g\mat R_g^{-1}\mat e_3\approx (\Id-\boldsymbol{\widehat \alpha}_m)\mat R_g^{-1}\mat e_3.$ The same approximation used before,    $\cos(I)=\cos(\iota_p)\approx 1$, 
gives $\mat R_g^{-1}\mat e_3=\mat e_3$ and $\mat p\approx(-\alpha_{m2},\alpha_{m1},1)$
(we are assuming
that the orientation of the Axis 1 of the guiding frame is  positive towards the Earth). So, we get the correspondence
\begin{equation}
  (p_1,p_2,\tau)=(-\alpha_{m2},\alpha_{m1},\alpha_{m3})\label{Eck2}
\end{equation}
between the libration
elements used by Eckhardt and ours.  The triple $(p_1,p_2,\tau)$ also appears in
Eckhardt's work, equation (5), where it is denoted as $\mat X$.

}. Several angular vectors used in the paper are listed
  in Table \ref{frames2}.

In  the real motion the moment of inertia of the extended body in the guiding frame is given by
\begin{equation}
  \mat I_{T,g}(t)= \ov {\mat I}+
  \boldsymbol{\dt}\mat I_{T,g}
  \label{IR3}
  \end{equation}
  where   $\boldsymbol{\dt}\mat I_{T,g}$ is  small (hypothesis (\ref{libhyp}))
  and its time average is null. Equation (\ref{apr1m}) implies
  that in the mantle frame, and up to first order in the small quantities $\|\boldsymbol \alpha_m\|$ and
  $\|\boldsymbol{\dt}\mat I_{T,g}\|$,
  the moment of inertia operator
  can be written as
  \begin{equation}
    \big(\Id-\boldsymbol{\widehat \alpha}_m\big)\big(\ov {\mat I}+
  \boldsymbol{\dt}\mat I_{T,g}\big)
    \big(\Id+\boldsymbol{\widehat \alpha}_m \big)\approx \ov {\mat I}+
    \boldsymbol{\dt}\mat I_{T,g}+\big[ \ov {\mat I},\boldsymbol{\widehat \alpha}_m\big]\,.
    \label{apr2m}
    \end{equation}
    Assuming that the time average of $\boldsymbol \alpha_m$ is either zero or of the order of the small terms
    that have already been neglected (if this statement were not true, then the definition of the
    guiding frame should have to be changed accordingly), then
    \begin{equation}
      \ov {\mat I}=   \lim_{\tau\to\infty}\frac{1}{\tau}\int_0^\tau
 \mat I_{T,g}(t)dt\approx
      \lim_{\tau\to\infty}\frac{1}{\tau}\int_0^\tau
      \mat I_{T,m}(t)dt,\label{apr3m}
  \end{equation}
  and so the time average of the moment of inertia tensor in the guiding frame coincides, up to terms of second
  order in small quantities,  with the time average of the  moment of inertia tensor in the mantle frame.
  The same reasoning implies that  $ \ov {\mat I}$
  can be understood as a time average of the moment of inertia operator in any  frame that oscillates 
  close to the guiding frame.

 \begin{table}
\begin{center}
  \caption{\label{frames2}
    List of: auxiliary reference frames, angular vectors, and   nominal (average)
    quantities associated with the  moment of inertia  used throughout the paper.
    The time  average unnormalized Stokes coefficients of the gravitational field of the extended
    body 
    in the guiding frame satisfy  
      $C_{21}=S_{21}=S_{22}=0$. $m$ is the mass and $R$ is the volumetric mean radius of the extended body.
  }
\renewcommand{\arraystretch}{1.5}
\begin{tabular}{|l |  l |}
  \hline
  $\mat R_g:\K_g\to \kappa$ & $\K_g$ Guiding frame\\
   $\mat R_s:\K_s\to \kappa$ & $\K_s$ Slow frame\\
   $\mat R_{pr}:\K_{pr}\to \kappa$& $\K_{pr}$ Precessional frame, see
                                    equation (\ref{Kpr})\\
  $\mat a$& $\Id+\mat{\widehat a}:\K_T\to\K_g$\\
  $\boldsymbol \alpha_m$&$\Id+\boldsymbol{\widehat \alpha}_m:\K_m\longrightarrow\K_g$\\
  $\boldsymbol \alpha_c$&$\Id+\boldsymbol{\widehat \alpha}_c:\K_c\longrightarrow\K_g$\\
  $\boldsymbol \alpha_p$&    $\Id+  \boldsymbol{\widehat \alpha}_p:\K_p\longrightarrow\K_g$
                          ($\K_p=$Principal axes frame)
  \\
  $\boldsymbol \beta$&   $\Id+  \boldsymbol{\widehat \beta}:\K_p\to\K_m$, see equation (\ref{ad3}),
                       ${\boldsymbol \beta}=\boldsymbol \alpha_p-\boldsymbol \alpha_m$\\ 
      $ \ov{\mat I}=$Diagonal$(\ov I_1,\ov I_2,\ov I_3)$
                            & Time average of $\mat I_{T,g}$ or nominal moment of inertia\\
   $ \ov{\mat I}_m$
                            & Time average of $\mat I_{m,g}$ (mantle)\\
     $ \ov{\mat I}_c$
                            & Time average of $\mat I_{c,g}$ (core), \  $\ov{\mat I}= \ov{\mat I}_m+\ov{\mat I}_c$\\
  $\ov B_{11}$, $\ov B_{22}$, $\ov B_{33}$& Mean deformation,  $\ov {\mat I}=
                                            {\rm I}_{\circ}(\Id-\ov {\mat B})$\\ 
   $\ov \alpha=\big(\ov I_{3}-\ov I_{2}\big)/\ov I_{1}$& $\ov \alpha=- \ov B_{33}+\ov B_{22}+
                                                            \Oc\big(|\mat B|^2)$\\
  
$\ov\beta=\big(\ov I_{3}-\ov I_{1}\big)/\ov I_{2}$& $\ov \beta=- \ov B_{33}+\ov B_{11}+ \Oc\big(|\mat B|^2)$\\
  $    \ov \gm=\big(\ov I_{2}-\ov I_{1}\big)/\ov I_{3}=\frac{4m R^2}{\ov I_3}C_{22}$
    &$\ov \gamma=- \ov B_{22}+\ov B_{11}+ 
                                                            \Oc\big(|\mat B|^2)=\ov \beta-\ov \alpha+\Oc\big(|\mat B|^2)$\\
  $\ov I_{e}=\big(\ov I_{1}+\ov I_{2}\big)/2$& Mean equatorial moment of inertia\\
  $\ov \alpha_e=\big(\ov I_{3}-\ov I_{e}\big)/\ov I_{3}=-\frac{m R^2}{\ov I_3}C_{20}$& $\ov \alpha_e=\big(\ov \alpha+\ov\beta)/2+ \Oc\big(|\mat B|^2)$\\
    $\alpha_{id}=\frac{\omega^2}{\gamma+\mu_0}$ & ideal flatness, equation (\ref{alphaid})\\
   $f_c=\big(\ov I_{c3}-\frac{\ov I_{c1}+\ov I_{c2}}{2}\big)/\ov I_{c3}$& Core oblateness\\
$   \ov B_{11}=\frac{\ov \beta+\ov \gamma}{3}$& Up to order $|\mat B|^2$\\
  $   \ov B_{22}=\frac{\ov \alpha-\ov \gamma}{3}$& Up to order $|\mat B|^2$\\
   $  \ov B_{33}=-\frac{\ov \alpha+\ov\beta}{3}$&Up to order $|\mat B|^2$\\
  $   \ov B_{11}=\frac{2}{3}\frac{m R^2}{{\rm I}_{\circ}}
  \left( 3C_{22}-\frac{1}{2}C_{20}\right)$& $\ov B_{11}$ in terms of unnormalized Stokes coefficients
  \\
   $   \ov B_{22}=\frac{2}{3}\frac{m R^2}{{\rm I}_{\circ}}
  \left( -3C_{22}-\frac{1}{2}C_{20}\right)$& $\ov B_{22}$ in terms of unnormalized Stokes coefficients
  \\
  $   \ov B_{33}=\frac{2}{3}\frac{m R^2}{{\rm I}_{\circ}}  C_{20} $& $\ov B_{33}$
                         in terms of unnormalized Stokes coefficients
  \\
   \hline
\end{tabular}
\end{center}
\end{table}
  
  \subsection{The guiding frame,  the average tidal force, and
     spin-orbit resonances.} \label{averageforce}

   The tidal force  operator transformed to  the  guiding frame
   $\mat J_g=\mat R_g^{-1}\mat J\mat R_g$ can be decomposed into a time-average part
   $\ov{\mat J}$ and
   an oscillatory part $\boldsymbol \delta \mat J_g(t)$. This implies that the torque matrix
   in $\K_g$ can be written as
 \begin{equation}
   [\mat I_g,\mathbf{J}_g]=
   [\ov {\mat I} ,\ov{\mat J}]+  [ \ov {\mat I}, \boldsymbol{\dt}\mathbf{J}_g]+
   [\boldsymbol{\dt}\mat I_g, \ov{\mathbf{J}}] +
 [\boldsymbol{\dt}\mat I_g\, , \boldsymbol{\dt}\mathbf{J}_g]\,.
 \label{torqueg}
\end{equation}
The libration hypothesis (\ref{libhyp})  implies that the last three terms
in the right-hand side of  equation (\ref{torqueg})  are small and the last one is  much smaller
than the others. The first term in the right-hand side
gives a constant  torque.
This term must be null (or very small)  otherwise the guiding motion would be displaced
and could be modified accordingly. Assuming that $\ov I_1<\ov I_2<\ov I_3$,
\begin{equation}
  [\ov {\mat I} ,\ov{\mat J}]=0\label{IJ}
  \end{equation}
implies
   \begin{equation}     \ov{\mat J}=
 \omega^2\left\{c_1\begin{bmatrix}
\frac{1}{3} & 0 & 0 \\
0 & \frac{1}{3} & 0 \\
0 & 0 & -\frac{2}{3}
\end{bmatrix}
+ c_2\begin{bmatrix}
1& 0 & 0 \\
0 & -1 & 0 \\
0 & 0 & 0
\end{bmatrix}
+c_3\begin{bmatrix}
1 & 0 & 0 \\
0 & 1 & 0 \\
0 & 0 & 1
\end{bmatrix}\right\}
\label{Jd3}
\end{equation}
where $c_1,c_2, c_3$ are nondimensional constants. Note that the term proportional to $c_3$ is isotropic
and does not generate any torque.

The mean force matrix transformed to the slow frame
$\ov {\mat J}_s(t)= \mat R_3(\omega t)\ov{\mat J}\mat R_3^{-1}(\omega t)$ is given by  
\begin{equation}
\ov {\mat J}_s(t) =  
\omega^2\left\{ c_1\begin{bmatrix}
\frac{1}{3} & 0 & 0 \\
0 & \frac{1}{3}& 0 \\
0 & 0 & -\frac{2}{3}
\end{bmatrix}
+ c_2\begin{bmatrix}
\cos(2\omega t) & \ \ \sin(2\omega t) & 0 \\
\sin(2\omega t) & -\cos(2\omega t) & 0 \\
0 & 0 & 0
\end{bmatrix}
+ c_3\begin{bmatrix}
1 & 0 & 0 \\
0 & 1 & 0 \\
0 & 0 & 1
\end{bmatrix}\right\}
\end{equation}
This equation implies that $c_2\ne 0$ if, and only if, the tidal force has a Fourier component with
frequency $2 \omega$ in the slow frame. The angular velocity of  $\K_g$
with respect to $\K_s$ is $\omega \mat e_3\in \K_s$, so the spin angular speed of $\K_g$
with respect to $\K_s$, which is the projection
of $\omega \mat e_3\in \K_s$ on $\mat e_3\in\K_s$, is equal to the sidereal angular speed  $\omega$. 
In conclusion, if  $c_2\ne 0$,  then  there must be some orbital  frequency $\omega_{orb}$ such that
$s\, \omega_{orb}= 2 \, \omega$ for some positive integer $s$, so  there is
an $s$-to-$2$ spin-orbit resonance.

In the Appendix \ref{hansen} we   compute the constants $c_1,c_2, c_3$  in the presence and in the absence of
spin-orbit resonances. In Table \ref{force} we list several quantities related to the force operator.
In Table \ref{eigtable} we list the symbols used to denote the free libration eigenfrequencies and
related quantities.
 
 \begin{table}
\begin{center}
  \caption{\label{force}
    List of quantities related to the dynamics,  force,  and torque.
    The names Jeans and Maclaurin associated with the  matrices $\mat S$ and $\mat C$, respectively,
    are due to the fact that  $\ov{\mat S}$ and $\ov{\mat C}$
    are responsible for  the ellipsoidal-hydrostatic deformations
    due to tidal and centrifugal forces, respectively, first studied by these authors (this nomenclature
    was inspired by \cite{ferraz2020tidal}).
    The vector
$\mat e_3\in\kappa$ is the normal to the invariable plane (``Laplace plane'') and
$\mat e_3\in\K_g$ is aligned with the mean axis of largest moment of inertia.}
\renewcommand{\arraystretch}{1.5}
\begin{tabular}{|l |  l |}
  \hline
  $\cal G$& Gravitational constant\\
  $m$& Total mass of the extended  body\\
    $\omega$& Sidereal angular velocity of the extended body ($time^{-1}$)\\
  $\boldsymbol \pi_T, \boldsymbol \pi_m, \boldsymbol \pi_c$& Total, mantle, and core angular momentum in
                                                             $\kappa$\\
  $\boldsymbol \omega_T, \boldsymbol \omega_m, \boldsymbol \omega_c$& Total, mantle, and core Tisserand
                                                                      angular
                                                                      velocities in
                                                             $\kappa$\\
  $m_\beta,\ \ \beta=1,\ldots$& Masses of the tide raising bodies (point masses)\\
  $\mat r_\beta$& Position in $\kappa$  of the point mass $\beta$\\
  $\iota_p$& Orbit inclination of a point mass with respect to $\mat e_3\in\kappa$\\ 
  $f_p$& True anomaly  of a point mass\\
  $\omega_p$&   Argument of the periapsis of a point mass\\
  $\Omega_p$& Longitude of the ascending node of a point mass\\
   $M_p$ & Mean anomaly of a point mass\\
  $a_p$& Semi-major axis of the orbit of a point mass\\
  $e$  & eccentricity of  the orbit of a point mass\\
   $\theta_g$& Inclination of  the mean body pole $\mat e_3\in\K_g$ to $\mat e_3\in\kappa$\\
$\psi_g$ & Longitude of the ascending  node of the
           body equator\\
  $\phi_g$ & Angle between the ascending  node and $\mat e_1\in\K_g$\\
$\chi=\iota_p+\theta_g$& Inclination  of $\mat e_3\in \K_g$  to the normal to the
                          orbital plane\\
   $X^{n,m}_k(e)$& Hansen coefficient, see equation (\ref{hanseneq})\\
  $\mat J=\sum_\beta
  \frac{3{\cal G}m_\beta}{r_\beta^5}\mathbf{r}_\beta\otimes\mathbf{r}_\beta$ &
                   tidal-force operator  ($time^{-2}$) in $\kappa$\\
  $\mat S=\mat J- \frac{\tr\mathbf{J}}{3}\Id $ &
  Jeans operator in $\kappa$ (see Section \ref{tidaldef})\\
  $\mat C= -\left(\boldsymbol{\omega}_m\otimes\boldsymbol{\omega}_m - \frac{\|\boldsymbol \omega_m\|^2}{3}\Id\right)$&
 Maclaurin operator  in $\kappa$  (see Section \ref{tidaldef})\\
  $\mat F = \mat C+\mat S$& Shear operator (traceless) in $\kappa$  (see Section \ref{tidaldef})\\
   $ {\mat J}_g=\mat R_g^{-1}\mat J\mat R_g$&  Jeans operator in $\K_g$\\
  $\ov {\mat J}=$Diagonal$(\ov J_1,\ov J_2,\ov J_3)$& Average-shear operator in $\K_g$\\
  $\ov {\mat S}=\!\!\frac{\omega^2}{3}$Diag({\tiny$c_1\!\!+\!\!3c_2,c_1\!\!-\!\!3c_2,\!\!-\!2c_1$})&
                                  Average-Jeans operator in $\K_g$\\
 $\ov {\mat C}=\frac{\omega^2}{3}$Diag$(1,1,-2))$& Average-Maclaurin operator in $\K_g$\\
 $\ov{\mat F} = \ov{\mat C}+\ov{\mat S}$& Average shear operator in $\K_g$, equation (\ref{ovF})\\
  $\boldsymbol \delta \mat J_g=\mat J_g-\ov {\mat J}$& Oscillatory part of $\mat J_g$ in $\K_g$\\
   $c_3=\frac{1}{3\omega^2}\tr\big(\ov {\mat J}\big)$ & Nondimensional coefficient of tidal compression\\
  $c_1=\frac{3}{2}(\ov S_{11}+\ov S_{22})/\omega^2$ & Tidal coefficient of polar flattening (nondimensional)\\
  $c_2=\frac{1}{2}(\ov S_{11}-\ov S_{22})/\omega^2$ & Spin-orbit-resonance coefficient of  equatorial flattening
                                                      \\
  $ \xi_1= c_1-c_2+1$ & Nondimensional coefficient of average  tidal force\\
  $ \xi_2= c_1+c_2+1$ & Nondimensional coefficient of average  tidal force\\
   \hline
\end{tabular}
\end{center}
\end{table}

 \begin{table} 
\begin{center}
  \caption{\label{eigtable}
    List of symbols used for the free-libration eigenvalues and related quantities (Sections
    \ref{free2}  and \ref{offsetsec}).  The eigenvalues are
    denoted as  $\lambda\in\C$,  where the imaginary part ${\rm  Im}\, \lambda$
   is the eigenfrequency ($2\pi/|{\rm Im} \lambda|$ is the ``libration period'') and $-{\rm Re}\lambda>0$
   is the damping
   rate ($1/|{\rm Re}\lambda|$ is 
   the ``damping time''). The FLL and NDFW modes have the same nature and  are not easily
   distinguishable, see equations (\ref{xx1}) and (\ref{xx2}). The $x_{dw}$ and $x_{\ell a}$ in this
 Table refer to the real part of these quantities that in general are complex numbers.}
\renewcommand{\arraystretch}{1.5}
\begin{tabular}{|l |  l |}
  \hline
  $ \lambda_{\ell o}=i\sigma_{\ell o}-\nu_{\ell o}$&
                                                     Libration in longitude, equation
                                                     (\ref{lambda32})\\
  $ \lambda_w=i\sigma_w-\nu_w$& Wobble, equation (\ref{lambdaw2})\\
  $\lambda_{\ell a}=i\sigma_{\ell a}-\nu_{\ell a}$& Free Libration in Latitude (FLL), eq.
                                                    (\ref{ldwella2})\\
  $ \lambda_{dw}=i\sigma_{dw}-\nu_{dw}$ & Nearly Diurnal Free Wobble (NDFW), eq. (\ref{ldwella2})\\
  $x_{\ell a}=\frac{\sigma_{\ell a}}{\omega}-1$ & FLL eigenfrequency in inertial space, Footnote
                                                  \ref{librationinertial}
  \\
    $x_{dw}=\frac{\sigma_{dw}}{\omega}-1$ & NDFW eigenfrequency in inertial space, Footnote
                                                  \ref{librationinertial}
  \\
  $z =c_1 \ov \alpha_e +\frac{c_2}{2}\ov \gamma-$\tiny{$ C(i\omega)\left(c_1^2+c_2^2\right)$} &
                     FLL eigenvalue for evanescent core, eq. (\ref{charpf})\\
  $y =f_c+i \frac{\eta_c}{\omega}\frac{{\rm I}_{\circ m}}{\Io}$ &
                     NDFW eigenvalue for evanescent core, eq. (\ref{charpf})
                    \\
      $ \boldsymbol \delta_{m}$ & Inertial offset of the mantle, equation (\ref{ofdef})\\
  $ \boldsymbol \delta_{c}$ & Inertial offset of the core, equation (\ref{ofdef})\\
  $A$& Amplitude of inertial librations, equation (\ref{inlib})\\
  \hline
\end{tabular}
\end{center}
\end{table}

 \subsection{Parameters of the rheology.}

 \label{rheosec}
 
 The structure of the mantle and the core may be
 complex and heterogeneous. In this paper we assume that the core mantle boundary
 moves as a rigid surface and the mantle is deformable. We  assume that the fluid
 inside the core is  Newtonian with an effective (eddy) viscosity  $\nu$ ($length^2/time$).
 In Table \ref{rheo}  $\nu$ is written in  different ways for reasons that will be explained
 in  forthcoming Sections.

 Although the mantle can be deformed along  infinitely many degrees of freedom,
 there are only five quantities, the elements of the traceless matrix $\mat B_T(t)$,   that
 are necessary for the  integration of  Euler's equation (\ref{eqrot1}).  So,
 the idea \cite{rr2015,rr2017} is to  phenomenologically construct   Lagrangian and  dissipation
 functions directly for the variables
 $\mat B_T$ and $\dot{\mat B}_T$, ignoring all other degrees of freedom of deformation,
 and from them to derive differential equations for the deformation variables.
 In \cite{rr2017} a method is presented to endow the body with an arbitrary
 linear viscoelastic rheology: the  ``Association Principle''(AP).

\begin{figure}[ptb]
\begin{center}
\begin{minipage}{.49\textwidth}
\centering
\begin{tikzpicture}[scale=0.9, transform shape]
\tikzstyle{spring}=[thick,decorate,decoration={zigzag,pre length=0.5cm,post length=0.5cm,segment length=6}]
\tikzstyle{damper}=[thick,decoration={markings,  
  mark connection node=dmp,
  mark=at position 0.5 with 
  {
    \node (dmp) [thick,inner sep=0pt,transform shape,rotate=-90,minimum width=15pt,minimum height=3pt,draw=none] {};
    \draw [thick] ($(dmp.north east)+(5pt,0)$) -- (dmp.south east) -- (dmp.south west) -- ($(dmp.north west)+(5pt,0)$);
    \draw [thick] ($(dmp.north)+(0,-5pt)$) -- ($(dmp.north)+(0,5pt)$);
  }
}, decorate]
\tikzstyle{ground}=[fill,pattern=north east lines,draw=none,minimum width=0.75cm,minimum height=0.3cm]

            \draw [thick] (0,0.7) -- (1,0.7);
            \draw [thick] (1,-0.7) -- (1,2.1);
            \draw [thick] (1.5,2.1) -- (1,2.1);
            \draw [thick] (1.5,-0.7) -- (1,-0.7);
            \draw [damper] (1.5,-0.7) -- (4,-0.7);
            \node at (2.8,-0.2) {$\eta$};
            \draw [thick] (4,-0.7) -- (4.5,-0.7);
            \draw [spring] (1.5,2.1) --  (4,2.1);
            \node at (2.8,2.35) {$\gamma$};
            \draw [spring] (1,0.9) -- (3,0.9);
            \draw [damper] (3,0.9) -- (4,0.9);
            \node at (2,1.15) {$\mu_0$};
            \node at (3.55,1.4) {$\eta_0$};
            \draw [thick] (4,2.1) -- (4.5,2.1);
            \draw [thick] (4.5,-0.7) -- (4.5,2.1);
            \draw [thick] (4,0.9) -- (4.5,0.9);
            \draw [latex-latex, thick] (1,-1.4) -- node[below] {$\varepsilon$} (4.5,-1.4);
            \draw [-latex, thick] (4.5,0.7) -- (5.5,0.7) node[below] {$\sigma$};
            \draw [latex-latex] (1,0.4) -- node[below] {$\varepsilon_0$} (2.8,0.4);
            \draw [latex-latex] (2.8,0.4) -- node[below] {$\tilde\varepsilon_0$} (4.5,0.4);

\node (wall) at (-0.15,0.5) [ground, rotate=-90, minimum width=3cm] {};
\draw [thick] (wall.north east) -- (wall.north west);
      \end{tikzpicture}
\end{minipage}

\end{center}
\caption{Example of a spring-dashpot model for the application of the Association Principle.
  The spring $\gamma$
represents the effect of gravity. The damper $\eta$  and the Maxwell element
$(\mu_0, \eta_0)$ represent the effect of the macroscopic (spatial average) rheology of the mantle;
$\varepsilon$, $\varepsilon_0$ and $\tilde\varepsilon_0$ denote strains and $\sigma$ the stress.
}
\label{simple-osc}
\end{figure}
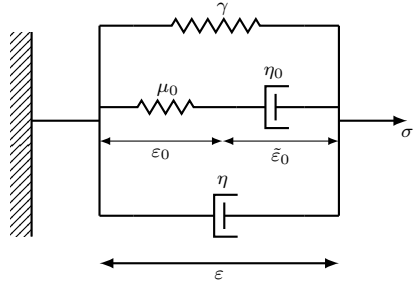

 The AP  was inspired
 in  the derivation of the Lam\'e coefficients in the theory of isotropic materials. For instance,
 we start with a general tensorial quadratic function $\mat B_T\to \sum_{ijkl}B_{Tij}B_{Tkl}\Gamma_{ijkl}$
 and using: the invariance under rotation (isotropy), the symmetry of $\mat B_T$, and  that  $\tr(\mat B_T)=0$; we obtain that the function must be equal to
 $\mat B_T\to  c \sum_{ij}B_{Tij}B_{Tij}$, where $c\in\R$ is a single parameter. In the case of a celestial body
 isotropy means that the body is spherically symmetric in the absence of centrifugal and
 external-gravitational stresses. So, isotropy implies that the Lagrangian and dissipation functions
 for the variable $\mat B_T$ must be a sum of identical Lagrangian and dissipation functions,
 one for each element of $\mat B_T$, and at the end we are lead to the construction of Lagrangian and
 dissipation functions for scalar variables.
 The last ingredient in the AP \cite{rr2017} comes from the fact that any linear
 viscoelastic rheology has a spring-dashpot representation   \citep{Bland} and from this representation
 we can obtain the desired Lagrangian and dissipation functions  for the deformation variables.
 Summarising we have:

 {\sl {\bf Association Principle (AP):}\footnote{There is a relation between the  AP and
    the Correspondence Principle \cite{efroimsky2012bodily}.
    The relation between both principles  is addressed in Section 4 of \citep{crr2018}.
    The main difference is that the AP is  defined directly in the time domain while
    the correspondence principle is defined in the frequency domain.
  }
  Any linear viscoelastic rheology 
   has always a  spring-dashpot representation 
   \citep{Bland}. If a  spring in parallel is added to 
   represent  self-gravity, then  the result is a
  spring-dashpot
  system as that represented in Figure  \ref{simple-osc}.
  If Lagrangian and dissipation functions are written
  for the  displacement  $\epsilon$ and for $\sigma=0$, then 
  the  {\bf AP} applied to deformations of the whole body consists in:
  \begin{itemize}
    \item[]
  ``To replace $\epsilon$ in the Lagrangian and dissipation functions by 
   ${\rm I}_{\circ}\mathbf{B}_T:\K_T\to\K_T$''.
\end{itemize}
}

The AP was originally formulated for a body whose average rotational dynamics could be well described by a
single rotation matrix, ``a one layer body''. This is not the case in this paper since the
fluid in the core may rotate almost independently of the mantle (in the
case of a round core with no CMB friction the motion of the core and mantle are uncoupled).
The AP  has to be slightly modified to be applied to a  body with several layers

The self-gravity term represented by the spring with elastic constant $\gamma$ refers to
the whole body. Indeed, $\gamma$ is a gravitational modulus for a  body made of a perfect fluid
(no rheology)
with  density stratification along concentric spherical shells. These are the conditions used to derive
Clairaut's equation whose solution gives the fluid Love number $k_f$, a quantity directly related to
our $\gamma$.\footnote{ The  gravitational modulus is related to
  the fluid Love number $k_f$  by means of $\frac{\omega^2}{\gamma}=\frac{R^5\omega^2}{3\Io G}k_f$.
  This is the same relation that appears in \cite{mathews2002modeling} (paragraph [21]) after
  we replace $\frac{\omega^2}{\gamma}$ by a compliance coefficient. So,  $\gamma^{-1}$ is  a dimensional
  gravitational compliance similar to those in  \cite{mathews2002modeling}.}

Each layer may have a different rheological model, so the  AP must be applied independently
to each layer.
In the case treated in this paper, in which the CMB is assumed rigid and the mantle deformable, the
AP can be restated as follows:
\begin{equation}\begin{split}&
\text{To replace the term $\gamma|\epsilon|^2/2$, which  corresponds to
      the}\\ & \text{self-gravitational term , by $\gamma {\rm I}_{\circ}\|\mat B_T\|^2/2$,
and all other }\\ & \text{terms in the Lagrangian and dissipation functions,
    which }\\ & \text{correspond to the rheology, by 
        ${\rm I}_{\circ m}\mathbf{B}_{m,m}$ and other auxiliary matrices.}
      \end{split}\label{AP}
    \end{equation}
    The same principle can be applied to a body with an arbitrary number of layers.
    In order to  understand the meaning of ``other auxiliary matrices'' see the  examples in Section
    \ref{tidaldef}.

In Table \ref{rheo} we present a list of symbols that we will use  to describe the rheology of the
mantle, of the fluid core, and of the whole body.

\begin{table}
\begin{center}
  \caption{\label{rheo} List of symbols used to describe physical and rheological  properties of the extended
    body. The  elastic ($\mu_j$) and viscosity ($\eta_j$)
      constants have the unusual dimensions  $time^{-2}$ and  $ time^{-1}$, respectively.
      The usual dimensions of shear modulus and viscosity  are obtained by means of multiplication
      of $\mu_j$ and $\eta_j$ 
      by the factor ${\rm I}_{\circ}/R_I^3=mass/length$. 
}
\renewcommand{\arraystretch}{1.5}
\begin{tabular}{|l |  l |}
  \hline
   $R$& Mean radius of the extended body\\
  $ R_{\Irm}:=\sqrt{\frac{5 {\rm I}_{\circ}}{2m}}$& Inertial radius ($length$)\\
  $ \gm\approx \frac{4}{5}\frac{Gm}{R_{\Irm}^3}$& Gravitational modulus ($time^{-2}$), see equation
                                                  (\ref{gm}) \\
  $ k_f\approx \frac{3}{2}\left(\frac{R_\Irm}{R}\right)^5$& Fluid Love number, see equation \ref{kf}\\
  $k(\sigma)$& Complex Love number at angular frequency $\sigma$ \\
  $k_2(\sigma)=|k(\sigma)|\approx$Re$\big(k(\sigma)\big)$& Real Love number \\
  $\delta(\sigma)=-\arctan\left(\frac{{\rm Im} k(\sigma)}{{\rm Re} k(\sigma)}\right)$& Phase lag\\
  $Q(\sigma)=1/\sin\big(\delta(\sigma)\big)$& Quality factor\\
  $J(\sigma)$&  Complex compliance of the rheology  ($time^{-2}$), Section \ref{parametersef}\\
  $ C(\lambda)=\left(\frac{3\Io G}{\omega^2 R^5}\right)^{-1}\, k(-i\lambda)$&  Complex compliance of the
          body (nondim.), Eq. (\ref{Ck})\\
    $\mu_0$& Prestress-elastic constant ($time^{-2}$) \\
 $\mu_j,\ j=1,\ldots$& Elastic constants  of the rheology ($time^{-2}$) \\
  $\eta_0$ &  Prestress-viscosity constant ($time^{-1}$)\\
  $\eta_j,\ j=1,\ldots$& Viscosity constants  of the rheology ($time^{-1}$) \\
   $\tau$& Characteristic time, Eqs. (\ref{tauKV}), (\ref{taugM}), (\ref{taugV}), (\ref{tauAndrade})\\
  $\tau_j=\eta_j/\mu_j,\ j=1,\ldots$& Characteristic times of the rheology\\
  $\ov{\mathbf \Lambda}:=\ov{\mathbf{F}}-\gamma \ov{\mathbf{B}}$
  &Prestress matrix, see Eqs. (\ref{ovLambda}),  (\ref{BKov1}) \\
$ \mathbf{B}_{0,m}:= \ov{\mathbf{B}}-
\frac{1}{\mu_0}\ov{\mathbf \Lambda}$&
 Fossil-deformation  matrix, see Eqs. (\ref{approx0}), (\ref{BKov2})\\
$R_c$& Mean radius of the core ($length$) \\
  $\nu$ &   viscosity (eddy) ($length^2/time$) of the fluid in the core\\
  $ k_c=\nu \frac{1}{R_c^2}\frac{{\rm I}_{\circ c}{\rm I}_{\circ m}}{{\rm I}_{\circ}}$&
                                                                            CMB coupling constant (\cite{2014Peal} equation (13))\\
  $\eta_c:=\frac{{\rm I}_{\circ}}{{\rm I}_{\circ c}{\rm I}_{\circ m}}k_c=\frac{\nu}{R_c^2}$& CMB viscosity constant
                                                                                ($time^{-1}$)\\
  $\ell_{c}:=R_c\sqrt{\frac{2\eta_c}{\om}}$& Viscous penetration depth ($length$) (see Section \ref{physicalkc})\\
  $E_k:= \frac{\eta_c}{\omega}=\frac{1}{2}\bigg(\frac{\ell_{c}}{R_c}\bigg)^2$& Ekman number  (see Section \ref{physicalkc})\\
      \hline
\end{tabular}
\end{center}
\end{table}

\section{Bodies that are spherically symmetric
  in the absence of external forces.}
\label{spherical}

In this section the extended  body is assumed  to   satisfy hypotheses
(\ref{hypotheses}) and  in  the absence of
centrifugal and tidal stresses is supposed to be spherically symmetric.
Since the CMB is rigid, the fluid core must remain spherical.
The equations for the rotation and deformation of the extended body
will be obtained within the Lagrangian formalism.

\subsection{Rotational  motion}\label{rotsec1}
The Lagrangian ${\cal L}_\mathrm{ROT}$ describing the rotation of the extended body  is
\begin{equation}\label{rotlag1}
{\cal L}_\mathrm{ROT} = \frac{1}{2}\boldsymbol{\omega}_m\cdot\mat I_m\boldsymbol{\omega}_m +
\frac{1}{2}\boldsymbol{\omega}_c\cdot\mat I_c\boldsymbol{\omega}_c -
\sum_\beta
  \frac{3{\cal G}m_\beta}{2 \|\mat r_\beta\|^5}\mathbf{r}_\beta\cdot\mat I_T\,\mathbf{r}_\beta \,.
\end{equation}
Since $\boldsymbol{\omega}_m$ and
$\boldsymbol{\omega}_c$ are  angular velocities
of  Tisserand frames related to  the mantle and to the  core, then 
$\boldsymbol{\pi}_m = \mat I_m\boldsymbol{\omega}_m$ and
$\boldsymbol{\pi}_c=\mat I_c\boldsymbol{\omega}_c$ are  the angular momentum
of the mantle and of the core.

For an infinitesimal rotation of vector $\delta \boldsymbol{\theta}$, we have
\begin{equation}
\delta \mat I_\alpha = [\widehat{\delta\boldsymbol{\theta}}, \mat I_\alpha]\,,\qquad\alpha\in\{m,c,T\}\,,
\end{equation}
and from this we obtain that the  Poincar\'e-Lagrange equations of motion are (see \cite{boue2017rotation}
for details)
\begin{subequations}
\label{eq.rot}
\begin{eqnarray}
\frac{\dd}{\dd t}\frac{\partial {\cal L}_\mathrm{ROT}}{\partial\boldsymbol{\omega}_m} &=& \boldsymbol{\omega}_m \times
\frac{\partial {\cal L}_\mathrm{ROT}}{\partial \boldsymbol{\omega}_m} + \frac{\partial {\cal L}_\mathrm{ROT}}{\partial \boldsymbol{\theta}}\,,
\\[0.3em]
\frac{\dd}{\dd t}\frac{\partial {\cal L}_\mathrm{ROT}}{\partial\boldsymbol{\omega}_c} &=& \boldsymbol{\omega}_c \times
\frac{\partial {\cal L}_\mathrm{ROT}}{\partial \boldsymbol{\omega}_c}\,.
\end{eqnarray}
\end{subequations}
 From these equations we get
\begin{equation}
 \dot{\boldsymbol{\pi}}_m = -3
   \sum_\beta
   \frac{{\cal G}m_\beta}{ \|\mat r_\beta\|^5}(\mat I_T\mathbf{r}_\beta)\times\mathbf{r}_\beta=
    -3
   \sum_\beta
   \frac{{\cal G}m_\beta}{ \|\mat r_\beta\|^5}(\mat I_m\mathbf{r}_\beta)\times\mathbf{r}_\beta\,,
\quad
  \dot{\boldsymbol{\pi}}_c =0\,.
 \label{eqrot1.2}
\end{equation}
If we add both equations
and use
 $\boldsymbol{\pi}_T=\boldsymbol{\pi}_m+\boldsymbol{\pi}_c$
we recover equation (\ref{eqrot1}).

\subsection{Tidal deformation}
\label{tidaldef}

As discussed in Section \ref{subinertia}   we will use the deformation matrices $\mat B_T$ and $\mat B_m$
to describe the variations of the moment of inertia of the body and the mantle, respectively. Since
the core is spherically symmetric for all times, $\mat B_c=0$ and
$\mat B_T=({\rm I}_{\circ m}/{\rm I}_{\circ})\mat B_m$.

We start assuming that the mantle  behaves according to the
effective  rheology described in Figure \ref{simple-osc}. In this case
the equation for the  displacement $\epsilon$
is
\begin{eqnarray}
\eta\dot\varepsilon+\gamma\varepsilon &=& \sigma-\lambda 
\label{aux1}\\
  \lambda=\mu_0\tilde\varepsilon_0=\eta_0\dot{\tilde\varepsilon}_0\Longrightarrow
\dot \varepsilon&=&\frac{\dot\lambda}{\mu_0}+\frac{\lambda}{\eta_0},\label{aux2}
\end{eqnarray}
where:  $\sigma$ is the stress upon the whole system,
$\lambda$ is the stress that acts upon the Maxwell element, and
$\tilde\varepsilon_0+\varepsilon_0=\varepsilon$. Lagrangian and dissipation functions
for this system with $\sigma=0$ can be easily obtained (see \cite{rr2017} for details)
and the application of the Association
Principle in equation (\ref{AP}) gives
\begin{equation}
{\cal L}_\mathrm{TID} = -\frac{1}{2}\gamma\,{\rm I}_{\circ}\, \|\,\mat B_{T,m}\,\|^2
-\frac{1}{2}\mu_{0m}\,{\rm I}_{\circ m}\,\|\,\mat B_{0,m}\,\|^2
-{\rm I}_{\circ m}\boldsymbol{\Lambda}_m\cdot(\mat B_{m,m} - \mat B_{0,m} - \tmat B_{0,m})
\end{equation}
where:  the  scalar product
between two matrices is
$\mat A\cdot \mat B = \frac{1}{2}\tr(\mat A^{\rm T}\mat B)$, 
$\boldsymbol{\Lambda}_m$ is an auxiliary  traceless matrix representing the stress upon the Maxwell element,
and 
$ \mat B_{0,m}$ and $\tmat B_{0,m}$ are other traceless auxiliary matrices representing the internal
variables $\epsilon_0$ and $\tilde \epsilon_0$ of the rheology. All matrices
represent operators in the mantle
frame $\K_m$ (note that $ \|\,\mat B_{T,m}\,\|^2= \|\,\mat B_{T}\,\|^2$).
To this Lagrangian function, we have to add a Rayleigh
dissipation function
\begin{equation}
{\cal D}_\mathrm{TID} = \frac{1}{2}\eta_{0m}\,{\rm I}_{\circ m}\,\|\,\dot{\tilde{\mat B}}_{0,m} 
\,\|^2 + \frac{1}{2}\eta_m\,{\rm I}_{\circ m}\,\|\,
\dmat B_{m,m}\,\|^2\,.
\end{equation}

The index $m$ in $\mu_{0m},\eta_{0m},\ldots$ is to indicate that the coefficients refer to the mantle.
Since in this work only the mantle can deform, it is possible to do the substitution
$\mat B_{m,m}=({\rm I}_{\circ}/{\rm I}_{\circ m})\mat B_{T,m}$ to eliminate $\mat B_{m,m}$ from 
 ${\cal L}_\mathrm{TID}$
 and ${\cal D}_\mathrm{TID}$ in favour of  $\mat B_{T,m}$. This is very convenient in the fit of the
 rheological parameters using  Love numbers. So, redefining the parameters of the rheology as
 \begin{equation}
\mu_0=\frac{{\rm I}_{\circ,T}}{{\rm I}_{\circ,m}}\mu_{0m}\,,\quad
\eta_0=\frac{{\rm I}_{\circ,T}}{{\rm I}_{\circ,m}}\eta_{0m}\,,\quad
\eta=\frac{{\rm I}_{\circ,T}}{{\rm I}_{\circ,m}}\eta_m\,,\label{mpar}
 \end{equation}
 and modifying  the auxiliary functions accordingly
 $\mat B_{0,m}\to({\rm I}_{\circ}/{\rm I}_{\circ m})\mat B_{0,m}$ and
 $\tilde{\mat B}_{0,m}\to({\rm I}_{\circ}/{\rm I}_{\circ m})\tilde{\mat B}_{0,m}$ we obtain
 the new Lagrangian function
\begin{equation}
{\cal L}_\mathrm{TID} = -\frac{1}{2}\gamma\,{\rm I}_{\circ}\, \|\,\mat B_{T,m}\,\|^2
-\frac{1}{2}\mu_{0}\,{\rm I}_{\circ}\,\|\,\mat B_{0,m}\,\|^2
-{\rm I}_{\circ}\boldsymbol{\Lambda}_m\cdot(\mat B_{T,m} - \mat B_{0,m} - \tmat B_{0,m})
\end{equation}
and the new dissipation function \footnote{\label{mol2} In the particular problem treated in  this paper
  we could have started directly with the effective parameters $\mu_0,\eta_0,\ldots$ and avoided
  the previous definition of the ``mantle parameters''  $\mu_{0m},\eta_{0m},\ldots$. We started with the
  mantle parameters for two reasons. At  first the simplification  is not possible for a body
  with more than one deformable layer. At  second the rescaling in equation (\ref{mpar}) shows that
  for a homogeneous mantle (see footnote \ref{mol}) the coefficients $\mu_0,\eta_0,\eta$ must change
  when the ratio $\frac{{\rm I}_{\circ,T}}{{\rm I}_{\circ,m}}$ is varied while the material
properties are preserved.}
 \begin{equation}
{\cal D}_\mathrm{TID} = \frac{1}{2}\eta_{0}\,{\rm I}_{\circ}\,\|\,\dot{\tilde{\mat B}}_{0,m} 
\,\|^2 + \frac{1}{2}\eta \,{\rm I}_{\circ}\,\|\,
\dmat B_{T,m}\,\|^2\,.
\end{equation}

Let ${\cal L} = {\cal L}_\mathrm{ROT} + {\cal L}_\mathrm{TID}$. The equations of motion for the
deformation variables are
\begin{equation}
\frac{\partial {\cal L}}{\partial \mat B_{T,m}} - \frac{\partial {\cal D}_\mathrm{TID}}{\partial \dmat
B_{T,m}} = \mat 0\,,
\qquad
\frac{\partial {\cal L}}{\partial \mat B_{0}} = \mat 0 \,,
\qquad
\frac{\partial {\cal L}}{\partial \tmat B_{0}} - \frac{\partial {\cal D}_\mathrm{TID}}{\partial \dot{\tilde{\mat B}}_{0}} = \mat 0\,\qquad
\frac{\partial {\cal L}}{\partial \boldsymbol{\Lambda}_m} = \mat 0\,.
\end{equation}
After simplifications we get
\begin{equation} \begin{split}
    \eta\dot{\mathbf{B}}_{T,m}+\gamma \mathbf{B}_{T,m}& = 
    -\mathbf{\Lambda}_{0,m}+\mathbf{F}_m\\
    \dot{\mathbf{B}}_{T,m} &=\frac{\dot{\mathbf{\Lambda}}_{0,m}}{\mu_{0}}+
    \frac{\mathbf{\Lambda}_{0,m}}{\eta_{0}}\,,
    \end{split}\label{eqLag1}
\end{equation}
 where: 
\begin{equation}\renewcommand{\arraystretch}{1.5}
  \begin{array}{l l l}
    \mathbf{F}_m&:=\mat C_m+\mat S_m \quad&\text{Shear matrix in}\ \K_m\\
    \mat C_m&:= -\left(\boldsymbol{ \omega}_{m,m}\otimes\boldsymbol{\omega}_{m,m} -
    \frac{\|\boldsymbol \omega_{m}\|^2}{3}\Id\right)\quad&\text{Maclaurin matrix in}\ \K_m\\
  \mat S_m&:=\mat J_m- \frac{\tr\mathbf{J}}{3}\Id\quad& \text{Jeans matrix in}\ \K_m\\
\mat J_m&=\sum_\beta
\frac{3{\cal G}m_\beta}{r_{\beta}^5}\mathbf{r}_{\beta,m}\otimes\mathbf{r}_{\beta,m}
\quad &\text{Tidal-force matrix in}\ \K_m\,.
\end{array}
\label{F}
\end{equation}

Note that equations (\ref{eqLag1}) and (\ref{F})  are equal to equations (\ref{aux1}) and (\ref{aux2}) after
the substitutions $\epsilon\to\mat B_{m,m}$, $\lambda\to \mat \Lambda_m$, and $\sigma\to\mat F_m$.
This is a direct consequence of the Association Principle and is not related to the special rheology
represented in Figure  \ref{simple-osc}. So, the same reasoning  applies to the  generalised
Voigt rheology represented in Figure \ref{genvoigt} and to the generalised Maxwell rheology represented
in Figure  \ref{genmax}.
\begin{figure}
\begin{center}
\begin{tikzpicture}[scale=1, transform shape]
\tikzstyle{spring}=[thick, decorate, decoration={zigzag, pre length=0.5cm, post length=0.5cm, segment length=6}]
\tikzstyle{damper}=[thick, decoration={markings,
  mark connection node=dmp,
  mark=at position 0.5 with
  {
    \node (dmp) [thick, inner sep=0pt, transform shape, rotate=-90, minimum width=15pt, minimum height=3pt, draw=none] {};
    \draw [thick] ($(dmp.north east)+(5pt,0)$) -- (dmp.south east) -- (dmp.south west) -- ($(dmp.north west)+(5pt,0)$);
    \draw [thick] ($(dmp.north)+(0,-5pt)$) -- ($(dmp.north)+(0,5pt)$);
  }
}, decorate]
\tikzstyle{ground}=[fill,pattern=north east lines,draw=none,minimum width=0.75cm,minimum height=0.3cm]

            \draw [latex-latex, thick] (0,-1) -- (7.9,-1);
            \node at (4,-0.8) {$B$};

            \draw [thick] (0,0.8) --  (0,2);

           \draw [spring] (0,0.8) -- node[below] {$\mu_1$} (1.5,0.8);
            \draw [damper] (1.5,0.8) -- (3,0.8);
            \node at (2.4,0.3) {$\eta_{1}$};

            \draw [thick] (3,0.4) --  (3,1.2);

           \draw [spring] (3,1.2) -- node[above] {$\mu_2$} (4.3,1.2);
            \draw [damper] (3,0.4) -- (4.3,0.4);
            \node at (3.75,-0.1) {$\eta_{2}$};

            \draw [thick] (4.3,0.4) --  (4.3,1.2);

            \draw [thick] (4.3,0.8) --  (4.8,0.8);
            \draw [thick,dashed] (4.8,0.8) -- (5.7,0.8);
            \draw [thick] (5.7,0.8) -- (6.2,0.8);

            \draw [thick] (6.2,0.4) --  (6.2,1.2);

           \draw [spring] (6.2,1.2) -- node[above] {$\mu_n$} (7.5,1.2);
            \draw [damper] (6.2,0.4) -- (7.5,0.4);
            \node at (6.95,-0.1) {$\eta_{n}$};

            \draw [thick] (7.5,0.4) --  (7.5,1.2);

            \draw [thick] (7.5,0.8) --  (7.9,0.8);

            \draw [thick] (7.9,0.8) --  (7.9,2);

            \draw [thick] (0,2) --  (2.5,2);
           \draw [spring] (2.5,2) -- node[above] {$\mu_0$} (4.5,2);
            \draw [damper] (4.5,2) -- (5.5,2);
            \node at (5.1,2.5) {$\eta_{0}$};
            \draw [thick] (5.5,2) --  (7.9,2);

            \draw [-latex,thick] (7.9,2) -- (8.5,2) node[right] {$\Lambda_{0}$};
            \draw [-latex,thick] (7.9,0.8) -- (8.5,0.8) node[right] {$\Lambda$};

      \end{tikzpicture}
\end{center}
\caption[]{  The model  in the figure  is a modification
  of the generalised Voigt  model in \cite{Bland} chapter 1
  eq. (27). The modification, the Maxwell element in parallel to the usual generalised
  Voigt model, is convenient for the introduction of  the prestress. 
}
\label{genvoigt}
\end{figure}
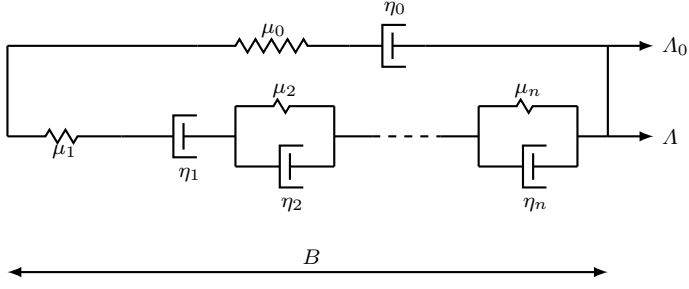
  \begin{figure}
\begin{center}
\begin{tikzpicture}[scale=1, transform shape]
\tikzstyle{spring}=[thick, decorate, decoration={zigzag, pre length=0.5cm, post length=0.5cm, segment length=6}]
\tikzstyle{damper}=[thick, decoration={markings,
  mark connection node=dmp,
  mark=at position 0.5 with
  {
    \node (dmp) [thick, inner sep=0pt, transform shape, rotate=-90, minimum width=15pt, minimum height=3pt, draw=none] {};
    \draw [thick] ($(dmp.north east)+(5pt,0)$) -- (dmp.south east) -- (dmp.south west) -- ($(dmp.north west)+(5pt,0)$);
    \draw [thick] ($(dmp.north)+(0,-5pt)$) -- ($(dmp.north)+(0,5pt)$);
  }
}, decorate]
\tikzstyle{ground}=[fill,pattern=north east lines,draw=none,minimum width=0.75cm,minimum height=0.3cm]

            \draw [latex-latex, thick] (0,0) -- node[left] {$B$} (0,2.8);
            \draw [damper] (1.5,2.8) --  (1.5,0);
            \node at (1,1.4) {$\eta$};

           \draw [spring] (3,2.8) -- node[left] {$\mu_0$} (3,1.2);
            \draw [damper] (3,1.2) -- (3,0);
            \node at (2.5,0.57) {$\eta_{0}$};

           \draw [spring] (4.5,2.8) -- node[left] {$\mu_1$} (4.5,1.2);
            \draw [damper] (4.5,1.2) -- (4.5,0);
            \node at (4,0.57) {$\eta_{1}$};

           \draw [spring] (8,2.8) -- node[left] {$\mu_n$} (8,1.2);
            \draw [damper] (8,1.2) -- (8,0);
            \node at (7.5,0.57) {$\eta_{n}$};

            \draw [thick] (1.5,2.8) -- (5.5,2.8);
            \draw [thick] (1.5,0) -- (5.5,0);
            \draw [thick,dashed] (5.5,2.8) -- (7,2.8);
            \draw [thick,dashed] (5.5,0) -- (7,0);
            \draw [thick] (7,2.8) -- (8,2.8);
            \draw [thick] (7,0) -- (8,0);
            \node at (6.25,1.4) {$\ldots\!\ldots$};

            \draw [-latex,thick] (1.5,2.8) -- (1.5,3.4) node[left] {$\eta\dot{B}$};
            \draw [-latex,thick] (3,2.8) -- (3,3.4) node[left] {$\Lambda_{0}$};
            \draw [-latex,thick] (4.5,2.8) -- (4.5,3.4) node[left] {$\Lambda_{1}$};
            \draw [-latex,thick] (8,2.8) -- (8,3.4) node[left] {$\Lambda_{n}$};

      \end{tikzpicture}
\end{center}
\caption[General oscillator with rheology]{The generalised Maxwell model.}
\label{genmax}
\end{figure}
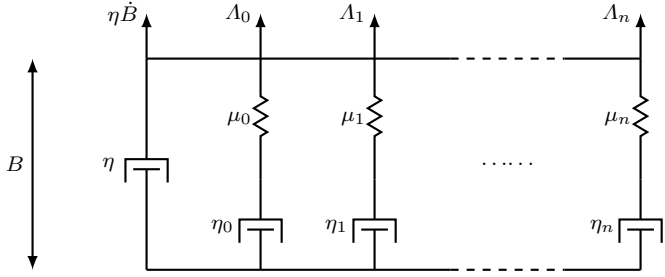
These models
are interesting because they can approximate \cite{gev2020} any other  viscoelastic model including those with infinitely
many internal variables as the Andrade model. 

The equations for the deformation variables of the generalised Voigt model  in Figure  \ref{genvoigt}
are 
\begin{equation}\label{maineq2}
\begin{split}
&\gamma \mathbf{B}_{T,m}+\mat \Lambda_{0m}+\mat \Lambda_m=\mathbf{F}_m\\
&\frac{1}{\mu_0}\dot{\mat\Lambda}_{0,m}+\frac{1}{\eta_0} \mat\Lambda_{0,m} =\dot{\mat B}_{T,m}\\
&\frac{1}{\mu_1}\dot{\mat\Lambda}_m+\frac{1}{\eta_1} \mat\Lambda_m =\dot{\mat B}_{1,m}\\
&\eta_j\dot {\mathbf{B}}_{j,m}+\mu_j\mat B_{j,m}=\mat\Lambda_m,\quad j=2,\ldots,n\\
&\mat B_{1,m}=\mat B_{T,m}-\big(\mat  B_{2,m}+\mat  B_{3,m}+\ldots+\mat  B_{n,m}\big)
\end{split}
\end{equation}
where $\mat F_m$ is given  in equation (\ref{F}).

The equations for the deformation variables of the generalised Maxwell model  in Figure  \ref{genmax}
are 
 {\renewcommand{\arraystretch}{1.5}  
 \begin{equation}
  \begin{split}
    & \eta\dot{\mathbf{B}}_{T,m}+ \gm\mat B_{T,m}+\mat\Lambda_{0m}+\mat\Lambda_{1m}\cdots+\mat
    \Lambda_{1n}= \mathbf{F}_m\\
   &  \dot{\mat\Lambda}_{jm}=-\frac{1}{\tau_j}\mat\Lambda_{jm}+
   \mu_{j}\dot{\mat B}_{T,m}\,,\qquad\tau_j=\frac{\eta_{j}}{\mu_{j}},\quad j=0,\ldots,n
       \end{split}\label{Beq}
\end{equation}
}
where $\mat F_m$ is given  in equation (\ref{F}).

\subsection{Core-mantle boundary}
\label{CMBfriction}
The last ingredient in the  model is a CMB  mechanical friction
that can be expressed as an additional dissipation function ${\cal
D}_\mathrm{CMB}$ given by
\begin{equation}
{\cal D}_\mathrm{CMB} = \frac{1}{2} k_c\,\|\boldsymbol{\omega}_m - \boldsymbol{\omega}_c\|^2\,.
\end{equation}
The rotation equations of motion (\ref{eq.rot}) should then be extended as follows
\begin{subequations}
\begin{eqnarray}
\frac{\dd}{\dd t}\frac{\partial {\cal L}_\mathrm{ROT}}{\partial\boldsymbol{\omega}_m} &=& -\frac{\partial {\cal D}_\mathrm{CMB}}{\partial \boldsymbol{\omega}_m} + \boldsymbol{\omega}_m \times
\frac{\partial {\cal L}_\mathrm{ROT}}{\partial \boldsymbol{\omega}_m} + \frac{\partial {\cal L}_\mathrm{ROT}}{\partial \boldsymbol{\theta}}\,,
\\[0.3em]
\frac{\dd}{\dd t}\frac{\partial {\cal L}_\mathrm{ROT}}{\partial\boldsymbol{\omega}_c} &=& -\frac{\partial {\cal D}_\mathrm{CMB}}{\partial \boldsymbol{\omega}_c} + \boldsymbol{\omega}_c \times
\frac{\partial {\cal L}_\mathrm{ROT}}{\partial \boldsymbol{\omega}_c}\,.
\end{eqnarray}
\end{subequations}
that implies an additional coupling term in equations (\ref{eqrot1})
\begin{equation}
  \dot{\boldsymbol{\pi}}_m = - k_c(\boldsymbol{\omega}_m - \boldsymbol{\omega}_c) - 3 \frac{{\cal G}m_E}
  {\|\mat r\|^5}(\mat I_T\mathbf{r})\times\mathbf{r}\,,\qquad
  \dot{\boldsymbol{\pi}}_c =  k_c(\boldsymbol{\omega}_m - \boldsymbol{\omega}_c)\,.
  \label{eqrot2.5}
\end{equation}

\subsection{A physical interpretation of $k_c$.}
\label{physicalkc}

The value of $k_c$ provided by the  INPOP19a  for the Moon  \cite{INPOPa}
was estimated   using  observational data. However, for most  bodies there is no
estimate of $k_c$ based on observations and   theoretical considerations have been used
to constrain $k_c$.
Fluid dynamic arguments (Section 2.3 of
\cite{2014Peal}) lead to the following  equation:
\begin{equation}
  k_c=\nu \frac{1}{R_c^2}\frac{{\rm I}_{\circ c}{\rm I}_{\circ m}}{\Io}\label{kc}
\end{equation}
where $\nu$ is the kinematic viscosity of the fluid in the core and
$R_c$ is the core mean radius. In   \cite{2014Peal} 
$\nu=35 \times 10^5$ cm$^2$/s is estimated for the liquid inside the core of   Mercury.
This viscosity is much larger than  the molecular
viscosity of iron, which  is about  $10^{-2}$ cm$^2$/s,
that is supposed to be the main
component in Mercury's fluid core. This situation
is similar to that found in the study of flows near the ocean
bottom:  molecular viscosity measured in the laboratory is much smaller
than the viscosity  estimated from field measurements. One way to overcome this difference
is to replace 
the molecular viscosity of water by  an effective value   often called eddy-viscosity
(see \cite{pedlosky2013geophysical} chapter 4). 
Our goal in this paragraph   is to associate this eddy viscosity, which
may depend on the temperature,  roughness of the core-mantle boundary (CMB), turbulence intensity, etc,
with a characteristic length of the flow.

In the forced libration problem the mantle  periodically oscillates
with respect to the fluid in the core with an angular frequency  $\omega_{osc}$.
The mantle oscillation  induces an  oscillation
in the fluid velocity that decays exponentially fast
with the distance from the mantle (see \cite{batchelor2000introduction} equation
(4.3.16)), under the assumption that the fluid is Newtonian. The decay length at the equator
is of  the order of $\sqrt{2 \nu/\om_{osc}}$, a quantity often called
``viscous penetration
depth''. Motivated by this we define the CMB penetration depth at frequency
$\omega_{osc}$ as 
 \begin{equation}
    \ell_c(\omega_{osc})=R_c\sqrt{\frac{2\eta_c}{\om_{osc}}}=\sqrt{\frac{2\nu}{\om_{osc}}}=R_c
    \sqrt{\frac{2}{\om_{osc}}\frac{\Io}{{\rm I}_{\circ c}{\rm I}_{\circ m}}k_c}
    \label{ellc}
  \end{equation}
where  $\eta_c$ is a   CMB-viscosity coefficient with dimension
1/time:
\begin{equation}
  \eta_c:=\frac{\Io}{{\rm I}_{\circ c}{\rm I}_{\circ m}}k_c=\frac{\nu}{R_c^2}\,.\label{etac}
  \end{equation}
  It is  implicit in the definition of the CMB penetration depth  that 
  $\ell_c(\omega_{osc})\ll R_c$. Note: If  there exists a solid core inside the fluid core
  then $\ell_c(\omega_{osc})$ must be much smaller than the distance from the CMB
  to the solid core, otherwise there may be a viscous interaction between the
  mantle and the solid core.

  It is convenient to associate $k_c$ or $\eta_c$ with a unique length scale
  equating the  oscillation frequency of the mantle to the body spin,
  $\omega_{osc}=\omega$. In this case we obtain
  \begin{equation}
    \ell_{c}:=\ell_c(\omega)=R_c\sqrt{\frac{2\eta_c}{\om}}\Longrightarrow
    \frac{1}{2}\bigg(\frac{\ell_{c}}{R_c}\bigg)^2=\frac{\nu}{\om R_c^2}=
    \frac{\eta_c}{\om}:=E_k\,,
      \label{Ekman}
    \end{equation}
   where  $E_k$ is equal to  the Ekman number
   (in meteorology the Ekman number is defined in a different way
   \cite{glickman2000glossary}).

   The characteristic length $\ell_c$ can be interpreted as the depth, measured from the CMB, that
   an oscillation  of the mantle with frequency  $\omega$ with
   respect to the guiding frame (or with respect to the average
   motion of the fluid inside the core) can perturb the mean flow of the fluid significatively
   (if the amplitude of the oscillation of the fluid at the CMB  is $a_f$,
   then at depth $\ell_c$ the amplitude is $\erm^{-1} a_f$). So, if we can constrain  $\ell_c$
   using the radius of the core or the distance of the CMB to a solid core, then we can
also constrain  $\eta_c$. 

 For Mercury,  using the values
  $\nu=87.5$ m$^2$/s, $R_c=2000$ km, $\Io/(mR^2)=0.346$, and
  ${\rm I}_{\circ m}/(mR^2)=0.149$ \cite{2014Peal} we obtain that  
  $E_k=\eta_c/\omega=1.76\times 10^{-5}$  and 
  $\ell_c=12$ km; for the Moon, using the values in INPOP19a with $R_c=381$ km
  we obtain
  $E_k=\eta_c/\omega=1.576\times 10^{-5}$  and 
  $\ell_c=2.1$ km; and  for the Earth,  using the values
  $R_c=3.48\times 10^3$ km, $\Io/(mR^2)=0.331$, and
  ${\rm I}_{\circ m}/(mR^2)=0.292$  \cite{zhang2020new}
   we obtain either\footnote{\label{uncertainty} There is a great
    uncertainty on the value of the viscosity (eddy) near the CMB of the Earth.
    Several values have been used in the literature: $\nu=10^{-6}$ m$^2$/s (molecular
    viscosity of iron),    $\nu=3.5\times 10^{-2}$ m$^2$/s
    ($E_k= 4\times 10^{-11}$)
    \cite{deleplace2006viscomagnetic}, $\nu=883$ m$^2$/s ($E_k=10^{-6}$)
    \cite{triana2019coupling}, $\nu=3.5 \times 10^5$ m$^2$/s ($E_k=4 \times 10^{-4}$)
    \cite{matsui2012large}, etc.}
 $E_k=\eta_c/\omega=4\times 10^{-11}$  and 
  $\ell_c=31$ m for   $\nu=3.5\times 10^{-2}$ m$^2$/s \cite{deleplace2006viscomagnetic}
or  $E_k=10^{-6}$ and $\ell_c=4.9$ km for $\nu=883$ m$^2$/s \cite{triana2019coupling}.

\section{The prestress frame.}
\label{prestress}

This section contains the concept of prestress frame and 
its physical interpretation.
Before entering into quantitative considerations we give an overview
of our reasoning and explain how it is related to old ideas.

The assumption that the Earth is in hydrostatic equilibrium leads to an estimate of the gravitational
Stokes coefficient  $C_{20}$ that is slightly smaller than that observed
(see \cite{Lambeck} section 2.4). 
The difference between the theoretical and the observed values of  $C_{20}$
has been interpreted by several authors
 as a  memory for the faster rotation rate of the
 past. \cite{mckenzie1966}, for instance, used this difference to
 estimate the viscosity of the mantle of the
 Earth as  $2.4 \times 10^{26}$Pa s. This idea was criticised in \cite{goldreich1969some}
 by means of the following argument. ``Imagine that the Earth's rotation were halted and that the present rotational bulge, but nothing else, were allowed to subside$\ldots$
 From the differences of the corresponding moments of inertia
 (of the halted Earth)$\ldots$,
 one would be hard pressed to describe the hypothetical non-rotating Earth as either
 an oblate or a prolate spheroid. In fact, it is no spheroid at all, but  a good example
 of triaxial body. This conclusion would have come as no surprise if we had instead been
 conditioned to regard the nonhydrostatic Earth as a collection of more or less random density inhomogeneities.''

 The starting point for our prestress frame comes from the two ideas
 in the paragraph above. We not only halt the body  rotation but also remove  all
 external gravitational field that acts upon the body
 and cause its permanent hydrostatic deformation.
 After relaxation the residual  nonhydrostatic deformation will be
 interpreted as a  fossil deformation of unknown origin
 that will define the orientation of the prestress frame.

 Clearly, our interpretation of the residual
 deformation as a transient state may be wrong and  density inhomogeneities is the correct cause.
 Fossil deformation being correct or not, our point is that it provides a physical argument to be used
 in the theory of   spherically symmetric bodies,
 presented in Section \ref{spherical} and on which we are confident in, to build a mathematical model
 to describe the librations of slightly aspherical bodies out of hydrostatic equilibrium.

 \cite{zanazzi2017triaxial}
 argued that the maximum triaxiality of a rock planet is determined
 by a critical yield strain. If the triaxiality of the body is beyond a certain
 limit this critical strain is exceeded and the rock begins
 to either plastically deform or fracture. The authors  estimate the critical
 triaxiality deforming a body from the  spherical shape. We apply their
 reasoning to the residual non hydrostatic body described above  
 and we  find a limit to the maximum strain of  the fossil deformation that
 is compatible to that  they found.

\subsection{Hydrostatic equilibrium.}
\label{hydeq}

The  force that acts upon the mantle of the extended body is the sum of the centrifugal  plus
the tidal forces as given in equation (\ref{F}). As argued in Section \ref{aerageI},
the average centrifugal and tidal forces in the mantle frame are close, up to second order in
small quantities, to  the same quantities in the guiding frame.
The Maclaurin operator in the guiding frame is
$\mat C_g= -\left(\boldsymbol{ \omega}_{m,g}\otimes\boldsymbol{\omega}_{m,g} -
  \frac{\|\boldsymbol \omega_{m}\|^2}{3}\Id\right)$. Due to equations (\ref{omegag}) and (\ref{apr1m}),
$\boldsymbol{\omega}_{m,g}=\boldsymbol{\omega}_{g,g}+ \dot{\boldsymbol{ \alpha}}_m=
\omega \mathbf{e_3}+ \boldsymbol {\omega}_{s,g}+ \dot{\boldsymbol{ \alpha}}_m$ and this implies that the
mean Maclaurin matrix (or mean centrifugal-force matrix) is
    \begin{equation}
      \ov{\mat C}=   \lim_{\tau\to\infty}\frac{1}{\tau}\int_0^\tau
 \mat C_g(t)dt= -\omega^2\left(\mat e_3\otimes \mat e_3 -
  \frac{1}{3}\Id\right)\,,\label{ovC}
  \end{equation}
 where terms of second order in small quantities were neglected.
This, the expression for $\ov{\mat J}$ in equation (\ref{Jd3}), and
$\ov{\mat S}=\ov{\mat J}- \frac{\tr\ov{\mathbf{J}}}{3}\Id $ imply 
 \begin{equation}
    \ov{\mathbf{F}}=\ov {\mat C}+\ov{\mat S}=
    \frac{\omega^2}{3}\begin{bmatrix}
1& 0 & 0 \\
0 & 1 & 0 \\
0 & 0 & -2
\end{bmatrix}+\frac{\omega^2}{3}
     \begin{bmatrix}
c_1+3c_2 & 0 & 0 \\
0 & c_1-3c_2 & 0 \\
0 & 0 & -2c_1
\end{bmatrix}\,.
\label{ovF}
\end{equation}

If  under static  forces the body were held together only by self-gravity,  as it happens 
in a body with  either one of the rheologies in Figures \ref{simple-osc}, \ref{genvoigt}, or
\ref{genmax},  then   either one of the equations (\ref{eqLag1}),
(\ref{maineq2}),
or (\ref{Beq}),  would give the equilibrium deformation matrix
\begin{equation}
 \mat B_{T,m}= \ov {\mat B}_{hyd}=\frac{1}{\gamma}\ov{\mat F}\,.\label{hyd}
  \end{equation}
If this  situation happens, then the body is said to be  in hydrostatic equilibrium.

If $0.2 \le {\rm I}_{\circ}/(mR^2) \le 0.4$, where 
$R$ is the volumetric mean radius of the body, then
 an extension of  the usual Darwin-Radau relation 
   for smaller values of
   $\Io /(mR^2)$ \cite{ragazzo2020theory}   gives 
  \begin{equation}
  \gm\approx \frac{4}{5}\frac{Gm}{R_{\Irm}^3}
  \quad\text{where}\quad R_{\Irm}:=\sqrt{\frac{5\Io}{2m}}:=\text{''inertial radius''.}
  \label{gm}
 \end{equation}

 The fluid Love number $k_f$
 is related to $\gm$ by means of $k_f=3\Io G/(R^5\gm)$ and the
 approximation above implies
 \begin{equation}
   k_f\approx \frac{3}{2}\left(\frac{R_\Irm}{R}\right)^5\,.
   \label{kf}
   \end{equation}

\subsection{The
  prestress and its physical interpretation.}
\label{secprestress}

If $\ov{\mat B}$ denotes the nominal (or average, see Section \ref{aerageI}) deformation matrix of the body,
then the nonhydrostatic deformation is defined according to   \cite{goldreich1969some}
     as  $\ov{\mathbf{B}}- \ov{\mathbf{B}}_{hyd}$.  The prestress $\ov{\mathbf \Lambda}$ is 
     the part of the mean centrifugal and tidal stresses
 that are not balanced  by the gravitational strength of the body, namely    
\begin{equation}
\ov{\mathbf \Lambda}:=\ov{\mathbf{F}}-\gamma \ov{\mathbf{B}}\,.\label{ovLambda}
\end{equation}
All  matrices in this equation represent average operators in the guiding frame $\K_g$ and,
up to second order in small quantities, in any other frame that oscillates sufficiently close to $\K_g$.

In order to give a physical interpretation to the prestress we consider a body
with a rheology as that represented in Figure   \ref{simple-osc} (the same argument
applies to the more general rheologies in Figures \ref{genvoigt} and 
\ref{genmax}). From equation (\ref{eqLag1}) we obtain that under static stress the deformation
variables satisfy
\[\begin{split}
    \eta\dot{\mathbf{B}}_{T,m}+\gamma \mathbf{B}_{T,m} &= 
    -\mathbf{\Lambda}_{0,m}+\ov {\mathbf{F}}\\
    \dot{\mathbf{B}}_{T,m} &=\frac{\dot{\mathbf{\Lambda}}_{0,m}}{\mu_{0}}+
    \frac{\mathbf{\Lambda}_{0,m}}{\eta_{0}}\,.\end{split}
\]
Suppose that the nondimensional viscosity coefficient $\eta_0/\omega$ is very large  such that
\begin{equation}\begin{split}
&      \dot{\mathbf{B}}_{T,m} =\frac{\dot{\mathbf{\Lambda}}_{0,m}}{\mu_{0}}+\frac{\mathbf{\Lambda}_{0,m}}{\eta_{0}}
  \approx \frac{\dot{\mathbf{\Lambda}}_{0,m}}{\mu_{0}}  \Longrightarrow
 \\ &
\mathbf{\Lambda}_{0,m}\approx  \mu_{0}\Big(\mathbf{B}_{T,m}-\mathbf{B}_{0,m}\Big)\,.
\end{split}
\label{approx0}
\end{equation}

The substitution of
\begin{equation}
  \mathbf{\Lambda}_{0,m}= \mu_{0}\Big(\mathbf{B}_{T,m}-\mathbf{B}_{0,m}\Big)
  \label{approx}
\end{equation}
into equations  (\ref{eqLag1}),
(\ref{maineq2}),
or (\ref{Beq}), leads to new equations that we call  equations with prestress.
For the simple rheology in
Figure \ref{simple-osc} the equation with prestress is
\begin{equation} 
    \eta\dot{\mathbf{B}}_{T,m}+\gamma \mathbf{B}_{T,m}+ \mu_{0}\Big(\mathbf{B}_{T,m}-\mathbf{B}_{0,m}\Big) = 
    \mathbf{F}_m\,.
  \label{def1}
\end{equation}
For the generalised Voigt rheology in Figure \ref{genvoigt} the equation with prestress is
\begin{equation}\label{def2}
\begin{split}
&\gamma \mathbf{B}_{T,m}+ \mu_{0}\Big(\mathbf{B}_{T,m}-\mathbf{B}_{0,m}\Big)+\mat \Lambda_m=\mathbf{F}_m\\
&\frac{1}{\mu_1}\dot{\mat\Lambda}_m+\frac{1}{\eta_1} \mat\Lambda_m =\dot{\mat B}_{1,m}\\
&\eta_j\dot {\mathbf{B}}_{j,m}+\mu_j\mat B_{j,m}=\mat\Lambda_m,\quad j=2,\ldots,n\\
&\mat B_{1,m}=\mat B_{T,m}-\big(\mat  B_{2,m}+\mat  B_{3,m}+\ldots+\mat  B_{n,m}\big)\,.
\end{split}
\end{equation}
For the generalised Maxwell rheology in Figure \ref{genmax} the equation with prestress is
 {\renewcommand{\arraystretch}{1.5}  
 \begin{equation}
  \begin{split}
    & \eta\dot{\mathbf{B}}_{T,m}+ \gm\mat B_{T,m}+\mu_{0}\Big(\mathbf{B}_{T,m}-\mathbf{B}_{0,m}\Big)
    +\mat\Lambda_{1,m}\cdots+\mat
    \Lambda_{n,m}= \mathbf{F}_m\\
   &  \dot{\mat\Lambda}_{j,m}=-\frac{1}{\tau_j}\mat\Lambda_{j,m}+
   \mu_{j}\dot{\mat B}_{T,m}\,,\qquad\tau_j=\frac{\eta_{j}}{\mu_{j}},\quad j=1,\ldots,n\,.
       \end{split}\label{def3}
\end{equation}
}
In equations (\ref{def1}), (\ref{def2}), and (\ref{def3}),  $\mat F_m$ is given  in equation (\ref{F}).

The body is at equilibrium  if its deformation is stationary and it does not
librate, namely it is in the guiding motion. The motion of $\K_m$ within $\K_g$ is given
by $\mat R_g^{-1}\mat R_m:\K_m\to\K_g$ and we assume that at equilibrium  $\mat R_g^{-1}\mat R_m=\Id$.
In this situation $\mat F_m=\ov{\mat F}$,
${\mat B}_{T,m}=\ov {\mat B}$, and the equilibrium condition for all the three equations
(\ref{def1}), (\ref{def2}), and (\ref{def3}) gives
\begin{equation}
  \mathbf{B}_{0,m}= \frac{\gm+\mu_0}{\mu_0}\ov{\mathbf{B}}-
\frac{1}{\mu_0}\ov{\mathbf{F}}\,.
\label{BKov2}
\end{equation}
This equation implies that 
\begin{equation}
\mu_0\big(   \ov{\mathbf{B}} -\mathbf{B}_{0,m}\big)= \ov{\mathbf{F}}-\gm\ov{\mathbf{B}}=\ov {\mat \Lambda}\,.
\label{BKov1}
\end{equation}
So, the elastic stress within the mantle at the equilibrium state is represented by
the prestress matrix $\ov{\mat\Lambda}$
(deviations from the equilibrium cause elastic stresses that must be added to the prestress).
We will call  $\mathbf{B}_{0,m}$ the {\sl fossil-deformation  matrix}.

If in equation (\ref{BKov2}) the elastic constant $\mu_0$ is known,
then all quantities in the right hand side of this equation are
known and it is possible to compute the fossil-deformation matrix $\mathbf{B}_{0,m}$.
Note that it is not necessary to know $\mu_0$ to compute 
the prestress matrix $\ov{\mat \Lambda}$.
Since both matrices $\ov{\mathbf{B}}$ and $\ov{\mathbf{F}}$ are diagonal in $\K_g$
(a consequence of equations (\ref{IJ}), (\ref{Jd3}), and (\ref{ovF})) and at the equilibrium
$\K_m$ coincides with $\K_g$, then $\mat B_{0,m}$ and $\ov {\mat \Lambda}$ are diagonal in $\K_m$.

In the following we assume that 
the  fossil-deformation matrix  $\mat B_{0,m}$ (or the prestress matrix
$ \ov{\mathbf{\Lambda}}$)  has three distinct eigenvalues,
which is generally true.
We call an orthonormal basis defined by the eigenvectors  of
 $\mat B_{0,m}$ as a  ``prestress frame''. The fossil-deformation  matrix 
does not move  in  the frame of the mantle $\K_m$ and
many times we will refer to  $\K_m$ as the prestress frame.

The prestress frame  breaks up the spherical symmetry of the body and, at the same time, it is a Tisserand
frame.
 This interesting fact
is a
consequence of the definition of a Tisserand
frame.  Indeed,  if   the prestress and its consequent deformation
were not fixed in $\K_m$ then as it would move within $\K_m$ it would carry 
angular momentum making the angular momentum relative  to
 $\K_m$ to be  non-null,
which is impossible by the definition of Tisserand frame.
The  fact that  $\mat B_{0,m}$
is frozen in $\K_m$ does not imply that the principal axes of the deformable body
are, or are even  close to be,
frozen in   $\K_m$. As discussed in the next section,
the principal axes of the deformable body will remain close to
the prestress frame only when  the deformations are small compared to
the mean triaxiality coefficients \footnote{
    At this point it would be interesting to relate the prestress frame 
       to analogous frames used by other authors.
      We restrict the discussion to the case of the Moon.
      \cite{eckhardt1981theory} defines 
 ``selenographic coordinates whose axes are the same as those of the
 Moon's principal moments of inertia in the absence of elastic deformation.''
 \cite{Viswanathan19} use the same frame as  \cite{Folkner14}: 
 ``The mantle coordinate system is defined by the principal axes of
 the undistorted mantle in which the moment of inertia matrix of
 the undistorted mantle is diagonal.'' In both Eckhardt and Folkner approaches,
 the body or the mantle reference frames, respectively, is defined using an 
 undistorted configuration on which  the real distorted situation is described.
   In our approach the undistorted
   configuration is replaced by the prestress frame 
   that is a Tisserand frame.   
   It seems that in Eckhardt and Folkner the undistorted
   frame is implicitly assumed to be a Tisserand frame, since 
   it is consistently used in this way.}.

  \subsection{The principal axes frame.}
   \label{frame}

There exists a frame $\K_p$ defined by the principal
axes of inertia of the body (the index $p$ stands for ``principal'').
This frame moves inside $\K_m$ according to $\mat R_m^{-1}\mat R_p:\K_p\to \K_m$,
where  $\mat R_p:\K_p\to \kappa$.

At a given time
the moment of inertia operator in $\K_m$ is given by
$\mat I_{T,m}=\ov{\mat I}+ \boldsymbol{\delta}\mat I_{T,m}$ with
$\ov I_j\gg |\big(\boldsymbol{\delta}\mat I_{T,m}\big)_{jj}|$, $j=1,2,3$.
In order to determine $\mat R_m^{-1}\mat R_p$ 
we further assume  that the variations of the moment of inertia due to tides and time-variable
  centrifugal forces are small enough such that
  \begin{equation}\label{principalaxes}
    \frac{|\dt I_{T,m23}|}{\ov I_3-\ov I_2}\ll 1\, , \quad
    \frac{|\dt I_{T,m13}|}{\ov I_3-\ov I_1}\ll 1\, ,\quad
    \frac{|\dt I_{T,m23}|}{\ov I_2-\ov I_1}\ll 1\,\quad\big(\text{Hypothesis}\big).
  \end{equation}
  In this case the matrix $\mat R_m^{-1}\mat R_p\approx \Id+\boldsymbol{\widehat \beta}$
  is close to the identity. By definition, the transformation $\mat R_m^{-1}\mat R_p$ must diagonalize
  $\mat I_{T,m}$ and this can be used to compute  $\boldsymbol{\widehat \beta}$.
 Since $\ov{\mat I}$ is diagonal,
  $\dt  I_{T,m\,ij}=-\Io  B_{T,m\, ij}$  for $i\ne j$ and  we obtain,  up to
  first order in small quantities, 
  \begin{equation}
    \beta_1=  \frac{\Io  B_{T,m\,23}}{\ov I_3-\ov I_2}\,
  , \quad
  \beta_{2}=  \frac{\Io  B_{T,m\,13}}{\ov I_1-\ov I_3}
  \, , \quad
  \beta_{3}=  \frac{\Io  B_{T,m\,12}}{\ov I_2-\ov I_1}\, .\label{ad}
\end{equation}
If we neglect small terms of the order of $\|\boldsymbol\delta \mat B\|^2
$, then  the usual mean ellipticity coefficients $\ov \alpha,\ov\beta,\ov\gamma$
can be written as
\begin{equation}\begin{split}
    \ov \alpha&:=\frac{\ov I_3-\ov I_2}{\ov I_1}\approx\frac{\ov I_3-\ov I_2}{\Io}\\
    \ov\beta&:=\frac{\ov I_3-\ov I_1}{\ov I_2}\approx \frac{\ov I_3-\ov I_1}{\Io}\\
    \ov \gm&:=\frac{\ov I_2-\ov I_1}{\ov I_3}\approx \frac{\ov I_2-\ov I_1}{\Io}\\
  \end{split}
    \label{apbtgm}
   \end{equation}
   Equations (\ref{ad}) and (\ref{apbtgm}) imply that
   \begin{equation}
     \beta_1(t) =\frac{ B_{T,m23}(t)}{\ov \alpha}\,
  , \
  \beta_2(t)=-\frac{ B_{T,m13}(t)}{\ov \beta}
  \, , \
  \beta_3(t)=
  \frac{ B_{T,m12}(t)}{\ov \gamma}\, ,\label{ad3}
\end{equation}
so, if the hypothesis in equation (\ref{principalaxes}) holds, then
the off diagonal elements of the deformation matrix $\mat B_{T,m}$ are related to the angular displacement
of the principal axes frame from the prestress frame.

\subsection{ Love numbers and rheology.}
\label{parametersef}

In this section we present equations that relate Love numbers to rheological models.
These equations can be used to  determine the parameters of the rheology.

For a given forcing frequency $\sigma$ 
the complex Love number  is given by 
\cite{rr2017} (equation (46))
\begin{equation}
  k(\sigma)=
  \frac{3\Io G}{R^5}\frac{1}{\gm+J^{-1}(\sigma)}\,,
 \label{kcomplex2}
\end{equation}
where $J^{-1}(\sigma)$ is the  complex rigidity of the rheology
 (we neglected  the inertia of deformation \citep{crr2018}).
As an example, we will compute  the  complex rigidity of the generalised Maxwell rheology represented in
Figure \ref{genmax}.

The linear equation (\ref{def3}) determines  the deformation variables of the generalised Maxwell rheology.
If we do the substitution  
\[
 \mat F_m\to\ov {\mat F}+ \mat F' \, \erm^{i\, \sigma \, t}\,,\quad
 \mat B_{T,m}\to \ov {\mat B}+\mat B'\, \erm^{i\, \sigma \, t}\,,\quad
 \mat \Lambda_{jm}\to \mat \Lambda_j'\,\erm^{i\, \sigma \, t}\,,
\]
 where $\mat F'$, $ \mat B'$, and  $\mat \Lambda_j'$
are understood as
constant complex amplitudes, into  equation (\ref{def3}) and use the equilibrium equation
(\ref{BKov2}); then we obtain
   {\renewcommand{\arraystretch}{1.1}  
 \begin{equation}
  \begin{split}
    &i \sigma\eta\mathbf{B}'+ (\gm+\mu_0)\mat B'
    +\mat\Lambda_1'\cdots+\mat \Lambda_n'=  \mat F'
 \\ & \\
   &  i\sigma\frac{1}{\mu_j}\mat\Lambda_j'+\frac{1}{\eta_j}\mat\Lambda_j'=
  i \sigma \mat B'\,\Rightarrow
\mat \Lambda_j'=\left(\frac{1}{\mu_j}+\frac{1}{i\sigma \eta_j}\right)^{-1}\mat B'\,,\\
    \end{split}\label{linch3}
  \end{equation}
  for $ j=1,\ldots,n\,$.
}
These equations imply
\begin{equation}
\bigg\{ \gamma+\mu_0+  \eta i\sigma+
  \sum_{j=1}^n\Bigl(\frac{1}{\mu_j}
  +\frac{1}{i\sigma \eta_j}\Bigr)^{-1}\Bigg\}\mat B'=
\mat F'\,.
   \label{love1.1}
  \end{equation}
  We define the complex rigidity  (this $J$ is unrelated to the force
  matrix $\mat J$)  of the generalised Maxwell model of Figure
\ref{genmax} with prestress as
  \begin{equation}
    J^{-1}(\sigma)=\mu_0+  \eta i\sigma+
  \sum_{j=1}^n\Bigl(\frac{1}{\mu_j}
  +\frac{1}{i\sigma \eta_j}\Bigr)^{-1}\,.\label{J-1genmax}
\end{equation}
The complex compliance  $J(\sigma)$ is the inverse of the complex rigidity.

  The combination of equations  (\ref{kcomplex2})  and (\ref{J-1genmax}) gives
\begin{equation}
  k(\sigma)=
  \frac{3\Io G}{R^5}\frac{1}{\gm+ J^{-1}(\sigma)}\,,\quad
 J^{-1}(\sigma)=
    \mu_0+  \eta i\sigma+
  \sum_{j=1}^n\left(\frac{1}{\mu_j}
  +\frac{1}{i\sigma \eta_j}\right)^{-1}
 \label{Lovegenmax}
\end{equation}    
that is the Love number of the body with the generalised Maxwell rheology represented in
Figure \ref{genmax} and with prestress  ($\eta_0\to\infty$).

A similar reasoning  gives the  Love number of the body with the generalised Voigt
rheology represented in Figure \ref{genvoigt} and with prestress ($\eta_0\to\infty$):
\begin{equation}
  k(\sigma)=
  \frac{3\Io G}{R^5}\frac{1}{\gamma+\mu_0+J_V^{-1}(\sigma)}\,,\quad
  J_V(\sigma)=\frac{1}{\mu_1}+\frac{1}{i\sigma\eta_1}+\sum_{j=2}^{n}\frac{1}{\mu_j+i\sigma\eta_j}\,.
\label{Lovegenvoigt}
\end{equation}
In this last expression $J_V(\sigma)$ is the complex compliance of the usual generalised Voigt rheology
(\cite{Bland} chapter 1
eq. (27)) while $J^{-1}(\sigma)=\mu_0+J^{-1}_V(\sigma)$ is the complex rigidity of the
 generalised Voigt
rheology represented in Figure \ref{genvoigt}  with prestress.

If we make $\eta_j=0$, $j=1,\ldots,n$, in equation (\ref{Lovegenmax}), then we
obtain the Love number of the body with the simple rheology represented in Figure \ref{simple-osc} and with 
prestress:
\begin{equation}
  k(\sigma)=\frac{3 \Io G}{R^5}\frac{1}{\gamma +\mu_0+i\sigma\eta}
  \,.\label{k}
\end{equation}
This Love number is equal to that obtained from a Kelvin-Voigt rheology with elastic coefficient
$\mu_0$ and viscosity coefficient $\eta$. For this reason we will usually refer to
the rheology represented in Figure  \ref{simple-osc}, after the limit $\eta_0\to\infty$
(prestress), as the Kelvin-Voigt rheology.

For the Kelvin-Voigt rheology, if $k(\sigma)$ is known at a certain frequency,
say $\sigma=\omega$, then $\mu_0$ and $\eta$ are given by
\begin{equation}
  \frac{3\Io G}{R^5}\frac{1}{ k(\omega)}-\gamma=
  \mu_0+i\omega\eta\,.\label{k2dt}
\end{equation}

The imaginary part of  $k(\sigma)$  is related to  the quality
factor (\cite{efroimsky2012bodily} Eq. (141))
as
\begin{equation}
  Q^{-1}(\sigma)=\sin \dt (\sigma)\quad \text{with}\quad k(\sigma)=|k(\sigma)|
  \big(\cos\dt(\sigma)-i\, \sin\dt(\sigma)\big)\,,\label{Qeq}
\end{equation}
where $\dt(\sigma)$ is the phase lag.
If $Q>10$ or $\delta< 1/10$, then $\delta/ \sigma$ is   called the ``time lag''
and
\begin{equation}
  \delta\approx \frac{1}{Q}
\,.\label{k2dt2} 
\end{equation}

\subsection{The fossil-deformation  matrix $\mathbf{B}_{0,m}$ and 
  the critical strain.}

The  prestress  at the mean configuration,
$\ov {\mathbf{\Lambda}}=\mu_0(\ov{\mathbf{B}}-\mathbf{B}_{0,m})$,
is caused by a fossil triaxial deformation $\ov{\mathbf{B}}-\mathbf{B}_{0,m}$.
As argued in   \cite{zanazzi2017triaxial}, there exists a critical   deformation
in which the  internal stress exceeds the yield stress  of von Mises and
the body
either plastically deform of fracture. The exact computation of the critical
deformation is complex, and essentially undoable, since it requires the computation of
the body internal stresses (the possible
presence of a liquid core may play an important role in this computation).
Associated with the critical stress there is a critical
strain that for the Earth is in the range $10^{-3}-10^{-5}$  \cite{zanazzi2017triaxial}.
The  strain in  the body is of the order of $\|\mat B-\mathbf{B}_{0,m}\|$
(see  \cite{rr2015}, in particular appendix 3). So,  we conclude that: 
for each body there exists a critical $B_{crit}$ such that $\mathbf{B}_{0,m}$
must satisfy
\begin{equation}
  \|\ov{\mathbf{B}}-\mathbf{B}_{0,m}\|\le B_{crit}\, ;\label{Bcrit}
\end{equation}
bodies with similar interior structure have the same $B_{crit}$; and,
 at least for terrestrial bodies,
$B_{crit}$ may be in the range $10^{-3}-10^{-5}$.

In tables \ref{table1} and \ref{table2} we present   the values
of the rheological parameters $\mu_0$ and $\eta$ of the Kelvin-Voigt rheology for some   terrestrial bodies.
These parameters were computed using  equation
(\ref{k2dt}) with the complex Love number at the diurnal frequency.
We also present   $\|\ov{\mathbf{B}}-\mathbf{B}_{0,m}\|$, where $\mathbf{B}_{0,m}$ is
obtained using equation (\ref{BKov2}).
As in  \cite{zanazzi2017triaxial},
we neglected the presence of a liquid core. The use
of $\mu_0$ 
calibrated at
the diurnal 
frequency in
the computation
of the fossil 
deformation
matrix is 
subjectable 
to criticism.

We point out that the values for   $\|\ov{\mathbf{B}}-\mathbf{B}_{0,m}\|$
in Table \ref{table2}
are within the limit $10^{-3}-10^{-5}$ proposed in
\cite{zanazzi2017triaxial}\footnote{\label{mol}
  Associated with  a {\sl nonhomogeneous} body with an effective Kelvin-Voigt rheology and with parameters
   $\Io$, $m$, $\mu_0$, and $\eta$ there is an {\sl equivalent homogeneous body} with the same parameters and the same rheology such that the molecular shear modulus $\mu_{0\,mol}$ and the 
   molecular viscosity $\eta_{0\,mol}$ of the later are given by
   \cite{crr2018}
   \[
     \mu_{0\,mol}  \ \text{(Pa)} =\frac{15}{152\pi }\frac{m}{R_\Irm}\, \mu_0\qquad
     \text{and}\qquad
     \eta_{mol} \ \text{(Pa sec)}=\frac{15}{152 \pi}\frac{m}{R_\Irm}\, \eta\,.
   \]
    Under this correspondence we find the following values for the pair
    $(\mu_{0\,mol},\eta_{mol})$ for each one of the  equivalent homogeneous
    bodies in Table \ref{table2}
    : Moon $\big($62 GPa, 5.2$\times 10^{14}$ Pa s$\big)$, Mercury
    $\big($8.9 GPa, 1.4$\times 10^{14}$ Pa s$\big)$, Earth
    $\big($120 GPa, 1.7$\times 10^{14}$ Pa s$\big)$, and Mars $\big($39 GPa,
    6.5$\times 10^{12}$ Pa s$\big)$. Note that the molecular characteristic time
    $\eta_{mol}/ \mu_{0\,mol}$ is equal to the characteristic time of the rheology
  $\eta/\mu_0$.}. 

\begin{table}
\begin{tabular}{llllllll}
  \hline\noalign{\smallskip}
  Body & $m\,${\tiny($\times\, 10^{24}\,$kg)}  & $R\,${\tiny(km)} &
  $\frac{\Io}{mR^2}$&$R_\Irm/R$ & $k_2$
  &Q& $k_f$\\
  \noalign{\smallskip}\hline\noalign{\smallskip}
  Moon &   0.07346 &   1737      &  0.393      &  0.992   & 0.0236 &   46&  1.43\\
  Merc &  0.3301   &   2439      &  0.346      &  0.930   &0.455&   89& 1.04  \\
  Earth &   5.974   &   6371      &  0.331      &  0.909   &0.280 &  14.5 &0.93  \\
  Mars &  0.6418    &   3389      &  0.365      &  0.955   &0.164 &   99.5& 1.19\\
  \noalign{\smallskip}\hline\noalign{\smallskip}
\end{tabular}
\caption{ Notation:
  $m=$mass, $R=$volumetric mean radius, $\Io/(mR^2)$=normalised
  mean moment of inertia,
  $R_\Irm/R=$inertial radius divided by $R$ ($R_\Irm$ is
  defined by $\Io=0.4 mR_\Irm^2$),
  $k_2$=$|k(\omega)|=$Love number, and  
  $Q$=quality factor. Both $k_2$ and $Q$ refer to a diurnal force period $T$
  given in Table
  \ref{table2}. Except for $k_2$ and $Q$ most values in  this table are
  well established, they were taken from:
  Moon  INPOP19a, ($\frac{\Io}{mR^2}$ is from \cite{yan2012cegm02}), 
  Mercury \cite{steinbrugge2018viscoelastic}
  ($k_2$ is from \cite{margot2018mercury} (section 3.4) and
   $Q=89$ is  a best fit for a quantity that may be in the range
  $25<Q<350$ according to \cite{baland2017obliquity}),
  Earth \cite{rr2017} Table 3 diurnal mode
  (the value $k=0.2803-0.01944 i$ takes into account the oceans and
  were obtained by means of an average using data from \cite{iers2010};
  in \cite{mathews2002modeling}
   (paragraph [21], Table 2, and Appendix D paragraph [134])
  $k=0.2810-i\, 0.035$),
  Mars \cite{jacobson2014martian} and  \cite{lainey2016quantification}
  (the tidal forcing frequency is not the diurnal but that of Phobos)
  .}
\label{table1}
\end{table}

\begin{table}
\begin{tabular}{llllllc}
  \hline\noalign{\smallskip}
Body& $T$ {\tiny (day)}& $\frac{2\pi}{\sqrt{\gm}}$ {\tiny (hour)}
 & $\frac{2\pi}{\sqrt{\mu_0}}$ {\tiny (hour)}&
$\frac{1}{\eta}$ {\tiny (sec)} & $\tau$ {\tiny (min)}
& $\|\ov{\mathbf{B}}-\mathbf{B}_{0,m}\|$  \\
  \noalign{\smallskip}\hline\noalign{\smallskip}
  Moon &   27.32 &   1.992      & 0.2575 & 2.57 &   136  & $0.70\times 10^{-5}$\\
  Merc &   58.65  &   1.421      &   1.249   &31.86 &   151.0 &    $1.0\times 10^{-4}$ \\
  Earth &  0.9973  &   1.363      &  0.8980   &194.1 &   15.85 &  $3.1\times 10^{-5}$ \\
  Mars &  1.026  &   1.736      & 0.6941   &960.4 &   2.363  & $9.3\times 10^{-5}$  \\  
   \noalign{\smallskip}\hline\noalign{\smallskip}
\end{tabular}
\caption{ Notation:
  $T=$sidereal rotation period, $\gm=$gravitational modulus,
 $\mu_0=$ modulus of elasticity, 
 $\eta=$ coefficient of viscosity, 
  $\tau=\eta/(\gamma+\mu_0)=$
  characteristic time, and
 $\|\ov{\mathbf{B}}-\mathbf{B}_{0,m}\|=
 \|\ov{\mathbf{F}}-\gamma \ov{\mathbf{B}}\| /\mu_0=$ deformation of the
 prestressed equilibrium configuration.
 $\ov{\mathbf{F}}$ and  $\ov{\mathbf{B}}$  were  computed using equations
 (\ref{ovF}), the relation between the Stokes coefficients and $\ov{\mat B}$
 given in Table \ref{frames2}, and the gravitational data from: Moon
 \cite{williams2014lunar}, Mercury \cite{mazarico2014gravity}, Earth
 \cite{yoder1995astrometric}, and Mars \cite{genova2016seasonal}.
 }
\label{table2}
\end{table}

\section{Equations for the libration
  of bodies that are out of hydrostatic equilibrium (prestressed).}
\label{equations}

In this section we consider a body that satisfies the hypotheses in equation
(\ref{hypotheses}). The mantle is supposed to be prestressed and 
  the eigendirections of the fossil-deformation  matrix  are fixed in
the Tisserand frame of the mantle
$\K_m$.
The core-mantle boundary (CMB) is assumed to be rigid and ellipsoidal with respect
to $\K_m$
The principal axes of the mantle
and the core do not need to be aligned. Nevertheless,
the centre of mass of the mantle and of the  core must coincide for all  time.
The mantle and the core do not need to be of constant density, but 
the layers of constant density of   the fluid in the core must be concentric and homothetic to  the CMB.
These hypotheses are further discussed  in Appendix \ref{Poinc}.

The Lagrangian function ${\cal L}_\mathrm{ROT}$ associated with the rotations
is the same as that in equation (\ref{rotlag1}), but
in this case $\mat I_c$ is not a multiple of the identity.
 As the cavity of the fluid core is fixed in the mantle frame, the
matrix of inertia $\mat I_c$
satisfies the same equation as $\mat I_m$ and $\mat I_T$ under
rotation of the mantle. 
The rotational part of the equations for the
librations is obtained by means
of a variational principle, as in Section \ref{rotsec1}.
If we also take into account
the CMB friction, as in Section \ref{CMBfriction}, then the  result is\footnote{The terms
  $\mat I_c\boldsymbol{\omega}_c\times\boldsymbol{\omega}_c
  - k_c(\boldsymbol{\omega}_m - \boldsymbol{\omega}_c)-3
   \sum_\beta
   \frac{{\cal G}m_\beta}{ \|\mat r_\beta\|^5}(\mat I_c\mathbf{r}_\beta)\times\mathbf{r}_\beta$
   in the first equation of system  (\ref{eqrot3}) represent  the torque of the  core upon the mantle.
    If the core is spherical, then 
   $\mat I_c\boldsymbol{\omega}_c\times\boldsymbol{\omega}_c=-3
   \sum_\beta
   \frac{{\cal G}m_\beta}{ \|\mat r_\beta\|^5}(\mat I_c\mathbf{r}_\beta)\times\mathbf{r}_\beta=0$
   and the torque of the core upon
   the mantle reduces to  $- k_c(\boldsymbol{\omega}_m - \boldsymbol{\omega}_c)$, which represents
   the shear-stress torque at the  CMB.  If the core is not spherical, then pressure can also
   produce torque. The term   $\mat I_c\boldsymbol{\omega}_c\times\boldsymbol{\omega}_c$ can be interpreted
   as the pressure torque due to the motion and the inertia of the fluid.    
   The term $-3
   \sum_\beta
   \frac{{\cal G}m_\beta}{ \|\mat r_\beta\|^5}(\mat I_c\mathbf{r}_\beta)\times\mathbf{r}_\beta$ 
   is the pressure torque from the core upon the mantle
   caused by the action of external gravity on the core.
   All the three terms  $\mat I_c\boldsymbol{\omega}_c\times\boldsymbol{\omega}_c$,
   $ - k_c(\boldsymbol{\omega}_m - \boldsymbol{\omega}_c)$, and $-3
   \sum_\beta
   \frac{{\cal G}m_\beta}{ \|\mat r_\beta\|^5}(\mat I_c\mathbf{r}_\beta)\times\mathbf{r}_\beta$
   produce   reactive counter-torques from the mantle upon the core. The first and second
   of these reactive counter-torques are present
   in the equation for $\dot{\boldsymbol{\pi}}_c$ in system
   (\ref{eqrot3}). The third pressure-reactive term, $3
   \sum_\beta
   \frac{{\cal G}m_\beta}{ \|\mat r_\beta\|^5}(\mat I_c\mathbf{r}_\beta)\times\mathbf{r}_\beta$, is absent
   because it is cancelled out by the external gravitational force that acts upon the core.

   Now, suppose the core is  spherical, so that the torque of the core upon the mantle is
   $ - k_c(\boldsymbol{\omega}_m - \boldsymbol{\omega}_c)=- \dot{\boldsymbol{\pi}}_c=-\frac{d}{dt}\mat I_c \boldsymbol{\omega}_c$. 
   If we take the limit as
   $k_c\to\infty$ while $\boldsymbol{\pi}_c$ remains bounded, then we obtain $\boldsymbol{\omega}_c\to\boldsymbol{\omega}_m$
 and  the mantle and core move as they formed a rigid body. In this case 
 the torque of the core upon the mantle becomes, as expected,
 $-\frac{d}{dt}\mat I_c \boldsymbol{\omega}_m$.
  }
  \begin{equation}\begin{split}
      \dot{\boldsymbol{\pi}}_m &=
      \mat I_c\boldsymbol{\omega}_c\times\boldsymbol{\omega}_c
      - k_c(\boldsymbol{\omega}_m - \boldsymbol{\omega}_c)
-3
   \sum_\beta
   \frac{{\cal G}m_\beta}{ \|\mat r_\beta\|^5}(\mat I_T\mathbf{r}_\beta)\times\mathbf{r}_\beta\\
  \dot{\boldsymbol{\pi}}_c &= \boldsymbol{\omega}_c\times\mat I_c\boldsymbol{\omega}_c
  + k_c(\boldsymbol{\omega}_m - \boldsymbol{\omega}_c)\\
  \dot {\mat I}_c&=[\boldsymbol{\widehat \omega}_m\,,\mat I_c]\,\quad\text{with:} \\
  \boldsymbol{\pi}_m&=\mat I_m \boldsymbol{\omega}_m\,,
  \quad \boldsymbol{\pi}_c=\mat I_c \boldsymbol{\omega}_c\,,\quad  \mat I_T= \mat I_m+\mat I_c\\
  \mat I_T&=\Io\big( \Id-\mat B_T\big)\,,\quad
  \mat I_m={\rm I}_{\circ m}\big( \Id-\mat B_m\big)\,,\quad
    \mat I_c={\rm I}_{\circ c}\big( \Id-\mat B_c\big)\,.\quad
   \end{split}
\label{eqrot3}
\end{equation}
If we add both equations
and use
 $\boldsymbol{\pi}_T=\boldsymbol{\pi}_m+\boldsymbol{\pi}_c$
we recover equation (\ref{eqrot1}).

Equations (\ref{eqrot3}) must be complemented by the equations that determine the deformation of the
mantle. The equations for the deformations depend on the
rheology of the mantle and for the three rheological models considered in this paper they are given
by equations (\ref{def1}), (\ref{def2}), and (\ref{def3}). In the following we rewrite these three equations
in the inertial frame.

For the Kelvin-Voigt rheology, represented in Figure \ref{simple-osc} with $\eta_0\to\infty$ (prestress),
\begin{equation}\begin{split} 
&  \eta\dot{\mathbf{B}}_{T}+\eta\big[\mat B_T\,,\boldsymbol{\widehat \omega}_m\big]
  +\gamma \mathbf{B}_{T}+ \mu_{0}\Big(\mathbf{B}_{T}-\mathbf{B}_{0}\Big) = 
  \mathbf{F}\\
&   \dot{\mathbf{B}}_{0}=\big[\boldsymbol{\widehat \omega}_m\,,\mat B_0\big]
  \end{split}
  \label{def1k}
\end{equation}
For the generalised Voigt rheology, represented in Figure  \ref{genvoigt}  with $\eta_0\to\infty$ (prestress),
\begin{equation}\label{def2k}
\begin{split}
&\gamma \mathbf{B}_{T}+ \mu_{0}\Big(\mathbf{B}_{T}-\mathbf{B}_{0}\Big)+\mat \Lambda=\mathbf{F}\\
&\frac{1}{\mu_1}\Big(\dot{\mat\Lambda}+\big[\mat \Lambda\,,\boldsymbol{\widehat \omega}_m\big]
\Big)+\frac{1}{\eta_1} \mat\Lambda =
 \dot{\mathbf{B}}_{1}+\big[\mat B_1\,,\boldsymbol{\widehat \omega}_m\big]\\
 &\eta_j\Big(\dot{\mathbf{B}}_{j}+\big[\mat B_j\,,\boldsymbol{\widehat \omega}_m\big]\Big)
 +\mu_j\mat B_{j}=\mat\Lambda,\quad j=2,\ldots,n\\
 &\mat B_{1}=\mat B_{T}-\big(\mat  B_{2}+\mat  B_{3}+\ldots+\mat  B_{n}\big)\\
 & \dot{\mathbf{B}}_{0}=\big[\boldsymbol{\widehat \omega}_m\,,\mat B_0\big]
\end{split}
\end{equation}
For the generalised Maxwell  rheology, represented in Figure  \ref{genmax}  with $\eta_0\to\infty$ (prestress),
 {\renewcommand{\arraystretch}{1.5}  
 \begin{equation}
  \begin{split}
    & \eta\Big(\dot{\mathbf{B}}_{T}+\big[\mat B_T\,,\boldsymbol{\widehat \omega}_m\big]\Big)
    + \gm\mat B_{T}+\mu_{0}\Big(\mathbf{B}_{T}-\mathbf{B}_{0}\Big)
    +\mat\Lambda_{1}\cdots+\mat
    \Lambda_{n}= \mathbf{F}\\
    &\frac{1}{\mu_j}\Big(\dot{\mat\Lambda}_j+\big[\mat \Lambda_j\,,\boldsymbol{\widehat \omega}_m\big]
\Big)+\frac{1}{\eta_j} \mat\Lambda_j =
\dot{\mathbf{B}}_{T}+\big[\mat B_T\,,\boldsymbol{\widehat \omega}_m\big]\,,\ j=1,\ldots,n\,,\\
 & \dot{\mathbf{B}}_{0}=\big[\boldsymbol{\widehat \omega}_m\,,\mat B_0\big]
       \end{split}\label{def3k}
\end{equation}
}
In equations (\ref{def1k}), (\ref{def2k}), and (\ref{def3k}),  $\mat F$ is given by
\begin{equation}
  \mat F = -\left(\boldsymbol{ \omega}_{m}\otimes\boldsymbol{\omega}_{m} -
    \frac{\|\boldsymbol \omega_{m}\|^2}{3}\Id\right) +\mat J- \frac{\tr\mathbf{J}}{3}\Id
  \label{Fk}
\end{equation}

\subsection{A characterisation of the core frame $\K_c$  in the absence of
 CMB friction.}

\label{Kc}

If  $k_c=0$,
then  equation (\ref{eqrot3}) becomes
$  \dot{\boldsymbol{\pi}}_c = \boldsymbol{\omega}_c\times\mat I_c\boldsymbol{\omega}_c$
that implies $ \dot{\boldsymbol{\pi}}_{c,c}=\frac{d}{dt}\big(\mat R_c^{-1}
\boldsymbol{\pi}_{c}\big)=0$, namely the angular momentum of the fluid is constant in the
Tisserand frame of the fluid core $\K_c$. If initially
$\mat R_c(0)\mat e_3=\boldsymbol{\pi}_c(0)/\|\boldsymbol{\pi}_c(0)\|$,
  then for all time
  $\mat R_c(t)\mat e_3=\boldsymbol{\pi}_c(t)/\|\boldsymbol{\pi}_c(0)\|$ and 
 the $\mat e_3$-axis
of $\K_c$ moves in the inertial space together
with the  angular momentum of the fluid.

  The archetypal example of an inviscid fluid motion inside an
  ellipsoid was provided by \cite{poincare1910precession} and 
  \cite{hough1895xii} (after related work
  by  \cite{poincare1885equilibre}).   We  remark that the results obtained from the Lagrangian function
(\ref{rotlag1}) are not bounded by the Poincar\'e-Hough model since no particular 
information from this model was used. On the contrary, being the Poincar\'e-Hough flow a particular
motion of a fluid inside an ellipsoidal cavity, the results obtained within this example must be in
agreement with those we obtained above. In the Appendix \ref{Poinc} we analyse the Poincar\'e-Hough flow
and recover the equation $\dot {\boldsymbol \pi}_{c,c}=0$ in this particular context.
In the Appendix \ref{Robert} we show that for $k_c>0$ the average mean vorticity of a viscous flow
\cite{stewartson1963motion}, \cite{roberts1965motion} is  equal to the core angular
velocity $\boldsymbol{\omega_c}$.

\subsection{The linearization of the equations about the guiding motion:
inertial part.}

\label{linsec}

In the libration problem we expect  the mantle frame $\K_m$ to
remain close to the guiding frame $\K_g$ and the angular velocity of 
the core $\boldsymbol \omega_c$ to remain close to both $\boldsymbol \omega_m$and $\boldsymbol \omega_g$.
 Under these two conditions the following approximations hold
 \begin{equation}\begin{split}
     &  \mat Y_m:=\mat R_g^{-1}\mat R_m=\exp\boldsymbol{\widehat \alpha}_m
     \approx\Id+\boldsymbol{\widehat \alpha}_m:\K_m\to\K_g\\
     &   \mat Y_c:=\mat R_g^{-1}\,\mat R_c=\boldsymbol{\widehat \alpha}_c=
     \Id+\boldsymbol{\widehat \alpha}_c\ldots:\K_c\to\K_g\\
      & \dot {\mat Y}_m\,\mat Y_m^{-1}= \boldsymbol {\widehat\omega}_{m,g}-\boldsymbol{\widehat \omega}_{g,g}
    \approx \dot{\boldsymbol{\widehat \alpha}}_m:\K_g\to\K_g\\
    & \dot {\mat Y}_c\,\mat Y_c^{-1}= \boldsymbol {\widehat\omega}_{c,g}-\boldsymbol{\widehat \omega}_{g,g}
    \approx \dot{\boldsymbol{\widehat \alpha}}_c:\K_g\to\K_g\,.
  \end{split}\label{apr1}
\end{equation}
where $\boldsymbol{ \alpha}_m$ and $\boldsymbol{ \alpha}_c$ are  angular vectors.
Although
$\dot{\boldsymbol{ \alpha}}_c$ is small, $\boldsymbol{ \alpha}_c$ can be large due to a possible drift.
As we will see, the oscillating part of $\boldsymbol{ \alpha}_c$ can be easily separated from  the
drift in the linearized equations in such a way that  we can pursue the linearization
assuming that   $\boldsymbol{ \alpha}_c$ is small. In this situation $\K_T$ also remains close to
$\K_g$ and we write
\begin{equation}\begin{split}
&      \mat Y_T:=\mat R_g^{-1}\mat R_T=\exp\mat {\widehat a}
\approx\Id+\mat{\widehat a}:\K_T\to\K_g\\
 & \dot {\mat Y}_T\,\mat Y_T^{-1}= \boldsymbol {\widehat\omega}_{T,g}-\boldsymbol{\widehat \omega}_{g,g}
 \approx \dot{\mat{\widehat a}}:\K_g\to\K_g\,.
 \end{split}
\label{apr1T}
\end{equation}

The three angular vectors $\boldsymbol{\widehat \alpha}_m$, $\boldsymbol{\widehat \alpha}_c$, and
$\mat{\widehat a}$ are not independent. Indeed,  the relations
$\mat I_{T,g}=\mat I_{m,g}+\mat I_{c,g}$  and
$\mat I_{T,g}\boldsymbol {\omega}_{T,g}=\mat I_{m,g}\boldsymbol {\omega}_{m,g}+\mat I_{c,g}
\boldsymbol {\omega}_{c,g}$ imply, up  to first order in the small angles, 
\[\begin{split}
    &     \mat I_{T,g}\big(\boldsymbol {\omega}_{T,g}-\boldsymbol {\omega}_{g,g}\big)=
  \mat I_{m,g}\big(\boldsymbol {\omega}_{m,g}-\boldsymbol {\omega}_{g,g}\big)
  +\mat I_{c,g}\big(\boldsymbol {\omega}_{c,g}-\boldsymbol {\omega}_{g,g}\big)
  \Longrightarrow\\ &
    \mat I_{T,g}\dot{\mat{ a}}=
  \mat I_{m,g}\dot{\boldsymbol{\alpha}}_m
  +\mat I_{c,g} \dot{\boldsymbol{\alpha}}_c \Longrightarrow
   \ov {\mat I}\dot{\mat{ a}}=
  \ov {\mat I}_{m}\dot{\boldsymbol{ \alpha}}_m
  +\ov {\mat I}_{c} \dot{\boldsymbol{ \alpha}}_c,
 \end{split} \]
where we  neglected the small time variations of $ \mat I_{T,g}$, $ \mat I_{m,g}$, and
$ \mat I_{c,g}$. The integration of this last equation gives the relation between the angular vectors
 \begin{equation}
   \ov {\mat I}\mat a= \ov{\mat I}_m\boldsymbol {\alpha}_m+\ov {\mat I}_c
   \boldsymbol{\alpha}_c\, .
   \label{alphamc}
 \end{equation}

Equations (\ref{apr1}), (\ref{apr1T}), and
$\boldsymbol{\omega}_{g,g}=\omega \mathbf{e_3}+ \boldsymbol {\omega}_{s,g}$, 
where $\|\boldsymbol {\widehat\omega}_{s,g}\|\ll \omega$  is a small quantity
(equation  (\ref{omegag})),
imply
  {\renewcommand{\arraystretch}{1.5}  
 \begin{equation}\begin{split}
     \boldsymbol {\widehat \omega}_{g,m}&
=\mat Y_m^{-1}  \boldsymbol{\widehat\omega}_{g,g}\mat Y_m\approx
(\Id-\boldsymbol{\widehat \alpha}_m)\boldsymbol{\widehat\omega}_{g,g}(\Id+\boldsymbol{\widehat \alpha}_m)
\\ & 
     \approx \boldsymbol{\widehat\omega}_{g,g}+
     \big[\boldsymbol{\widehat\omega}_{g,g},\boldsymbol{\widehat \alpha}_m\big]
     \approx  \boldsymbol{\widehat\omega}_{g,g}+
\omega \big[\mat{\widehat e}_3,\boldsymbol{\widehat \alpha}_m\big]\\
&\approx \omega \mat{\widehat e}_3+\boldsymbol {\widehat\omega}_{s,g}
+
\omega \big[\mat{\widehat e}_3,\boldsymbol{\widehat \alpha}_m\big]\\
     \boldsymbol {\widehat \omega}_{m,m}&
=\mat Y_m^{-1}  \boldsymbol{\widehat\omega}_{m,g}\mat Y_m\approx
\mat Y_m^{-1}\big\{\boldsymbol{\widehat\omega}_{g,g}+\dot{\boldsymbol{\widehat \alpha}}_m
\big\} \mat Y_m \approx \boldsymbol{\widehat\omega}_{g,m}+\dot{\boldsymbol{\widehat \alpha}}_m\\  
\boldsymbol {\widehat \omega}_{c,m}&=\mat Y_m^{-1}  \boldsymbol{\widehat\omega}_{c,g}\mat Y_m\approx
\mat Y_m^{-1}\big\{\boldsymbol{\widehat\omega}_{g,g}+\dot{\boldsymbol{\widehat \alpha}}_c
\big\} \mat Y_m \approx\boldsymbol{\widehat\omega}_{g,m}+\dot{\boldsymbol{\widehat \alpha}}_c\\ 
\boldsymbol{\widehat \omega}_{T,m}&=\mat Y_m^{-1}  \boldsymbol{\widehat\omega}_{T,g}\mat Y_m\approx
\mat Y_m^{-1}\big\{\boldsymbol{\widehat\omega}_{g,g}+\dot{\mat{\widehat a}}
\big\} \mat Y_m \approx\boldsymbol{\widehat\omega}_{g,m}+\dot{\mat {\widehat a}}
  \end{split}\label{apr2}
\end{equation}
}

As in Section \ref{averageforce}, 
the force operator $\mat J_g:\K_g\to\K_g$   can be decomposed into a constant
  part $\ov {\mat J}$  plus a time oscillating part $\boldsymbol{\dt}\mat J_g$, 
 $\mat J_g=\ov {\mat J}+\boldsymbol{\dt}\mat J_g$.
  In the mantle frame this decomposition becomes
\begin{equation}
  \mat J_m=\mat Y_m^{-1}\mat J_g\mat Y_m\approx \ov{ \mat J}+
  [ \ov {\mat J}\, ,\boldsymbol { \widehat \alpha}_m] +
  \boldsymbol{\dt}\mat J_g+[\boldsymbol{\dt}\mat J_g\,
  ,\boldsymbol { \widehat \alpha}_m]\,.
  \label{Jprime2}
\end{equation}
  In the following we assume the 
  \begin{equation}
    \text Hypothesis:\ \text{the effect of}\ [\boldsymbol{\dt}\mat J_g\,
    ,\boldsymbol { \widehat \alpha}_m]\  \text{is negligible.}
    \label{hypsmall2}
  \end{equation}
This simplifies a lot the mathematical
 analysis of the problem  since the homogeneous part of the linear equation to be obtained
 is of constant coefficients. The drawback is that we exclude
 the possibility of parametric resonances
 that may exist. In some situations  $[\boldsymbol{\dt}\mat J_g\,
 ,\boldsymbol { \widehat \alpha}_m]$ can be  at least partially 
 eliminated by means of an averaging procedure.

 The total  moment of inertia operator in $\K_m$ can be decomposed as
 \begin{equation}
   \mat I_{T,m}=
   \ov{\mat I}-\Io \boldsymbol\delta \mat B_{T,m}\ \text{where}\  \ov{\mat I}=\Io(\Id-\ov{\mat B})\
   \text{and}\
   \boldsymbol\delta \mat B_{T,m}= \mat B_{T,m}-\ov{\mat B}\,.\label{B1}
   \end{equation}
  Since 
  $\mat I_{c,m}=\ov{\mat  I}_c$ is constant,
 \begin{equation}
   \mat I_{m,m}=
   \ov{\mat I}-\ov{\mat I}_c-\Io \boldsymbol\delta \mat B_{T,m}\,,\quad
   \ov {\mat I}_m=  \ov{\mat I}-\ov{\mat I}_c\,,\quad 
   \text{and}\quad
   \Io\boldsymbol\delta \mat B_{T,m}={\rm I}_{\circ m}\boldsymbol\delta \mat B_{m,m}\,.
   \end{equation}

At first we will consider the case in which the principal axes of the core are not aligned with those of the 
mantle. In this case it is convenient to choose a $\K_m$ in which  the  average moment of inertia of the whole 
body $\ov {\mat I}$ is diagonal.
Note that $\mat I_{c,m}=\ov{\mat I}_c$ is constant but it is 
not diagonal.  In the frame of the mantle equation (\ref{eqrot3}) becomes
\begin{equation}
  \begin{split}
           &\frac{d}{dt}\left(\mat I_{m,m}\boldsymbol{\omega}_{m,m}\right)
    +\boldsymbol{\omega}_{m,m}\times \mat I_{m,m}\boldsymbol{\omega}_{m,m}
+\boldsymbol{\omega}_{c,m}
\times\ov{ \mat I}_c\boldsymbol{\omega}_{c,m}\\
&\qquad \qquad+  k_c(\boldsymbol{\omega}_{m,m} - \boldsymbol{\omega}_{c,m})
  =\big[{\mat I}_{T,m},\mat J_m\big]^\vee\\
  & \ov{ \mat I}_c\dot{\boldsymbol{\omega}}_{c,m}
  +\big(\boldsymbol{\omega}_{m,m}-\boldsymbol{\omega}_{c,m}\big)
  \times\ov{ \mat I}_c\boldsymbol{\omega}_{c,m} + k_c(\boldsymbol{\omega}_{c,m} - \boldsymbol{\omega}_{m,m})=0 
 \\
 &
  \dot {\ov {\mat I}}_c=0\,,
  \end{split} \label{eqrot4}
\end{equation}
where we used the check map $^\vee$ (the inverse of the hat map)
to represent the torque, see Section \ref{sechat}.
The linearized equations are obtained by means of the substitution of the relations
previously obtained  into equations (\ref{eqrot4}) and then by neglecting small quantities of second order,
where the small quantities are: $\boldsymbol\alpha_m$,  $\boldsymbol\alpha_c$,
$\boldsymbol{\dt}\mat J_g$, $\boldsymbol\delta \mat B_{T,m}$, and $ \boldsymbol{\omega}_{s,g}$.
The expression obtained from the equation for $\dot{\boldsymbol \pi}_m$ is long and will
be omitted. The expression obtained from the equation for
 $\dot{\boldsymbol \pi}_c$ is
\begin{equation}
 \ov{\mat I}_{c}\ddot{\boldsymbol\alpha}_{c}+\omega \ov{\mat I}_{c}\big(\mat e_3\times 
  \dot{\boldsymbol\alpha}_m\big)+\omega\big( \dot{\boldsymbol\alpha}_m
  -\dot{\boldsymbol\alpha}_c\big)\times  \ov {\mat I}_{c}\mat e_3+    k_c\big(\dot{\boldsymbol\alpha}_c
  -  \dot{\boldsymbol\alpha}_m\big)= - \ov{\mat I}_{c}\dot{\boldsymbol{\omega}}_{s,g}\,.
\label{lincore2}\end{equation}
Note that if $\mat I_c$ is not diagonal then the three components of this equation are coupled.
If $\mat I_c$ is diagonal, which means that the principal axes of the core are aligned with the principal
axes of $\ov {\mat I}$, then the equations simplify a lot.

In the following we assume that $\ov {\mat I}_c$, $\ov {\mat I}_m$ and $\ov {\mat I}_T$ are all diagonal
in $\K_m$. In this case the linearization of equations (\ref{eqrot4}) gives, 
 for the angular motion of the mantle:
{\renewcommand{\arraystretch}{1.5}
\begin{equation}
  \begin{split}
    &\left(\begin{array}{lll}
       \ov I_{m1}\ddot{\alpha}_{m1}
    &-\omega (\ov I_{m1}+\ov I_{m2}-\ov I_{m3})\dot{\alpha}_{m2}
            & +\omega^2\xi_1(\ov I_3-\ov I_2)\alpha_{m1}\\
  \ov I_{m2}\ddot{\alpha}_{m2}
   & +\omega (\ov I_{m1}+\ov I_{m2}-\ov I_{m3})\dot{\alpha}_{m1}
            & +\omega^2\xi_2(\ov I_3-\ov I_1)\alpha_{m2}\\
            \ov I_{m3}\ddot{\alpha}_{m3}& &+ \omega^2(\xi_2-\xi_1)(\ov I_2-\ov I_1)\alpha_{m3}
           \end{array}\right)\\
         &+\left(\begin{array}{r}\omega  (\ov I_{c3}-\ov I_{c2})\dot\alpha_{c2}
                   +k_c(\dot\alpha_{m1}-\dot\alpha_{c1})\\
  -\omega  (\ov I_{c3}-\ov I_{c1})\dot\alpha_{c1}
                   +k_c(\dot\alpha_{m2}-\dot\alpha_{c2})\\
                   k_c(\dot\alpha_{m3}-\dot\alpha_{c3})\end{array}\right)\\
              & +\Io\left(\begin{array}{r}
                           \omega^2\xi_1\delta B_{T,m23}-\omega \dot{\delta B}_{T,m13}\\
                            - \omega^2\xi_2\delta B_{T,m13}-\omega \dot{\delta B}_{T,m23}\\
                            +\omega^2(\xi_2-\xi_1)\delta B_{T,m12}-\omega \dot{ \delta B}_{T,m33}
                          \end{array}\right)\\
 & =  \left(\begin{array}{l}
\big(\ov I_3-\ov I_2\big)\dt J_{g23} \\
\big(\ov I_1-\ov I_3\big)\dt J_{g13} \\
\big(\ov I_2-\ov I_1\big)\dt J_{g12}
                        \end{array}\right)
                      - \left(\begin{array}{l}
\ov I_{m1}\dot{{\omega}}_{s,g1} +\omega\big(\ov I_3-\ov I_2\big){\omega}_{s,g2} \\
\ov I_{m2}\dot{{\omega}}_{s,g2} -\omega\big(\ov I_3-\ov I_1\big){\omega}_{s,g1} \\
0   \end{array}\right)\, ,
                              \end{split}\label{lineq}                        
  \end{equation}
}
where we used 
\footnote{\label{c1c2comments} If we  assume
that the body is in
a Cassini state in a $s$-to-2 spin-orbit resonance, $s$ integer,
then the coefficients $c_1$ and
$c_2$ are explicitly given in terms of Hansen coefficients in equations
(\ref{c1c3}) and (\ref{c2}). These expressions imply:
for the Moon $\xi_1=0.9976$, $\xi_2=3.927$; for  Enceladus
$\xi_1=1.00012$, $\xi_2=3.99351$; and for Mercury $\xi_1=1.27513$, $\xi_2=2.14751$;
and all these constants are of order of one.
If there is no spin-orbit  resonance then $c_2=0$ and $\xi_1=\xi_2$, for the Earth $c_1=0.000027$ and
$\xi_1=1.000027$.

For bodies that are out of spin-orbit resonance and are not close to massive bodies,  the average
external gravitational upon them is small and, so   
$c_1\approx 0$ and $\xi_1=\xi_2\approx 1$.

For a body in {\bf 1:1} spin orbit resonance, if the inclination
      of the body spin axis  to the normal to the orbital plane is small and 
      the eccentricity of the orbit is small, then equations (\ref{c1c3}), (\ref{c2}), and
      (\ref{X030}) give $c_1\approx c_2 \approx \frac{3}{2} \frac{m_p}{m_p+m}$, where $m_p$ is the mass
      of the point mass (the tidal raising body) and $m$ is the mass of the extended body. In the case
      of the Moon or Enceladus $c_1\approx c_2\approx 3/2$ and $(\xi_1,\xi_2)\approx (1,4)$.
}
\begin{equation}
  \xi_1= c_1-c_2+1,\qquad
\xi_2=c_1+c_2+1\,.
\label{xi}  \end{equation}
The terms in the left hand side of  equation  (\ref{lineq})
are related to the body free librations,
the term in  the first line represents the rigid part, the one in  the second line
the coupling with the fluid core, and the one in the third line
the first order correction due to the
mantle deformations.
The first term in the right hand side of  equation (\ref{lineq})
represents  the tidal torque due to the orbiting point masses  and
the second  the inertial (or ``fictitious'')  torque that appears when 
$\boldsymbol{\omega}_s\ne 0$.

For the motion of the core, the linearization of equations (\ref{eqrot4}) gives equation
(\ref{lincore2}) that  in coordinates becomes 
\begin{equation}
  \begin{array}{rcl}
    \ov I_{c1}\ddot\alpha_{c1}-\omega\ov I_{c1} \dot\alpha_{m2}
    - \omega\ov I_{c3}(\dot\alpha_{c2}-\dot\alpha_{m2})+
    k_c(\dot\alpha_{c1}-\dot\alpha_{m1})&=&-\ov I_{c1}\dot{\omega}_{s,g1}\\
     \ov I_{c2}\ddot\alpha_{c2}+\omega\ov I_{c2} \dot\alpha_{m1}
    + \omega\ov I_{c3}(\dot\alpha_{c1}-\dot\alpha_{m1})+
    k_c(\dot\alpha_{c2}-\dot\alpha_{m2})&=&-\ov I_{c2}\dot{\omega}_{s,g2}\\
    \ov I_{c3}\ddot\alpha_{c3}+ k_c(\dot\alpha_{c3}-\dot\alpha_{m3})&=&
    0\,.
\end{array}\label{lincore}\end{equation}

If the guiding motion is a   good
approximation  to the  real motion of the body,
then the torque terms that contain $\boldsymbol \omega_{s,g}$,
which are  due  to the non-inertial character of $\K_s$, are mostly cancelled out 
by true torque terms in  $\boldsymbol \delta \mat J_g$, see an example in Appendix \ref{simple}.

\subsection{The  case in which the  fluid core is an oblate spheroid.}
\label{eqs}

In  the following sections we will restrict our attention to the case
where the core is an oblate ellipsoid of revolution with:
 $\ov I_{c12}=\ov I_{c13}=\ov I_{c23}=0$, and 
\begin{equation}
  \ov I_{c1}= \ov I_{c2}=\ov I_{c3}(1-f_c)\,,\quad\text{with}\quad f_c\ge 0
  \label{fc}
\end{equation}

In this case it is convenient to rewrite the equations using the following set
of nondimensional and positive parameters (the size of the parameters was suggested by 
 data in  the literature for the: Earth,  Moon, Mercury,
and Enceladus):
\begin{equation}
  \begin{split}
&  
  f_{\circ}:=\frac{{\rm I}_{\circ c}}{{\rm I}_{\circ m}}\,,
  \quad\text{significance of the fluid core
    (possibly large);}\\
&
\ov \alpha=\frac{\ov I_3-\ov I_2}{\ov I_1}\,,\quad
\ov\gamma=\frac{\ov I_2-\ov I_1}{\ov I_3}\,,\quad f_c\,, \quad  
\quad\text{ellipticity coefficients (small, $\ll 1$);}
\\
&
 \frac{\eta_c}{\om}  \ \bigg(\eta_c=\frac{{\rm I}_{\circ}}{{\rm I}_{\circ c}{\rm I}_{\circ m}}k_c\bigg),
  \quad \text{Ekman number (\ref{Ekman}) (very   small, $\ll\ll 1$).}
\end{split}
\label{parlist}
\end{equation}

The equations have other parameters: all the parameters in the rheology that have either dimension of
$time^{-2}$ (elastic constants) or $time^{-1}$ (viscosity constants), the nondimensional parameters of
average tidal force $c_1$ and $c_2$ (or $\xi_1$ and $\xi_2$), and the amplitudes of the time periodic
tidal force of dimension $time^{-2}$. The sidereal angular frequency $\omega$ is used to nondimensionalize
all these parameters.  Since the beginning we have neglected quantities of order two with respect to
the deformation variables $\mat B$, so we can use  identities that are valid up to
first order in $\|\ov{\mat B}\|$ as, for instance:
\begin{equation}\begin{split}
    \ov \beta&=\ov \alpha+\ov \gamma\\
    \frac{\ov I_3-\ov I_2}{\ov I_{m1}}&=\ov \alpha\,\frac{\ov I_1}{\ov I_{m1}}=
    \ov \alpha \,\frac{\Io}{{\rm I}_{\circ m}}\\
\frac{\Io}{\ov I_{m1}}\delta B_{T,m23}&=\frac{\Io}{{\rm I}_{\circ,m}}\delta B_{T,m23}\\
    \frac{\ov I_{c3}}{\ov I_{c1}}&=\frac{\ov I_{c3}}{\ov I_{c2}}=1+f_c,\ \ov I_{c3}=
{\rm I}_{\circ m}\left(1+\frac{2f_c}{3}\right)\\
\frac{k_c}{\ov I_{c1}}&=\frac{k_c}{\ov I_{c2}}=\frac{k_c}{\ov I_{c2}}=\eta_c\frac{{\rm I}_{\circ m}}{{\rm I}_{\circ}}\,.
\end{split}\label{II}
\end{equation}
In order to simplify some expressions we will  use the ratios
${\rm I}_{\circ c}/\Io$ and  ${\rm I}_{\circ m}/\Io$ that can   be written in terms
of $ f_{\circ}=\frac{{\rm I}_{\circ c}}{{\rm I}_{\circ m}}$ as 
\begin{equation}
\frac{{\rm I}_{\circ}}{{\rm I}_{\circ m}}=1+f_\circ \quad\text{and}\quad
\frac{{\rm I}_{\circ c}}{{\rm I}_{\circ}}=\frac{f_\circ}{(1+f_\circ)}\,.
\label{fo}
\end{equation}

Using the approximations above equation (\ref{lineq})  can be written as

\nd{\bf Equations for the motion of the mantle:}
{\renewcommand{\arraystretch}{2.1}
\begin{equation}
  \begin{split}
  & 
  \left(\begin{array}{l}
          \ddot{\alpha}_{m1}-\omega\dot \alpha_{m2} +
          \omega\left(\frac{\Io}{{\rm I}_{\circ m}}\ov \alpha\right)\dot  \alpha_{m2}
 +   \omega^2\left(\frac{\Io}{{\rm I}_{\circ m}}\ov \alpha\right)\xi_1  \alpha_{m1}\\
\ddot \alpha_{m2}+\omega\dot  \alpha_{m1} -\omega\left(\frac{\Io}{{\rm I}_{\circ m}}
          \ov \beta\right)\dot \alpha_{m1} 
    +\omega^2\left(\frac{\Io}{{\rm I}_{\circ m}}
                                             \ov \beta\right) \xi_2  \alpha_{m2}\\
  \ddot \alpha_{m3} +\omega^2\left(\frac{\Io}{{\rm I}_{\circ m}}
          \ov \gamma\right) (\xi_2-\xi_1) \alpha_{m3}
        \end{array}\right)
    + \omega f_c\frac{{\rm I}_{\circ c}}{{\rm I}_{\circ m}}
      \left(\begin{array}{c}
       \dot\alpha_{c2}- \dot\alpha_{m2} \\
     \dot\alpha_{m1}      
              -\dot\alpha_{c1} \\ 0
            \end{array}\right)        
      \\
 &     
 + \eta_c\frac{{\rm I}_{\circ c}}{{\rm I}_{\circ}}\,    \left(\begin{array}{l}
  \dot\alpha_{m1}    -\dot\alpha_{c1}\\
   \dot\alpha_{m2}  -\dot\alpha_{c2}\\
 \dot\alpha_{m3}  -\dot\alpha_{c3}\end{array}\right)+ \omega\frac{\Io}{{\rm I}_{\circ m}}
 \left(\begin{array}{r}
                           \omega\xi_1\delta B_{T,m23}- \dot{\delta B}_{T,m13}\\
                            - \omega\xi_2\delta B_{T,m13}- \dot{\delta B}_{T,m23}\\
                            +\omega(\xi_2-\xi_1)\delta B_{T,m12}- \dot{ \delta B}_{T,m33}
                          \end{array}\right)                                               
    \\
               &= \frac{\Io}{{\rm I}_{\circ m}} \left(\begin{array}{l}
\ \ \ov\alpha\,\dt J_{g23} \\
-\ov\beta\,\dt J_{g13} \\
\ \ \ov\gamma\,\dt J_{g12}
                        \end{array}\right)
                      - \left(\begin{array}{l}
\dot{{\omega}}_{s,g1}\\
\dot{{\omega}}_{s,g2}\\
0
    \end{array}\right)\,
+\omega\frac{{\rm I}_{\circ}}{{\rm I}_{\circ m}}
\left(\begin{array}{l}                  
  -\ov\alpha\,{\omega}_{s,g2} \\
   \ \ \ov\beta\, {\omega}_{s,g1} \\
  \ \   0      \end{array}\right)\, ,
                              \end{split}\label{eq1}
  \end{equation}
}
and equation (\ref{lincore2})

\nd {\bf Equations for the motion of the core:}
{\renewcommand{\arraystretch}{2.1}
   \begin{equation}
  \begin{array}{llll}
    \ddot\alpha_{c1}- \omega\dot\alpha_{c2}&
  +\omega\,f_c(\dot\alpha_{m2}-\dot\alpha_{c2})&+\eta_c\frac{{\rm I}_{\circ m}}{{\rm I}_{\circ}}\,(\dot \alpha_{c1}
  -\dot \alpha_{m1})&=-\dot{\omega}_{s,g1}
\\
    \ddot\alpha_{c2}+ \omega\dot\alpha_{c1}&-\omega\,f_c(\dot\alpha_{m1}-\dot\alpha_{c1})
 & + \eta_c\frac{{\rm I}_{\circ m}}{{\rm I}_{\circ}}\,(\dot\alpha_{c2} -\dot \alpha_{m2})&=-\dot{\omega}_{s,g2}\\
   \ddot\alpha_{c3}& 
    &+ \eta_c\frac{{\rm I}_{\circ m}}{{\rm I}_{\circ}}\,(\dot \alpha_{c3}-\dot \alpha_{m3})&=
  0\\   
        \end{array}
\label{eq2}
\end{equation}
}

\subsection{The linearization of the equations about the guiding motion:
deformation part.}

\label{linsec2}

In order to close the system  of equations (\ref{lineq}) and (\ref{lincore}) it is necessary
to include equations for the time evolution of the deformation variables. 
The  linearization of the equations for the deformation is  
 almost  restricted to the linearization of the shear operator $\mat F_m$.

The expression for $\mat F_m$ in equation (\ref{F}) implies
\begin{equation}\begin{split}
    \mat F_m&=\ov {\mat F}+\boldsymbol\delta \mat F_m\quad\text{where:}\\
\boldsymbol\delta \mat F_m    &=\boldsymbol\delta \mat C_m+
\boldsymbol\delta \mat S_m\,,
\end{split}\label{dtFm}
  \end{equation}
  $ \mat C_m= -\left(\boldsymbol{ \omega}_{m,m}\otimes\boldsymbol{\omega}_{m,m} -
    \frac{\|\boldsymbol \omega_{m}\|^2}{3}\Id\right)$, and $ \mat S_m=\mat J_m- \frac{\tr\mathbf{J}}{3}\Id$.
  The average shear operator $\ov {\mat F}$ is that given in equation
(\ref{ovF}).
Using the   expression for $\boldsymbol \omega_{m,m}$ in equation (\ref{apr2}) we obtain
    \begin{equation}
      \boldsymbol\delta \mat C_m=\omega   \begin{bmatrix}
  \frac{2}{3}\dot \alpha_{m3} & 0 & -\dot \alpha_{m1}\\
  0 & \frac{2}{3}\dot \alpha_{m3} &  -\dot \alpha_{m2} \\
    -\dot \alpha_{m1} &  -\dot \alpha_{m2} & -\frac{4}{3}\dot \alpha_{m3}
 \end{bmatrix}+
       \omega^2
   \begin{bmatrix}
  0 & 0& \alpha_{m2}\\
  0 &0 & -\alpha_{m1} \\
  \alpha_{m2} & -\alpha_{m1}& 0
\end{bmatrix}
-\omega   \begin{bmatrix}
  0 & 0 & \omega_{s,g1}\\
  0 & 0 & \omega_{s,g2}  \\
     \omega_{s,g1} &  \omega_{s,g2} & 0
 \end{bmatrix}\,.
\end{equation}
Using the expression  for
$\mat J_m$ in equation (\ref{Jprime2}) we obtain that $ \boldsymbol\delta \mat S_m$ is the traceless
part of $
 [ \ov {\mat J}\, ,\boldsymbol { \widehat \alpha}_m] +
 \boldsymbol{\dt}\mat J_g$ and using the expression for $\ov {\mat J}$ in equation (\ref{Jd3})
\begin{equation}
  \boldsymbol\delta \mat S_m=
 \omega^2 \begin{bmatrix}
  0 & -2c_2\, \alpha_{m3}& (c_1+c_2)\alpha_{m2}\\
  -2c_2\, \alpha_{m3} &0 & (-c_1+c_2)\alpha_{m1} \\
   (c_1+c_2)\alpha_{m2} &  (-c_1+c_2)\alpha_{m1} & 0
 \end{bmatrix}+\boldsymbol{\dt}\mat J_g- \frac{\tr
  \boldsymbol{\dt}\mat J_g}{3}\Id\,.
    \end{equation}
    The combination of these two expressions gives
    {\renewcommand{\arraystretch}{1.5}
  \begin{equation}\begin{split}
  \boldsymbol{\delta}{\mat F}_m=
 &\omega   \begin{bmatrix}
  \frac{2}{3}\dot \alpha_{m3} & 0 & -\dot \alpha_{m1}\\
  0 & \frac{2}{3}\dot \alpha_{m3} &  -\dot \alpha_{m2} \\
    -\dot \alpha_{m1} &  -\dot \alpha_{m2} & -\frac{4}{3}\dot \alpha_{m3}
 \end{bmatrix}
 +\omega^2
   \begin{bmatrix}
  0 & -(\xi_2-\xi_1)\alpha_{m3}& \xi_2\alpha_{m2}\\
   -(\xi_2-\xi_1)\alpha_{m3}&0 & -\xi_1\alpha_{m1} \\
  \xi_2\alpha_{m2} & -\xi_1\alpha_{m1}& 0
\end{bmatrix}
\\ & \\
&
-\omega   \begin{bmatrix}
  0 & 0 & \omega_{s,g1}\\
  0 & 0 & \omega_{s,g2}  \\
     \omega_{s,g1} &  \omega_{s,g2} & 0
 \end{bmatrix}
 +\boldsymbol{\dt}\mat J_g- \frac{\tr
  \boldsymbol{\dt}\mat J_g}{3}\Id
\, .
\end{split}
\label{dtF}
  \end{equation}
}

If we  substitute 
  $ \mat B_{T,m}=\ov{\mat B}+\boldsymbol\delta \mat B_{T,m}$
  and $ \mat F_m=\ov {\mat F}+\boldsymbol\delta \mat F_m$ in any one of the equations
  (\ref{def1}), (\ref{def2}), and (\ref{def3}), then the constant terms $\ov{\mat B}$,
  $\ov {\mat F}$, and $\mu_{0}\mathbf{B}_{0,m}$ cancel out due to the equilibrium equation
  (\ref{BKov2}). At the equilibrium all auxiliary variables $\mat \Lambda_m, \mat B_{1,m},\ldots$
  are null, so we may represent their variations by the same letter (if $i\ne j$, then
  $\mat B_{T,mij}=\boldsymbol\delta
  \mat B_{T,mij}$ is also true).
  In the following we list, for  each one of the rheologies
  considered in this paper, the  linearized  equations for the deformations.

For the Kelvin-Voigt rheology (Figure \ref{simple-osc}):
\begin{equation} 
  \eta\dot{\boldsymbol \delta \mathbf{B}}_{T,m}+(\gamma+\mu_0) \boldsymbol \delta\mathbf{B}_{T,m} = 
    \boldsymbol \delta\mathbf{F}_m\,.
  \label{def1d}
\end{equation}
For the generalised Voigt rheology (Figure \ref{genvoigt})
\begin{equation}\label{def2d}
\begin{split}
&(\gamma+\mu_0) \boldsymbol \delta\mathbf{B}_{T,m}+\mat \Lambda_m=\boldsymbol \delta\mathbf{F}_m\\
&\frac{1}{\mu_1}\dot{\mat\Lambda}_m+\frac{1}{\eta_1} \mat\Lambda_m =\dot{\mat B}_{1,m}\\
&\eta_j\dot {\mathbf{B}}_{j,m}+\mu_j\mat B_{j,m}=\mat\Lambda_m,\quad j=2,\ldots,n\\
&\mat B_{1,m}=\boldsymbol \delta\mat B_{T,m}-\big(\mat  B_{2,m}+\mat  B_{3,m}+\ldots+\mat  B_{n,m}\big)\,.
\end{split}
\end{equation}
For the generalised Maxwell rheology (Figure \ref{genmax})
 {\renewcommand{\arraystretch}{1.5}  
 \begin{equation}
  \begin{split}
    & \eta\dot{\boldsymbol \delta\mathbf{B}}_{T,m}+ (\gm+\mu_0)\boldsymbol \delta\mat B_{T,m}
        +\mat\Lambda_{1,m}\cdots+\mat
    \Lambda_{n,m}= \boldsymbol \delta\mathbf{F}_m\\
   &  \dot{\mat\Lambda}_{j,m}=-\frac{1}{\tau_j}\mat\Lambda_{j,m}+
   \mu_{j}\dot{\boldsymbol \delta\mathbf{B}}_{T,m}\,,\qquad\tau_j=\frac{\eta_{j}}{\mu_{j}},\quad j=1,\ldots,n\,.
       \end{split}\label{def3d}
\end{equation}
}
In all these equations    $\boldsymbol \delta\mat F_m$ is given  in equation (\ref{dtF}).

\section{Free librations and Love numbers.}
\label{free2}

The homogeneous part of the linearized equations (or the equations
without the forcing terms $\boldsymbol\delta \mat J$ and $\boldsymbol \omega_{s,g}$)
will be called ``free-libration equations''.
This denomination   follows the literature  on the librations
of the Moon and other satellites in spin-orbit resonance but not that on the librations of the Earth,
where ``free-libration''  
means torque-free libration 
(the  motion  in the absence of any external gravitational torque).
Equations (\ref{eq1}) and (\ref{eq2}) become the equations for the torque-free-librations only
if $c_1=c_2=0$ (or $\xi_1=\xi_2=1$), i.e. the average gravitational coefficients are zero.
For the Earth  $c_2=0$ (there is no spin-orbit resonance) and  $c_1=0.000027$ (due to the average
gravitational field of the Moon and Sun),
so the difference between
the eigenvalues of torque-free librations ($c_1=0$)  and  of free librations is small.

    Our  goal in this  Section is to
   compute and    to present formulas  for  free libration modes.
   The number of eigenvalues can be very large depending
   on the  complexity of the rheology. Here we will be interested only in those eigenvalues that
   are related to the rotational motion. These are the eigenvalues that continue to exist when the rheology
   of the mantle is continuously deformed to that of a rigid mantle. These eigenvalues can be computed
   perturbatively from those of the rigid motion.

The most general problem  that we will consider is that of a 
body with  a deformable mantle and a rigid   oblate  fluid core.
The rheology may be either  the  generalised Voigt or
generalised Maxwell, the Kelvin-Voigt being a particular case of the generalised Maxwell.
 The following argumentation is based
on the generalised Maxwell rheology but it could be done equally well using the generalised Voigt rheology
with the same result.

In order to find the equation for the eigenmodes we do the substitution
\begin{equation}\label{subs}
  \boldsymbol\alpha_m\to \boldsymbol\alpha_m \erm^{\lambda \, t}\,,\
    \boldsymbol\alpha_c\to  \boldsymbol\alpha_c\, \erm^{\lambda \, t}\,,\
\boldsymbol\delta \mat B_{T,m}\to \boldsymbol\delta \mat B\, \erm^{\lambda \, t}\,,\
 \mat \Lambda_{mj}x=\mat \Lambda_j\,\erm^{\lambda \, t}\,,
\end{equation}  
where $ \boldsymbol\alpha_m, \boldsymbol\alpha_c, \boldsymbol\delta \mat B$, and  $\mat \Lambda_j$
in the right hand side of the substitutions are understood as
constant complex vectors or matrices. This notation will be used
only in this section. Equation (\ref{def3d}) for the  deformation variables gives the following relation
   {\renewcommand{\arraystretch}{1.1}  
 \begin{equation}
  \begin{split}
    & \lambda\eta\boldsymbol\delta\mathbf{B}+ (\gm+\mu_0)\boldsymbol \delta\mat B
    +\mat\Lambda_1\cdots+\mat \Lambda_n= \boldsymbol \delta \mat F_{\boldsymbol \alpha} \qquad\text{where}\\
    & \\
    & \boldsymbol \delta \mat F_{\boldsymbol \alpha}:=
 \omega \lambda  \frac{2}{3}  \alpha_{m3} \begin{bmatrix}
1 & 0 & 0\\
  0 & 1 &  0 \\
   0& 0 & -2
 \end{bmatrix}\\ &
 + 
   \begin{bmatrix}
  0 &\  & -\omega^2(\xi_2-\xi_1)\alpha_{m3}&\ &-\lambda \omega \alpha_{m1}+ \omega^2\xi_2\alpha_{m2}\\
   -\omega^2(\xi_2-\xi_1)\alpha_{m3}&\ &0 &\ &-\lambda\omega \alpha_{m2} -\omega^2\xi_1\alpha_{m1} \\
 -\lambda\omega\alpha_{m1} + \omega^2 \xi_2\alpha_{m2} &\ &  -\lambda\omega\alpha_{m2}-\omega^2\xi_1\alpha_{m1}
   &\ & 0
\end{bmatrix}
 \\ & \\
   &  \lambda\frac{1}{\mu_j}\mat\Lambda_j+\frac{1}{\eta_j}\mat\Lambda_j=
   \lambda\boldsymbol\delta \mat B\,\Rightarrow
\mat \Lambda_j=\left(\frac{1}{\mu_j}+\frac{1}{\lambda \eta_j}\right)^{-1}\delta \mat B\,,\\
    \end{split}\label{linch3.1}
  \end{equation}
  for $ j=1,\ldots,n\,$.
}
These equations imply
\begin{equation}
 \underbrace{\bigg\{ \gamma+\mu_0+  \eta \lambda+
  \sum_{j=1}^n\Bigl(\frac{1}{\mu_j}
  +\frac{1}{\lambda\eta_j}\Bigr)^{-1}\Bigg\}}_{\gamma+J^{-1}(-i\lambda)}\boldsymbol\delta\mat B=
\boldsymbol \delta \mat F_{\boldsymbol \alpha}
   \label{love1}
  \end{equation}
  where $J^{-1}(\sigma)$ is the  complex rigidity (equation (\ref{J-1genmax}))
  of the generalised Maxwell rheology.

  Equations  (\ref{Lovegenmax}) and (\ref{love1})   lead to  a definition of the
  nondimensional compliance of the whole body $C(\lambda)$,
  which includes the effects of  both   rheology and
  self-gravity, that is equal to the one  given in \cite{mathews2002modeling} (paragraph [21])
\begin{equation}
  C(\lambda):= \frac{\omega^2}{\gm+J^{-1}(-i\lambda)}=
  \left(\frac{3\Io G}{\omega^2 R^5}\right)^{-1}\, k(-i\lambda)\,.
   \label{Ck}
\end{equation}
In this way  equation (\ref{love1}) becomes
\begin{equation}
\boldsymbol\delta\mat B= C(\lambda)\frac{1}{\omega^2} 
\boldsymbol \delta \mat F_{\boldsymbol \alpha}\,.
   \label{love2}
  \end{equation}
Equation (\ref{love2}) can be used to eliminate $\boldsymbol\delta\mat B$ from  the homogeneous part
of equation (\ref{eq1}),
  after the substitution (\ref{subs}).
  As a result we obtain a linear system that has only the variables
  $\boldsymbol \alpha_m$ and
  $\boldsymbol \alpha_c$ as unknowns. Moreover, the system  can be split into two uncoupled systems:
  one for the polar motion
  of the form
  $\mat A_P(\alpha_{m1},\alpha_{m2},\alpha_{c1},\alpha_{c2})^{\rm T}=0$ and another for the libration in longitude
  of the form   $\mat A_L(\alpha_{m3},\alpha_{c3})^{\rm T}=0$. The matrices $\mat A_P(\lambda)$ and
  $\mat A_L(\lambda)$ have entries  that are rational functions in $\lambda$.
  The eigenvalues associated
  with polar motion  (libration in longitude) are the roots of the characteristic equation
  ${\rm Det}(\mat A_P)=0$ (${\rm Det}(\mat A_L)=0$).
  After a reduction of the  terms in ${\rm Det}(\mat A_P)$ (${\rm Det}(\mat A_L)$)
  over a common denominator
  the characteristic equation can be written in polynomial form as  $P_P(\lambda)=0$ ($P_L(\lambda)=0$).

  The characteristic polynomials may have high degree depending on the complexity of the rheology.
  If the goal is to find all
  eigenvalues of the problem, then the best approach is to substitute numbers and to do the computations
  numerically. In the following we  show that for the rotational eigenvalues
  it is possible to obtain approximate mathematical formulas that
   are valid for any rheology.
  The following result  is crucial:
   \begin{proposition}[ \citep{crr2018}(Proposition 3.1)]
  \label{bound}
  For any rheology associated with a generalised Voigt model, equation {\rm (\ref{Lovegenvoigt})},
   or to a generalised Maxwell model,
    equation {\rm (\ref{Lovegenmax})}, the following inequality
    holds:
    \[
|k(\sigma)|\le|k(0)|= \frac{3\Io G}{R^5}\frac{1}{\gm+\mu_0}, \quad\text{for all}\quad \sigma\in\R.
\]
\end{proposition}
This shows that for $\lambda$ in the imaginary axis,
$\lambda=i\, \sigma$, the maximum of the modulus of the complex compliance $|C(\lambda)|$
is $C(0)=\frac{\omega^2}{\gamma+\mu_0}$.

The quantity $\frac{\omega^2}{\gamma+\mu_0}$ is equal to the flattening coefficient
$\ov\alpha=\ov\beta= \alpha_{id}$
of a body: with steady uniform rotation about the $\mat e_3$-axis, that is free from gravitational
interaction  ($\xi_1=\xi_2=1$), and has no prestress $\mat B_{0,m}=0$;
see  equations (\ref{ovF}), (\ref{BKov2}) and the
expression for  $\ov \alpha$
in   Table \ref{frames2}. We denote this ``ideal flattening coefficient'' as (if $\mu_0=0$, then
this would be the hydrostatic flattening)
\begin{equation}
  \alpha_{id}:=\frac{\omega^2}{\gamma+\mu_0}\,\quad \text{and}\quad |C(i\sigma)|\le   \alpha_{id}
   \quad\text{for all}\quad \sigma\in\R
  \label{alphaid}
  \end{equation}
Since  $\alpha_{id}$  is of the order of magnitude of the  ellipticity parameters
$\ov\alpha,\ov\beta,\ov\gamma$ and $f_c$, we conclude that $|C(\lambda)|$ is a small quantity provided
that $\lambda$ is restricted to a small neighbourhood of the imaginary axis
($|C(\lambda)|$ becomes  large if $\lambda$ is far from the imaginary axis).

In order to keep track of all the small quantities in the equations, namely
$\ov\alpha,\ov\beta,\ov\gamma, f_c, \eta_c,$ and $C(\lambda)$, we will multiply them  by a scaling
variable $\epsilon>0$.
The forthcoming analysis is restricted  to a strip $|{\rm Re}(\lambda)|<\epsilon$. With the  introduction
of $\epsilon$ and for fixed values of $\ov\alpha,\ov\beta,\ov\gamma, f_c$, and $\eta_c$
the characteristic polynomial can be considered as  a function of $\lambda$ and $\epsilon$.
Note that the value of $\lambda$ in $C(\lambda)$ cannot be fixed a priori.

\subsection{Libration in Longitude}
\label{liblongsec}

At first we will analyse the eigenvalues associated with the libration in longitude determined
by  $P_L(\lambda,\epsilon)=0$. This equation has an eigenvalue $\lambda=0$ that is trivial and
can be factored out\footnote{\label{trivialroot} Our libration equations depend only on the
     time derivatives of the angles  $\alpha_{c1}$,  $\alpha_{c2}$, or $\alpha_{c3}$
     and not on the angles themselves. This gives rise to degenerated eigenmodes where
     all variables are zero but the angles  $\alpha_{c1},\alpha_{c2},\alpha_{c3}$.
     These degenerated eigenmodes could be easily
     removed if we had considered the angular velocities of the core as variables of
     the problem instead of the angles themselves. We decided to keep the angles
     because they are the variables which are more easily visualised.}.
   After factorisation the equation becomes
   \begin{equation}\begin{split}
   &  \frac{\lambda}{\omega}  \epsilon  \left(\frac{\Io}{{\rm I}_{\circ m}}  \left(-3 \ov \gamma
           (\xi_1-\xi_2)+C(\lambda) \left(4 \left(\frac{\lambda}{\omega}\right)^2-3 (\xi_1-\xi_2)^2\right)\right)+3
         \frac{\eta_c}{\omega}
         \frac{\lambda}{\omega} \right)\\
    &   +\frac{\eta_c}{\omega }  \epsilon ^2 \left(-3 \ov \gamma
      (\xi_1-\xi_2)+C(\lambda) \left(4 \left(\frac{\lambda}{\omega}\right)^2
        -3 (\xi_1-\xi_2)^2\right)\right)+
     3 \left(\frac{\lambda}{\omega}\right)^3=0
     \end{split}
     \label{charalong}
   \end{equation}
   For $\epsilon=0$ we obtain  $\lambda^3=0$. We  use a Newton polygon to  determine  the leading terms
   in the series expansions of $\lambda$ in fractional exponents of  $\epsilon$
   (Puiseux series). At
     first we obtain a pair of roots, for which  $\lambda\sim \epsilon^{1/2}$, that
        are determined by
     the equation
     \begin{equation}\label{C0}\left(\frac{\lambda}{\omega}\right)^2+\frac{\Io}{{\rm I}_{m0}}(\xi_2-\xi_1) \big(\ov \gamma -C(0)(\xi_2-\xi_1)
       \big)=0.
     \end{equation}
     This equation determines up to leading order in the small parameters the eigenvalue of libration
     in longitude:
     \begin{equation}
         \lambda_{\ell o}=i\,\omega \sqrt{ \frac{\Io}{{\rm I}_{m0}}(\xi_2-\xi_1) \,\,
       \Big(\ov \gamma -C(0)\, (\xi_2-\xi_1)\Big)}:=i\, \sigma_{\ell o}
       \label{lambda3}
       \end{equation}
       where\footnote{\label{van}
         In the same way $C(0)$ is related to an ideal flattening $\alpha_{id}$ due to centrifugal
       forces, described before equation
       (\ref{alphaid}), $C(0)\, (\xi_2-\xi_1)$ is related to an ideal ellipticity coefficient
       $\gamma_{id}=\frac{ \ov I_{id2}-\ov I_{id1}}{\ov I_{id3}}$ due to tidal deformations.
       For simplicity suppose that the extended body is in 1-to-1
       spin orbit resonance with an orbiting point mass with  circular orbit. In this case
       $c_1=c_2=3/2$, which implies $\xi_2-\xi_1=3$ (see Footnote \ref{c1c2comments}).
       As in the definition of $\alpha_{id}$ we also assume that the body has  no prestress $\mat B_{0,m}=0$.
       In this ideal situation
       the equilibrium equations (\ref{ovF}) and  (\ref{BKov2}) and  the
expression for  $\ov \gamma$
in   Table \ref{frames2} imply that  $\gamma_{id}=C(0)\, (\xi_2-\xi_1)=3C(0)$. In the absence of elastic
rigidity ($\mu_0=0$)  $\gamma_{id}$ would correspond to  the hydrostatic equilibrium.

It is of note that $\ov \gamma=\gamma_{id}$ implies $\sigma_{\ell o}=0$. This fact is well explained in
\cite{VANHOOLST2013299} Sections 1 and 2. In a simplified way their explanation
is the following.
``If the tidal response for static tides were to be as for the short-periodic tides, the sum of all tides
would be aligned with the satellite-planet axis.$\ldots$ Therefore, there would be no gravitational
torque on the satellite, unless a frozen-in asymmetry unrelated to
tides would be present.'' The ``frozen-in asymmetry'' is what we called prestress. Since we are using
the approximation $C(0)\approx C(i\sigma_{\ell o})$ in equation (\ref{lambda3}), we are indeed
assuming that the tidal response to  static tides is the same as that at frequency $\sigma_{\ell o}$.
Equation (\ref{lambda3}) is equivalent to
equation (18) in \cite{VANHOOLST2013299}.}
       \[
         C(0)=\frac{\omega^2}{\gamma+\mu_0}=  \left(\frac{3\Io G}{\omega^2 R^5}\right)^{-1}\, k(0)=
          \alpha_{id}\,.
       \]

Using the parameters in Tables \ref{table2}
and \ref{tab.INPOP} we find from equation (\ref{lambda3})
that for the Moon $\sigma_{\ell o}=2.889$ years, which is close to the value $2.892$ years
estimated from observations in    \cite{rambaux2011moon} (and almost the same value $2.887$ years
we would obtain
if we considered the mantle as rigid).

     We started the computation using the approximation $\lambda=0$ and for this
     reason we used  $k(-i\lambda)\approx k(0)$ (or $C(\lambda)\approx C(0)$)
     in equation (\ref{C0}). 
           Now, assume the   frequency of free longitudinal librations $ \sigma_{\ell o}$ is known.
           Then we can replace  $C(0)$ in equation (\ref{C0})
           by the correct value $C(i  \sigma_{\ell o})$
           and solve the equation for $C(i  \sigma_{\ell o})$.
       So,  if the frequency of free longitudinal libration  $ \sigma_{\ell o}$ is known,
       then we obtain 
       \begin{equation}
           {\rm Re}\big(k(\sigma_{\ell o})\big)\approx 
         \left(\frac{3\Io G}{\omega^2 R^5}\right)\bigg\{\frac{\ov \gamma}{\xi_2-\xi_1}-
         \frac{\Io}{{\rm I}_{m0}}\frac{1}{(\xi_2-\xi_1)^2}\left(\frac{\sigma_{\ell o}}{\omega}\right)^2\bigg\}
         \,.
         \label{klambda3}
       \end{equation}

       In order to obtain the second  term in the  Puiseux expansion of  $\lambda_{\ell o}$
     in powers of $\epsilon^{1/2}$ 
 it is necessary to use the Taylor expansion of $C(\lambda)$ about $\lambda=0$, i.e. 
 \begin{equation}
   C(\lambda)=C(0) (1-\lambda  \tau)+\Oc(\lambda^2)\,,\quad\text{where}\quad \tau=-\frac{1}{C(0)}
   \frac{\partial C}{\partial \lambda}(0)\label{tau}
 \end{equation}
 Note that $\tau$ is a characteristic time and it is not necessarily a small quantity.
 For the Kelvin-Voigt rheology
 \begin{equation}
   \tau=\frac{\eta}{\gamma+\mu_0}\,,\label{tauKV}
 \end{equation}
 for the Generalised Maxwell rheology
  \begin{equation}
   \tau=\frac{\eta +\sum_{j=1}^n\eta_j}{\gamma+\mu_0}\,,\label{taugM}
 \end{equation}
 and for the generalised Voigt rheology
 \begin{equation}
   \tau=\frac{\eta_1}{\gamma+\mu_0}\,.\label{taugV}
 \end{equation}
       The second  term in the  Puiseux expansion of $\lambda_{\ell o}$ is of order $\epsilon$ and gives the
       decay rate of the libration in longitude:
   \begin{equation}
         \lambda_{\ell o}=i\sigma_{\ell o}-\frac{\eta_c}{2}\,
         \frac{{\rm I}_{\circ c}}{\Io} -
         \tau \omega^2 C(0)\frac{\Io}{{\rm I}_{\circ m}}\frac{(\xi_2-\xi_1)^2}{2}
         :=i\sigma_{\ell o}-\nu_{\ell o}\,.
                      \label{lambda32aux}
       \end{equation}

       The characteristic equation (\ref{charalong}) for the longitudinal motion
       has an additional  pure real eigenvalue that  up to leading
       order in small quantities is given by $-\eta_c\frac{{\rm I}_{\circ m}}{\Io}$. This mode corresponds
       to a steady decay of the difference $|\dot\alpha_{m3}-\dot\alpha_{c3}|$ due to
       the core-mantle friction.

       \subsection{Wobble}
       \label{wobblesec}

       We now consider the polar motion. The characteristic equation $P_P(\lambda,\epsilon)$
       can be easily computed but it is  long and it will not be shown. There is
       a trivial double root $\lambda=0$ that can be factored out (see Footnote \ref{trivialroot}).
       For $\epsilon=0$, the characteristic equation becomes 
       $ \lambda ^2 \left(\lambda ^2+\omega^2\right)^2 =0$.

       The root $\lambda=0$ will give rise to the
       ``wobble'',  which is a free precession (it happens in the absence of external forces)
         of the angular velocity vector about the axis of largest moment of inertia,
         as viewed in the body frame
          (the Chandler's wobble in the case of the Earth). As in the libration of longitude case,
       a Newton polygon analysis shows that the equation for the leading order expansion of the Puiseux
       series  is:
 \begin{equation}
   \frac{\lambda^2}{\omega^2}+\xi_1\xi_2\left(\frac{\Io}{{\rm I}_{\circ m}}\right)^2\Big(\ov \alpha-\xi_1C(0)\Big)
     \Big(\ov \beta-\xi_2C(0)\Big)=0
     \label{charw}
     \end{equation}
The solution to this equation gives the eigenfrequency of free wobble
  \begin{equation}
         \lambda_w=i\,\omega\,\frac{\Io}{{\rm I}_{\circ m}}\sqrt{\xi_1\xi_2\Big(\ov \alpha-\xi_1C(0)\Big)
     \Big(\ov \beta-\xi_2C(0)\Big)}:=i\,\sigma_w
       \label{lambdaw}
     \end{equation}
     where $C(0)=\frac{\omega^2}{\gamma+\mu_0}$.
   
        In the particular case of a  body
        of revolution $\ov \alpha=\ov \beta=\ov\alpha_e$
        and with negligible  gravitational torque  $\xi_1=\xi_2=1$
       (or $c_1=c_2=0$), which is the case of the Earth,  then equation (\ref{lambdaw})
       becomes
       \begin{equation}
        \lambda_w=i\,\omega\,\frac{\Io}{{\rm I}_{\circ m}}\Big(\ov \alpha_e-C(0)\Big)\,.
          \label{lambdawrev}
        \end{equation}
       If in this last equation we replace the approximation $C(0)$ by the value of the complex compliance
       at the real wobble frequency, then we obtain
       $ \lambda_w=i\,\omega\,\frac{\Io}{{\rm I}_{\circ m}}\Big(\ov \alpha_e-C(i\sigma_w)\Big)$ that is
        the formula  
 for $\sigma_1$ in \cite{mathews2002modeling}
equation (37)\footnote{Our complex compliance $C(i\sigma_w)$ is
  equivalent to the complex compliance
  $\tilde\kappa$ in equation (37) of \cite{mathews2002modeling}.  Our generalised Maxwell rheology aims to describe
  the rheology of the mantle in a generalised sense including oceans, atmosphere,
  and other effects,   as far as these effects  can be considered in
  a   spherically average sense.
}.

If $\sigma_w$ is known, then we can replace $C(0)$ in equation (\ref{lambdawrev})
by $C(\lambda_w)$ and to  solve the equation for this quantity.
In this way we obtain the following  estimate for  the
Love number at the Chandler's wobble frequency for a body
of revolution
        and with negligible  gravitational torque  $\xi_1=\xi_2=1$
    \begin{equation}
         {\rm Re}\big( k(\sigma_w)\big)\approx  \frac{3\Io G}{\omega^2 R^5}\left(\ov \alpha
           -\frac{{\rm I}_{m0}}{\Io}\frac{\sigma_w}{\omega}\right)\,.
       \label{ksigmaw}
 \end{equation}

 Using the parameter $\tau$, given in equation (\ref{tau}),  and  the second  term
 in the  Puiseux expansion of $\lambda_{w}$
 we compute
 the decay rate of the wobble, 
   \begin{equation}\begin{split}
     \lambda_w=&i\sigma_w-\frac{\eta_c}{2}\frac{{\rm I}_{\circ c}}{{\rm I}_{\circ m}}\Big(
   \ov \alpha  \xi_1+\ov\beta\xi_2- C(0)(\xi_1^2+\xi_2^2)\Big)\\ & -
   \tau\omega^2 C(0)\frac{\Io^2}{{\rm I}^2_{\circ m}}\frac{\xi_1\xi_2}{2}  (\ov \beta  \xi_1+
   \ov\alpha\xi_2-2 C(0)\xi_1\xi_2):=i\sigma_w-\nu_w
\end{split}
                      \label{lambdaw2aux}
                    \end{equation}
Note:  The conditions $\ov \alpha>\xi_1C(0)$ and $\ov \beta>\xi_2C(0)$ that imply the oscillatory nature
of the solution also imply that both terms in the real part of $\lambda_w$ are negative (this condition
is related to that discussed in Footnote \ref{van}).

\subsection{Libration in Latitude  and Nearly
  Diurnal Free Wobble (NDFW)} 
    \label{llndfwsec}

  We will now study the roots of the  equation  $P_P(\lambda,\epsilon)=0$
  near $(\lambda,\epsilon)=(i\,\omega, 0)$ (the results near the  root
        $(\lambda,\epsilon)=(-i\,\omega,0)$ are obtained by complex conjugation).
        After a  change of variables $\lambda=i\,x\,\omega +i\, \omega$, the equation for the dominant terms
        of the Puiseux expansion of $x (\epsilon)$, which is obtained  using a Newton's polygon, is
        \begin{equation}
          x^2-x(f_0+1) (y+z )+(f_0+1)z\,y=0\label{charp}
        \end{equation}
        where
        \begin{equation}\begin{split}
            z&=\frac{1}{2} \Big(\ov \alpha  (\xi_1-1)+\ov \beta  (\xi_2-1)-
            C(i\omega)\left((\xi_1-1)^2+(\xi_2-1)^2\right)
          \Big)\\
          &=c_1 \ov \alpha_e +\frac{c_2}{2}\ov \gamma-C(i\omega)\left(c_1^2+c_2^2\right)\,,\\
          y&=f_c+i \frac{\eta_c}{\omega(1+f_0)}=f_c+i \frac{\eta_c}{\omega}\frac{{\rm I}_{\circ m}}{\Io}
        \end{split}
        \label{zpole}
\end{equation}
and where we used $\xi_1-1=c_1-c_2$, $\xi_2-1=c_1+c_2$, $\ov\beta-\ov\alpha\approx\ov\gamma$,
and $\ov \alpha_e=(\ov\alpha+\ov\beta)/2$.

If we assume the conditions
$\ov \alpha>\xi_1C(0)$ and $\ov \beta>\xi_2C(0)$, which imply the stability of the Chandler wobble motion,
and $\ov \gamma \ge  2 C(0) c_2$,
which in the case
$\ov\gamma>0$ and $c_2>0$ implies the stability of free librations in longitude, then a computation using that
$|C(i\sigma)|\le C(0)$ and the assumption $c_1>0$ shows that ${\rm Re}\,z> 0$. So:
\begin{equation}
  {\rm Re}\,z:=z_r>0\,,\   {\rm Im}\,z:=z_i\ge 0\,,\  {\rm Re}\,y:=y_r > 0\,,\ 
  {\rm Im}\,y:=y_i\ge 0\,.
  \label{inyz}
  \end{equation}

If $f_0=0$, then equation (\ref{charp}) has  two roots:
$y=f_c+i \frac{\eta_c}{\omega}$ and $z$.  The  first root  depends on the ellipticity of the core
  $f_c$ and it is related to the   ``nearly diurnal free wobble'' (NDFW),   a term  used in
  the literature about free librations of the Earth. The second root depends only on the properties of the
  mantle  and it is related  to the ``free librations in latitude'' (FLL), a term  used in
  the literature on   free librations of the Moon
  \cite{rambaux2011moon}.

  Equation (\ref{charp}) depends on the parameters $(y,z)$ and $f_0$.
  If  the discriminant $\Delta$ of equation (\ref{charp}) is different from zero, then the equation
  has two solutions. One of them   will be denoted as $x_{nd}$ and will be related to the
  NDFW eigenvalue  $\lambda_{dw}=i\omega(1+x_{dw})$. The other
    will be denoted as  $x_{\ell a}$ and will be related to the 
   FLL eigenvalue $\lambda_{\ell a}=i\omega(1+x_{\ell a})$.
In order to classify a given solution one must deform $f_0>0$ from
its current value to  $f_0=0$ keeping $(y,x)$ constant. If the function $f_0\to\Delta$  remains different
from zero 
during the deformation, then the root will move in the complex plane as a smooth function of $f_0$ and it
will eventually become either $y$, and the root will be classified as $x_{nd}$, or $z$, and
the root will be classified as $x_{\ell a}$. An explicit algorithm for the classification of the two
roots is given in Appendix \ref{class}.  In this Appendix not only the limit of an evanescent core
is studied, $f_0\to 0$, but also that of an evanescent mantle, $f_0\to\infty$.

  If we decompose $x_{dw}=x_{dwr}+ix_{dwi}$ and   $x_{\ell a}=x_{\ell ar}+ix_{\ell ai}$
  into real and imaginary parts, then we can write
  the eigenvalues associated with the NDFW and FLL, respectively, as 
  \begin{equation}
    \begin{split}
      \lambda_{dw}&=i\omega(1+x_{dw})=i(1+x_{dwr})- x_{dwi}:=i\sigma_{dw}-\nu_{dw}\\
      \lambda_{\ell a}&=i\omega(1+x_{\ell a})=i(1+x_{\ell ar})- x_{\ell ai}:=i\sigma_{\ell a}-\nu_{\ell a}\,.
      \end{split}
  \label{ldwella}
  \end{equation}
  A computation using the algebraic manipulator  Mathematica shows that
  if the inequalities (\ref{inyz}) hold, then:
  \[
  x_{dwr}>0\,,\quad x_{\ell a r}>0\,,\quad x_{dwi}\ge 0\,,\quad \text{and}\quad x_{\ell a i}\ge 0\,.
  \]

   The eigenfrequencies  $\sigma_{dw}= \om(1+x_{dwr})$ and  $\sigma_{\ell a}= \om(1+x_{\ell ar})$
   are close to $\omega$, since $|x_{dwr}|$ and $|x_{\ell ar}|$ are much smaller than one.
   For this reason the approximations $C(i \sigma_{dw})\approx C(i \sigma_{\ell a})\approx  C(i \omega)$
   are  good. The positive quantities $x_{dwr}$ and $x_{\ell a r}$
        are angular frequencies of cycles per sidereal  day ($2\pi/\omega$)
     associated with the NDFW and the FLL, respectively,
     in the slow frame $\K_s$\footnote{
  \label{librationinertial}
  The motion of  the pole  of the mantle ($\mat e_3\in\K_m$)  in the guiding frame is given by
   \[
     (\Id+ \boldsymbol{\widehat \alpha}_m)\mat e_3=
     \mat e_3+ \boldsymbol{\alpha}_m\times\mat e_3\in\K_g
   \]
  If $\boldsymbol{\alpha}_m$ oscillates as an eigenmode with  eigenvalue
  $\lambda= i\,\omega(1+x)$ and eigenvector  $\boldsymbol \alpha_m=\epsilon (1,i,0)$ (this is the case of the
  NDFW and the FLL modes) then   
   \[\boldsymbol \alpha_m=\epsilon{\rm Re}\left(\exp[t\lambda]
           \begin{bmatrix}1
       \\
    i\\
0
\end{bmatrix}\right)= \epsilon\left(\begin{matrix}\ \ \cos\big(t \omega(1+ x)\big)
       \\
-  \sin\big(t\omega(1+ x)\big)\\
0
\end{matrix}\right)=\epsilon\mat R_3^{-1}\big( t\omega \left(1+x\right)\big)\mat e_1\,.
     \]
    Since the transition
     from the guiding frame to the slow frame is given by $\mat R_3(\omega t):\K_g\to\K_s$ we obtain that
     the image of the pole of the mantle in $\K_s$ is
     \[
       \mat R_3(\omega t)\Big(    \mat e_3+ \boldsymbol{\alpha}_m\times\mat e_3\Big)=
        \mat e_3+\epsilon\Big( \mat R_3^{-1}\big( t\omega x\big)\mat e_1\Big)\times\mat e_3\,.
      \]
      So,  the period associated with $\lambda$  in the slow frame is $2 \pi/(\omega x)$
      and the motion is retrograde if $x>0$ and prograde if $x<0$.
    }.
    If the obliquity of the
    spin axis with respect to the normal to the invariable plane is zero, i.e. $\theta_g=0$, then
    the same statement of the previous period is valid after replacing the slow frame by the inertial frame 
\footnote{ \label{inframe} Suppose  the motion of the slow frame is a precession
  about the normal to  the invariable plane (see equations (\ref{Rs}) and (\ref{zeta})) with 
  $\mathbf{R}_s=\mathbf{R_3}(\psi_g)\mathbf{R_1}(\theta_g)\mathbf{R_3}(\zeta):\K_s\to\kappa$,
  where: $\psi_g=\dot \psi_g\,  t $, $\zeta=-\dot \psi_g\cos\theta_g \,t$,
  and $\dot \psi_g$ and $\theta_g$ are constants. 
  From the last equation in Footnote \ref{librationinertial} we obtain that  the motion 
        of the pole in the inertial frame is given by the sum of the usual precessing
      vector $\mat R_3(\dot \psi_g t )\mat R_1(\theta_g)\mat e_3$ plus the small ``physical libration''
      \[
        \epsilon
        \mat R_3(\dot \psi_g\, t)\mat R_1(\theta_g)\Big\{\Big(\mat R_3^{-1}
        \big( t(-\dot\psi_g \cos \theta_g+\omega x)\big)\mat e_1\Big)\times\mat e_3\Big\}\,. \]
      
      If $\theta_g= 0$, then the libration of the spin axis in the inertial frame is equal to that
      in the slow frame. If $\theta_g\ne 0$, then the three components of the physical libration
      of the spin axis are
      \[\epsilon\left[\begin{array}{ll}
    + \cos ^2(\theta_g/2) \sin \Big(t \big(2 \dot\psi \sin^2(\theta_g/2)-x \omega \big)\Big)&
    -\sin ^2(\theta_g/2) \sin \Big(t \big(2 \dot\psi \cos^2(\theta_g/2)+x \omega \big)\Big)
                \\
                - \cos ^2(\theta_g/2) \cos \Big(t \big(2 \dot\psi \sin^2(\theta_g/2)-x \omega \big)\Big)&
                + \sin ^2(\theta_g/2) \cos \Big(t \big(2 \dot\psi \cos^2(\theta_g/2)+x \omega \big)\Big)              \\
                & -\sin\theta_g \cos \big(t (\dot\psi_g\cos \theta_g+x \omega )\big)
                                \end{array}\right]\,.\]                                
                          }. The two free modes are retrograde precession modes when viewed in the
                          slow  frame.

\begin{figure}[hptb!]
\centering
\begin{minipage}{0.5\textwidth}
\centering
\includegraphics[width=0.95\textwidth]{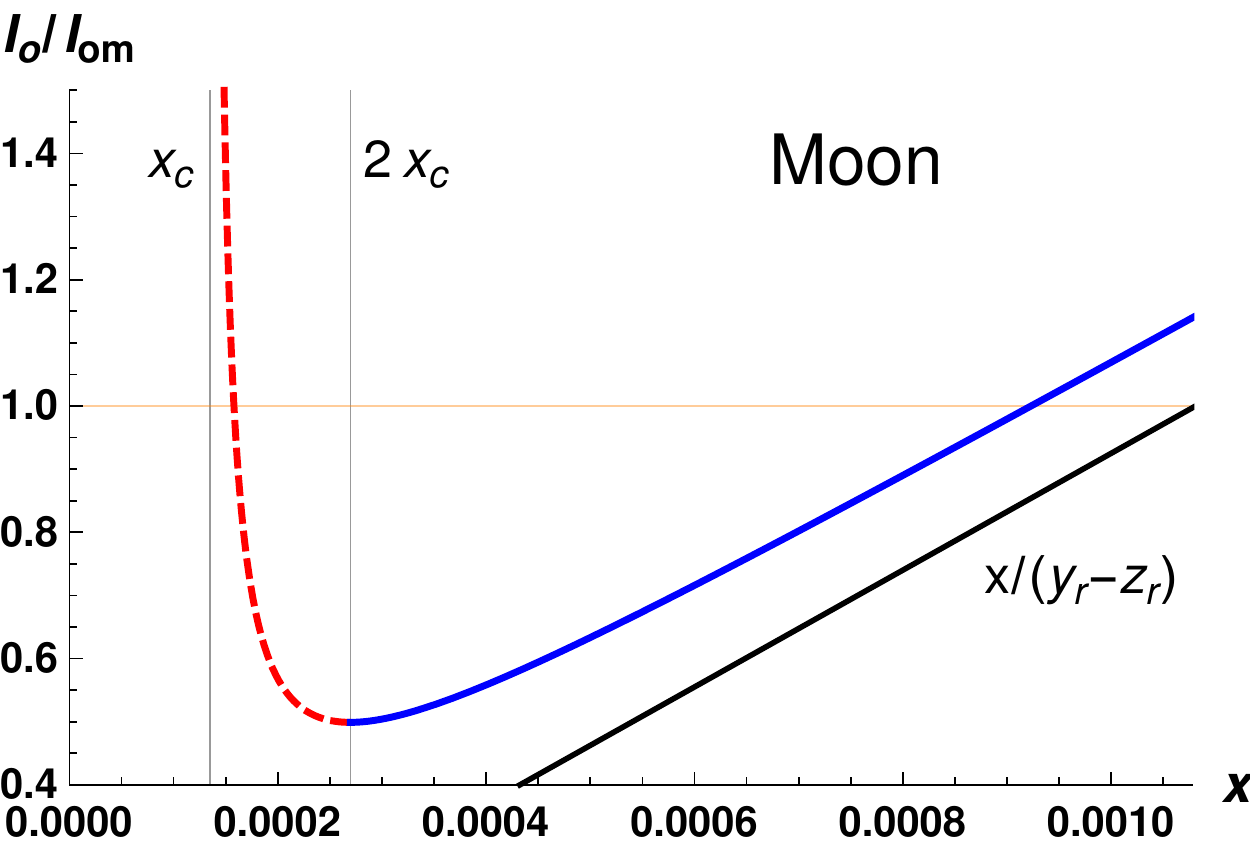}
\includegraphics[width=0.95\textwidth]{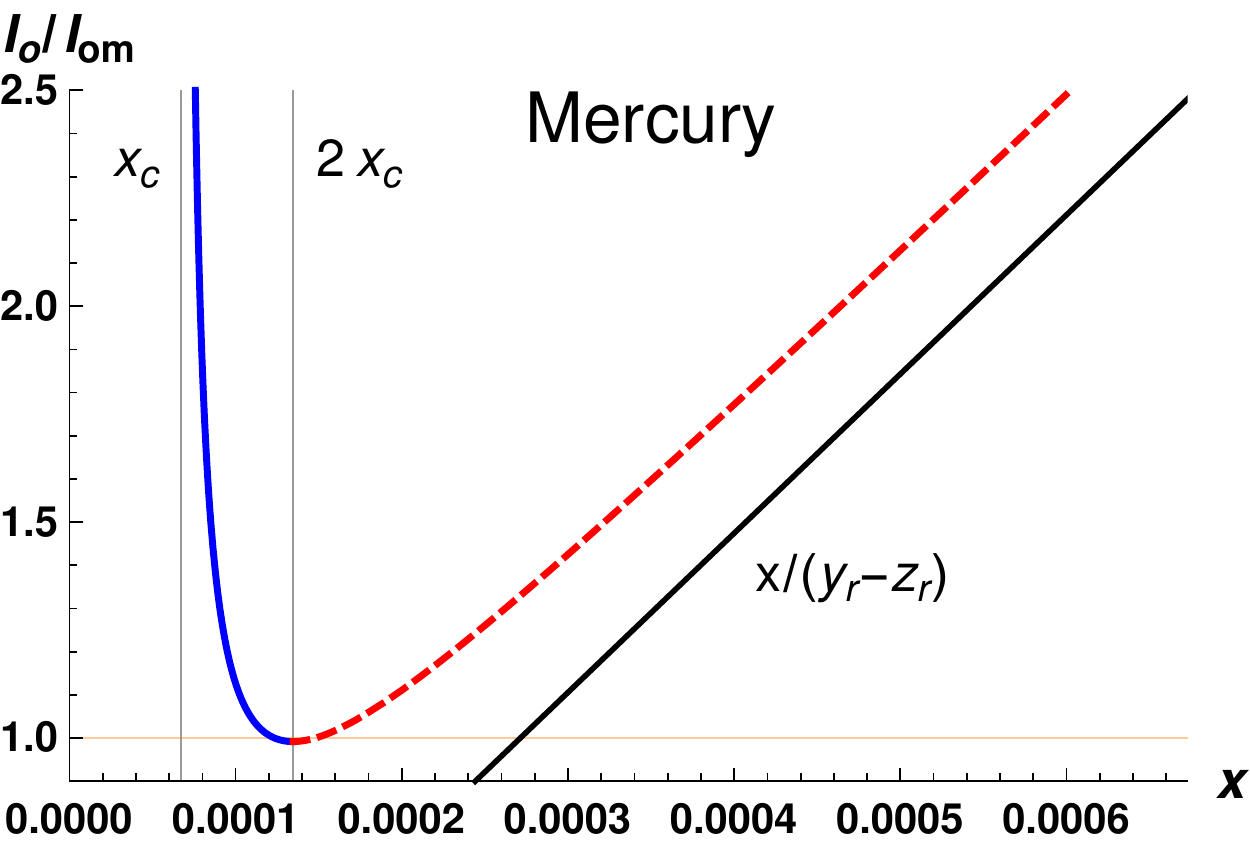}
\end{minipage}\hfill
\begin{minipage}{0.5\textwidth}
\centering
\includegraphics[width=0.95\textwidth]{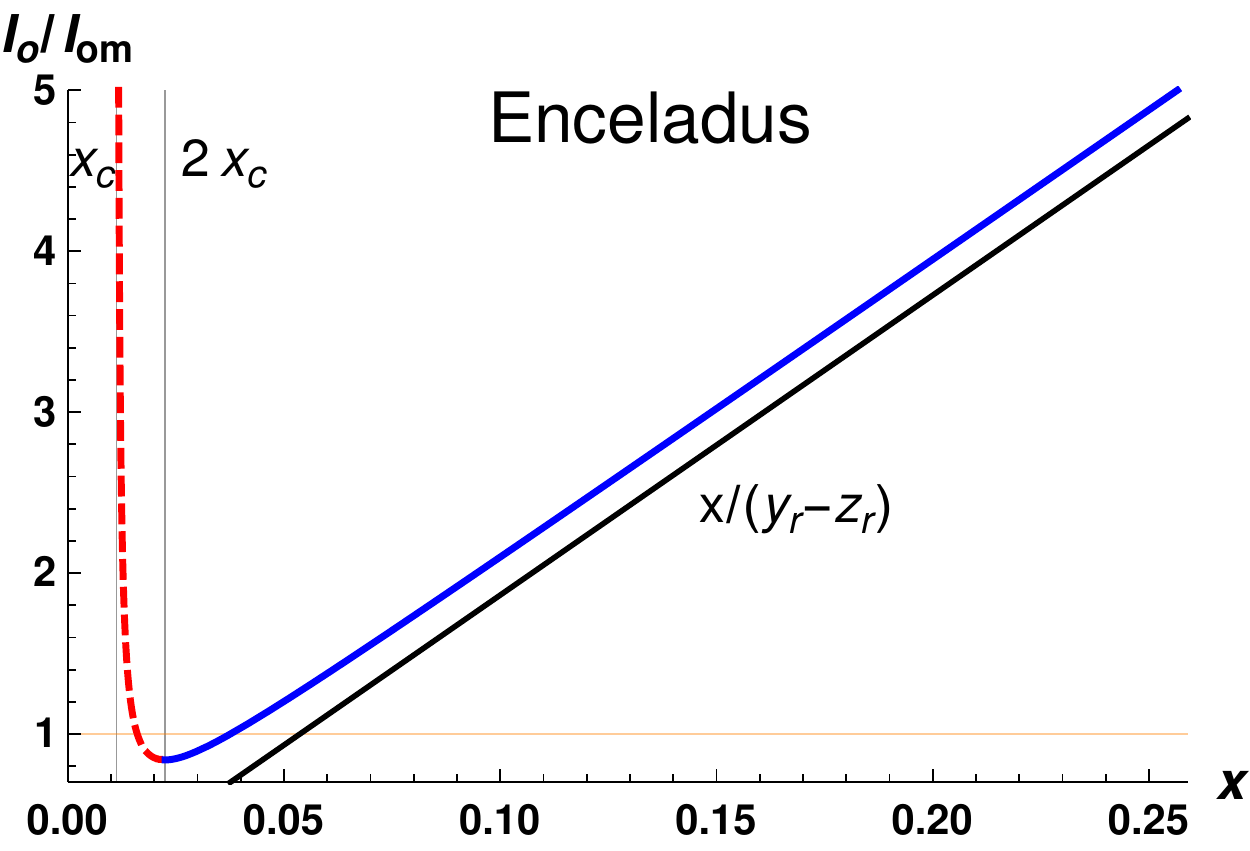}
\includegraphics[width=0.95\textwidth]{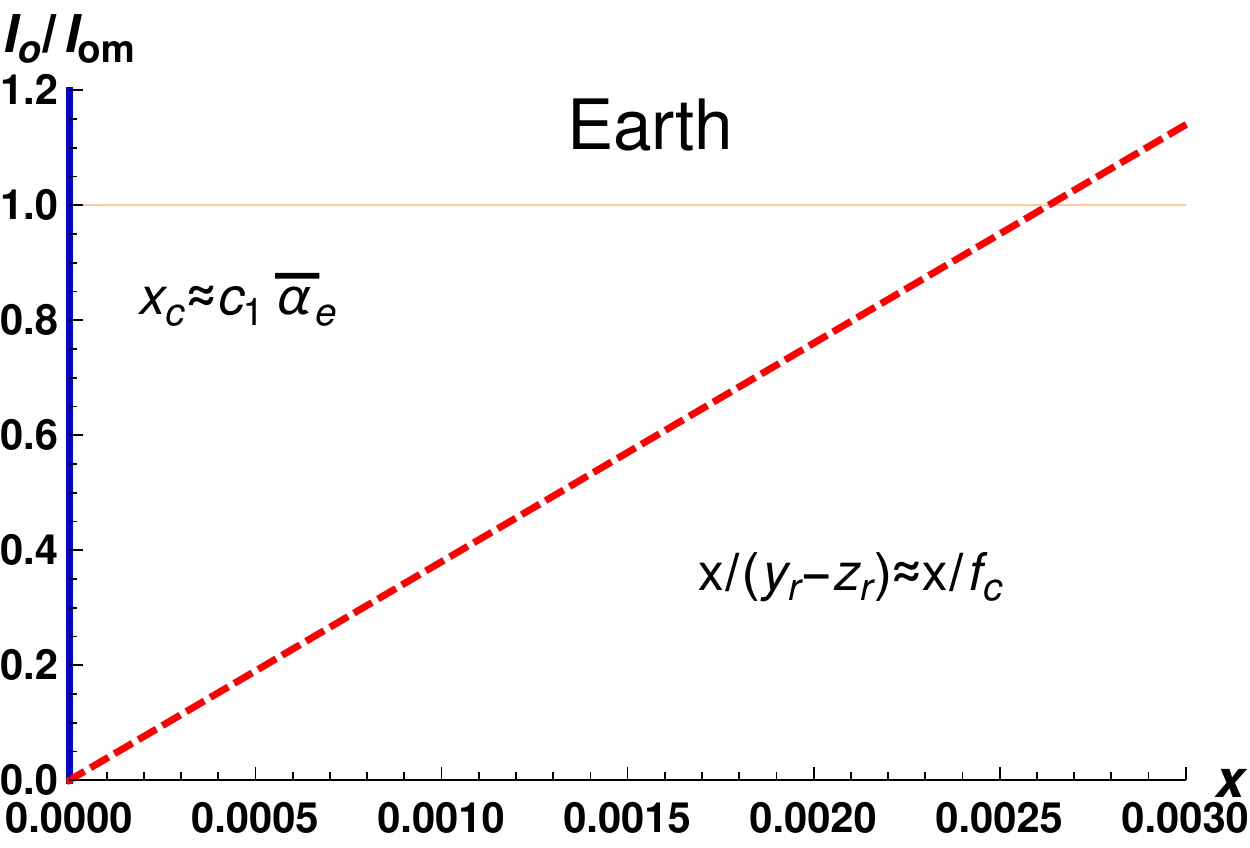}
\end{minipage}\hfill
\caption{Diagram:Vertical axis $\frac{\Io}{{\rm I}_{\circ m}}$ and
  horizontal  axis $x_{\ell a r}$ (solid-blue line) and  $x_{dw r}$ (dashed-red line).
  $x_{dw r}$ (NDFW) and  $x_{\ell a r}$ (FLL)
  are angular frequencies of cycles per sidereal  day ($2\pi/\omega$)
  in the slow frame $\K_s$ (see Footnote \ref{inframe}
  for the relation to angular frequencies in the inertial frame).
  The values of $c_1$ and $c_2$ are from  Footnote \ref{c1c2comments}.Other  data for the Earth, Mercury, and
  Moon  were taken from Tables \ref{table1} and \ref{table2} and references cited in their captions.
  The $k_2=0.014$ of Enceladus is from \cite{Marie2017}, the values $\ov \alpha_e=0.0162$ and
  $\ov \gamma = 0.0183$ were computed using the gravitational coefficients in \cite{iess2014gravity},
  and other data are from  \cite{THOMAS201637}. The approximated
  values of $1+f_0=\frac{\Io}{{\rm I}_{\circ m}}$ are:
  $1+7\times 10^{-4}$ (Moon), $4.2$ (Enceladus), $2.35$ (Mercury), $1.13$ (Earth).}
\label{examples}
\end{figure}

As an illustration, 
in Figure \ref{examples} we present the eigenfrequencies as  function of the ratio
$1+f_0=\frac{\Io}{{\rm I}_{\circ m}}$ for four different set of parameters $(y,z)$ with
$y_i=z_i=0$. The  four pairs ($z_r$,$y_r$) were computed
    with the parameters of:
    Moon and  Enceladus (both in {\bf 2:2} spin-orbit resonance), Mercury ({\bf 3:2}
    spin-orbit resonance),
    and the Earth (no spin orbit resonance). Note that  for the Earth and Mercury (Moon and Enceladus)
    the value of $x_{dwr}$ increases (decreases) as the size of the core 
    increases. In the opposite way,
    for the Earth and Mercury (Moon and Enceladus)  $x_{\ell a r}$ decreases (increases)
    as  the core increases. These results are in agreement with the statements in  (\ref{conj}).

  We will present explicit formulas in two cases. 
  At first suppose that there is no spin-orbit resonance $c_2=0$, the extended body is axisymmetric
  $\ov \gamma=0$, and the gravitational coefficient $c_1\ll 1$ (for the Earth $c_1=0.000027$).
  In this case $p_r\approx (f_c +c_1 \ov \alpha_e)/2$, $q_r\approx (f_c -c_1 \ov \alpha_e)/2$, and
  the eigenfrequencies are  approximately 
  \begin{equation}
    \sigma_{dw}= \omega \left(1+\frac{\Io}{{\rm I}_{\circ m}}f_c\right)\quad\text{and}\quad 
    \sigma_{\ell a}=  \omega \left(1+c_1 \ov \alpha_e\right)\,.\label{earthlib}
    \end{equation}
    The formula we obtained for $\sigma_{dw}$ agrees with that for $\sigma_2$ in equation
    (37) in \cite{mathews2002modeling}  \footnote{\label{NDFW}
      Equation (37, $\sigma_2$) in \cite{mathews2002modeling} has
      an additional  term $\tilde \beta$ that is
        due to the deformability  of the mantle
        and the core-mantle   magnetic coupling. These two effects are known to be
        important in the dynamics of the NDFW. Since they were not taken into account
        in our modelling,  our formula, with the accepted  value of $f_c$ for the Earth,
        does not give the observed value of  $\sigma_{dw}$.}. 
      In the case of the Earth  the  motion associated with $\sigma_{\ell a}$
      ($2\pi/( \omega c_1 \ov \alpha_e)\approx 30834$ years)
    can be understood as  a ``precession on precession'': the mantle spin 
    precesses in the inertial frame, this precessing vector defines the $\mat e_3$-axis of the slow frame,
    and   libration in latitude is a secondary precession of the mantle spin with respect
    to the $\mat e_3$-axis of the  slow frame.

    The second case we will analyse is that of a body in {\bf 1:1}
    spin orbit resonance and we will take the Moon
    as an example. In this case both $c_1$ and $c_2$ are of the order of  one (see footnote
    \ref{c1c2comments}).
   The moment of inertia of the core of the Moon is small,
   $f_0={\rm I}_{\circ c}/{\rm I}_{\circ m}\approx 7\times 10^{-4}$, such that, up to small corrections,
   $x_{dw}=y_r$,
   $x_{\ell a}= z_r$ and 
    \begin{equation}
      \sigma_{dw}=i\, \omega (1+f_c)\quad \sigma_{\ell a}= i\, \omega\Big(1
       +c_1 \ov \alpha_e +\frac{c_2}{2}\ov \gamma-C_r\left(c_1^2+c_2^2\right)\Big)\,
       \label{lambdamoon}
       \end{equation}
(for the Moon the term proportional to $C_r$ in $ \sigma_{\ell a}$ can also be neglected).
The expression for $\sigma_{dw}$ in equation (\ref{lambdamoon})
is the same as the one due to Petrova and Gusev as reported in
\cite{williams2008lunar} (p.111) and \cite{rambaux2011moon} (equation (16)).

Using the parameters in Tables \ref{table2}
and \ref{tab.INPOP} we find from equation (\ref{lambdamoon})
that in the inertial frame the NDFW has a period equal to
469 years   (this period is considerably
    larger than that reported in \cite{williams2008lunar}, which is $197$ years,
    but is almost  within the range of the most recent estimate $367\pm 100$ years
    \cite{Viswanathan19}). In the same way,  we obtain from equation (\ref{lambdamoon})
that in the inertial frame the FLL mode
    has a period of $80.84$ years (the value obtained from observations in \cite{rambaux2011moon}
    is 80.86 years).

    \subsection{Eigenmodes and Summary.}
    \label{eigsum}

In this paragraph we  present  the rotational eigenmodes
in  time domain and we summarise the results obtained in this Section. 

For each one of  the vectors $\boldsymbol \alpha_m$, $\boldsymbol \alpha_c$, and $\boldsymbol \delta \mat B$,
     only the components that are different from zero at leading order in the small quantities will be shown.
 The components of  $\boldsymbol \delta \mat B$ are determined by equation (\ref{love2}) and 
     only the components $(1,2),(1,3)$, and $(2,3)$ will appear in the expressions of the eigenmatrix.
   We will  assume  that the variations of the moment of inertia due to tides and time-variable
centrifugal forces are small enough (hypothesis  (\ref{principalaxes})),
so that we can use the vector $\boldsymbol \beta$ (equation (\ref{ad3})) with components 
   \[
     \beta_1(t) =\frac{ B_{T,m23}(t)}{\ov \alpha}\,
  , \
  \beta_2(t)=-\frac{ B_{T,m13}(t)}{\ov \beta}
  \, , \
  \beta_3(t)=
  \frac{ B_{T,m12}(t)}{\ov \gamma}\,
\]
to represent  the eigenmatrix $\boldsymbol \delta \mat B$.  
We recall that $\boldsymbol \beta$ gives the angular displacement
of the principal axes frame from the prestress frame, $(\Id+\boldsymbol{\widehat \beta}):\K_p\to\K_m$.

The $(\boldsymbol \alpha_m,\boldsymbol \alpha_c)$ components of the  eigenmodes above
represent oscillations with respect to the
guiding frame. The analysis of  these oscillations using  other frames
follows the  ideas in Footnote
\ref{librationinertial}.

The motion of the frame of the  principal axes of inertia with respect to the guiding frame is given by
$(\Id+\boldsymbol{\widehat \alpha}_p):\K_p\to\K_g$, where
$\boldsymbol \alpha_p=\boldsymbol \alpha_m+
\boldsymbol \beta$.

All the eigenmodes below are composed by two oscillating vectors. Only one of these vectors
will be given, because the second one can be obtained by exchanging
$\big(\cos(\sigma t),\sin(\sigma t)\big)\to\big(\sin(\sigma t),-\cos(\sigma t)\big)$,
which corresponds to the time translation $t\to t-\frac{\pi}{2\sigma}$.

The complex compliance $C$ is related to the complex Love number as (equation (\ref{Ck})):
\[
  C(i\sigma)=
  \frac{\omega^2 R^5}{3\Io G}\, k(\sigma)\,.
  \]
The ``characteristic time''  $\tau$ is defined as (equation (\ref{tau})): 
 \[
   C(\lambda)=C(0) (1-\lambda  \tau)+\Oc(\lambda^2)\,,\quad\text{where}\quad \tau=-\frac{1}{C(0)}
   \frac{\partial C}{\partial \lambda}(0)
 \]
 Note that the definition of $\tau$ does not depend on the particular model used for the rheology of
 the mantle. In terms of the parameters of the rheology, $\tau$ is given in equation:
 (\ref{tauKV}) Kelvin-Voigt,  (\ref{taugM}) generalized Maxwell,  (\ref{tauKV}) generalized Voigt, and
 (\ref{tauAndrade}) Andrade. 

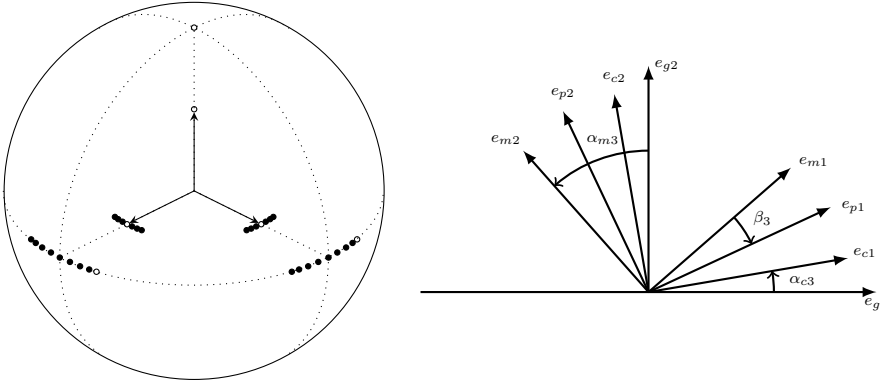
\begin{figure}[hptb!]
\centering
\begin{minipage}{0.5\textwidth}
\centering
\begin{tikzpicture}[tdplot_main_coords, scale = 2.5]
 
\coordinate (P) at ({1/sqrt(3)},{1/sqrt(3)},{1/sqrt(3)});

  \draw (0,0) circle (1cm);
 
\tdplotsetrotatedcoords{0}{0}{0};
\draw[dotted,
    tdplot_rotated_coords,
    black
] (0.71,-0.71,0) arc (-45:135:1);

 \tdplotsetrotatedcoords{0}{90}{90};
\draw[dotted,
    tdplot_rotated_coords,
    black
] (0.707,-0.707,0) arc (-45:135:1);

 \tdplotsetrotatedcoords{90}{90}{90};
\draw[dotted,
    tdplot_rotated_coords,
    black
] (0.707,0.707,0) arc (45:225:1);
 
\draw[dotted] (0,0,0) -- (1.0,0,0);
\draw[dotted] (0,0,0) -- (0,1.0,0);
\draw[dotted] (0,0,0) -- (0,0,1.0);

\draw[-stealth] (0,0,0) -- (0.48,0,0);
\draw[-stealth] (0,0,0) -- (0,0.48,0);
\draw[-stealth] (0,0,0) -- (0,0,0.48);

\draw[] (0.5,0,0) circle[radius=0.4pt];
\draw[] (0,0.5,0) circle[radius=0.4pt];
\draw[] (0,0,0.5) circle[radius=0.4pt];

\draw[] (0.97,0.243,0) circle[radius=0.4pt];
\draw[fill=black] (0.98,0.199,0) circle[radius=0.4pt];
\draw[fill=black] (0.99,0.141,0) circle[radius=0.4pt];
\draw[fill=black] (0.995,0.0707,0) circle[radius=0.4pt];
\draw[fill=black] (1,0,0) circle[radius=0.4pt];
\draw[fill=black] (0.98,-0.199,0) circle[radius=0.4pt];
\draw[fill=black] (0.99,-0.141,0) circle[radius=0.4pt];
\draw[fill=black] (0.995,-0.0707,0) circle[radius=0.4pt];
\draw[fill=black] (0.97,-0.243,0) circle[radius=0.4pt];

\draw[] (-0.243,0.97,0) circle[radius=0.4pt];
\draw[fill=black] (-0.199,0.98,0) circle[radius=0.4pt];
\draw[fill=black] (-0.141,0.99,0) circle[radius=0.4pt];
\draw[fill=black] (-0.0707,0.995,0) circle[radius=0.4pt];
\draw[fill=black] (0,1,0) circle[radius=0.4pt];
\draw[fill=black] (0.199,0.98,0) circle[radius=0.4pt];
\draw[fill=black] (0.141,0.99,0) circle[radius=0.4pt];
\draw[fill=black] (0.0707,0.995,0) circle[radius=0.4pt];
\draw[fill=black] (0.243,0.97,0) circle[radius=0.4pt];
\draw[] (0,0,1.0) circle[radius=0.4pt];

\draw[fill=black] (0.4975,0.03535,0) circle[radius=0.4pt];
\draw[fill=black] (0.495,0.0705,0) circle[radius=0.4pt];
\draw[fill=black] (0.49,0.0995,0) circle[radius=0.4pt];
\draw[fill=black] (0.4975,-0.03535,0) circle[radius=0.4pt];
\draw[fill=black] (0.495,-0.0705,0) circle[radius=0.4pt];
\draw[fill=black] (0.49,-0.0995,0) circle[radius=0.4pt];

\draw[fill=black] (-0.0995,0.49,0) circle[radius=0.4pt];
\draw[fill=black] (-0.0705,0.495,0) circle[radius=0.4pt];
\draw[fill=black] (-0.03535,0.4975,0) circle[radius=0.4pt];
\draw[fill=black] (0.0995,0.49,0) circle[radius=0.4pt];
\draw[fill=black] (0.0705,0.495,0) circle[radius=0.4pt];
\draw[fill=black] (0.03535,0.4975,0) circle[radius=0.4pt];
 
\end{tikzpicture}
\end{minipage}\hfill
\begin{minipage}{0.5\textwidth}
\centering
\begin{tikzpicture}[scale=0.75, transform shape]
\tikzstyle{spring}=[thick, decorate, decoration={zigzag, pre length=0.5cm, post length=0.5cm, segment length=6}]
\tikzstyle{damper}=[thick,decoration={markings,
  mark connection node=dmp,
  mark=at position 0.5 with
  {
    \node (dmp) [thick, inner sep=0pt, transform shape, rotate=-90, minimum width=15pt, minimum height=3pt, draw=none] {};
    \draw [thick] ($(dmp.north east)+(5pt,0)$) -- (dmp.south east) -- (dmp.south west) -- ($(dmp.north west)+(5pt,0)$);
    \draw [thick] ($(dmp.north)+(0,-5pt)$) -- ($(dmp.north)+(0,5pt)$);
  }
}, decorate]
\tikzstyle{ground}=[fill, pattern=north east lines, draw=none, minimum width=0.75cm, minimum height=0.3cm]

            \draw [-latex, thick] (0,0) -- (0,4) node[right] {$e_{g2}$};
            \draw [-latex, thick] (-4,0) -- (4,0) node[below] {$e_{g1}$};

            \draw [-latex, thick] (0,0) -- (2.5,2.2);

            \draw [-latex, thick] (0,0) -- (3.2,1.5);

            \draw [-latex, thick] (0,0) -- (3.5,0.6);
            \node at (2.9,2.3) {$e_{m1}$};
            \node at (3.6,1.5) {$e_{p1}$};
            \node at (3.8,0.7) {$e_{c1}$};

            \draw [-latex, thick] (0,0) -- (-2.2,2.5);

            \draw [-latex, thick] (0,0) -- (-1.5,3.2);

            \draw [-latex, thick] (0,0) -- (-0.6,3.5);

            \node at (-2.5,2.7) {$e_{m2}$};
            \node at (-1.5,3.5) {$e_{p2}$};
            \node at (-0.6,3.8) {$e_{c2}$};

            \node at (-0.8,2.7) {$\alpha_{m3}$};
            \node at (2.,1.3) {$\beta_{3}$};
            \node at (2.7,0.2) {$\alpha_{c3}$};

  \coordinate (o) at (0,0);
  \coordinate (x) at (1,0);
  \coordinate (y) at (0,1);
  \coordinate (z1) at (2.5,2.2);
  \coordinate (z2) at (3.5,0.6);
  \coordinate (z3) at (3.2,1.5);
  \coordinate (z4) at (-2.2,2.5);
  \draw pic[draw, ->, angle eccentricity=1.2,angle radius = 2.2cm,thick] {angle = x--o--z2}; 
  \draw pic[draw, <-, angle eccentricity=1.2,angle radius = 2cm,thick] {angle = z3--o--z1};
  \draw pic[draw, ->, angle eccentricity=1.2,angle radius = 2.5cm,thick] {angle = y--o--z4};

      \end{tikzpicture}
\end{minipage}\hfill
\caption{Left:3-D representation of the oscillations of free librations in longitude.
  Positions at constant time intervals of the frame of the mantle $\K_m$ (of the core $\K_c$)
  are depicted by black dots on the external sphere (on the internal sphere) that
  is fixed in  the guiding-frame
  $\K_g$.  The white dots indicate the initial
  condition. Right: Relative positions of the several vectors on the equatorial plane of
  $\K_g$. Note: Deformations decrease the amplitude of librations of $\K_p$, the frame of principal
  axes of the mantle,  in comparison to the librations of $\K_m$. The angle $\beta_3$
  between these two  frames tends to zero  in the limit as  the mantle becomes rigid.
}
\label{figlong}
\end{figure}

{\bf Free libration in longitude.}
\begin{equation}\begin{split}
         \lambda_{\ell o}=&i\,\omega \sqrt{ \frac{\Io}{{\rm I}_{m0}}(\xi_2-\xi_1) \,\,
       \Big(\ov \gamma -C(0)\, (\xi_2-\xi_1)\Big)} \\ &-\frac{\eta_c}{2}\,
         \frac{{\rm I}_{\circ c}}{\Io} -
         \tau \omega^2 C(0)\frac{\Io}{{\rm I}_{\circ m}}\frac{(\xi_2-\xi_1)^2}{2}\\
         :=&i\sigma_{\ell o}-\nu_{\ell o}\,.
         \end{split}
                      \label{lambda32}
       \end{equation}     
       Only the coordinates $\alpha_{m3}$, $\alpha_{c3}$, and $\beta_3$ of the eigenmodes are nonnull.
       The eigenvector  is:
       {\renewcommand{\arraystretch}{1.5}
         \begin{equation}\begin{split}&
             \begin{bmatrix} \alpha_{m3} \\ \alpha_{c3}\\ \beta_3\end{bmatrix}=
        \erm^{-t \nu_{\ell o }}\begin{bmatrix} &\cos(\sigma_{\ell o}t)\\
            \frac{\eta_c}{\sigma_{\ell o}}\frac{{\rm I}_{\circ m}}{\Io}    &\sin(\sigma_{\ell o}t)\\
-\frac{C(0)(\xi_2-\xi_1)}{\ov \gamma}&\cos(\sigma_{\ell o}t)
\end{bmatrix}\\
\end{split}\label{ellomode}
         \end{equation}
       }
       The free libration in longitude eigenmode is represented in Figure \ref{figlong}.

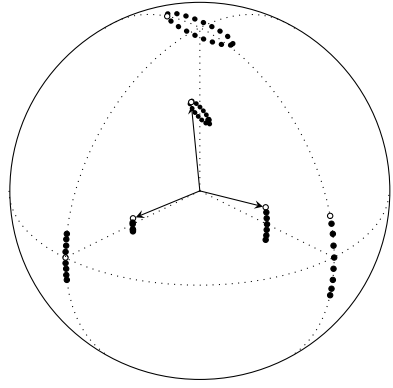
\begin{figure}
\begin{center}
\begin{tikzpicture}[tdplot_main_coords, scale = 2.5]
 
\coordinate (P) at ({1/sqrt(3)},{1/sqrt(3)},{1/sqrt(3)});

  \draw (0,0) circle (1cm);

 \draw[black, dot diameter=2pt, dot spacing=3.4pt, dots, rotate around={-24:(0,0,0.99)}] (0,0,0.99) ellipse (0.19cm and 0.04cm);

 \draw[black, dot diameter=1.8pt, dot spacing=1.8pt, dots,  rotate around={-50:(0,0,0.48)}] (0,0,0.48) ellipse (0.08cm and 0.02cm);
 
\tdplotsetrotatedcoords{0}{0}{0};
\draw[dotted,
    tdplot_rotated_coords,
    black
] (0.71,-0.71,0) arc (-45:135:1);

 \tdplotsetrotatedcoords{0}{90}{90};
\draw[dotted,
    tdplot_rotated_coords,
    black
] (0.707,-0.707,0) arc (-45:135:1);

 \tdplotsetrotatedcoords{90}{90}{90};
\draw[dotted,
    tdplot_rotated_coords,
    black
] (0.707,0.707,0) arc (45:225:1);
 
\draw[dotted] (0,0,0) -- (1.0,0,0);
\draw[dotted] (0,0,0) -- (0,1.0,0);
\draw[dotted] (0,0,0) -- (0,0,1.0);

\draw[-stealth] (0,0,0) -- (0.4776,0,0.0328755);
\draw[-stealth] (0,0,0) -- (0,0.4704,0.09552);
\draw[-stealth] (0,0,0) -- (-0.0328755,-0.09552,0.469152);

\draw[] (0.4975,0,0.03535) circle[radius=0.4pt];
\draw[] (0,0.49,0.0995) circle[radius=0.4pt];
\draw[fill=white] (-0.03535,-0.0995,0.4887) circle[radius=0.4pt];

\draw[] (1,0,0) circle[radius=0.4pt];

\draw[fill=black] (0.99,0,0.141) circle[radius=0.4pt];
\draw[fill=black] (0.994,0,0.1094) circle[radius=0.4pt];
\draw[fill=black] (0.995,0,0.0707) circle[radius=0.4pt];
\draw[fill=black] (0.9994,0,0.03464) circle[radius=0.4pt];

\draw[fill=black] (0.99,0,-0.141) circle[radius=0.4pt];
\draw[fill=black] (0.994,0,-0.1094) circle[radius=0.4pt];
\draw[fill=black] (0.995,0,-0.0707) circle[radius=0.4pt];
\draw[fill=black] (0.9994,0,-0.03464) circle[radius=0.4pt];

\draw[] (0,0.97,0.243) circle[radius=0.4pt];

\draw[fill=black] (0,0.98,0.199) circle[radius=0.4pt];
\draw[fill=black] (0,0.99,0.141) circle[radius=0.4pt];
\draw[fill=black] (0,0.995,0.0707) circle[radius=0.4pt];
\draw[fill=black] (0,1,0) circle[radius=0.4pt];
\draw[fill=black] (0,0.98,-0.199) circle[radius=0.4pt];
\draw[fill=black] (0,0.99,-0.141) circle[radius=0.4pt];
\draw[fill=black] (0,0.995,-0.0707) circle[radius=0.4pt];

\draw[fill=black] (0,0.97,-0.243) circle[radius=0.4pt];

\draw[fill=white] (0,-0.243,0.97) circle[radius=0.4pt];

\draw[fill=black] (0.4975,0,-0.03535) circle[radius=0.4pt];
\draw[fill=black] (0.498,0,-0.04468) circle[radius=0.4pt];
\draw[fill=black] (0.499,0,-0.0316) circle[radius=0.4pt];
\draw[fill=black] (0.5,0,0) circle[radius=0.4pt];
\draw[fill=black] (0.4999,0,0.01) circle[radius=0.4pt];

\draw[fill=black] (0,0.49,-0.0995) circle[radius=0.4pt];

\draw[fill=black] (0,0.495,0.0705) circle[radius=0.4pt];
\draw[fill=black] (0,0.4975,0.03535) circle[radius=0.4pt];
\draw[fill=black] (0,0.5,0) circle[radius=0.4pt];

\draw[fill=black] (0,0.495,-0.0705) circle[radius=0.4pt];
\draw[fill=black] (0,0.4975,-0.03535) circle[radius=0.4pt];

\draw[fill=black] (0.03535,0.0995,0.4887) circle[radius=0.4pt];
 
\end{tikzpicture}

\end{center}
\caption{Representation of the oscillations of the wobble. The Figure shows the motion of $\K_m$ and
  $\K_c$ with respect to $\K_g$ as explained in the caption of Figure \ref{figlong}. The $\mat e_3$-axis
  of $\K_m$ traces an elliptical path. The minor axis of the ellipsis is along the $\mat e_1$-axis of $\K_g$
  (the principal axis of minimum moment of inertia $\ov I_1$), which 
  in the case of the Moon is towards the Earth. Using the formulas given in this paper with
  $\xi_1$ and $\xi_2$ given in Footnote \ref{c1c2comments}, $C(0)=\omega^2/(\mu_0+\gamma)$ with $\mu_0$
  and $\gamma$ given in 
  Table \ref{table2}, $\ov \alpha$ and $\ov \beta$ obtained from the data in Table \ref{tab.INPOP} (Moon)
  and from  \cite{mazarico2014gravity} (Mercury, $\ov \alpha=0.000099$, $\ov \beta=0.000192$), we obtain  
  (minor axis)/(major axis)=$\sqrt{\frac{\xi_1(\ov \alpha-C(0)\xi_1)}{\xi_2(\ov \beta-C(0)\xi_2)}}$=
  0.402 (Moon) and 0.553 (Mercury). The value for the Moon  reported in \cite{williams2008lunar} is 0.406.
  }
\label{figwobble}
\end{figure}

{\bf Wobble.}
 \begin{equation}\begin{split}
     \lambda_w=&i\,\omega\,\frac{\Io}{{\rm I}_{\circ m}}\sqrt{\xi_1\xi_2\Big(\ov \alpha-\xi_1C(0)\Big)
     \Big(\ov \beta-\xi_2C(0)\Big)}\\ & 
   \\ &-\frac{\eta_c}{2}\frac{{\rm I}_{\circ c}}{{\rm I}_{\circ m}}\Big(
   \ov \alpha  \xi_1+\ov\beta\xi_2- C(0)(\xi_1^2+\xi_2^2)\Big)\\ & -
   \tau\omega^2 C(0)\frac{\Io^2}{{\rm I}^2_{\circ m}}\frac{\xi_1\xi_2}{2}  (\ov \beta  \xi_1+
   \ov\alpha\xi_2-2 C(0)\xi_1\xi_2)\\
   :=&i\sigma_{w}-\nu_{w}\,.
\end{split}
                      \label{lambdaw2}
                    \end{equation}
                    The coordinates $\alpha_{m3}$, $\alpha_{c3}$, and $\beta_3$ of the eigenmodes are null.
    The eigenvector is:
           {\renewcommand{\arraystretch}{1.5}
         \begin{equation}\begin{split}&
             \begin{bmatrix}\alpha_{m1}\\\alpha_{m2}
             \end{bmatrix}=  \erm^{-t \nu_{w}}\begin{bmatrix}&\cos(\sigma_{w}t)\\
               \alpha_{w2}&\sin(\sigma_{w}t)
               \end{bmatrix}\,,
             \begin{bmatrix}\beta_{1}\\\beta_{2}
             \end{bmatrix}=   - \erm^{-t \nu_w} C(0)
\begin{bmatrix} \frac{\xi_1}{\ov \alpha}&\cos(\sigma_{w}t)\\
              \frac{\xi_2}{\ov \beta}\alpha_{w2}&\sin(\sigma_{w}t)
            \end{bmatrix}\\
            & \\ &
  \begin{bmatrix}\alpha_{c1}\\\alpha_{c2}
             \end{bmatrix}=f_c \begin{bmatrix}\alpha_{m1}(t)\\\alpha_{m2}(t)
             \end{bmatrix}
+\erm^{-t \nu_{w}}\frac{\eta_c}{\omega}\frac{{\rm I}_{\circ m}}{\Io}
\begin{bmatrix}
\alpha_{w2}&\ \ \sin(\sigma_{w}t)\\
 & - \cos(\sigma_{w}t)
\end{bmatrix}\,,
\end{split}
\end{equation}
}
where
\[
 \alpha_{w2}:= \sqrt{\frac{\xi_1(\ov \alpha-C(0)\xi_1)}{\xi_2(\ov \beta-C(0)\xi_2)}}\,.
\] 
      The wobble  eigenmode is represented in Figure \ref{figwobble}.

{\bf Nearly Diurnal Free Wobble (NDFW) and Free Libration in Latitude (FLL).}
  \begin{equation}
    \begin{split}
      \lambda_{dw}&=i\omega(1+x_{dw})=i\sigma_{dw}-\nu_{dw}\\
      \lambda_{\ell a}&=i\omega(1+x_{\ell a})=i\sigma_{\ell a}-\nu_{\ell a}\,,
      \end{split}
  \label{ldwella2}
\end{equation}
where  $x_{dw}$ and $x_{\ell a}$ are the solutions to the equation
       \begin{equation} \begin{split}
       &   x^2-x(f_0+1) (y+z )+(f_0+1)z\,y=0\,,\\ & \text{and}\\
       &z=c_1 \ov \alpha_e +\frac{c_2}{2}\ov \gamma-C(i\omega)\left(c_1^2+c_2^2\right) \ 
      \Big( \text{FLL eigenvalue for} \ {\rm I}_{\circ c}=0 \Big)\,,\\
          &y =f_c+i \frac{\eta_c}{\omega}\frac{{\rm I}_{\circ m}}{\Io}
 \quad\qquad\qquad\qquad \ \ \,
      \Big( \text{NDFW eigenvalue for} \ {\rm I}_{\circ c}=0 \Big)\,.       \end{split}
        \label{charpf}
\end{equation}
 The FLL and NDFW eigenvalues  have the same nature and  are not easily
   distinguishable, see equations (\ref{xx1}) and (\ref{xx2}).

 The coordinates $\alpha_{m3}$, $\alpha_{c3}$, and $\beta_3$ of the eigenmodes are null.
  We will choose a normalisation of the eigenvectors such that for both modes NDFW and FLL the components
  of $\boldsymbol \alpha_m$ and $\boldsymbol \beta$ are the same and given by:                  
          {\renewcommand{\arraystretch}{1.5}
         \begin{equation}
             \begin{bmatrix}\alpha_{m1}\\\alpha_{m2}
             \end{bmatrix}=  \erm^{-t \nu}\begin{bmatrix}\ \ \cos(\sigma t)\\
             - \sin(\sigma t)
               \end{bmatrix}\,,
             \begin{bmatrix}\beta_{1}\\\beta_{2}
             \end{bmatrix}= \erm^{-t \nu}
\begin{bmatrix} \frac{\xi_1-1}{\ov \alpha}&\big(C_i \sin(\sigma t) - C_r \cos(\sigma t)\big)\\
              \frac{\xi_2-1}{\ov \beta}&\big(C_r \sin(\sigma t) + C_i \cos(\sigma t)\big)\,,
            \end{bmatrix}
 \label{lambdadwla}
\end{equation}
}
where: $(\nu,\sigma)=(\nu_{dw},\sigma_{dw})$ for the NDFW,
$(\nu,\sigma)=(\nu_{\ell a},\sigma_{\ell a})$ for the FLL, and $C(i\omega)=C_r+iC_i$. 

For the NDFW
  {\renewcommand{\arraystretch}{1.5}
  \begin{equation}
\boldsymbol{\alpha}_c={\rm Re}\left\{ \exp\big(i\,\sigma_{dw}t-\nu_{dw}t\big)
\frac{1}{f_0}\left(-1+\frac{x_{\ell a}}{ y}\right) 
\begin{bmatrix} 1  \\ i \\ 0
\end{bmatrix}\right\}\label{eigdwc}\end{equation}
}
and
for the FLL
 {\renewcommand{\arraystretch}{1.5}
  \begin{equation}
\boldsymbol{\alpha}_c={\rm Re}\left\{ \exp\big(i\,\sigma_{\ell a}t-\nu_{\ell a}t\big)
\frac{1}{f_0}\left(-1+\frac{x_{dw}}{ y}\right) 
\begin{bmatrix} 1  \\ i \\ 0
\end{bmatrix}\right\}\label{eiglac}\end{equation}
}
where: $y=f_c+i \frac{\eta_c}{\omega}\frac{{\rm I}_{\circ m}}{\Io}$,
$x_{\ell a}=x_{\ell a r}+i \,\nu_{\ell a}/\omega$, and
$x_{dw}=x_{dw r}+i \,\nu_{dw}/\omega$.

\begin{figure}
\begin{center}

\begin{tikzpicture}[tdplot_main_coords, scale = 2.5]
 
\coordinate (P) at ({1/sqrt(3)},{1/sqrt(3)},{1/sqrt(3)});

  \draw (0,0) circle (1cm);

  \draw[black, dot diameter=2pt, dot spacing=3pt, dots] (0,0,0.99) ellipse (0.14cm and 0.07cm);

  \draw[black, dot diameter=2pt, dot spacing=3.2pt, dots] (0,0,0.45) ellipse (0.22cm and 0.11cm);
 
\tdplotsetrotatedcoords{0}{0}{0};
\draw[dotted,
    tdplot_rotated_coords,
    black
] (0.71,-0.71,0) arc (-45:135:1);

 \tdplotsetrotatedcoords{0}{90}{90};
\draw[dotted,
    tdplot_rotated_coords,
    black
] (0.707,-0.707,0) arc (-45:135:1);

 \tdplotsetrotatedcoords{90}{90}{90};
\draw[dotted,
    tdplot_rotated_coords,
    black
] (0.707,0.707,0) arc (45:225:1);
 
\draw[dotted] (0,0,0) -- (1.0,0,0);
\draw[dotted] (0,0,0) -- (0,1.0,0);
\draw[dotted] (0,0,0) -- (0,0,1.0);

\draw[-stealth] (0,0,0) -- (0.48,0,0);
\draw[-stealth] (0,0,0) -- (0,0.432,0.20274);
\draw[-stealth] (0,0,0) -- (0,-0.21146,0.4365);

\draw[] (0.5,0,0) circle[radius=0.4pt];
\draw[] (0,0.45,0.218) circle[radius=0.4pt];
\draw[fill=white] (0,-0.218,0.45) circle[radius=0.4pt];

\draw[] (1,0,0) circle[radius=0.4pt];

\draw[] (0,0.99,0.141) circle[radius=0.4pt];
\draw[fill=black] (0,0.994,0.1094) circle[radius=0.4pt];
\draw[fill=black] (0,0.995,0.0707) circle[radius=0.4pt];
\draw[fill=black] (0,0.9994,0.03464) circle[radius=0.4pt];
\draw[fill=black] (0,1,0) circle[radius=0.4pt];
\draw[fill=black] (0,0.99,-0.141) circle[radius=0.4pt];
\draw[fill=black] (0,0.994,-0.1094) circle[radius=0.4pt];
\draw[fill=black] (0,0.995,-0.0707) circle[radius=0.4pt];
\draw[fill=black] (0,0.9994,-0.03464) circle[radius=0.4pt];

\draw[fill=black] (0.99,0,0.141) circle[radius=0.4pt];
\draw[fill=black] (0.994,0,0.1094) circle[radius=0.4pt];
\draw[fill=black] (0.995,0,0.0707) circle[radius=0.4pt];
\draw[fill=black] (0.9994,0,0.03464) circle[radius=0.4pt];

\draw[fill=black] (0.99,0,-0.141) circle[radius=0.4pt];
\draw[fill=black] (0.994,0,-0.1094) circle[radius=0.4pt];
\draw[fill=black] (0.995,0,-0.0707) circle[radius=0.4pt];
\draw[fill=black] (0.9994,0,-0.03464) circle[radius=0.4pt];

\draw[fill=white] (0,-0.141,0.99) circle[radius=0.4pt];

\draw[] (0,-0.218,0.45) circle[radius=0.4pt];

\draw[] (0,0.45,0.218) circle[radius=0.4pt];

\draw[fill=black] (0,0.47,0.171) circle[radius=0.4pt];

\draw[fill=black] (0,0.487,0.1133) circle[radius=0.4pt];

\draw[fill=black] (0,0.497,0.055) circle[radius=0.4pt];

\draw[fill=black] (0,0.5,0) circle[radius=0.4pt];

\draw[fill=black] (0,0.45,-0.218) circle[radius=0.4pt];

\draw[fill=black] (0,0.47,-0.171) circle[radius=0.4pt];

\draw[fill=black] (0,0.487,-0.1133) circle[radius=0.4pt];

\draw[fill=black] (0,0.497,-0.055) circle[radius=0.4pt];

\draw[fill=black] (0.45,0,0.218) circle[radius=0.4pt];

\draw[fill=black] (0.47,0,0.171) circle[radius=0.4pt];

\draw[fill=black] (0.487,0,0.1133) circle[radius=0.4pt];

\draw[fill=black] (0.497,0,0.055) circle[radius=0.4pt];

\draw[fill=black] (0.45,0,-0.218) circle[radius=0.4pt];

\draw[fill=black] (0.47,0,-0.171) circle[radius=0.4pt];

\draw[fill=black] (0.487,0,-0.1133) circle[radius=0.4pt];

\draw[fill=black] (0.497,0,-0.055) circle[radius=0.4pt];

\end{tikzpicture}
\hspace{1cm}
\begin{tikzpicture}[tdplot_main_coords, scale = 2.5]
 
\coordinate (P) at ({1/sqrt(3)},{1/sqrt(3)},{1/sqrt(3)});

  \draw (0,0) circle (1cm);

  \draw[black, dot diameter=2pt, dot spacing=3pt, dots] (0,0,0.99) ellipse (0.14cm and 0.07cm);

  \draw[black, dot diameter=2pt, dot spacing=3.2pt, dots] (0,0,0.463) ellipse (0.14cm and 0.07cm);
 
\tdplotsetrotatedcoords{0}{0}{0};
\draw[dotted,
    tdplot_rotated_coords,
    black
] (0.71,-0.71,0) arc (-45:135:1);

 \tdplotsetrotatedcoords{0}{90}{90};
\draw[dotted,
    tdplot_rotated_coords,
    black
] (0.707,-0.707,0) arc (-45:135:1);

 \tdplotsetrotatedcoords{90}{90}{90};
\draw[dotted,
    tdplot_rotated_coords,
    black
] (0.707,0.707,0) arc (45:225:1);
 
\draw[dotted] (0,0,0) -- (1.0,0,0);
\draw[dotted] (0,0,0) -- (0,1.0,0);
\draw[dotted] (0,0,0) -- (0,0,1.0);

\draw[-stealth] (0,0,0) -- (0.48,0,0);
\draw[-stealth] (0,0,0) -- (0,0.4704,0.09552);
\draw[-stealth] (0,0,0) -- (0,-0.09552,0.4704);

\draw[] (0.5,0,0) circle[radius=0.4pt];
\draw[] (0,0.49,0.0995) circle[radius=0.4pt];
\draw[fill=white] (0,-0.0995,0.49) circle[radius=0.4pt];

\draw[] (1,0,0) circle[radius=0.4pt];

\draw[] (0,0.99,0.141) circle[radius=0.4pt];
\draw[fill=black] (0,0.994,0.1094) circle[radius=0.4pt];
\draw[fill=black] (0,0.995,0.0707) circle[radius=0.4pt];
\draw[fill=black] (0,0.9994,0.03464) circle[radius=0.4pt];
\draw[fill=black] (0,1,0) circle[radius=0.4pt];
\draw[fill=black] (0,0.99,-0.141) circle[radius=0.4pt];
\draw[fill=black] (0,0.994,-0.1094) circle[radius=0.4pt];
\draw[fill=black] (0,0.995,-0.0707) circle[radius=0.4pt];
\draw[fill=black] (0,0.9994,-0.03464) circle[radius=0.4pt];

\draw[fill=black] (0.99,0,0.141) circle[radius=0.4pt];
\draw[fill=black] (0.994,0,0.1094) circle[radius=0.4pt];
\draw[fill=black] (0.995,0,0.0707) circle[radius=0.4pt];
\draw[fill=black] (0.9994,0,0.03464) circle[radius=0.4pt];

\draw[fill=black] (0.99,0,-0.141) circle[radius=0.4pt];
\draw[fill=black] (0.994,0,-0.1094) circle[radius=0.4pt];
\draw[fill=black] (0.995,0,-0.0707) circle[radius=0.4pt];
\draw[fill=black] (0.9994,0,-0.03464) circle[radius=0.4pt];

\draw[fill=white] (0,-0.141,0.99) circle[radius=0.4pt];

\draw[fill=black] (0,0.49,-0.0995) circle[radius=0.4pt];

\draw[fill=black] (0,0.495,0.0705) circle[radius=0.4pt];
\draw[fill=black] (0,0.4975,0.03535) circle[radius=0.4pt];
\draw[fill=black] (0,0.5,0) circle[radius=0.4pt];

\draw[fill=black] (0,0.495,-0.0705) circle[radius=0.4pt];
\draw[fill=black] (0,0.4975,-0.03535) circle[radius=0.4pt];

\draw[fill=black] (0.49,0,-0.0995) circle[radius=0.4pt];

\draw[fill=black] (0.495,0,0.0705) circle[radius=0.4pt];
\draw[fill=black] (0.4975,0,0.03535) circle[radius=0.4pt];

\draw[fill=black] (0.495,0,-0.0705) circle[radius=0.4pt];
\draw[fill=black] (0.4975,0,-0.03535) circle[radius=0.4pt];

\draw[fill=black] (0.49,0,0.0995) circle[radius=0.4pt];

\end{tikzpicture}

\end{center}
\caption{Representation of the oscillations of the NDFW (LEFT) and FLL (RIGHT).
  The Figures show the motion of $\K_m$ and
  $\K_c$ with respect to $\K_g$ as explained in the caption of Figure \ref{figlong}. The
  only difference between the NDFW and the FLL is the relative
  amplitude of the oscillations of the mantle and the core.
  For the Moon the amplitude of librations of the core
  is larger for the NDFW and the opposite is true for the FLL. Using
  the parameters given in  Table \ref{tab.INPOP} (Moon)  we obtain
  (amplitude  core)/(amplitude  mantle)=
  7107 (NDFW) and  0.2010 (FLL).
   }
\label{figNDFWFLL}
\end{figure}
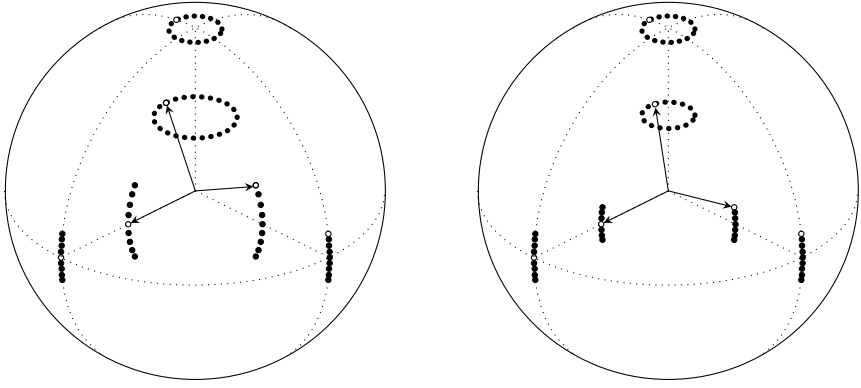

The NDFW and FLL  eigenmodes are represented in Figure \ref{figNDFWFLL}.

\section{The  offset of the
  rotation axes of mantle and core
  from the  Cassini state of the rigid-body approximation.}
 \label{offsetsec}

 The definition of  guiding motion  is based on the  hypothesis 
 that the extended body rotates almost steadily as a rigid body. It is therefore natural
 to assume in the theoretical determination of a guiding motion that the extended body is rigid.
 The slow frame determined from the rigid-body motion is in general non inertial due to  axial precession.
 In this Section we investigate the effect of inertial forces in the slow frame when 
 we replace a rigid body  in guiding motion for  a body 
 made of  a rigid mantle and a fluid core.

 The theoretical determination of the  Cassini state 
 of a rigid body  in  s-to-2 spin-orbit resonance (e.g. Moon)  \cite{peale1969generalized} is
 done  by means of   equations in which  the torque
 is  averaged  after being written in the orbital frame $\K_{or}$, which is a frame
 that 
  precesses with the orbit of the point mass and is 
  given by
  $\mathbf{R}_{or}=\mathbf{R_3}(\Omega_p)\mathbf{R_1}(\iota_p):\K_{or}\to\kappa$ .
   In the following we 
  consider the case of an extended body in s-to-2 spin-orbit resonance with a point mass.
  The case of no spin-orbit resonance  is easier and will be treated in Section \ref{subwithout}.

  In  Appendix \ref{hansen}
  (equations (\ref{Rgapp}), (\ref{Rs}),  and (\ref{zeta})) we parameterized the guiding motion
  associated to a Cassini state as
$\mathbf{R_g}=\mathbf{R_3}(\psi_g)\mathbf{R_1}(\theta_g)\mathbf{R_3}(\phi_g)$,
where: $\theta_g$ ($\dot\theta_g=0$) is the inclination of the body equator to the
$(\mat e_1,\mat e_2)$-plane
in $\kappa$
(the Laplace plane or invariable plane),
$\psi_g$  ($\ddot\psi_g=0$) is the longitude of the ascending  node of the
body equator, and  $\phi_g$  ($\ddot\phi_g=0$) is
the angle between the ascending  node and Axis 1 (a principal
axis of smallest moment of inertia). The  
 slow frame $\mat R_s:\K_s\to\kappa$
 is defined by $\mathbf{R}_s= \mathbf{R}_g\mathbf{R_3}^{-1}(\omega\, t)=
 \mathbf{R_3}(\psi_g)\mathbf{R_1}(\theta_g)\mathbf{R_3}(\zeta)$
 where  $\zeta=\phi_g-\omega t$  and  (due to equation (\ref{omegag}))
 $\dot\zeta=-  \dot \psi_g\cos\theta_g$. These definitions  imply  (see equation (\ref{simple4.5}))
 \begin{equation}
   \boldsymbol \omega_{s,g}= \dot \psi_g \sin \theta_g\left(
\begin{array}{c}
 \sin\phi_g \\
  \cos\phi_g\\
 0 
\end{array}
\right)
\,,\quad
\dot{\boldsymbol \omega}_{s,g}= \dot \phi_g\dot \psi_g \sin \theta_g
\left(
\begin{array}{c}
\ \ \, \cos\phi_g \\
- \sin\phi_g\\
 0 
\end{array}
\right)\,.
\label{osg}
 \end{equation}

The guiding motion becomes simpler in the  ``precessional frame'' $\K_{pr}$ defined
as\footnote{ \label{footorpr} The transformation
between the precessional frame
and the orbital frame is
\[
  \mat R^{-1}_{or}  \mat R_{pr}=
  \mat R^{-1}_1(\iota_p)\mat R_3(\psi_g-\Omega_p)\mat R_1(\theta_g)=\mat R_1(\theta_g-\iota_p)\
  \text{or} \ 
{\scriptsize \left(
\begin{array}{ccc}
 -1 & 0 & 0 \\
 0 & -\cos \left(\theta _g+\iota _p\right) & \sin \left(\theta _g+\iota _p\right) \\
 0 & \ \ \sin \left(\theta _g+\iota _p\right) & \cos \left(\theta _g+\iota _p\right) \\
\end{array}
\right)}\,,
\]
where we used
$  \psi_g=\Omega_p $,  for Cassini state 1 (e.g. Mercury), or 
 $\psi_g=\Omega_p+\pi$, for Cassini state 2  (e.g. Moon), see Eq.
(\ref{psiOmega}). Note:
$  \mat R^{-1}_{or} \mat R_{pr}:\K_{pr}\to\K_{or}$
does not depend on time.
}:
\begin{equation}
  \mat R_3(\psi_g)\mat R_1(\theta_g):\K_{pr}\to\kappa\Longrightarrow
  \left\{ \begin{array}{ll}
  \mat R_3(\phi_g)&:\K_g\to \K_{pr} \\
            \mat R_3(\zeta)&:\K_s\to\K_{pr}\,. \end{array}\right.
  \label{Kpr}
\end{equation}
Note that $\K_s$ ($\K_g$) rotates  slowly (fastly) with angular velocity
$\dot \zeta=-\dot \psi_g\cos\theta_g$ ($\dot\phi_g$)
inside $\K_{pr}$,
in such a way that the   $\mat e_3$-axis of $\K_{s}$ ($\K_g$) remains aligned 
with the $\mat e_3$-axis of $\K_{pr}$. The angular velocity
of the precessional frame $\boldsymbol\omega_{pr}= \dot \psi_g \mat e_3\in\kappa$
is represented by a constant vector $\boldsymbol\omega_{pr,pr}=
\dot \psi_g \sin\theta_g\mat e_2+\dot \psi_g\cos\theta_g \mat e_3$ in $\K_{pr}$, and 
the angular velocity of the guiding frame, which  is given by
\begin{equation}
  \boldsymbol{\omega}_{g,pr}= \mat R_3(\phi_g)
  \boldsymbol{\omega}_{g,g}= \omega \mat e_3+\mat R_3(\phi_g)
  \boldsymbol{\omega}_{s,g}=\omega \mat e_3+\dot\psi_g\sin\theta_g\mat e_2\,,\label{ogpr}
\end{equation}
is also constant in  $\K_{pr}$.

By definition, the guiding motion is a solution to Euler's
equation after time-averaging in $\K_{pr}$. If we use: the notation $\langle f\rangle =
\lim_{T\to\infty}\frac{1}{T}\int_0^Tf(t)dt$,  that  $\dot\phi_g$ is constant,
$[\ov{\mat I},\ov {\mat J}]=0$,  and $\mat R_3(\phi_g):\K_g\to\K_{pr}$, then we obtain
\begin{equation}
  \begin{split}
  & \langle\mat I_{pr}\rangle=
    \frac{1}{2\pi}\int_0^{2\pi}
    \mat R_3^{-1}(\phi_g)\ov{\mat I}\mat R_3(\phi_g)d\phi_g=
    \left( \begin{array}{ccc} \frac{\ov I_1+\ov I_2}{2} & 0 & 0\\0 &\frac{\ov I_1+\ov I_2}{2} & 0\\
             0&0&\ov I_3\end{array}\right)\ \ \text{and} \\
        &\Big\langle
         \frac{d}{dt}\mat I_{pr}\boldsymbol\omega_{g,pr}
  +\boldsymbol\omega_{pr,pr}\times \mat I_{pr}\boldsymbol\omega_{g,pr}\Big\rangle=
  \boldsymbol\omega_{pr,pr}\times \langle\mat I_{pr}\rangle\boldsymbol\omega_{g,pr}=\\
  &=\left( \begin{array}{c}\ov I_3  \dot\phi_g  \dot \psi_g \sin\theta_g +\Oc(\dot\psi_g^2)\\0\\0
           \end{array}\right)=
      \Big\langle\mat R_3^{-1}(\phi_g)[\ov{\mat I},\ov {\mat J}+
 \boldsymbol\delta\mat J(t)]\mat R_3(\phi_g)\Big\rangle^\vee=\\ & =
  \left(
         \begin{array}{c}
 (\ov I_3-\ov I_2)\big\langle \delta J_{23}(t)\cos (t \dot\phi_g)\big\rangle
 + (\ov I_3-\ov I_1)\big\langle\delta J_{13}(t)\sin (t \dot\phi_g)\big\rangle\\
 (\ov I_1-\ov I_3) \big\langle\delta J_{13}(t)\cos (t \dot\phi_g)\big\rangle
 + (\ov I_3-\ov I_2) \big\langle\delta J_{23}(t)\sin (t \dot\phi_g)\big\rangle\\
           (\ov I_2-\ov I_1)\big\langle\delta J_{12}(t)\big\rangle
           \end{array}
\right)\,.
\end{split}\label{Kprav}
\end{equation}
By definition, the average value of $\delta J_{12}$ is zero.
The average $ \big\langle\delta J_{23}(t)\sin (t \dot\phi_g)\big\rangle$   is one  half of
the Fourier coefficient associated with $\sin\phi_g$
of the Fourier expansion of     $ \delta  J_{23}$.  The computation
of this  Fourier coefficient, and of the others,
is similar to the computation of the  constants $c_1$ and $c_2$ done
 in the Appendix
\ref{hansen}. The result is:
{\footnotesize \begin{equation}
  \begin{split}
    \big\langle \delta J_{23}(t)\cos (t \dot\phi_g)\big\rangle &=\left(\frac{3G m_p}{2a_p^3}\right)
   \frac{1}{4}\Big( X^{-3,2}_s(e) +\big( X^{-3,2}_s(e)-2X^{-3,0}_0(e)\big) \cos\chi_p\Big) \sin\chi_p\\
 \big\langle\delta J_{13}(t)\sin (t \dot\phi_g)\big\rangle &=
 -\left(\frac{3G m_p}{2a_p^3}\right)\frac{1}{4}\Big( X^{-3,2}_s(e) +\big( X^{-3,2}_s(e)+2X^{-3,0}_0(e)\big) \cos\chi_p\Big) \sin\chi_p\\
 \big\langle \delta J_{23}(t)\sin (t \dot\phi_g)\big\rangle &=
 \big\langle\delta J_{13}(t)\cos (t \dot\phi_g)\big\rangle=0\,,
 \end{split}\label{avJ}
\end{equation}}
where  $ X^{-3,0}_0(e)$ and $X^{-3,2}_s(e)$ are the Hansen coefficients
given  in Appendix \ref{hansen} and  $\chi_p$ is the obliquity, namely
the angle between the body   axis of largest moment of inertia
 ($\mat e_3\in \K_g$) and 
 the normal to the orbital plane.
 We remark  that $\mat e_3\in\K_g$ 
  is the spin axis of the guiding  motion  with respect to the orbital  frame
 ($\mat  R_{or}^{-1}\mat R_g:\K_g\to \mat \K_{or}$). According to Footnote
 \ref{footorpr}, $\chi_p=\theta_g-\iota_p$  for Cassini state 1 (e.g. Mercury), or 
  $\chi_p=\theta_g+\iota_p$ for Cassini state 2  (e.g. Moon).

  Equations (\ref{Kprav}) and (\ref{avJ}) imply that the Cassini state is equal to  the guiding
  motion $\mathbf{R_g}=\mathbf{R_3}(\psi_g)\mathbf{R_1}(\theta_g)\mathbf{R_3}(\phi_g)$
  if
   \begin{equation}
     \dot\psi_g\dot\phi_g\sin\theta_g=-
     \frac{3}{2}\frac{G m_p}{a_p^3}\sin\chi_p\left(\ov\alpha_e
       X^{-3,0}_0\cos\chi_p+\frac{\ov\gamma}{4}X^{-3,2}_s
    (1+\cos\chi_p)\right):=\cal P
\,,\label{peale}
  \end{equation}
  where $\ov\alpha_e=\frac{\ov I_3-(\ov I_1+\ov I_2)/2}{\ov I_3}$ and
  $\ov \gamma=\frac{\ov I_2-\ov I_1}{\ov I_3}$.
Equation (\ref{peale}) is that   obtained by  \cite{peale1969generalized}, as presented in    
    \cite{boue2020cassini}.
The signs of the angles in equation (\ref{peale}) have
been a source of mistakes
(\cite{baland2017obliquity} Section 6.1.1). Our  sign conventions are: $\dot\psi_g<0$,
$\dot \phi_g=\omega-\dot\psi_g\cos\theta_g>0$,
$\theta_g>0$ ,  $\chi_p>0$ , and $\iota_p>0$.
There are two possible relations among $\theta_g$,  $\chi_p$,
and $\iota_p$: $\theta_g=\chi_p+\iota_p$ for the Cassini state 1 (e.g. Mercury)
and $\theta_g=\chi_p-\iota_p$ for Cassini state 2 (e.g. Moon). For a given  $\iota_p$,
the solutions $\chi_p$ to  equation (\ref{peale})
in case 1 are the same as those in   case 2 after a change of sign.
The Cassini state we are interested in is obtained in the following way. Given $\iota_p$
choose $\theta_g=\chi_p+\iota_p$ and solve for $\chi_p$. Among the solutions with
$-\pi<\chi_p\le \pi$ choose the one, denoted as  $\tilde \chi_p$,
with the smallest absolute value. If  $\tilde \chi_p>0$ ($\tilde\chi_p<0$),  then
$\chi_p=\tilde \chi_p$ ($\chi_p=-\tilde \chi_p$) and
we have a Cassini state 1 with $\theta_g=\chi_p+\iota_p$
(a Cassini state 2 with $\theta_g=\chi_p-\iota_p$).

In the follwowing we
restrict the tidal force matrix  $\boldsymbol \delta \mat J_g$ to
its Fourier modes in $\cos\phi_g$ and $\sin\phi_g$, and write
$ \dt J_{g12}(t)=0$,
\begin{equation}
  \begin{split}
 \dt J_{g13}(t)&= 2
 \big\langle\delta J_{13}(t)\sin (t \dot\phi_g)\big\rangle\sin (\phi_g)=
 \frac{\ov I_3}{\ov I_3-\ov I_1}({\cal P - R}) \sin(\phi_g)\\
   \dt J_{g23}(t)&=2
   \big\langle \delta J_{23}(t)\cos (t \dot\phi_g)\big\rangle \cos\phi_g
   = \frac{\ov I_3}{\ov I_3-\ov I_2} ({\cal P  + R})\cos\phi_g\,,
 \end{split}\label{Fourier}
\end{equation}
where we have used equations (\ref{avJ}) and (\ref{peale}) and the definition
\begin{equation}
  {\cal R}:=
    \frac{3}{2}\frac{G m_p}{a_p^3}\sin\chi_p\left(\frac{\ov\alpha_e}{2}
       X^{-3,2}_s(1+\cos\chi_p) +\frac{\ov\gamma}{2}X^{-3,0}_0\cos\chi_p
    \right)\,.\label{calR}
  \end{equation}
  
The guiding motion solves the  forced  rigid-body equations only in an average sense.
The residue we obtain  after the substitution
of the guiding motion into  the rigid-body equation
written in the guiding frame is, after
using equations  (\ref{omegag}), (\ref{osg}),
and (\ref{Fourier}), and   ${\cal P}= \dot\psi_g\dot\phi_g\sin\theta_g$ (Peale's equation):
{\footnotesize\begin{equation}
   \begin{split} & [\ov {\mat I},\boldsymbol  \delta \mat J_g]^\vee-\ov {\mat I}\dot{\boldsymbol \omega}_{g,g}
    -\dot{\boldsymbol \omega}_{g,g}\times\ov{\mat I}\dot{\boldsymbol \omega}_{g,g}=
   \left(\begin{array}{c}\big(\ov I_3-\ov I_2\big)\dt J_{g23}
             -\ov I_{1}\dot{{\omega}}_{s,g1} -\omega\big(\ov I_3-\ov I_2\big){\omega}_{s,g2}\\
\big(\ov I_1-\ov I_3\big)\dt J_{g13}-
             \ov I_{2}\dot{{\omega}}_{s,g2} -\omega\big(\ov I_3-\ov I_1\big){\omega}_{s,g1}\\0
            \end{array}\right)=\\  &
          \left(\begin{array}{c}\ov I_3 ({\cal P+ R})\cos\phi_g
                  -(\ov I_1 - \ov I_2 + \ov I_3)\dot\phi_g\dot\psi_g\sin\theta_g\cos\phi_g\\
\ov I_3 ({\cal R- P})\sin\phi_g
                    -(\ov I_1 - \ov I_2 - \ov I_3)\dot\phi_g\dot\psi_g\sin\theta_g\sin\phi_g
\\0
                \end{array}\right)=
   \left(\begin{array}{c}\big((\ov I_2-\ov I_1) {\cal P}+ \ov I_3 {\cal R}\big)\cos\phi_g\\
 \left(  (\ov I_2-\ov I_1) {\cal P}+ \ov I_3 {\cal R } \right)\sin\phi_g         
\\0
\end{array}\right)\,\end{split}\label{ofeq1}\end{equation}}
where terms of order $\dot\psi_g^2$ were neglected.

Now, we will investigate  the motion in the guiding frame 
of a body with  a deformable
mantle and a fluid core. The analysis will be restricted to the   frequency $\dot \phi_g$.

The mean moment of inertia of the whole  body can be  decomposed 
as  $\ov{\mat I}=\ov{\mat I}_m+\ov{\mat I}_c$. If we write $\ov{\mat I}\dot{\boldsymbol \omega}_{s,g}=
  (\ov{\mat I}_m+\ov{\mat I}_c)\dot{\boldsymbol \omega}_{s,g}$ and use equation
  (\ref{ofeq1}) and  $  \mat I_c \dot{\boldsymbol \omega}_{s,g}=\dot \phi_g\dot \psi_g \sin \theta_g
   \mat I_c\mat R^{-1}_3(\phi_g)\mat e_1= {\cal P} \mat I_c\mat R^{-1}_3(\phi_g)\mat e_1 $,
  then  we can write the right hand side of equation (\ref{lineq}) (the torque) as
{\footnotesize  \begin{equation}
   \begin{split}
 &  \big(\ov I_3-\ov I_2\big)\langle\dt J\rangle_{g23}   -
 \overbrace{\ov I_{m1}}^{\ov I_{1}-\ov I_{c1}}\dot{{\omega}}_{s,g1} -\omega\big(\ov I_3-\ov I_2\big){\omega}_{s,g2}=
 \Big((\ov I_2-\ov I_1) {\cal P}+ \ov I_3 {\cal R}+{\cal P} \ov I_{c1}\Big)\cos\phi_g
 \\
   & \big(\ov I_1-\ov I_3\big)\langle\dt J\rangle_{g13}-
   \underbrace{\ov I_{m2}}_{\ov I_2-\ov I_{c2}}\dot{{\omega}}_{s,g2} +\omega\big(\ov I_3-\ov I_1\big){\omega}_{s,g1} =\Big((\ov I_2-\ov I_1) {\cal P}+ \ov I_3 {\cal R}-{\cal P} \ov I_{c2}\Big)\sin\phi_g
  \,.
 \end{split}\label{ofeq2}
\end{equation}}
The torque  has a term of forced librations
$\big((\ov I_2-\ov I_1) {\cal P}+ \ov I_3 {\cal R}\big)\mat R^{-1}_3(\phi_g)\mat e_1$ that is responsible
for the oscillations  of the rigid body (and  of the mantle) about the Cassini state, and
a term $  \mat I_c \dot{\boldsymbol \omega}_{s,g}$.   This last term, which 
is due to the non-inertial character of the slow frame,  will be called  the
inertial torque. It is partially responsible for the displacement of the  spin axis of the mantle
from  the rigid-body  state of Cassini.

 The inertial torque that acts upon the mantle  is
 exactly the opposite of that in the right-hand side of equation  (\ref{lincore2}),
 i.e.
 \begin{equation}
   -\ov{\mat I}_c\dot{\boldsymbol \omega}_{s,g}=-{\cal P} \mat I_c\mat R^{-1}_3(\phi_g)\mat e_1\,,
   \label{tcore}
   \end{equation}
 that acts upon the core. Moreover,  $-\ov{\mat I}_c\dot{\boldsymbol \omega}_{s,g}$
 is the unique direct  torque upon the core (it does not depend on the relative position of the mantle)
 regardless all possible  tidal-forcing torques that act
 upon the mantle, and this feature distinguishes the inertial torque
 $\ov {\mat I}_c\dot{\boldsymbol \omega}_{s,g}$ from
  other torque terms. 

  The two nonhomogeneous terms in equation  (\ref{dtF}) show that the  deformation variables
  $\boldsymbol \delta \mat B$ are under the effect of both  tidal and inertial
  torques.   Equations (\ref{Fourier}) and
 (\ref{tcore}) 
  can be used to write  the  two non-homogeneous terms in equation  (\ref{dtF}) as
   \begin{equation}\begin{split}
      & -\omega \omega_{s,g1} +\dt J_{g13}=
       \frac{ (\ov I_1 {\cal P}-\ov I_3 {\cal R})}{\ov I_3-\ov I_1}\sin \phi_g\\ & \qquad =
       \frac{3 G m_p}{2 a_p^3} 
       \left(-X^{-3,0}_0\cos\chi_p- X^{-3,2}_s\cos^2(\chi_p/2)\right)\sin \chi_p\sin \phi_g\\
     &  -\omega \omega_{s,g2} +\dt J_{g23}
     =\frac{ (\ov I_2 {\cal P}+\ov I_3 {\cal R})}{\ov I_3-\ov I_2}\cos \phi_g\\
     &\qquad=
     \frac{3 G m_p}{2 a_p^3} 
       \left(-X^{-3,0}_0\cos\chi_p+ X^{-3,2}_s\cos^2(\chi_p/2)\right)\sin \chi_p\cos \phi_g
       \end{split}\label{ofeq4}
\end{equation}
where terms of order $\dot\psi_g^2$ were neglected.

Equations:
(\ref{lineq}) for  $\boldsymbol\alpha_m$,  (\ref{lincore2}) for  $\boldsymbol \alpha_c$ ,
and one of the equations
(\ref{def1d}), (\ref{def2d}), or (\ref{def3d}) for the deformation variables
$\boldsymbol \delta\mat B$, 
 with the forcing terms given respectively by equations  (\ref{ofeq2}), (\ref{tcore}), and 
 (\ref{ofeq4}) can be easily solved numerically in any particular problem. Nevertheless,
 in order to understand the motion at frequency $\dot \phi_g$ in the guiding frame and its
 consequences in the precessional frame 
 it is interesting to further treat
    the problem in general form.

 To simplify the following analysis  we will assume that the principle axes of the
 core cavity are aligned to those of the mantle, so that both $\mat I_m$ and $\mat I_c$ are diagonal
 in the frame of the mantle. In this case the equations for
 $\alpha_{m1},\alpha_{m2},\alpha_{c1},\alpha_{c2}$ decouple from those for
 $\alpha_{m3},\alpha_{c3}$ and due to $\delta J_{12}=0$ and $\omega_{s,g3}=0$  we can make
 $\alpha_{m3}=\alpha_{m3}=\delta B_{T,m12}=\delta B_{T,m33}=0$.
 The solution for the angular variables of the mantle and core can be
 written as
\begin{equation}
  \boldsymbol \alpha_m=\underbrace{\left(\begin{array}{ccc} x_{11}&x_{12}&0
                                           \\ x_{21}&x_{22}&0\\
                               0&0&1\end{array}\right)}_{:=\mat X}
\left(
\begin{array}{c}
- \cos\phi_g \\
\,  \ \ \sin \phi_g\\
 0 
\end{array}
\right)=-\mat X\mat R_3^{-1}(\phi_g)\mat e_1\ \ \text{and}\ \
\boldsymbol \alpha_c=-
\mat Y\mat R_3^{-1}(\phi_g)\mat e_1\,,
\label{XY}
\end{equation}
where $\mat X$ and $\mat Y$ are matrices with constant coefficients. The solution for the deformation
variables can  be written in a similar way.

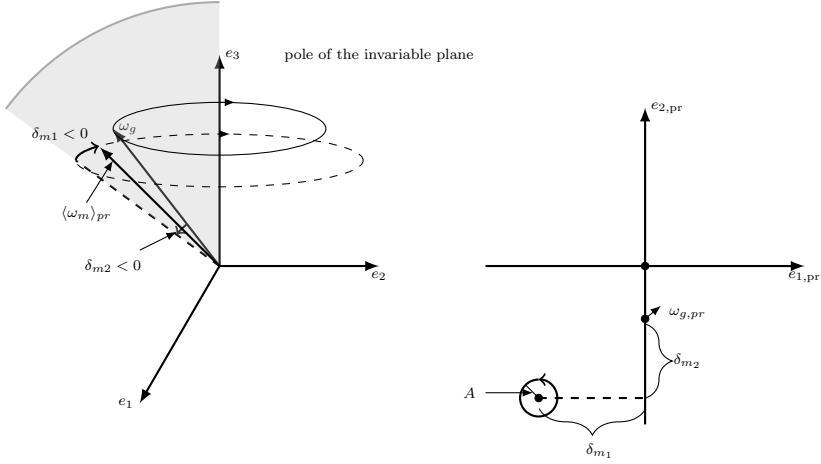
\begin{figure}
\begin{center}
\begin{tikzpicture}[scale=0.7, transform shape]
\tikzstyle{spring}=[thick, decorate, decoration={zigzag, pre length=0.5cm, post length=0.5cm, segment length=6}]
\tikzstyle{damper}=[thick,decoration={markings,
  mark connection node=dmp,
  mark=at position 0.5 with
  {
    \node (dmp) [thick, inner sep=0pt, transform shape, rotate=-90, minimum width=15pt, minimum height=3pt, draw=none] {};
    \draw [thick] ($(dmp.north east)+(5pt,0)$) -- (dmp.south east) -- (dmp.south west) -- ($(dmp.north west)+(5pt,0)$);
    \draw [thick] ($(dmp.north)+(0,-5pt)$) -- ($(dmp.north)+(0,5pt)$);
  }
}, decorate]
\tikzstyle{ground}=[fill, pattern=north east lines, draw=none, minimum width=0.75cm, minimum height=0.3cm]

            \draw [-latex, thick] (0,0) -- (0,4) node[right] {$e_3$};
            \draw [-latex, thick] (0,0) -- (3,0) node[below] {$e_2$};
            \draw [-latex, thick] (0,0) -- (-1.5,-2.6) node[left] {$e_1$};

            \draw [-latex, thick] (0,0) -- (-2.,2.6) node[right] {$\omega_{g}$};
            \draw [thick, dashed] (0,0) -- (-2.7,2.);
  \coordinate (o) at (0,0);
  \coordinate (x) at (1,0);
  \coordinate (y) at (0,1);
  \coordinate (z1) at (-2.,2.6);
  \coordinate (z2) at (-2.7,2.);
  \coordinate (z3) at (-2.25,2.25);
  \draw pic[draw, ->, angle eccentricity=1.2,angle radius = 1cm,thick] {angle = z1--o--z2};

  \draw pic[draw,angle radius = 5cm,thick,fill=gray!50, opacity=0.3] {angle = y--o--z2}; 

  \draw[draw, ->, black] (0,2.6) ellipse (2cm and 0.5cm);
  \draw[black, dashed] (0,2.) ellipse (2.7cm and 0.5cm);

            \draw [-latex, thick] (0,0) -- (-2.25,2.25);
            \draw [-latex] (0,3.1) -- (0.3,3.1);
            \draw [-latex] (0,2.5) -- (0.2,2.5);
             \node at (-3,2.5) {$\delta_{m1}<0$};
             \node at (-2.,0) {$\delta_{m2}<0$};
            \draw [-latex] (-1.5,0.3) -- (-0.8,0.65);
             \node at (-2.5,1) {$\langle\omega_m\rangle_{pr}$};
            \draw [-latex] (-2.5,1.2) -- (-2,2);

\tdplotdrawarc[->,color=black,thick,xscale=5.4]{(0,2.)}{0.5}{-180}{-213}{anchor=south west,color=black}

             \node at (3,4) {pole of the invariable plane};

            \draw [-latex, thick] (8,-3) -- (8,3) node[right] {$e_{2,\rm{pr}}$};
            \draw [-latex, thick] (5,0) -- (11,0) node[below] {$e_{1,\rm{pr}}$};

\draw[fill=black] (8,0) circle[radius=2pt];
\draw[fill=black] (8,-1) circle[radius=2pt];

            \node at (8.8,-0.9) {$\omega_{g,pr}$};

\draw[fill=black] (6,-2.5) circle[radius=2pt];
            \draw [thick, dashed] (6,-2.5) -- (8,-2.5);
\tdplotdrawarc[->,color=black,thick,xscale=1]{(6,-2.5)}{0.35}{-270}{90}{anchor=south west,color=black}

\draw [decorate,decoration={brace,amplitude=10pt},xshift=-4pt,yshift=0pt]
(8.15,-2.7) -- (6.15,-2.7);

\draw [decorate,decoration={brace,amplitude=10pt},xshift=-4pt,yshift=0pt]
(8.15,-1.1) -- (8.15,-2.5);

            \node at (7.15,-3.5) {$\delta_{m_1}$};
            \node at (8.8,-1.8) {$\delta_{m_2}$};

            \draw [] (6,-2.5) -- (5.75,-2.25);

            \draw [-latex] (8,-1) -- (8.3,-0.75);

            \draw [-latex] (5,-2.4) -- (5.9,-2.4);

            \node at (4.7,-2.4) {$A$};

      \end{tikzpicture}
\end{center}
\caption{LEFT: We denote by $\Sigma\in\kappa$
    the shaded plane that contains $\mat e_3$ (the normal to the invariable plane)
    and  $\boldsymbol \omega_g$ (the  angular velocity of the rigid-body approximation). The circles indicate
    that  the precession of $\boldsymbol \omega_g$ is retrograde, $\dot \psi_g<0$. 
                The inertial torque rotates $\boldsymbol \omega_g$  of a small  angle $\delta_{m2}$  inside $\Sigma$
                and  rotates $\boldsymbol \omega_g$  of a small  angle $\delta_{m1}$  perpendicular to $\Sigma$.
      If $\delta_{m2}>0$ ($\delta_{m2}<0$),
      then the displacement is towards (opposite to)  $\mat e_3$.
      If $\delta_{m1}<0$ ($\delta_{m1}>0)$, then
       the displacement is towards (opposite to) the direction
        of the retrograde axial precession.
        The angle  $|\delta_{m1}|>0$  is entirely
        due to dissipative effects. RIGHT: The projection of
        $\boldsymbol \omega_{g,pr}/\|\boldsymbol \omega_{g,pr}\|$
        onto the $(\mat e_1,\mat e_2)$-plane of the
        precessional frame $\K_{pr}$.
        This plane moves in the inertial space always orthogonal to the
        axis of largest moment of inertia of the rigid-body approximation, with
        $\mat e_1\in \K_{pr}$
        in the direction of the ascending node of the equator of the rigid-body approximation,  and
        $\mat e_2\in\K_{pr}$ pointing  towards the pole of the invariable plane. The circle  in
        the Figure represents the inertial libration of the normalised angular velocity of the mantle.
        The amplitude $A$ is given in equation (\ref{inlib}).}
\label{delta}
\end{figure}

The relations in  equation (\ref{apr2}) imply that 
$\boldsymbol {\omega}_{m,g} =\boldsymbol{\omega}_{g,g}+  \dot{\boldsymbol\alpha}_m\in \K_g$,  and
so $
\boldsymbol \omega_{m,pr}=  \boldsymbol{\omega}_{g,pr}+\mat R_3(\phi_g)
  \dot{\boldsymbol\alpha}_m$.
Using equation (\ref{XY}) we obtain 
\[
  \dot {\boldsymbol\alpha}_m=-\dot \phi_g\mat X\dot{\mat R}_3^{-1}(\phi_g)\mat e_1 =
 - \dot \phi_g\mat X\mat R_3^{-1}(\phi_g)\underbrace{\mat R_3(\phi_g)
    \dot{\mat R}_3^{-1}(\phi_g)}_{-\mat{\widehat e}_3}\mat e_1 =
   \dot \phi_g\mat X\mat R_3^{-1}(\phi_g)\mat e_2 
\]
and using equation (\ref{ogpr}) and
the same reasoning for the core angular velocity we obtain, up to small errors,
\begin{equation}
  \begin{split}
     \boldsymbol \omega_{m,pr} &=   \omega \mat e_3+\dot\psi_g\sin\theta_g\mat e_2+
    \dot \phi_g\mat R_3(\phi_g)\mat X\mat R_3^{-1}(\phi_g)\mat e_2\\
      \boldsymbol \omega_{c,pr} &=  \underbrace{ \omega \mat e_3+\dot\psi_g\sin\theta_g\mat e_2}_{\boldsymbol{\omega}_{g,pr}}+
      \dot \phi_g\mat R_3(\phi_g)\mat Y\mat R_3^{-1}(\phi_g)\mat e_2\,.
    \end{split}
    \label{alphamcdif}
  \end{equation}
Note that    $\boldsymbol \omega_{m,pr}$
and $\boldsymbol \omega_{c,pr}$ are not  stationary in  $\K_{pr}$ unless $\mat X$ and $\mat Y$ commute
with $\mat R_3(\phi_g)$. However, the images of
$\phi_g\to \mat R_3(\phi_g)\mat X\mat R_3^{-1}(\phi_g)\mat e_2$  and
$\mat R_3(\phi_g)\mat Y\mat R_3^{-1}(\phi_g)\mat e_2$ are circles in the $(\mat e_1,\mat e_2)$ plane
of $\K_{pr}$ that are centred, respectively, at the points
\begin{equation}
\mat x:=\begin{bmatrix}\frac{x_{12}-x_{21}}{2}\\
    \frac{x_{11}+x_{22}}{2}\\0\end{bmatrix}\quad\text{and}\quad
  \mat y:=\begin{bmatrix}\frac{y_{12}-y_{21}}{2}\\
     \frac{y_{11}+y_{22}}{2}\\0\end{bmatrix}\,.
   \label{xycenters}
 \end{equation}
  The expression for the angular velocity of the mantle
 can be written as (see Figure \ref{delta}):
 \begin{equation}\begin{split}
 &    \boldsymbol \omega_{m,pr} =   \omega \mat e_3+\dot\psi_g\sin\theta_g\mat e_2+
    \dot \phi_g  \mat x+  \dot \phi_g A \begin{bmatrix} \cos(2 \phi_g+h)\\  \sin(2\phi_g+h)\end{bmatrix}\\
    &A=\frac{1}{2}\sqrt{(x_{12}+x_{21})^2+(x_{11}-x_{22})^2}\,,\quad
    \tan h=\frac{x_{22}-x_{11}}{x_{12}+x_{21}}\,.
 \end{split}\label{inlib}
 \end{equation}
 Similar expressions hold for $ \boldsymbol \omega_{c,pr}$.

The offsets of the normalised-angular velocity of the mantle and of the core
from the angular velocity of the Cassini state of the  rigid-body
are given by the nondimensional vectors
   \begin{equation}\begin{split}
     \boldsymbol \delta_m&:=
     \frac{\langle \boldsymbol \omega_{m,pr}\rangle-\boldsymbol{\omega}_{g,pr}}{\dot\phi_g}=
       \mat x\in\K_{pr}
        \quad \text{Mantle offset}\\
          \boldsymbol \delta_c&:=
        \frac{\langle \boldsymbol \omega_{c,pr}\rangle-\boldsymbol{\omega}_{g,pr}}{\dot\phi_g}=
        \mat y\in\K_{pr}
        \quad \, \text{ Core  offset}\,,
      \end{split}
      \label{ofdef}
    \end{equation}
    where    the brackets indicate that
    we are neglecting the oscillatory parts of $ \boldsymbol \omega_{m,pr}$ and
    $ \boldsymbol \omega_{c,pr}$. We recall that terms of the order $(\dot\psi_g/\omega)^2$
    were neglected during the computations.
  The angles $\dt_m$ have the  geometric interpretation given in Figure \ref{delta}.

    In the following
    paragraphs we will present different approximations 
     for the inertial offsets of  bodies with and without  spin-orbit resonances.
     These approximations will depend on the complex compliance at frequency $\dot\phi_g$,
     $C=C_r+i\,C_i=C(i\, \dot \phi_g)\approx C(i\, \omega)$, that is introduced in the problem
     by means of  equation $\boldsymbol\delta\mat B=
    C(i\, \dot \phi_g)\frac{1}{\omega^2} \boldsymbol \delta \mat F_{\boldsymbol \alpha}$
    (equation (\ref{love2}) with $\lambda=i \dot \phi_g$). The equation  $\boldsymbol\delta\mat B=
   \frac{C}{\omega^2} \boldsymbol \delta \mat F_{\boldsymbol \alpha}$
is used
    to eliminate the deformation variables from
    the problem.

 \subsection{Axial precession without  spin-orbit resonance}   
 \label{subwithout}

 In this Section we investigate the  mantle offset
  of a body without
  spin-orbit resonance. The situation is that described  below equation
  (\ref{c2}), in which several point masses may orbit an extended body with precessing
  Keplerian orbits and no spin-orbit
  resonances.  

  As before  the  guiding motion
  $\mat R_g=\mathbf{R_3}(\psi_g)\mathbf{R_1}(\theta_g)\mathbf{R_3}(\phi_g):\K_g\to\kappa$
  will be  that of a rigid body with body frame $\K_g$. The theoretical determination
  of the guiding motion may be done  as in the case where a  spin-orbit resonance exists.
  The averaging of the equations of motion in the  precessional frame and the requirement that the guiding
  frame is a solution to these equations lead again to equation (\ref{Kprav}). In this case, 
the  average value of $\delta J_{12}$ is zero and, for a single orbiting point mass,
{\footnotesize \begin{equation}
  \begin{split}
    \big\langle \delta J_{23}(t)\cos (t \dot\phi_g)\big\rangle &=
 \big\langle\delta J_{13}(t)\sin (t \dot\phi_g)\big\rangle=
-
 \left(\frac{3G m_p}{2a_p^3}\right)X^{-3,0}_0(e)\frac{1+ 3 \cos(2\iota_p)}{16}\sin(2\theta_g)\\
  \big\langle \delta J_{23}(t)\sin (t \dot\phi_g)\big\rangle &=
 \big\langle\delta J_{13}(t)\cos (t \dot\phi_g)\big\rangle=0\,,
 \end{split}\label{avJ2}
\end{equation}}
If there are several point masses, then  we must add the contribution of each one. 
 The non-null average quantities become
\begin{equation}
     \big\langle \delta J_{23}(t)\cos (t \dot\phi_g)\big\rangle =
 \big\langle\delta J_{13}(t)\sin (t \dot\phi_g)\big\rangle=
-\omega^2\frac{\sin(2\theta_g)}{4}\sum s_\beta\,,
 \label{avJ3}
\end{equation}
 where $s_\beta$  expresses the contribution
   of the point mass $\beta$ with $s_\beta$ defined as  in equation (\ref{c1nores2}), namely
\[ s_\beta=
     \frac{3{\cal G}m_\beta}{\omega^2 a_\beta^3} (1-e_\beta^2)^{-3/2}\,
     \frac{1+3 \cos (2 \iota_\beta)}{8}\,.
  \]
In the case of no spin-orbit resonance, equation (\ref{Kprav}) is verified if 
   \begin{equation}
   \dot\phi_g \dot\psi_g\sin\theta_g=
    -\ov\alpha_e\frac{\sin(2\theta_g)}{2}\omega^2\sum s_\beta:={\cal P}\,,
    \label{pealeanalogue}
  \end{equation}
  where terms of order $\dot\psi_g^2$ were neglected.
  This equation:   substitutes the equation (\ref{peale}) of Peale for the case of spin-orbit
  resonance, 
   coincides with equation  (\ref{dotpsi}) up to first order in $\dot\psi_g/\omega$, and 
   can be found in
  \cite{williams1994contributions}.

Since our main example is the Earth with its  axial
 precession due to 
 the  Moon and  Sun, we assume that $\ov I_1=\ov I_2$ and  $I_{c1}=I_{c2}=I_{c3}(1-f_c)$.
 In this case it can be checked that the equations for the forcing terms,
  namely (\ref{ofeq2}), (\ref{tcore}), and 
  (\ref{ofeq4}), hold provided  ${\cal P}$ is that  in equation
  (\ref{pealeanalogue}) and ${\cal R}=0$. In particular, equation (\ref{ofeq2})
  implies that the only torque that acts upon the mantle is the inertial torque
  $\ov{\mat I}_c\dot{\boldsymbol \omega}_{s,g}$.

  If we assume that $\sum s_\beta$ is a small quantity
  \footnote{\label{c1comments} For a point mass of mass $m_\beta$,
with  a low eccentricity and low
   inclination ($\iota_\beta$) orbit of anomalistic mean motion 
   $n_\beta$, $ s_\beta \approx \frac{3}{2}( \frac{n_{\beta}}{\omega})^2
   \frac{m_\beta}{m_\beta+m}$.
 If we assume that  the orbit inclinations
      $\iota_\beta$ of the tidal generating point masses satisfy  $0<\iota_\beta<\frac{\arccos (-1/3)}{2}$,
      then  $s_\beta>0$  and $\dot\psi_g<0$.}, then equation
    (\ref{pealeanalogue}) implies $\frac{|\dot\psi_g|}{\omega \ov\alpha_e}\ll 1$.
     Motivated by this,
    we will assume  the   hypothesis:
 \begin{equation}
     \epsilon:= \frac{\dot\psi_g}{\omega }
    \max\left\{\frac{1}{f_c},\frac{1}{\ov\alpha_e}\right\}\ll 1
     \label{epsilondef0}
   \end{equation}
   (for  the Earth  $\epsilon=3.5\times 10^{-10}$).
   We recall that the compliance
   $|C|$ is comparable to the  flattening coefficients (equation (\ref{alphaid})) and,
   as before, we assume that $\frac{\eta_c}{\omega}$ is at most of the order of $f_c$ (for the Earth
   $\frac{\eta_c}{\omega f_c}\ll1$).
    
   The vector  $\boldsymbol \delta_{m}$  can be computed using an algebraic manipulator (Mathematica)
   but the expressions are too long to be useful. However, up  to first order in quantities of order
   $\epsilon$  the expression becomes simpler and we obtain 
 \begin{equation}\begin{split}
     \delta_{m1}&=2 \cot \theta_g \left(c_1\, \frac{C_i}{\ov\alpha_e}+\frac{\dot\psi_g  \cos\theta_g}{\omega}
 \frac{{\rm I}_{\circ c}}{\Io}\frac{\frac{\eta_c}{\omega}\frac{{\rm I}_{\circ m}}{\Io}}{|y|^2}\right)\\
       \delta_{m2}&=2 \cot \theta_g \left(c_1\, \frac{C_r}{\ov\alpha_e}-\frac{\dot\psi_g  \cos\theta_g}{\omega}
         \frac{{\rm I}_{\circ c}}{\Io}\frac{f_c}{|y|^2}\right)\,,
       \end{split}\label{ofnon}
     \end{equation}
      where $y=f_c+i \frac{\eta_c}{\omega}\frac{{\rm I}_{\circ m}}{\Io}$ ($\approx f_c$ if
      $\frac{\eta_c}{\omega f_c}\ll 1$) is the complex parameter given in equation (\ref{charpf})
      and $ c_1=\frac{1+3 \cos (2 \theta_g)}{4}\sum s_\beta$ is given in equation (\ref{c1nores2}).
       Within the approximations we made, the amplitude of the
      librations $A$ given in equation (\ref{inlib}) is
      zero.

Using that $c_1>0$,  $\dot\psi_g<0$, and $C_i\le 0$ we obtain
      \begin{equation}
        \delta_{m1}\le 0\quad \text{and}\quad  \delta_{m2}>0\,.
        \label{dtin1}
        \end{equation}
        See Figure \ref{delta} for a geometric interpretation of these angles.
         For the Earth, with $\eta_c/\omega=10^{-6}$ \cite{triana2019coupling}, and
        other  parameter values as given in  Section \ref{chandler}, 
        we obtain $\delta_{m1}=-1.96\times 10^{-7}=-0.040''$ and $\delta_{m2}=5.82\times 10^{-5}=12.0''$.

\subsection{Axial precession with  spin-orbit resonance}   
 \label{subwith}

 In this Section we present  approximated formulas for the  offsets $\boldsymbol \delta_m$ and
 $\boldsymbol\delta_c$ for a body in s-to-2 spin-orbit resonance.
  We assume that   the core is an oblate ellipsoid with   $I_{c1}=I_{c2}=I_{c3}(1-f_c)$.
  Our main example is  the Moon in spin-orbit resonance with the Earth.

  In this case equation (\ref{peale}) implies that $|\dot \psi_g|/\omega$  is
  of the order of magnitude of $\ov\alpha_e$. Using this scaling and the algebraic manipulator
  Mathematica, we computed  $\boldsymbol \delta_m$ and  $\boldsymbol \delta_c$
  up to corrections of second order in the ellipticity coefficients.
  The result we obtained is:
 $\boldsymbol \delta_c$
  up to corrections of second order in the ellipticity coefficients.
  The result we obtained is
\begin{equation}\begin{split}
     \delta_{m2}+i\, \delta_{m1}&=\bigg\{C\, (x+y) U_1-\frac{{\rm I}_{\circ c}}{\Io}x^2\tan\theta_g
     +\frac{{\rm I}_{\circ m}}{\Io} (x+y) U_2 \bigg\}\bigg/ den\\ & \\
     \delta_{c2}+i\, \delta_{c1}&=\bigg\{C\, y \, U_1+
     \left(\frac{{\rm I}_{\circ m}}{\Io}\,x +z\right) x\,\tan\theta_g
     +\frac{{\rm I}_{\circ m}}{\Io}\,y\, U_2 \bigg\}\bigg/den\,,
   \end{split}
 \label{ofres}
  \end{equation}
  where: $C$ is the complex compliance (equation (\ref{Ck})), 
  $x:=\frac{\dot\psi_g}{\omega}\cos\theta_g<0$,  $y$ and $z$ are the complex parameters
  given in  equation  (\ref{charpf}),
  \begin{equation}
    den=\frac{{\rm I}_{\circ m}}{\Io}x^2
  +x(y+z)+ y z=\frac{{\rm I}_{\circ m}}{\Io}(x_{\ell a}+x)
(x_{dw}+x) \label{chareq}
\end{equation}
is the characteristic  equation  (\ref{charpf}) evaluated at $x$, 
$x_{\ell a}$ and $x_{dw}$   are the roots of the characteristic equation, 
  {\footnotesize\begin{equation}
      U_1=\left(\frac{3 G m_p}{2 a_p^3\omega^2}\right)^2 \sin \chi_p
      \left\{\big(X^{-3,0}_0\big)^2\cos\chi_p\left(-1+\frac{3}{2}\sin^2\chi_p\right)-
        \big(X^{-3,2}_s\big)^2 \cos ^6\left(\frac{\chi_p }{2}\right)\right\}\,,\label{U1}
    \end{equation}}
  and
  \begin{equation}
      U_2=\frac{{\cal R}}{\omega^2} \left(\frac{\ov \alpha_e}{2} c_2+
        \frac{\ov\gamma }{4} c_1-(c_1+1)c_2 \,C\right)\,,
    \label{U2}\end{equation}
where $\cal R$ is the function in equation (\ref{calR}).
  We recall that $\omega x_{dw}$ and $\omega x_{\ell a}$
  are essentially the eigenfrequencies of the nearly diurnal free wobble (NDFW) and the
  free libration in latitude (FLL)
  in the inertial space. The amplitude of the  offset may become
  large when the precession frequency $-\dot\psi_g\cos\theta_g$ is close to one of these resonance
  frequencies. The function $U_1$ is related to the force upon the deformation variables
  in equation (\ref{ofeq4}) and $U_2$ is a small factor, of second order  in the parameters
  $\ov \alpha_e$ and $\ov\gamma$,
  that becomes relevant only when $\frac{{\rm I}_{\circ c}}{{\rm I}_{\circ m}}\ll 1$,
  which is the case of the Moon.

  The expressions in equation (\ref{ofres}) are complicated enough to be analysed in general, and so
  we restrict the following discussion to the case of the Moon and Mercury.
  The NDFW and FLL periods of free libration in inertial space were computed solving equation
  (\ref{charpf}).
  
  The parameters used for the Moon are those from INPOP and are given in
  Tables \ref{table1}, \ref{table2} and \ref{tab.INPOP}.
  For the Moon the period: of  the NDFW is $469$ years, of the FLL is $80.8$, and of the precession
  ($2 \pi/(-\dot \psi \cos \theta_g)$) is 18.6 years (27.21 days in the guiding frame).
  So, the real part of both quantities
  $x_{\ell a}+\frac{\dot\psi_g}{\omega } \cos\theta_g$ and
  $x_{dw}+\frac{\dot\psi_g}{\omega } \cos\theta_g$ are negative.

  Using equation  (\ref{ofres}) we find  $\delta_{m1}=-8.08\times 10^{-7}$rad
  and  $\delta_{m2}=-1.69\times 10^{-5}$rad (the relative error to the solution  we find  solving
  numerically the linear system for $\mat X$ and $\mat Y$ is about $0.3\%$).
  The angle  $|\delta_{m2}|=3.5^{\prime\prime}$ corresponds to
  displacement of  the angular velocity of the
  mantle away from the pole to the ecliptic, see Figure \ref{delta}.

  The angular displacement  $\delta_{m1}=\,rad=-0.17^{\prime\prime}$
  is different from the observed value $-0.26^{\prime\prime}$
  (\cite{williams2001lunar}   Figure 1)
  but  both  have  the same direction. 
  There is a great uncertainty in the determination of the value of $k_c$
  for the Earth  (the values of the Ekman number vary from  $10^{-11}$ to $10^{-4}$, see
  Footnote \ref{uncertainty}) and the same may happen for the Moon.
  If we multiply by 2.7 the value of $k_c$  we have used for the Moon,
  which is given in
  Table \ref{tab.INPOP}, while  keeping all other parameters with the same values we will
  get the observed
  value $\delta_{m1}=-0.26^{\prime\prime}$. We believe that this change of $k_c$ would not change
  considerably any  other dynamical property of the Moon.

  The orientation of the spin axis of Mercury was studied in several papers. Here we will focus 
   on the results presented in
  \cite{baland2017obliquity}, which ``are valid only if the rotational dynamics
  leading to Mercury's equilibrium spin axis orientation is similar to that of a solid body''.
  We will use equation (\ref{ofres}) to show the importance of a fluid  core in this
  problem. The following parameter values were used in  \cite{baland2017obliquity}
  and will also  be   used here: $C_{20}=-5.03216\times 10^{-5}$, $C_{22}=0.80389\times 10^{-5}$,
  $\iota_p=8.533^\circ$, $\omega= \pi/(58.646\, {\rm day})$, $\dot\psi= -2 \pi/(325513 \, {\rm year})$,
  $e=0.2056$,  $a_p=5.791\times 10^7$km,  $\ov I_3/m R^2= 0.3433$, $m=3.30414\times 10^{23}$kg,
  $R=2440$ km, $k_2=0.5$ and $k_2/Q= 0.00563$.
  From these values  and Peale's equation we obtain $\tilde \chi=5.893\times 10^{-4}$ rad
  ($2.0258^\prime$), which is the angle between Mercury's spin axis and the normal to its orbit. 
  Up  to a negligible error  $\theta_g=\iota_p+\tilde \chi_p$  (see Footnote \ref{thetagfoot}).

\begin{figure}[hptb!]
\centering
\begin{minipage}{0.5\textwidth}
\centering
\includegraphics[width=0.95\textwidth]{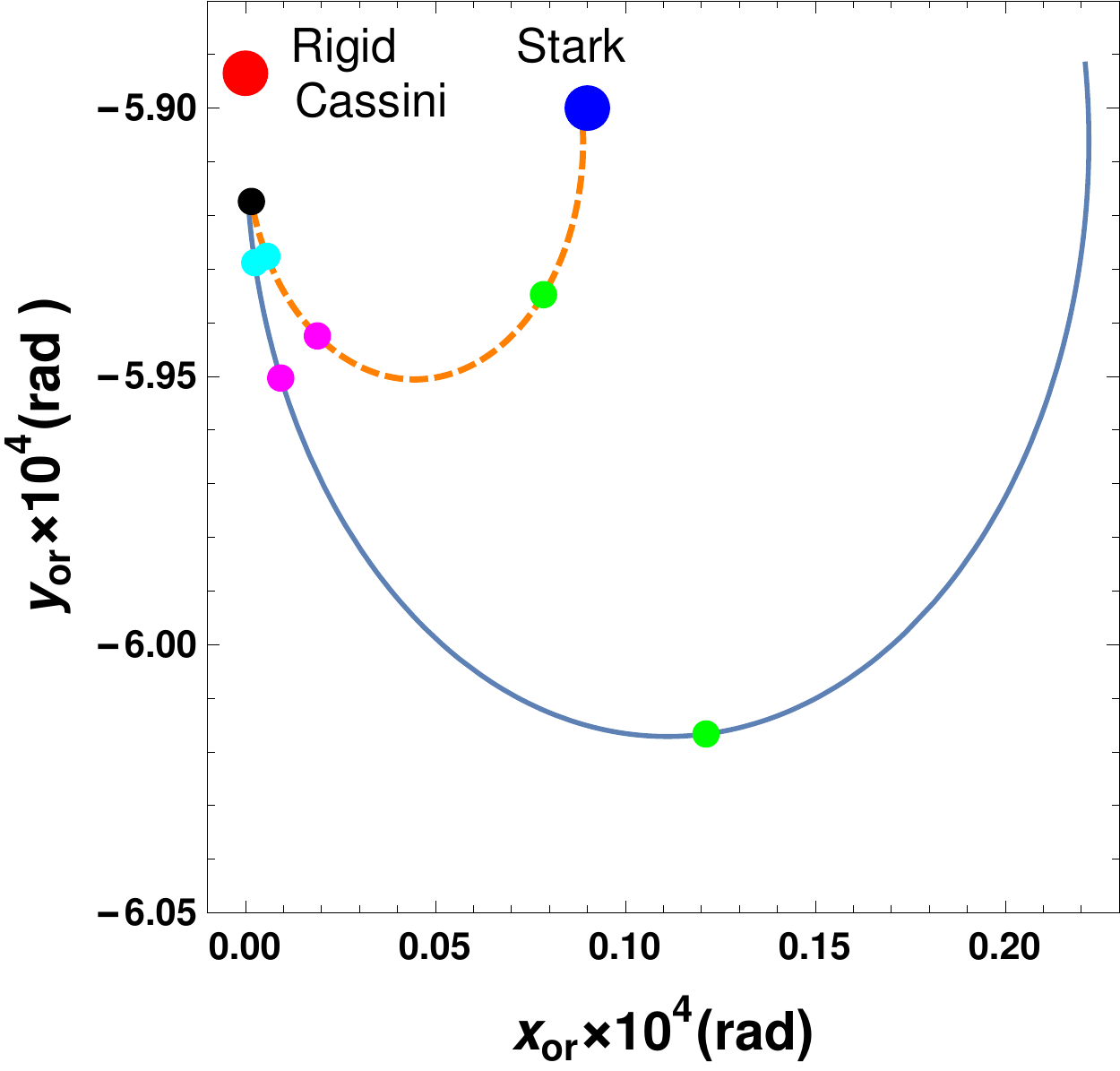}
\end{minipage}\hfill
\begin{minipage}{0.5\textwidth}
\centering
\includegraphics[width=1.0\textwidth]{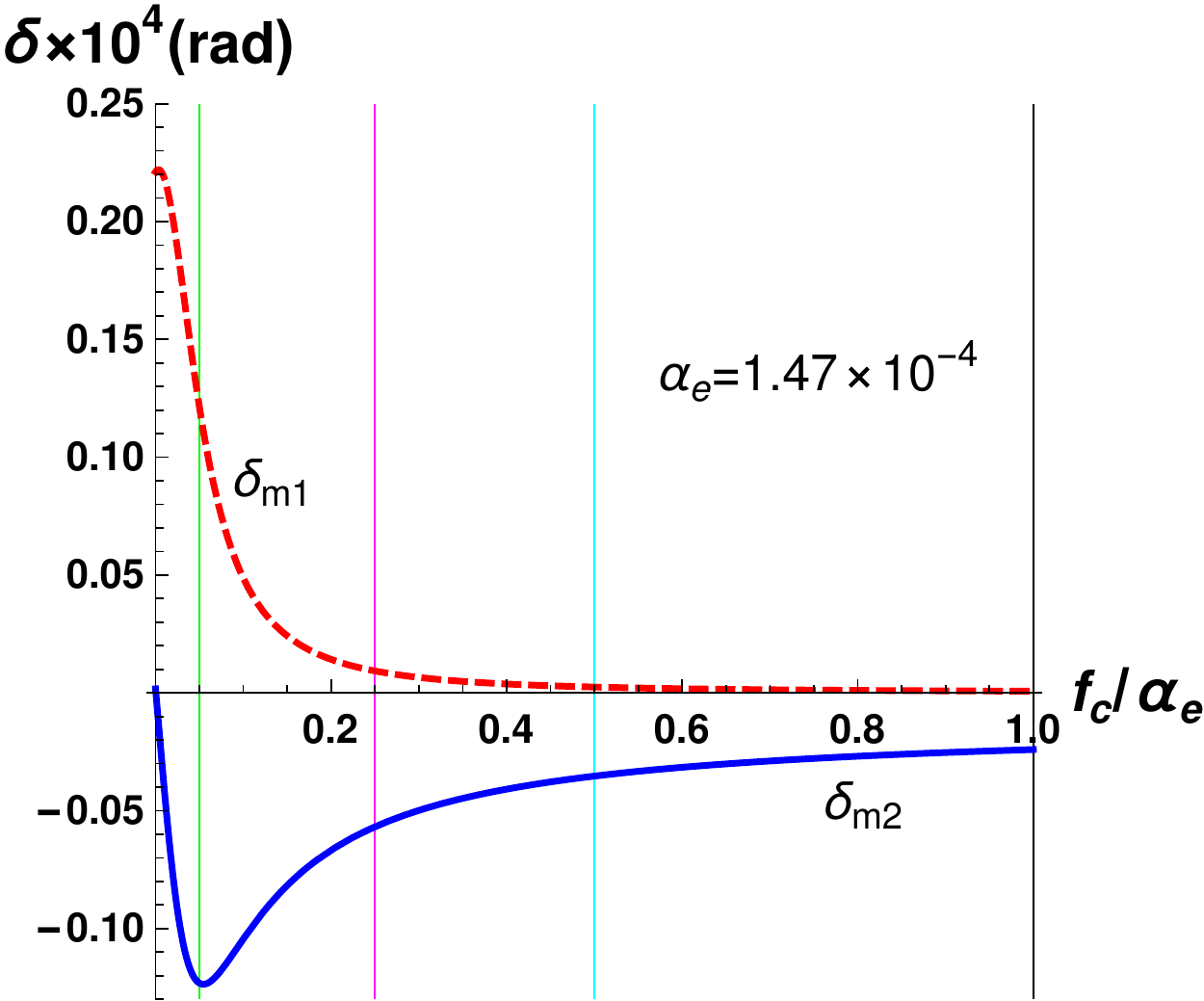}
\end{minipage}\hfill
\caption{LEFT: Diagram showing  average positions of the spin axis of Mercury (normalised to unit norm)
  projected
  on the orbital plane. The $\mat e_1$-axis is in the direction of   the ascending
  node of the orbit on the invariable plane and
  the  $\mat e_2$-axis points towards the pole of the invariable plane. This Figure is the analogue
  of Figure 4 in \cite{baland2017obliquity}. The big red point  ``Rigid Cassini'' indicates
  the position of the spin axis at the Cassini state of the rigid Mercury.
  The big blue point ``Stark'' 
  represents the spin orientation at a certain date
  obtained from observations in \cite{stark2015a}, as reported in
  \cite{baland2017obliquity}. Each point on a curve represents  the average position
  of the spin of the mantle
  $\langle \boldsymbol \omega_{m,pr}\rangle$ as given in equation (\ref{ofdef}). The vector
  from  the red-rigid-Cassini point to a point on a curve is the offset that corresponds to that
  point. The solid-blue curve corresponds to the reference  value $k_c=\tilde k_c$
   while the dashed-orange
   curve  corresponds to the value $k_c=2.5\tilde k_c$. The value of $f_c$ varies from
   $\tilde f_c=1.466\times 10^{-4}$ at the black point  to zero at the other end of the curve.
   The points of colours cyan, magenta, and green corresponds  to $f_c=0.5 \tilde f_c$,
   $f_c=0.25 \tilde f_c$, and  $f_c=0.05 \tilde f_c$, respectively.
    For all the
  range of variation of $f_c$,
  the amplitude of the librations of the mantle  ($A$ in equation (\ref{inlib})) are less than
  3 \% of the magnitude of the average offset. RIGHT: The Figure contains
   the graphs of $f_c\to \delta_{m1}$ (dashed-red curve) and  $f_c\to \delta_{m2}$ (solid-blue curve)
   for $k_c=\tilde k_c$. The vertical lines correspond to the values of $f_c$ of the corresponding
   coloured
   points  in the figure at the left-hand side.
   }
\label{Mercury}
\end{figure}

   Our model requires three parameters for the core. We will fix
   ${\rm I}_{\circ m}/\Io=0.425$ \cite{margot2018mercury}, and choose two reference values
   $\tilde k_{c}/\Io=5.35\times 10^{-12}\, s^{-1}$  \cite{2014Peal}
   and $\tilde f_{c}=\ov\alpha_e=1.466\times 10^{-4}$.
   We computed the inertial offset of the mantle for  several values of $k_c$ but decided to
   present the results  only for two: $k_c=\{\tilde k_{c}, 2.5\,\tilde k_{c}\}$.
   The value of the core oblateness $f_c$ was varied
   continuously from zero to its reference value. For the reference values the two periods
   of free-libration $2\pi/(\omega x_{dw})=287$ years and $2\pi/(\omega x_{\ell a})=2100$ years 
    are much smaller than $2\pi/(-\dot \psi_g\cos\theta_g)=330$ kyr. As $f_c$ decreases
    to zero the period of the NDFW increases and eventually it reaches  $330$ kyr.
    At the resonance
   the amplitude of the offset is controlled by the viscosity parameter $k_c$. The  inertial offsets
   we obtained are shown in Figure \ref{Mercury}. The results in this figure show that:
   both parameters $f_c$ and $k_c$ are important in the determination of the inertial offset,
   and for $f_c$ small the magnitude of the inertial offset may be enough to explain the displacement
   the position of Mercury' spin observed in   \cite{stark2015a}\footnote{ \label{baland}
It is enough to make ${\rm I}_{\circ c}=0$ in 
equation (\ref{ofres}) to obtain the inertial offset of a deformable solid body, which is the case
considered in \cite{baland2017obliquity}. Equation (\ref{ofres}) with the parameters in ibid. 
gives 
$\delta_{m1}=0.003''$ and $\delta_{m2}=-0.0055'$. The values obtained in ibid.
(see Figure 4 and Table C3) are: $\delta_{m1B}=\epsilon_\zeta=0.995''$ and
$\delta_{m2B}=\Delta \epsilon_\Omega=-0.006'$. The signs of the angles agree, 
$\delta_{m2}\approx 0.92 \, \delta_{m2B}$, but the value we obtained for
$\delta_{m1}\approx \delta_{m1B}/260$ is much smaller.}.

\section{The Chandler wobble period of the Earth and the
 necessity of   complex rheological models.}
\label{chandler}

Our simplified model for the Earth, which ignores the existence of a solid inner core, admits
only two modes of  torque-free librations: the nearly diurnal free wobble (NDFW) and the Chandler's wobble.
As mentioned in the previous section (see footnote \ref{NDFW})  our model
cannot provide the correct period of the NDFW.
 The Chandler wobble period seems to be mostly determined by the rheology
 of the mantle (paragraph [61] of \cite{mathews2002modeling}) and
 not  sensitive to  other variables, e.g to the deformation of the core-mantle boundary.
So, the Chandler's
wobble  is within the range of applicability of our model.

 The Chandler's wobble eigenvalue can be estimated using equation (\ref{lambdaw2}):  
\begin{equation}\begin{split}
     \lambda_w&=i\underbrace{\,\omega\,\frac{\Io}{{\rm I}_{\circ m}}\Big(\ov \alpha_e-C(0)\Big)}_{:=\sigma_w}
-\eta_c\frac{{\rm I}_{\circ c}}{{\rm I}_{\circ m}}\Big(\ov \alpha_e-C(0)\Big)
-   \tau\omega^2 C(0)\frac{\Io^2}{{\rm I}^2_{\circ m}}\Big(\ov \alpha_e-C(0)\Big)\\
&=\bigg(i
-E_k\frac{{\rm I}_{\circ c}}{{\rm I}_{\circ}}
-   \tau\omega C(0)\frac{\Io}{{\rm I}_{\circ m}}\bigg)\sigma_w
\end{split}
                      \label{lambdaw3}
                    \end{equation}
                    where: $\sigma_w$ is the Chandler's wobble frequency,
                    $E_k=\eta_c/\omega$ is the Ekman number of the flow at the core mantle boundary, and
 $C(0)=\omega^2/(\gamma+\mu_0)$.

  Following \cite{zhang2020new} we use the following values for the inertial coefficients
  \[
    \frac{\Io}{{\rm I}_{\circ m}}=1.13213\qquad \ov \alpha_e= 0.0032845\,.
  \]
  The values of $\gamma$ and $\omega$ for the Earth are given in  Table  \ref{table2} and
  so, $\sigma_w$ is determined by the prestress elastic constant $\mu_0$. If we use the value of $\mu_0$
  provided in Table \ref{table2}, then we obtain $\sigma_\omega=1.90\times 10^{-7}\,sec^{-1}$
  that corresponds to the period 382.5 days.
This value is far from that obtained from observations
 ($\approx 433$ days   \cite{vondrak2017new})
 but it is within the interval (381.9,385)  obtained in \cite{mathews2002modeling}
 Table 3A.

In   \cite{mathews2002modeling}, the explanation for the difference  
 between the estimated frequency and
 the observed one
 is that the rheological behaviour of the Earth, encoded in
 its complex Love numbers,   is frequency dependent and the Love number they
 used to obtain $\sigma_w$ was at the 
 diurnal frequency and not  at the Chandler wobble frequency.
 Using our notation, the equation used in
 \cite{mathews2002modeling} to obtain the wrong Chandler's wobble period was
 $  \sigma_w=\omega\,\frac{\Io}{{\rm I}_{\circ m}}\Big(\ov \alpha_e-C(\omega)\Big)$
 while, as explained in their Appendix D, they would get the correct period if they
 had used $  \sigma_w=\omega\,\frac{\Io}{{\rm I}_{\circ m}}\Big(\ov \alpha_e-C(\sigma_w)\Big)$.
 
\begin{figure}[hptb!]
\centering
\begin{minipage}{0.5\textwidth}
\centering
\includegraphics[width=1.0\textwidth]{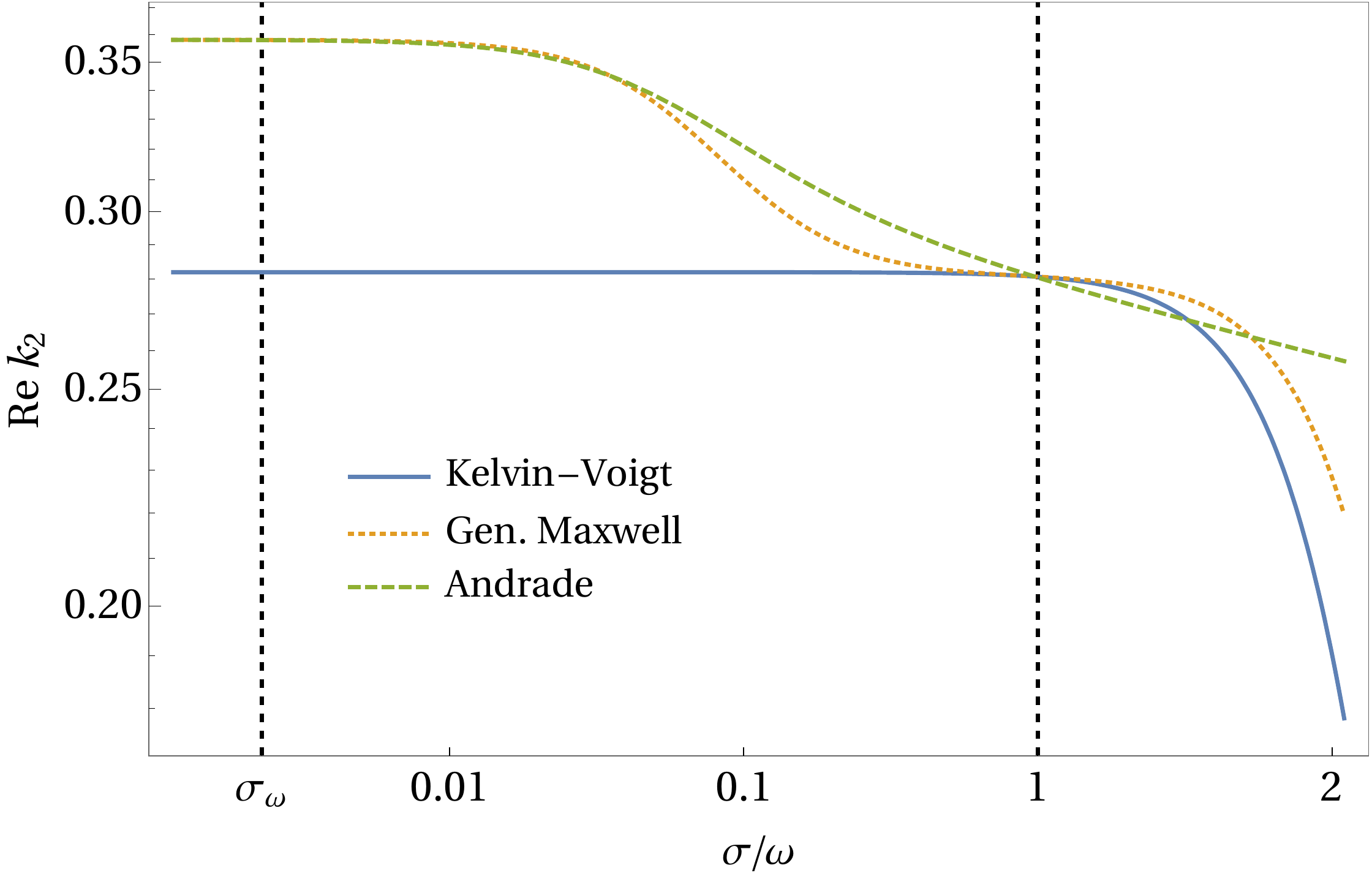}
\end{minipage}\hfill
\begin{minipage}{0.5\textwidth}
\centering
\includegraphics[width=1.0\textwidth]{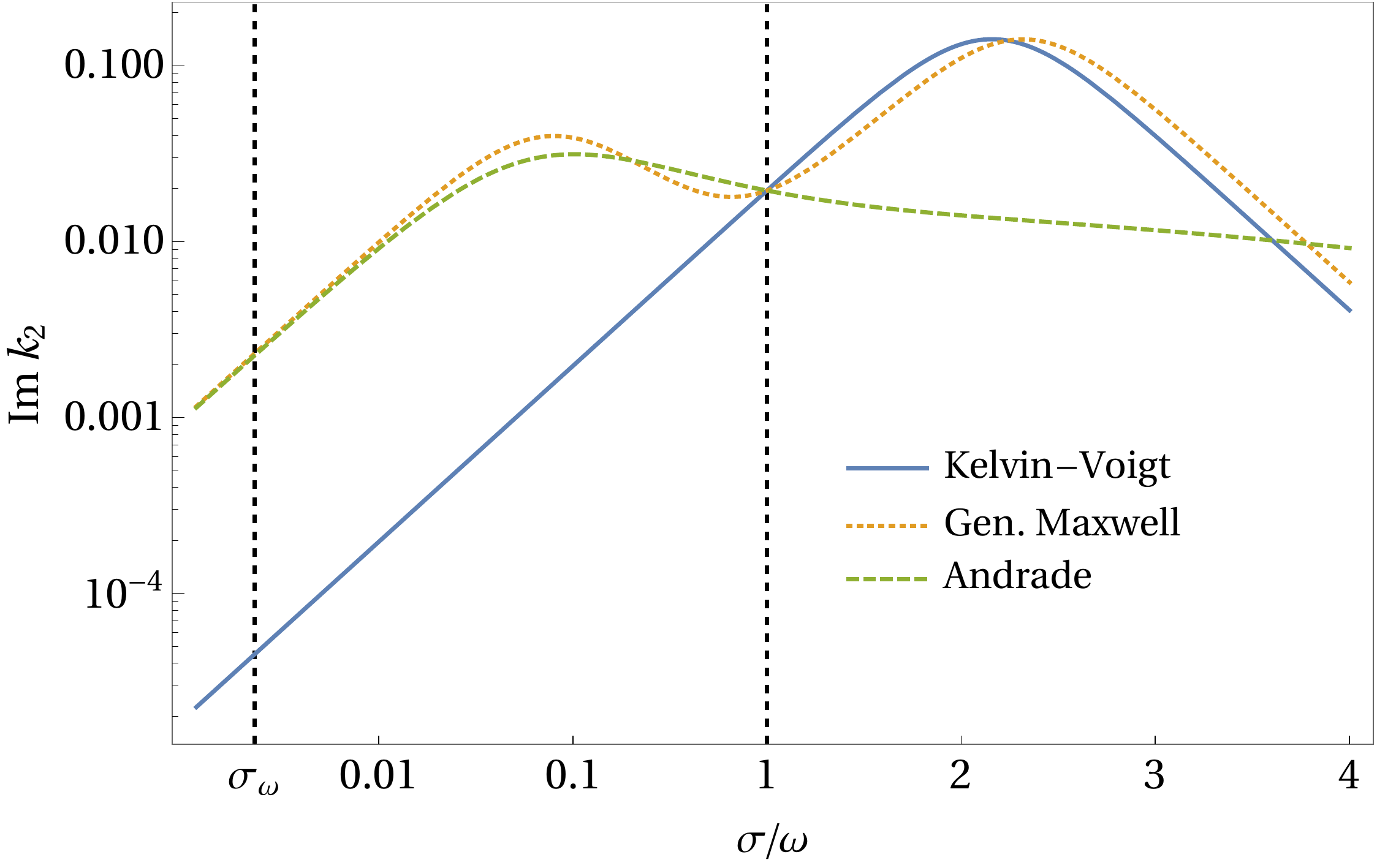}
\end{minipage}\hfill
\caption{Comparison between the Love number (real part in the left and imaginary part in the right)
  generated by: a Kelvin-Voigt rheology
  ($\frac{\mu_0}{\omega ^2}=712\,,\frac{\eta}{\omega}=70.8$),
a generalised Maxwell rheology ($\frac{\mu_0}{\omega ^2}=495\,,\frac{\mu_1}{\omega ^2}=219\,,
\frac{\eta}{\omega}=49.1\,,\frac{\eta_1}{\omega}=2200$), and an Andrade rheology
($\frac{\mu_0}{\omega ^2}=495\,,\frac{\mu_1}{\omega ^2}=728\,,\frac{\eta_1}{\omega}=2250
\,,\tau_{A}\omega=0.0151\,,\alpha=0.2$).
The Kelvin-Voigt model was calibrated with the Love number at diurnal frequency and
the generalised Maxwell and the Andrade models were
calibrated with the Love number at  both diurnal and  Chandler's wobble
frequencies. In the left figure  and for the Kelvin-Voigt rheology
the ratio $ {\rm Re} \left( \frac{k(\sigma)}{k(0)}\right)$
 decreases from
1 at $\sigma=0$ to $0.995$ at $\sigma=\omega$, which shows that  real part of the Love number is almost
constant for $\sigma\le \omega$.}
\label{Lovefig}
\end{figure}

 The explanation for the failure of our estimate of the Chandler's wobble period
 is similar but not the same as  that in \cite{mathews2002modeling}.
 As in  ibid.,
 we also calibrated $\mu_0$ using the Love number at the diurnal frequency. But we do have a model for the
 rheology that allows for the variation of the Love number $k(\sigma)$ with the frequency
 and so,  instead of using the approximation $k(\sigma_w)\approx k(\omega)$, as \cite{mathews2002modeling}
 did, we used  $k(\sigma_w)\approx k(0)$, as suggested by the argumentation in Section
 \ref{free2}. Our failure occurred because
  the variation of the Real part of the Love number
  of the Earth  for $\sigma\in(0,\omega)$ is very small when we use
  the  Kelvin-Voigt rheology for  the mantle, see Figure \ref{Lovefig} left.
  This is also the reason for us to have obtained the same wrong
  period for the Chandler's wobble as \cite{mathews2002modeling}.

The above argumentation shows that the mantle of the Earth
requires a more complex rheological model than that of Kelvin-Voigt.
In the following  we use two different rheologies
to fit the Love numbers at  both diurnal and Chandler's wobble frequencies:
a generalised Maxwell rheology with four parameters ($\mu_0,\eta,\mu_1,\eta_1$)
and an Andrade rheology with prestress, for which the complex Love number is  
\begin{equation}
  k(\sigma)=
  \frac{3\Io G}{R^5}\frac{1}{\gamma+\mu_0+J_A^{-1}(\sigma)}\,,\quad
  J_A(\sigma)=\frac{1}{\mu_1}+\frac{1}{i\sigma\eta_1}+\frac{\Gamma(1+\alpha)}{\mu_1(i\omega \tau_A)^\alpha}\,.
\label{LoveAndrade}
\end{equation}
If we set  $\alpha=0.2$, as in \cite{efroimsky2012tidal}, then the Andrade rheology
will have four free parameters
($\mu_0,\eta_1,\mu_1,\tau_A$).

The period of the Chandler's wobble  $\approx 433$ days and  equation 
(\ref{ksigmaw}) imply that the real part of the Love number at the Chandler's wobble
frequency 
is approximately  $0.358$.
If we assume  that ${\rm Re}\,  k(\sigma_w)\approx k(0)$, as we did in Section \ref{free2},
then $  C(0)= \frac{\omega^2}{\gm+\mu_0}=
\frac{\omega^2 R^5}{3\Io G}\, k(0)$ implies $\mu_0/\omega^2=495$. There are several
estimates of the real part
of $\lambda_w$ using observations, here we use 
${\rm Re}\, \lambda_w=- \sigma_w/(2\times Q_w)$ where $Q_w=127$ \cite{nastula2015chandler}
\footnote{In \cite{nastula2015chandler}, and in many other references, we
  find the value of the ``quality factor'' $Q_w$ associated with the Chandler's wobble and not
  ${\rm Re}\, \lambda_w$. The two quantities are related by ${\rm Re}\, \lambda_w=- \sigma_w/(2\times Q_w)$.
  Note: the quality factor $Q_w$ is not $Q(\sigma_w)$. The reason for this difference is the following.
  In equation (\ref{k2dt}) the quality factor is given by  $Q^{-1}(\sigma)=\sin(\delta(\sigma))$,
  where $\delta(\sigma)$ is the phase lag
  of the body response to a tidal force with frequency $\sigma$. The Chandler's wobble eigenvalue
  $\lambda_w$ is related to  a natural mode of oscillation of the body in the absence of
  oscillatory-external forces
  and so, it cannot be associated with a phase lag. The relation
  ${\rm Re}\, \lambda_w=- \sigma_w/(2\times Q_w)$ comes from the definition
  $Q_w^{-1}=\frac{1}{2\pi E}\int\dot E dt$, where $E$ is the energy stored and
  $\frac{1}{2\pi}\int\dot E dt$ is the energy dissipated per cycle, applied to the free oscillations
  of an underdamped harmonic
  oscillator. In \cite{gross2007earth} (Table 11) there is a review of several estimates
  of $Q_w$, which are in  the range $(30,1000)$, and in \cite{vondrak2017new})
  we find the estimate $Q_w=35$.}. If we neglect the contribution of the friction at the  CMB to the
real part of $\lambda_w$, then we can use  equation (\ref{lambdaw3}) to estimate
$\tau \om=\frac{{\rm I}_{\circ m}}{\Io}C^{-1}(0) Q_w^{-1}/2=2.79$, or $\tau\approx 10.6$ hours.

For the generalised Maxwell model with four parameters ($\mu_0,\eta,\mu_1,\eta_1$)
$\tau=(\eta+\eta_1)/(\gamma+\mu_0)$  (equation (\ref{taugM})). This relation, the value   $\mu_0/\omega^2=495$, 
  the Love number at the diurnal frequency in
Table  \ref{table1} ($k(\omega)=0.2803 - 0.01944\, i$), and  equation
(\ref{Lovegenmax}) imply the values of $\mu_1$ and $\eta_1$ given in the
caption of Figure  \ref{Lovefig}.

For the Andrade model with prestress
\begin{equation}
  \tau=\frac{\eta_1}{\gamma+\mu_0}\,,
  \label{tauAndrade}
\end{equation}
where  $ \tau=-\frac{1}{C(0)}
   \frac{\partial C}{\partial \lambda}(0)$ (equation   (\ref{tau}))  and 
$ C(\lambda)=  \left(\frac{3\Io G}{\omega^2 R^5}\right)^{-1}\, k(-i\lambda)$
  (equation (\ref{Ck})). 
If we fix $\alpha=0.2$, then a computation similar to that done for  the generalised Maxwell model gives
the parameter values listed in the caption of Figure \ref{Lovefig}.

\section{Forced libration within the linear approximation.}

\label{forcedlib}

Formulas for forced librations can be obtained by substitution of the Fourier
decomposition of   force (known) and   angles (unknown)  in the linearized equations.
In this way the Fourier coefficient of each angle is written as a
complex rational function of  the forcing
frequency. While the numerator of these rational functions depend on the
type of forcing, the denominator is always the characteristic polynomial
of the homogeneous system. Since the characteristic polynomial can be factored
using the eigenvalues,  general features of
the body response to  an external force are determinated by the eigenfrequencies and their associated
eigenvectors.

In this Section all the quantitative results are for the  Kelvin-Voigt rheology but qualitative aspects
of the analysis also  apply to  other rheologies.

\subsection{Rigidity$\times$deformation.}
\label{rigiddef}

\begin{figure}[hptb!]
\centering
\begin{minipage}{0.5\textwidth}
\centering
\includegraphics[width=1.0\textwidth]{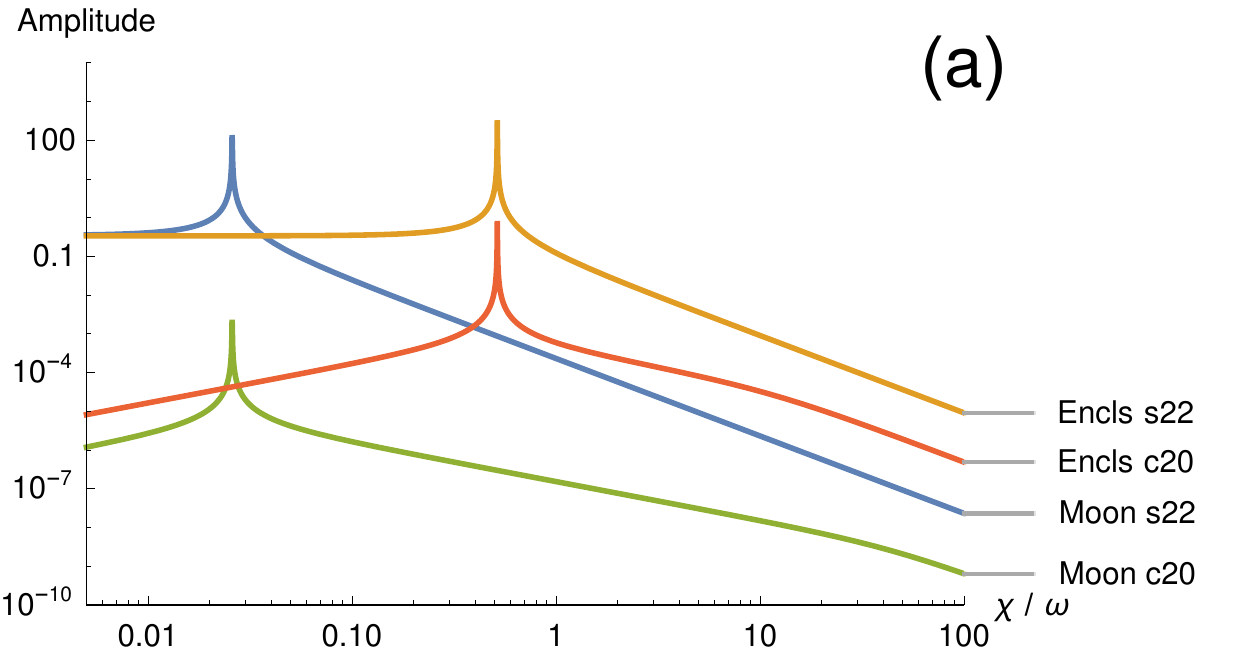}
\includegraphics[width=1.0\textwidth]{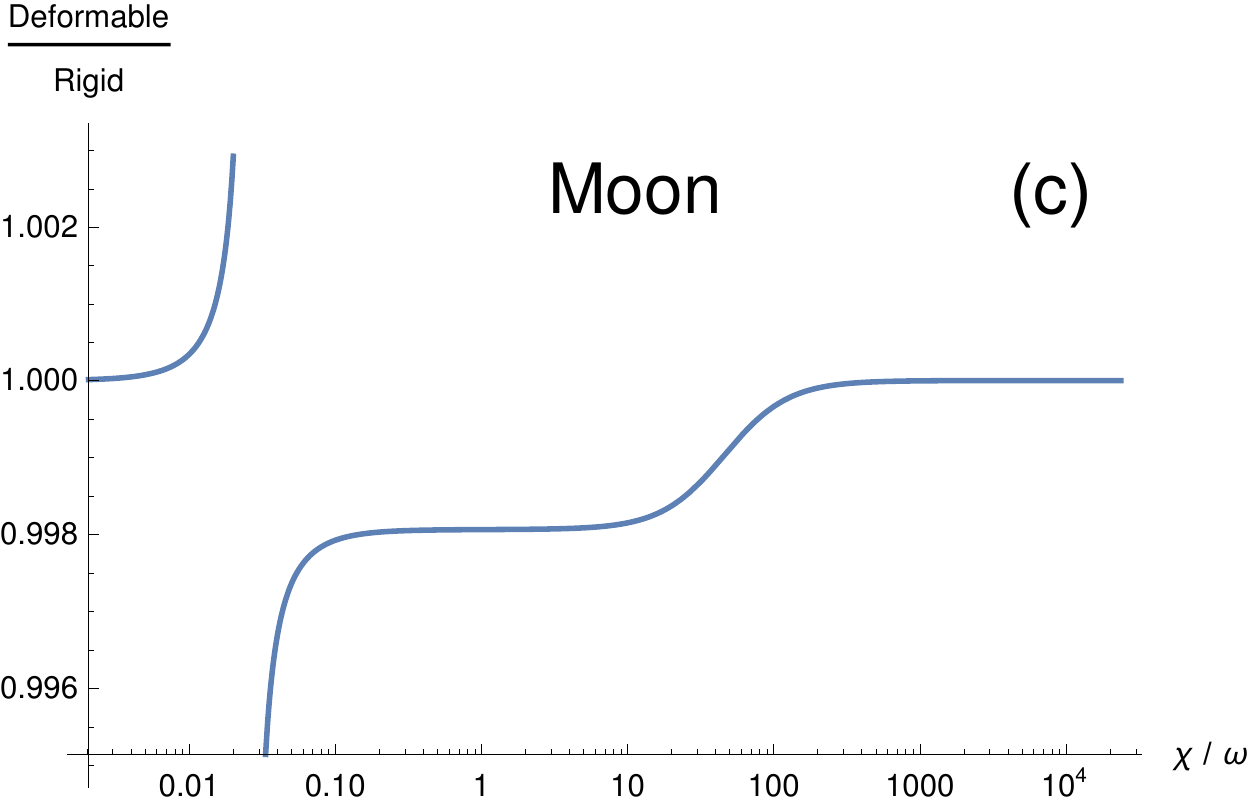}
\end{minipage}\hfill
\begin{minipage}{0.5\textwidth}
\centering
\includegraphics[width=1.0\textwidth]{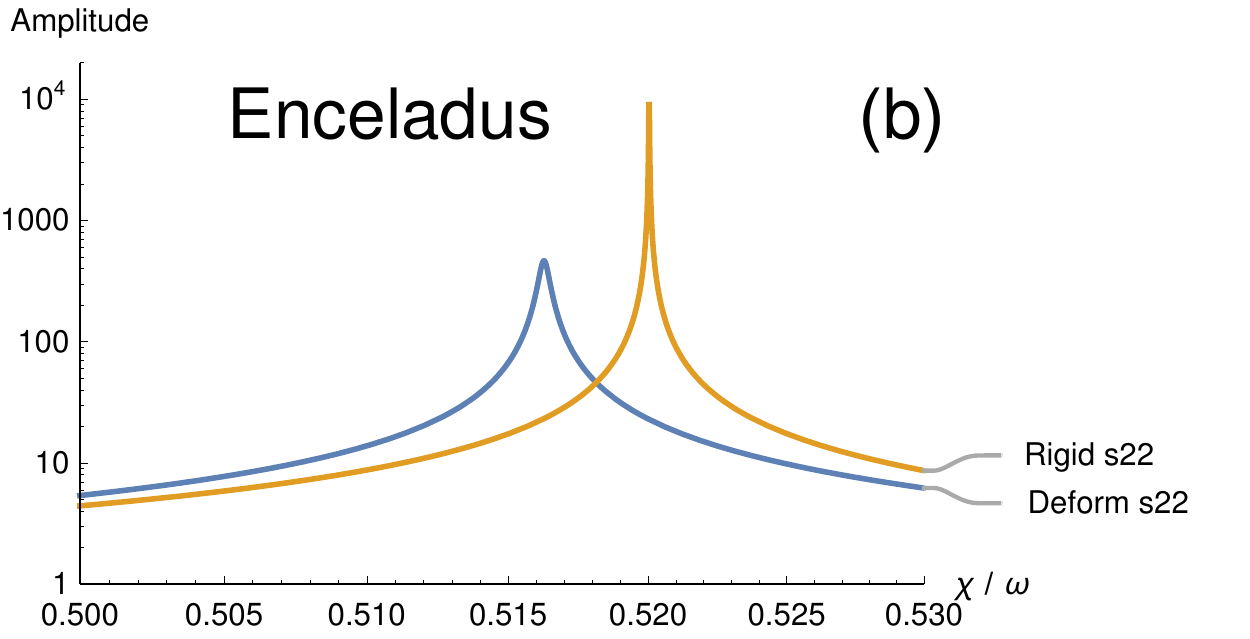}
\includegraphics[width=1.0\textwidth]{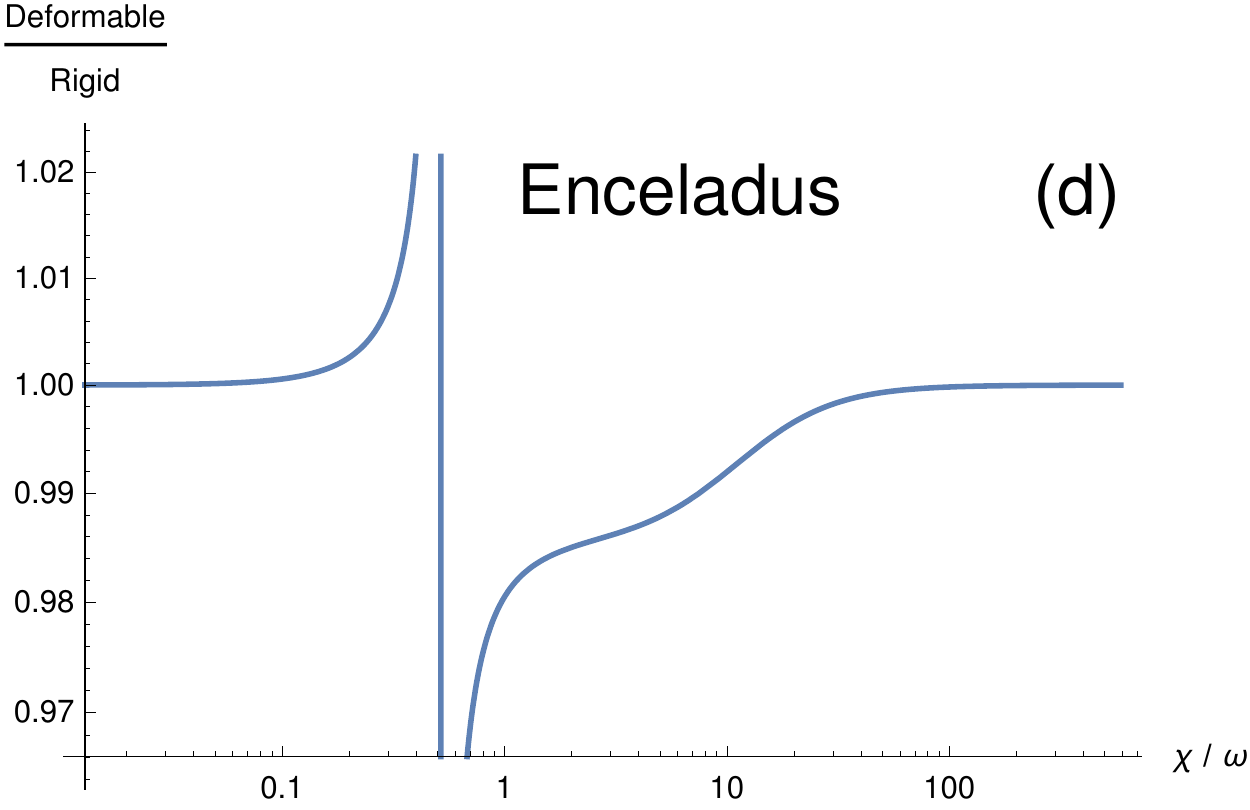}
\end{minipage}\hfill
\caption{Rigidity$\times$deformation. (a) Amplitude of
  longitudinal libration of the Moon and Enceladus, both with  deformable mantles,
for forcing terms of  the form
$(s_{22},c_{20})=(1,0)\exp(i \chi)$ and $(s_{22},c_{20})=(0,1)\exp(i \chi)$. In Figures
(b), (c), and (d) the forcing term is $(s_{22},c_{20})=(1,0)\exp(i \chi)$.
(b) Amplitude of longitudinal libration of Enceladus near its
resonant frequency
($\approx 0.516\, \omega$): comparison between  deformable and rigid mantle.
(c) (Moon) and (d) (Enceladus): Ratio between the amplitude of longitudinal libration
obtained from a model with deformable mantle over that with rigid mantle.
The parameters (Kelvin-Voigt rheology) used to generate the Figure were the following:
  Moon: $\omega =2.662\times 10^{-6}\,s^{-1},\
  f_0=0.000700, \ \ov \alpha =0.000402,\ \ov \beta =0.000636,\
  \ov \gamma =0.000229,\ f_c=0.000158,\ \tau\omega=0.0217 ,\,\ \eta_c/\omega=0.0000158,\ 
  \eta^{-1}\omega=6.99\times 10^{-6}$.
  Enceladus:
  $\omega =5.308\times 10^{-5}\,s^{-1},\
  f_0=3.96, \ \ov \alpha =0.00707,\ \ov \beta =0.0254,\
  \ov \gamma =0.0183,\ f_c=0.0161,\ \tau\omega= 0.0889,\,\ \eta_c/\omega=0.0000320,\ 
  \eta^{-1}\omega=0.000985$.}
\label{Moonamp}
\end{figure}

If tidal variations of the total moment of inertia
 are small compared to the average values, then forced librations are well described by
the rigid-mantle approximation. The deformations of the mantle are important
in both  dissipation of energy  and close to   resonance frequencies.
In   Figure \ref{Moonamp} we illustrate this fact showing  the
amplitude of longitude libration of the Moon and of Enceladus
as a function of the forcing
frequency.  We use two rheological models for the mantle: deformable with the Kelvin-Voigt rheology
and rigid.
The librations  are forced by a hypothetical  point mass with an almost circular
orbit  in the equatorial plane of the extended body.
If the Jeans operator $\mat S_g$ is  written 
using Stokes variables,  
 then  the only terms that
 appear in the right hand side of equations 
 (\ref{eq1}) and (\ref{def1d})  are $s_{22}$ and $c_{20}$\footnote{
   \begin{equation}\renewcommand\arraystretch{2}
       \mat S_g =\mat J_g- \frac{\tr\mathbf{J}}{3}\Id=
       \omega ^2 \left(
\begin{array}{ccc}
 -\frac{1}{2}c_{20}+c_{22} &  s_{22} & c_{21} \\
   s_{22} &  -\frac{1}{2}c_{20}-c_{22} & s_{21} \\
  \ \ \ c_{21} &\ \ \  s_{21} &\ \ \  c_{20} 
\end{array}
\right)\,,
\label{Stokes2}
\end{equation}
Up to first order in the eccentricity:
 \[
   \begin{split}
     c_{20}&=- \frac{G m_p}{a_p^3 \omega^2}\big(1+ e\cos(M_p))\\
     s_{22}&=\frac{3}{2}\frac{G m_p}{a_p^3 \omega^2}\bigg(\sin(2 M_p-2\om t) +\frac{e}{2}
     \Big(-\sin( M_p-2\om t)+7\sin(3M_p-2\om t)\Big)\bigg).\end{split}
 \]
 The notation follows that in Appendix \ref{hansen}.}.
The term $s_{22}$ is proportional to  the longitudinal  torque and $c_{20}$ acts
to flat the body down to the equator,  so it does not affect a rigid body.

In Figure \ref{Moonamp} (a) we show the amplitude of longitudinal libration
of the mantle ($\alpha_{m3}$) for forcing terms of  the form
$(s_{22},c_{20})=(1,0)\exp(i \chi)$ and $(s_{22},c_{20})=(0,1)\exp(i \chi)$,
 where  $\chi>0$ is the 
 forcing frequency. In the scale of the Figure it is not possible to distinguish
 between the response function  of a body with a deformable mantle
 from that of a body with
 a rigid
 mantle. For the same frequency the angular response  to
 $s_{22}$ is much larger than  that to $c_{20}$ (approximately
 1000 times larger for the Moon
 and 200 times for Enceladus). This gives an idea on the relative importance
 of deformation for forced librations in longitude, since $c_{20}$  affects only a body with 
 a deformable mantle.

 In Figures \ref{Moonamp} (b), (c), and (d) the forcing term is of the form
 $(s_{22},c_{20})=(1,0)\exp(i \chi)$.
 In Figure  \ref{Moonamp} (b) the domain of variation of $\chi$ is restricted
 to a small neighbourhood of the resonance frequency of Enceladus.
 In this domain it is possible
 to observe the difference between the amplitude of libration of Enceladus
 with a deformable mantle from that with  a rigid mantle. The graphs for the Moon
 are similar.   
In Figures \ref{Moonamp} (c) and (d) we show the ratio between the amplitude of
 libration  of a body with deformable mantle over that with a rigid mantle. The ratio
 is very close to one except near the resonant frequency.
 The rigid mantle
 approximation is better for the Moon than for Enceladus because the Moon
 is more rigid.

 \subsection {Eigenmodes and forced librations.}
\label{eigfluidsec}
 
 The phase space of
 a forced linear system with constant coefficients
 can be  decomposed into eigenspaces of the homogeneous system.
 If the forcing terms are decomposed into  eigenspace components, then the forced linear
 system decouples into a system of independent oscillators, one for each eigenmode.
 The geometry of the oscillations in a particular  eigenmode is determined by the respective
 eigenvectors.

 The  idea in the previous paragraph can be applied to the rotational eigenmodes of Section \ref{free2}
 with caution, as we now explain.
 In the computation of the rotational eigenvalues and eigenvectors  we were able to
 reduce all the information in the  internal variables of the rheology (represented by extensions of
 the springs and dash-pots of their representation) into the Love number and the deformation variables
 $\boldsymbol \delta\mat B$. In this way we were able to compute the $\boldsymbol \alpha_m$,
 $\boldsymbol \alpha_c$, and $\boldsymbol \delta\mat B$ without having to compute the components
 associated with the internal variables. This approximation  is good close to the frequencies of the free eigenmodes, namely at low
 frequencies for the libration in longitude and wobble modes and close to the diurnal frequency
 for the nearly diurnal free wobble and librations in latitude modes.
 For forced librations where the forcing
 frequency is  far from the natural frequencies the part of the eigenvectors that depends
   of the internal variables may be relevant. In the following discussion we will
 disregard this issue.

 Libration in longitude are oscillations in the two dimensional
 eigenspace associated with the  eigenvector in equation (\ref{ellomode}).
 The ratio between the $\beta_3$-component and the  $\alpha_{m3}$-component of this eigenvector
 is $\beta_3/\alpha_{m3}=-\frac{C(0)}{\ov \gamma}(\xi_2-\xi_1)$.
 As shown in Figure  \ref{figlong} (Right),
 $\beta_3(t)$ is the angle, measured in the  plane orthogonal to $\mat e_3\in\K_m$,
 between the $I_1$-principal axis of the deformable body and its average position, which
 is the $\mat e_1$-axis
 of $\K_m$ (the prestress frame). So,
 $|\beta_3/\alpha_{m3}|=\frac{C(0)}{\ov \gamma}(\xi_2-\xi_1)$\footnote{We remark that the complex compliance
   $C(0)$ in this equation appears because we used the  approximation $C(i\,\sigma_{\ell o})\approx C(0)$ in
   order to obtain an estimate of $\sigma_{\ell o}$. It seems natural to replace $C(0)$ by $C(i\chi)$
   in the case of a forced oscillation.} gives the relative importance of 
 deformation in longitudinal librations\footnote{For the libration in longitude of  the Moon
   and for the Kelvin-Voigt rheology with the data in Table \ref{table2},
 $| \beta_{3}| /| \alpha_{m3}|=3 \frac{\omega^2}{\gamma+\mu_0}/\ov \gamma=2 \times 10^{-3}$.
      In the notation of  \cite{eckhardt1981theory} and \cite{Folkner14}
      $\alpha_{m3}=\tau$ is the longitudinal  angle
      between the ideal Cassini state and the principal axes of 
      `` the undistorted mantle'' and  $\beta_{3}$
      is the longitudinal angle between the
       principal axes of 
       `` the undistorted mantle'' and the principal axes of the ``distorted mantle''.
          From \cite{rambaux2011moon} Table 3 the angle of longitudinal
      libration is
      $1.8^{\prime\prime}$, or 15.23 meters at the equator. 
      According to equation  $|\beta_3/\alpha_{m3}|=\frac{C(0)}{\ov \gamma}(\xi_2-\xi_1)$,
      the part due to the Moon's
       deformation is $0.03$ meters, which is of the order of magnitude
       of the typical horizontal
       displacement of $0.05$ meters due to tidal variations detected by the Lunar Laser
       Ranger
       \cite{williams2008lunar} (p.109).}. The same idea can be applied to the other rotational modes.

 \subsection { Parametric Resonances: limitations of the forced libration model.}

 In our  derivation of the linearized equations  we assumed the hypothesis (\ref{hypsmall2}):
The effect of $ [\boldsymbol{\dt}\mat J_g\,
,\boldsymbol { \widehat \alpha}_m]$ is negligible.  In some situations
an averaging procedure justifies this assumption.
 The convenience of this hypothesis is that the homogeneous
 linearized equations for librations are of constant coefficients and the analysis
 of their solutions is easy. The drawback of this approach is that it prevents the occurrence of parametric
 resonances.

 The equations for free librations with constant coefficients, on the contrary to those for forced librations,
 are useful even when   $ [\boldsymbol{\dt}\mat J_g\,
 ,\boldsymbol { \widehat \alpha}_m]$ is not negligible.
 Indeed,  if
 $\lambda=i \omega_0+\nu_0$ is an eigenvalue of the equation with constant coefficients,
 then parametric resonances may occur  for
 forcing frequencies $\chi$ that satisfy: $\omega_0/\chi\approx j/2$, $j=1,2,\ldots$. So, the
 eigenfrequencies of the equations
 with constant coefficients determine the  parametric-resonance frequencies. Note that 
even a small dissipation of energy, $\nu_0>0$,
 may have an effect on preventing  the parametric resonance to occur, specially for large values
 of $j$  (see, e.g., \cite{arnoldode}).

\section{The Moon in INPOP and in JPL ephemerides}\label{INPOP} 

In the main international ephemerides,
the Moon is assumed to be composed of a homogeneous liquid
core surrounded by a prestressed viscoelastic mantle. This model that we will call INPOP19a
was the main motivation for the present paper.
The geometry of the Moon's core cavity is
fixed and the tidal response of the viscoelastic mantle is represented by the {\em constant time lag
  model}  \cite{INPOPa} \cite{Viswanathan19},  \citep{Folkner14}, and  \cite{JPL}.
The INPOP19a
is formally equivalent to the model we presented in Section \ref{equations}, equations (\ref{eqrot3})
and (\ref{def1k}), except for the gravitational modulus $\gamma$ and for the fact that INPOP19a
incorporate additional physical effects, as for instance figure-figure torques and higher
order gravitational moments.
Since $\gamma$ appears always added to $\mu_0$ and for the Moon self-gravity seems to be much less
important than elastic rigidity $\mu_0\gg\gamma$, in the following we set $\gamma=0$.

\subsection{INPOP19a parameters}

The parameters used in INPOP19a
are summarised in Table \ref{tab.INPOP}. They are related to the parameters
of the present model as follows
\begin{eqnarray}
\frac{{\rm I}_{\circ,T}}{\mathcal{M}\mathcal{R}_T^2} &=&\frac{ C_T}{\mathcal{M}\mathcal{R}_T^2} +
\frac{2}{3}C_{20T} \,,
\\
\frac{{\rm I}_{\circ,c}}{\mathcal{M}\mathcal{R}_T^2} &=& \frac{C_c}{\mathcal{M}\mathcal{R}_T^2} +
\frac{2}{3}C_{20c}\,,
\\
\frac{{\rm I}_{\circ,m}}{\mathcal{M}\mathcal{R}_T^2} &=& 
  \frac{{\rm I}_{\circ,T}}{\mathcal{M}\mathcal{R}_T^2}
- \frac{{\rm I}_{\circ,c}}{\mathcal{M}\mathcal{R}_T^2}\,,
\\
 \mat B_0(0)&=&\frac{2}{3}\frac{\mathcal{M}\mathcal{R}_T^2}{{\rm I}_{\circ,T}}
  \begin{bmatrix}
  3C_{22T}-\frac{1}{2}C_{20T} & 0 & 0 \\
  0 & -3C_{22T}-\frac{1}{2}C_{20T} & 0 \\
  0 & 0 & C_{20T}
  \end{bmatrix} \\ \nonumber
&+& \frac{n^2}{3\mu_0}\begin{bmatrix}
-1& 0 & 0 \\
0 & -1 & 0 \\
0 & 0 & 2
\end{bmatrix}=\mat B_{0,m}\ ,\label{prestress_inpop}
\\
\mat B_c(0) &=&\frac{2}{3}\frac{\mathcal{M}\mathcal{R}_T^2}{{\rm I}_{\circ,c}}
  \begin{bmatrix}
  -\frac{1}{2}C_{20c} & 0 & 0 \\
  0 & -\frac{1}{2}C_{20c} & 0 \\
  0 & 0 & C_{20c}
  \end{bmatrix}=\mat B_{c,m} \, \ ,
\\
\mu_0 &=& 3\frac{\GMEMB}{1+EMRAT}\frac{{\rm I}_{\circ,T}}{\mathcal{M}\mathcal{R}_T^2}
  \frac{1}{\mathcal{R}^3_T\,k_2}\,,
\\
\eta &=& \mu_0 \tau_M\,,
\\
\bar{k}_c &=& \frac{k_c}{C_T}\frac{C_T}{\mathcal{M}\mathcal{R}_T^2}\frac{\mathcal{M}\mathcal{R}_T^2}{{\rm I}_{\circ,T}}\,,
\\
  \mat F &=& 
              -\left(\boldsymbol{\omega}_m\otimes\boldsymbol{\omega}_m-\frac{\omega_m^2}{3}\Id\right)
                      +3\frac{\GMEMB}{1+1/EMRAT}\frac{1}{r^5}\left(\mathbf{r}\otimes\mathbf{r} - \frac{r^2}{3}\Id\right)\,.\nonumber
\end{eqnarray}

\begin{table}
\begin{center}
\caption{\label{tab.INPOP}Parameters taken from INPOP19a  \cite{INPOPa}.}
\renewcommand{\arraystretch}{1.1}
\begin{tabular}{p{6.5cm}ccc}
\hline
Parameter & Notation & Value & Unit \\ \hline
Earth-Moon mass ratio & $\qquad{}EMRAT\qquad{}$ & 81.300\,566\,772\,76764 & \\
Gravitational mass of E-M barycenter & \GMEMB & $\quad{}8.997\,011\,394\,021\,228\times10^{-10}\quad{}$ & au$^3$/day$^{2}$ \\
Lunar time delay for tide & $\tau_M$ & $0.094\,332\,332\,227\,022\,27$ & day \\
Lunar potential Love number & $k_2$ & $0.023\,559$ & \\
Polar moment of inertia of the Moon & $C_T/\mathcal{M}\mathcal{R}^2_T$ & $0.393\,140\,294\,559\,018$ & \\
Polar moment of inertia of the core & $C_c/\mathcal{M}\mathcal{R}^2_T$ & $0.000\,275$ & \\
Lunar gravity field & $C_{20T}$ & $-0.000\,203\,212\,558\,851\,8901$ & \\
--- & $C_{22T}$ & $2.238\,295\,071\,767\,246\times10^{-5}$ & \\
Lunar core gravity field & $C_{20c}$ & $-4.342\,243\,760\,334\,537\times10^{-8}$ & \\
--- & $C_{22c}$ & 0 & \\
Coefficient of viscous friction at CMB & $k_c/C_T$ & $6.443\,479\,383\,181\,008\times10^{-9}$ & rad/day \\
Lunar radius & $\mathcal{R}_T$ & 1738 & km \\
Astronomical  unit & au & 149\,597\,870.7 & km \\
\hline
\end{tabular}
\end{center}
\end{table}

\subsection{Comparison between INPOP19a and our model.}
\label{integration}

In this section we   integrate
numerically equations (\ref{eqrot3})
and (\ref{def1k}) to compute  the librations of the Moon forced by the  Earth and the Sun. 
The positions of the Earth and the Sun relative to the Moon were obtained
from the orbits of the Earth and Moon as  given  in \cite{yoder1995astrometric} (Table 4a).
The parameterization of the orbits is that  in equation (\ref{xapp}) with
$r=\left(\frac{a_p(1-e^2)}{1+e \cos f}\right)$ and
\begin{equation}
  f = M_p + e\left(2-\frac{1}{4}e^2\right) \sin M_p + \frac{5}{4}e^2 \sin 2 M_p
  +\frac{13}{12}e^3 \sin 3 M_p+ \Oc(e^4).
\label{true.an}
\end{equation}

\begin{figure}[hptb!]
\centering
\includegraphics[width=0.6\textwidth]{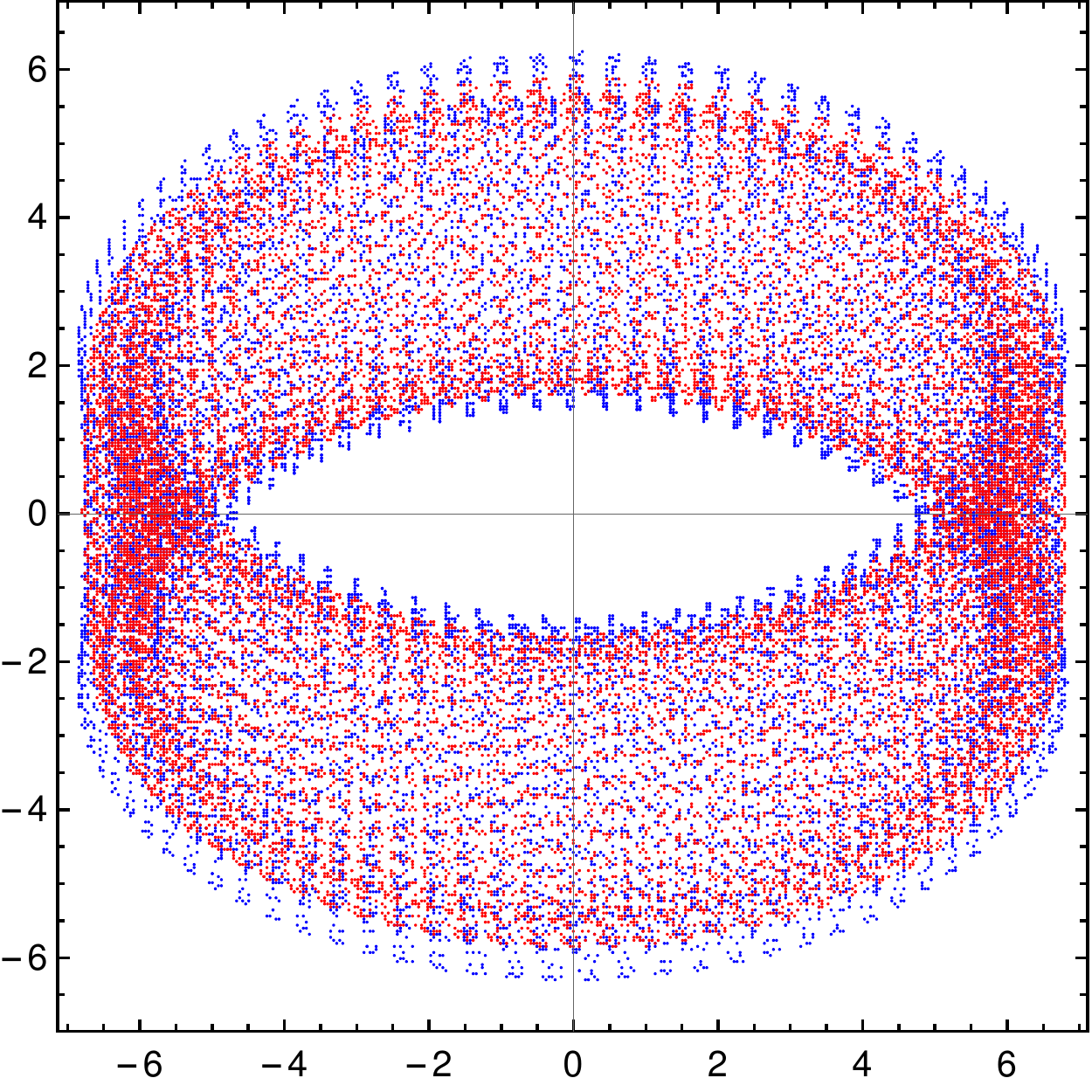}
\caption{Path of the normalised angular velocity vector
  $(\omega_1, \omega_2)/|\boldsymbol{\omega}|$ ($\times 10^{-4}$) in the $(\mat e_1,\mat e_2)-$plane of $\K_m$,
  $\boldsymbol{\omega}_m$ (blue) and $\boldsymbol{\omega}_T$ (red).} 
\label{omegaxybf}
\end{figure}

In Figure \ref{omegaxybf}, we plot  the first and second components of the
angular velocity of the mantle
$\boldsymbol \omega_{m,m}/|\boldsymbol \omega_{m,m}|$ and   
of  the Tisserand angular velocity of the   whole Moon
$\boldsymbol \omega_{T,m}/|\boldsymbol{\omega}_T|$.
As expected,  the mantle angular velocity has a larger amplitude of oscillations.

\begin{figure}[hptb!]
\centering
\begin{minipage}{0.33\textwidth}
\centering
\includegraphics[width=1.0\textwidth]{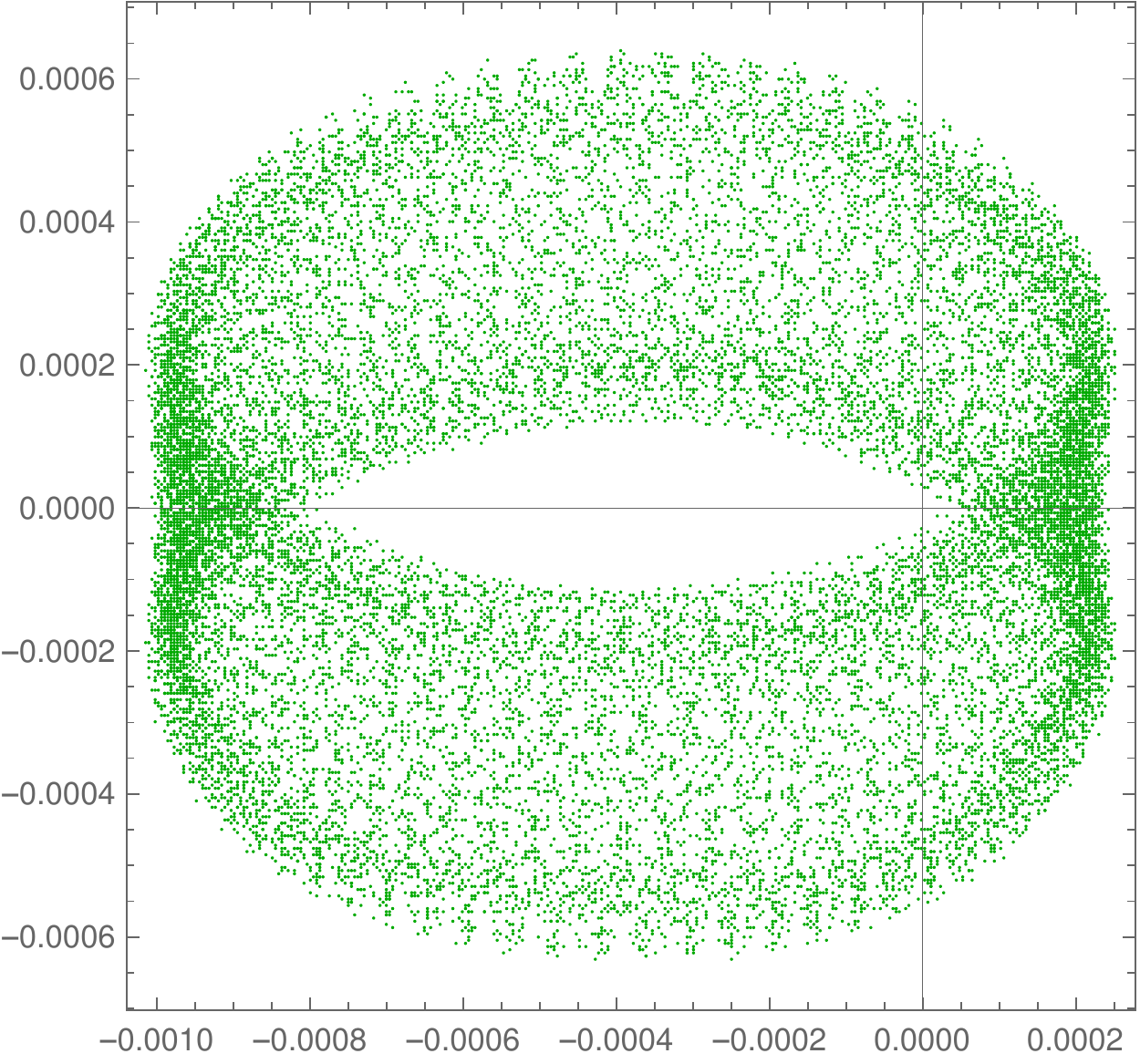}
\end{minipage}\hfill
\begin{minipage}{0.31\textwidth}
\centering
\includegraphics[width=1.0\textwidth]{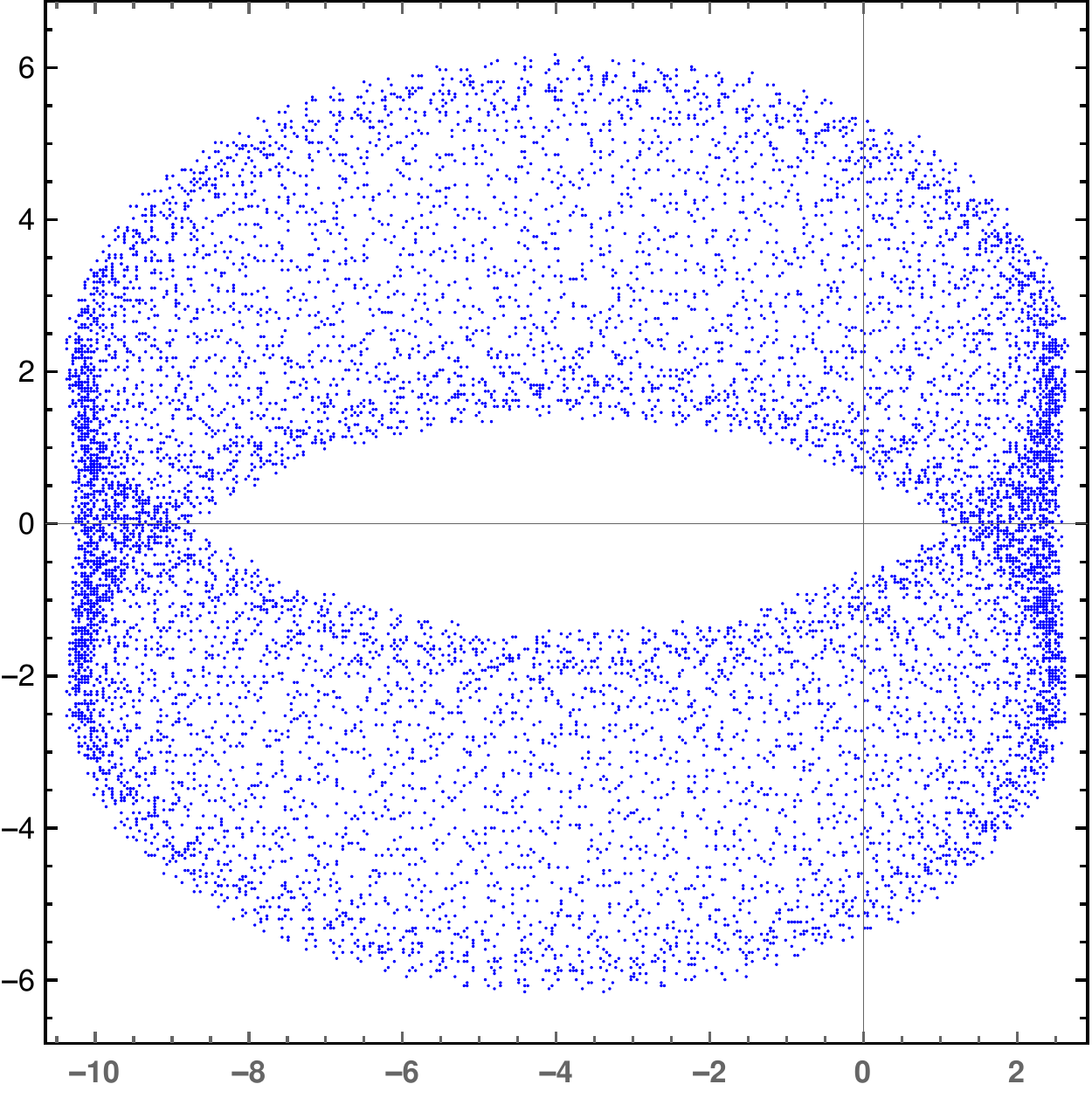}
\end{minipage}
\begin{minipage}{0.31\textwidth}
\centering
\includegraphics[width=1.0\textwidth]{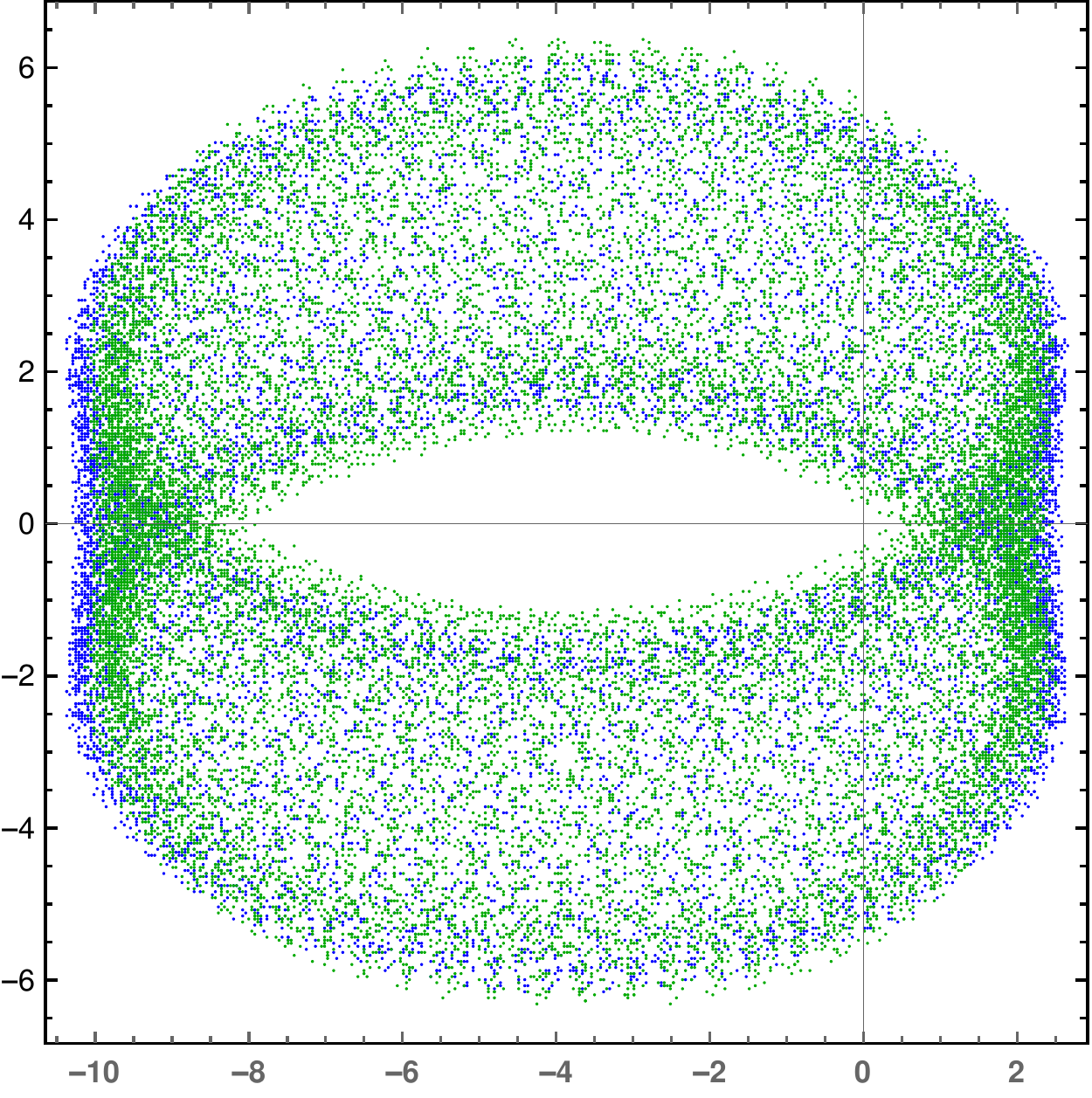}
\end{minipage}\hfill
\caption{
Path of the normalised angular velocity vector
$(\omega_{m,m1}, \omega_{m,m2})/|\boldsymbol{\omega}_m|$ ($\times 10^{-4}$): {\bf Left} INPOP19a,
{\bf Center} our  model with a tilted core cavity, as illustrated in Figure \ref{draw_tilt}, and
{\bf Right} superposition of both plots INPOP19a (green) and our model with a tilted cavity (blue).
}
\label{omegaxytilt}
\end{figure}

\begin{figure}[hptb!]
\centering
\includegraphics[width=0.8\textwidth]{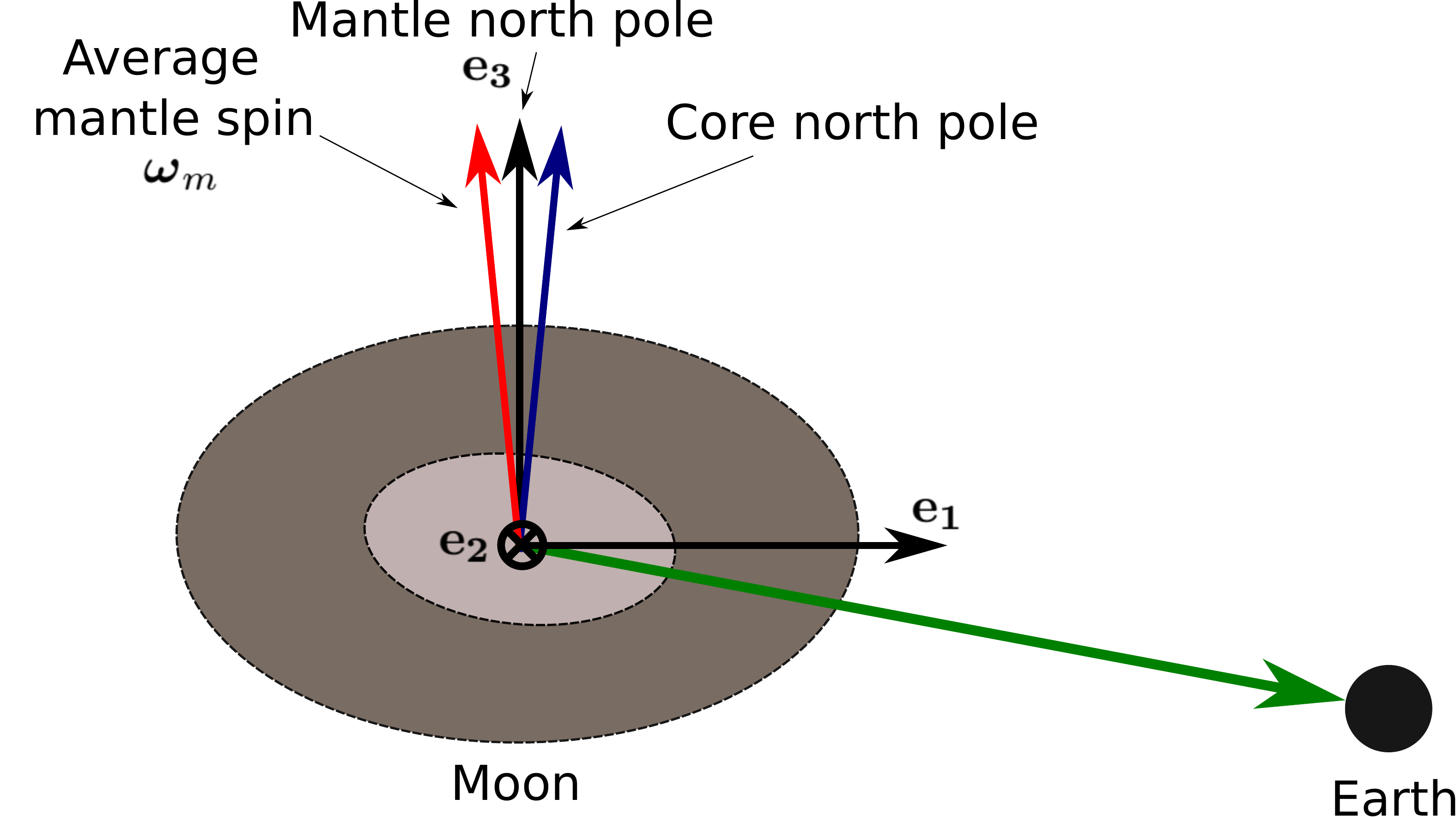}
\caption{Inclination between  the symmetry axis of the core cavity and the mantle frame $\K_m$.} 
\label{draw_tilt}
\end{figure}

In Figure \ref{omegaxytilt} - Left, we plot
$\boldsymbol \omega_{m,m}/|\boldsymbol \omega_{m,m}|$
using the data  from INPOP19a 
(the data for the Figure was generously  provided by Prof. Herve Manche).
In the figure generated with the INPOP19a data  the oscillations of
$\boldsymbol\omega_m$ are  centred at  the point $(-0.0004,0)$ while in Figure \ref{omegaxybf}
the oscillations are symmetric with respect to the origin.
This discrepancy lead us to investigate the role played by a possible small
tilt of the symmetry axis of the core with respect to the $\mat e_3$-axis of $\K_m$.
In Figure \ref{omegaxytilt} - Right we show that the same offset obtained with the INPOP19a
data can be
obtained from our model 
if we set the tilt equal to  $0.325^\circ$, see Figure \ref{draw_tilt} \footnote{The inclination
  of the symmetry axis of the core cavity with respect to $\K_m$ implies that 
  $\mat I_{c,m}$ is not diagonal in $\K_m$. If the density of the fluid is constant throughout the cavity,
  then $\mat I_{c,m}$ will be clearly  constant in time because the geometry of the cavity is fixed
  in $\K_m$. In Appendix \ref{Poinc} we show that under certain hypothesis  $\mat I_{c,m}$ can be
  constant in time  even
  when the density of the fluid inside the cavity is not constant.}.
As pointed out by Prof. V. Viswanathan (personal communication),
  the offset observed in INPOP19a  does not come
from any tilt of the core cavity with respect to the mantle frame but from other physical effects
which were not taken into account in our model, e.g. higher order gravitational moments.

Figures \ref{omegaxybf} and  \ref{omegaxytilt} show that the agreement between the results obtained with
INPOP19a
and those obtained with our model is excellent, except for the offset of the centre of libration of
$\boldsymbol\omega_{m,m}$
that in INPOP19a is due to physical effects not taken into account in our model.


\section{Conclusions}
\label{conclusions}

The main result in this paper is the set of  equations 
(\ref{eqrot3}), for the librations of a body made of a mantle and a fluid core, and  
 either equation (\ref{def1k}), or (\ref{def2k}), or (\ref{def3k}), for the deformations
of the mantle. The equations for the deformations are so general that
the mantle can have essentially any linear viscoelastic rheology (models with infinite memory, as the Andrade,
can
be approximated with arbitrary accuracy \cite{gev2020}).
 While  the (rigid) inertial part of these equations is well known and the deformation part already
appeared in \cite{rr2017}, the combination of the two parts is new.

In Section \ref{prestress} we introduced the important concept  of   prestress frame.
There is an old idea that part of the deformation of bodies   out of hydrostatic
equilibrium is due to an almost static transient state associated with a large viscosity of the mantle
(fossil deformation).
We reformulated this idea mathematically using rheological models of bodies that in the absence
of external forces eventually become spherical. With this concept we linked  the triaxiality
of the body to the parameters of the rheology, equation (\ref{BKov2}), and we separated
the time scale of the librations (years) from the time scale of variations of the 
triaxiality (thousands of years or more). Our point is:
fossil deformation,  being correct or not, provides a physical argument to be  used
 in the theory of   spherically symmetric bodies,
on  which we are confident in, to build a mathematical model
 to describe the librations of slightly aspherical bodies out of hydrostatic equilibrium.

 Most of the present paper is related to the validation of our equations for the librations.
 In order to do this,  in Section \ref{free2},  we present equations for the  rotational
 eigenvalues and eigenvectors of free librations. These equations apply to both
 bodies in and out-of spin orbit resonance and are formulated in terms of Love numbers,
 so they are independent on the rheological model used for the mantle. We are not aware of other
 formulas for free librations with such  level of generality, including our definition of the
 characteristic time $\tau$ in equation (\ref{tau}) that does not depend on the choice
 of a rheological model. In particular cases our formulas
 agree with others found in the literature. We summarised and illustrated  all our results about
 free librations in Section \ref{eigsum}.

 Another important result in this paper are the formulas for the inertial offsets, in particular
 that in equation (\ref{ofres}) for a body  in s-to-2 spin-orbit resonance. This formula is new
 and may
 be particularly useful in the estimation of the core parameters  $k_c$ and $f_c$.

 Some of our  formulas for the eigenvalues of free librations, e.g.
 (\ref{xx1}) and (\ref{xx2}), are quite complicated. If
 other features are included to the model; as for instance, a  solid core, deformation
 of the fluid cavity, and  core-mantle-magnetic coupling; then  these formulas
 will probably become much more complicated. In this case  the analytical treatment of the
 problem may be  doable but not  elucidative. In contrast, the model for librations we presented in
 this paper is quite simple and the addition of few other effects,  by means of the Lagrangian formalism,
 will not complicate it a lot. So, we conclude that 
  the most promising
 approach to the libration problem 
 is the numerical simulation of models with a minimum of  physical parameters that can be 
  calibrated
  directly from observations (see, e.g., \cite{noyelles2018rotation}).
  \footnote{Data sharing not applicable to this article as no datasets were generated or analysed during the current study.}

\begin{acknowledgements}
  We thank Prof.  Herve Manche (Observatoire de Paris, IMCCE) for
  the data used to generate Figure \ref{omegaxytilt}-Left and also for helping with the interpretation
  of the results. We also thank 
  Prof. V. Viswanathan (NASA Goddard Space Flight Center; University of Maryland)
  for the discussions about the INPOP19a model
  and about the offset of the
  centre of libration in Figure  \ref{omegaxytilt}-Left.
  CR is partially supported by FAPESP grant 2016/25053-8. YG is partially supported by FAPESP grants 2015/26253-8 and 2018/02905-4.
  LR was partially supported by CFisUC projects (UIDB/04564/2020 and UIDP/04564/2020), and ENGAGE SKA (POCI-01- 0145-FEDER-022217), funded by COMPETE 2020 and FCT, Portugal.
\end{acknowledgements}

\appendix

\section{The value of the coefficients $c_1,c_2,c_3$ in
  equation  (\ref{Jd3})
  for a  body in a Cassini state.}
\label{hansen}

The theoretical determination of the  almost equilibrium orientation of the
spin axis of an extended body,
or the guiding motion  as  described in Section \ref{preliminaries},
is a difficult task that requires  some knowledge about the rheological behaviour of
the body. An important situation, which will be the one considered in this appendix,
is that of a rigid body under the influence
of a single point mass that moves according to a Keplerian precessing orbit.
  In this
case
the guiding motion is called ``a Cassini state'' \cite{peale1969generalized}.
We remark that different Cassini states can be obtained under different rheological
hypotheses (see, for instance,  \cite{boue2020cassini} for bodies with a rigid mantle
and a fluid core).

Let $\kappa$  be the inertial frame and  $\K_g$ be the guiding frame
 described in Section
\ref{preliminaries}.
We recall that the common origin of both frames is the centre of mass of the extended
body. The position of the point mass with respect to the extended body
in the frame  $\kappa$
 is given by 
\begin{equation}
  \mathbf{r}=\mathbf{R_3}(\Omega_p)\mathbf{R_1}(\iota_p)\mathbf{R_3}(\omega_p)
  \begin{bmatrix}
r \cos(f_p)  \\
r \sin(f_p) \\
0 
\end{bmatrix}
\label{xapp}
\end{equation}
where:   $\iota_p$, $f_p$, $\omega_p$, and $\Omega_p$
are  the inclination with respect to an invariable plane (the ``Laplace plane''),
the true anomaly, the argument of the periapsis and the longitude of the ascending node,
respectively, of a classical Keplerian orbit that has  semi-major axis $a_p$, 
eccentricity $e$, and mean anomaly $M_p$ (the index $p$ stands for point-mass).

We recall that the three axes of $\K_g$ are those of the body's 
principal moments of inertia, $\ov I_{1}< \ov I_{2}< \ov I_{3}$,
in the absence of deformation.
The orientation of the guiding frame  $\mathbf{R_g}:\K_g\to\kappa$ is
given by three Euler angles
\begin{equation}
 \mathbf{R_g}=\mathbf{R_3}(\psi_g)\mathbf{R_1}(\theta_g)\mathbf{R_3}(\phi_g)
 \label{Rgapp}
 \end{equation}
 where: $\theta_g$ is the inclination of the body equator to the Laplace plane\footnote{\label{thetagfoot}
   In this paper we  slightly
   misused the angle $\theta_g$.
   For instance, in the case of the Earth $\tilde\theta_g=23.44^\circ$  is the inclination
   of the Earth equator (the plane that is  perpendicular to the average Earth's rotation axis)
   to the ecliptic and we use it as if it were $\theta_g$, namely  the inclination of the Earth
   conventional north pole (the  axis of largest moment of inertia of a rigid Earth) to the pole of
   the ecliptic. For the guiding motions in this appendix 
   $\tilde \theta_g=\theta_g-\frac{\dot\psi_g}{\omega}\sin\theta_g$ implies
   $\frac{|\tilde\theta_g-\theta_g|}{\theta_g}<\frac{|\dot\psi_g|}{\omega}$, so the relative error
 between the two angles is small.},
$\psi_g$ is the longitude of the ascending  node of the
body equator, and  $\phi_g$  is the angle between the ascending  node and Axis 1.
 There are two possible orientations of Axis 1, so $\mat R_g$ is not uniquely
defined. In the case of the Moon it is usual to choose the positive orientation of
Axis 1 as that pointing towards the Earth
\citep[see, e.g.,][]{eckhardt1981theory}. To change from one orientation
to the other it is enough to change the signs of both
unit vectors $\mat e_1$ and $\mat e_2$.

The orientation of the slow frame is given by
$\mathbf{R}_s(t)= \mathbf{R}_g(t)\mathbf{R_3}^{-1}(\omega\, t):\K_s\to \kappa$,
equation (\ref{Rg}), that in the present case
 becomes
\begin{equation}
  \mathbf{R}_s=\mathbf{R_3}(\psi_g)\mathbf{R_1}(\theta_g)\mathbf{R_3}(\zeta)\quad\text{where}\quad
  \zeta=\phi_g-\omega t\,.\label{Rs}
\end{equation}
The identity  $\langle\boldsymbol{\omega}_{s,g}\,,\mat e_3\rangle=0$ implies
\begin{equation}
\dot\zeta=-  \dot \psi_g\cos\theta_g\,.\label{zeta}
  \end{equation}

The position of the point mass in the guiding frame $\K_g$
is given by $\mathbf{r_g}=\mathbf{R_g}^{-1}\mathbf{r}$ and 
 equations (\ref{xapp}) and
(\ref{Rgapp}) imply 
  \begin{equation}
        \mathbf{r_g}=r \Big( \cos \big(f_p\big) \mat u +
 \sin \big(f_p\big) \mat v \Big)
\label{xgapp0}
\end{equation}
where $\mat u$  and $\mat v$ are the first and second columns, respectively,
of the matrix
$\mathbf{R_g}^{-1}\mathbf{R_3}(\Omega_p)\mathbf{R_1}(\iota_p)\mathbf{R_3}(\omega_p)$.
The  force matrix in the guiding frame $\mathbf{J}_{g}$, as defined in
Section \ref{preliminaries}, is given by
\begin{equation}\begin{split}
   & J_{gij}=\bigg(\frac{3{\cal G}m_p}{|\mathbf{r_g}|^5}\mathbf{r_g}\otimes\mathbf{r_g}
    \bigg)_{ij}\\
    &\  \quad =
 \frac{3{\cal G}m_p}{2r^3}\bigg(\big(u_iu_j+v_iv_j\big)+ \cos\big(2f_p\big)\big(u_iu_j-v_iv_j\big)+
\sin\big(2f_p\big)\big(u_iv_j+v_iu_j\big)\bigg)\, .\end{split}
 \label{J2}
\end{equation}
Notice that  the vectors $\mat u$ and $\mat v$
   in equation (\ref{xgapp0}) are the rows of an orthogonal matrix, which implies
   $|\mat u|=|\mat v|=1$ and $\langle \mat u,\mat v\rangle=0$. So, 
   \begin{equation}
     \tr \mat J_g=\frac{3{\cal G}m_p}{r^3}\, .
     \label{trapp}
   \end{equation}

Now we impose that the
extended body is in a Cassini state that is characterised by three laws
\cite{colombo1966cassini},  \cite{peale1969generalized}, \cite{peale1974possible}, which are generalisations
of the  original
laws of Cassini (1693) (aimed to describe
the motion of the Moon, see \cite{eckhardt1981theory}).  
\begin{itemize}
\item[1-] The body spin 
  is in $s$-to-2 resonance with
  the orbital mean motion, where $s$ is a positive integer (for the Moon
  $s=2$ and for  Mercury $s=3$);
\item[2-] The   inclination of the body equator $\theta_g$ with respect to the
  Laplace plane is constant;
\item[3-] Either the ascending node (state 1) or  the descending node (state 2)
   of the body equator on the Laplace plane
  precesses in coincidence
  with the ascending node of the  orbit on the Laplace plane (the 
  spin axis, orbit normal, and Laplace plane  normal are coplanar).
\end{itemize}

The third law implies that either
\begin{equation}\begin{array}{rll}
  \psi_g&=\Omega_p \quad&\text{state 1  (e.g. Mercury) or}\\
  \psi_g&=\Omega_p+\pi \quad &\text{state 2 (e.g. Moon),}
\end{array}
\label{psiOmega}
\end{equation}
then the second law
implies that the inclination  of the body
spin axis with respect to the normal to the
orbital plane, given by ($\theta_g\ge 0$, $\iota_p\ge 0$) \begin{equation}
  \begin{array}{rll}
    \chi_p&=\theta_g-\iota_p\ge 0\quad& \text{state 1}\\
    \chi_p&=\theta_g+\iota_p\quad &\text{state 2}\end{array}
  \label{chi}
  \end{equation}
  is constant (the inequality in state 1 is of dynamical origin,
  see \cite{peale1969generalized} paragraph below equation (18)); and then the first law implies that
$s M_p=2(\phi_g-\omega_p)$\footnote{This resonance condition is due to
  \cite{peale1969generalized},  here presented  in a form given in
 \cite{boue2020cassini}. Notice that 
  the sidereal rotation period (the spin period) of the guiding motion is
  $2\pi/\omega$, where $\omega= \dot\phi_g+\dot \psi_g\cos\theta_g$. As discussed
  in the cited references, the angular
  velocity  $ \dot\phi_g$ that  appears in the resonance condition is
  the spin of the extended body
  relative to the frame  $\K_{or}$  that precesses with the orbit of the point mass,
  given by
  $\mathbf{R_3}(\Omega_p)\mathbf{R_1}(\iota_p):\K_{or}\to\kappa$.}
\footnote{Assume $\iota_p=\theta_g=0$. If the orbital mean motion is synchronous
  with the constant spin then $s=2$ and it is 
  clear that the   resonance  $M_p+\omega_p =\phi_g$ has the meaning stated
  in the first law. If $s\ne 2$ then the relation  $s M_p=2(\phi_g-\omega_p)$
  means that after  a time interval equal to 
  the anomalistic period, which is $T=2\pi/\dot M_p$, the orbit 
  has an angular displacement of $2\pi+\, T \dot\omega_p$ rad while the smallest
  axis of inertia of the body has an angular displacement of
  $s\, \pi +\, T\dot\omega_p$ rad. So, if initially the periapsis occurs on
the smallest axis of inertia, then  all other periapsis will occur on the same axis
(if $s$ is even, then always at the same side of the extended body).}.

For a  body in a Cassini state 2 (e.g. Moon)
equation (\ref{xgapp0}) becomes\footnote{\label{Eck1}
  In \cite{eckhardt1981theory},
  the orientation
  of the body  frame of the rigid Moon $\K$ with respect to an inertial frame
  $\kappa$
is given by  $\mat R_3(\psi)\mat R_1^{-1}(\theta)\mat R_3(\phi):\K\to\kappa$, where:
  $\psi$ is the longitude of the descending node of the
  lunar equator, $\theta$ is the inclination of the lunar equator to the ecliptic, and
  $\phi$ is the angle between the descending node and the axis of smallest moment of
  inertia  pointing towards the Earth.
 The Cassini's laws  are equivalent to  $(\psi,\theta,\phi)=
  (\Omega_p, \iota_p, \pi+M_p+\omega_p)$, so 
  Eckhardt's guiding frame is
  $\mat R_3(\Omega_p)\mat R_1^{-1}(\iota_p)\mat R_3(\pi+M_p+\omega_p):\K_g\to\kappa$,
  which coincides with ours, namely
  $\mat R_3(\Omega_p+\pi)\mat R_1(\iota_p)\mat R_3(M_p+\omega_p):\K_g\to\kappa$.}  
{\scriptsize
  \begin{equation}
        \mathbf{r_g}=r  \cos f_p\underbrace{\begin{bmatrix} 
    -\cos\phi_g \cos\omega_p - \cos\chi_p \sin\phi_g \sin\omega_p\\ 
    \cos\omega_p \sin\phi_g - \cos\phi_g \cos\chi_p \sin\omega_p\\ \sin\chi_p \sin\omega_p
  \end{bmatrix}}_{ \mat u}+
r \sin f_p\underbrace{\begin{bmatrix}
-\cos\chi_p \cos\omega_p \sin\phi_g + \cos\phi_g \sin\omega_p\\ -\cos\phi_g \cos\chi_p \cos\omega_p - 
  \sin\phi_g \sin\omega_p\\ \cos\omega_p \sin\chi_p
\end{bmatrix}}_{\mat v}
\label{xgapp}
\end{equation}
}  
and for a body in a Cassini state 1 we must replace $\mat u\to-\mat u$ and $\mat v\to-\mat v$.
Since  $J_{gij}$ given in equation (\ref{J2})
depends only on the products of components of $\mat u$ and $\mat v$, 
the matrix $\mat J_g$ has the same expression for both states 1 and 2. Therefore,
the following computation of $c_1$, $c_2$ and $c_3$
holds in both cases.

The
 coefficients $c_1,c_2,$ and $c_3$  we aim  to compute are  determined
by the constant term in the Fourier expansion of $J_{gij}$ given in equation (\ref{J2}).
 In order to compute the Fourier
 expansion of the factors involving $r$ and $f_p$ we use the Hansen coefficients
 defined by \footnote{
     A table of the Hansen coefficients $ X_s^{-3,2}(e)$, for $-4\le s\le 8$ and
   eccentricity up to order $e^6$, can be found in \cite{correia2014deformation}.
   In this paper we use
      \begin{equation}\begin{split}
          X_0^{-3,0}(e)&=(1-e^2)^{-3/2}\\
          X_2^{-3,2}(e)&= 1 - \frac{5}{2} e^2 + \frac{13}{16} e^4 - \frac{35}{288} e^6\\
          X_3^{-3,2}(e)&= \frac{7}{2} e - \frac{123}{16} e^3 + \frac{489}{128} e^5
          \end{split}\,.\label{X030}
     \end{equation}}
 \begin{equation}
 \bigg(\frac{r}{a_p}\bigg)^n\erm^{i m f}=\sum_{k=-\infty}^\infty X^{n,m}_k(e) \erm^{i k M}
 \label{hanseneq}
 \end{equation}

   The computation of the coefficients  $c_1,c_2,$ and $c_3$ in  equation (\ref{Jd3})
   can be done in the following way. At first we use equation (\ref{trapp}) to obtain 
   \begin{equation}
c_3=\frac{{\cal G}m_p}{\omega^2 a_p^3} X_0^{-3,0}(e)\, .
     \label{c3}
   \end{equation}
   The term
   $ \left( \frac{3{\cal G}m_p}{\omega^2|\mathbf{r_g}|^5}\mathbf{r_g}\otimes\mathbf{r_g}\right)_{33}$
   can be easily computed using equations
   (\ref{J2}) and (\ref{hanseneq}) and the
   result is $\frac{3{\cal G}m_p}{2 \omega^2 a_p^3} X_0^{-3,0}(e)\sin^2(\chi_p)=c_3-2c_1/3$.
   So, we obtain
that
   \begin{equation}
     c_1=\frac{3\,{\cal G}m_p}{2\,\omega^2 a_p^3} X_0^{-3,0}(e)\bigg(1-\frac{3}{2}\sin^2(\chi_p)\bigg)\,.
     \label{c1c3}
   \end{equation}

   In order to compute the constant $c_2$ 
   associated with an $s-to-2$ spin-orbit resonance it is enough to compute the constant
   term in the Fourier expansion of
 $\left( \frac{3{\cal G}m_p}{\omega^2|\mathbf{r_g}|^5}\mathbf{r_g}\otimes\mathbf{r_g}\right)_{11}$ and then to use that this term is equal to
 $c_1/3+c_2+c_3$, equation (\ref{Jd3}). Assuming that the precession of the periapsis is
 different from zero a computation gives
 \begin{equation}
   c_2=\frac{3\,{\cal G}m_p}{2\,\omega^2 a_p^3} X_s^{-3,2}(e)   \cos^4(\chi_p/2)
   \label{c2}
   \end{equation}

   Another case of interest is that of a point mass that moves in a precessing
   Keplerian orbit, as that parameterised in equation (\ref{xapp}), with constant
   inclination, $\iota_p=$constant, and  with no spin-orbit resonance.
   By no spin-orbit resonance we mean that
   the constant frequencies $\dot \psi_g$, $\dot\Omega_p$,  $\omega$, $\dot\omega_p$,
   and $\dot M$ are noncomensurable.
   Equation (\ref{trapp}) implies that
   $c_3$ is again given by equation (\ref{c3}) and the condition of
   no  spin-orbit resonance
   implies $c_2=0$. As before, to compute
   $c_1$ we look for a term in $J_{g33}$ that is constant in time. Using that
   $X^{-3,\pm 2}_0=0$ and the  no spin-orbit hypothesis we obtain that
   \[
    \frac{3{\cal G}m_p}{2r^3}\bigg(\cos\big(2f_p\big)\big(u_3^2-v_3^2\big)+
    \sin\big(2f_p\big)2 u_3v_3\bigg)
  \]
  does not contain any term that is  constant.
  Using that   $-\dot \psi_g+\dot\Omega_p\ne 0$, an analysis
  of the remaining term, $\frac{3{\cal G}m_p}{2r^3}\big(u_3^2+v_3^2\big)$, shows that
 the constant term
 of $J_{g33}$ is
 \[
    \frac{3{\cal G}m_p}{2 a_p^3} X^{-3,0}_0  \frac{1}{8}
    \Big(-\cos (2 \theta_g)-(3 \cos (2 \theta_g)+1) \cos (2 \iota_p)+5\Big)=
    \omega^2\left(c_3-\frac{2}{3}c_1\right)\]
  So, using equation (\ref{c3}) for $c_3$ we obtain that
  for no spin-orbit resonance:
  \begin{equation}\begin{split}
      c_1&=\frac{3{\cal G}m_p}{\omega^2 a_p^3}  X^{-3,0}_0
      \frac{1}{32} \Big(3 \cos (2 \theta_g)+1\Big) \Big(3 \cos (2 \iota_p)+1\Big)\\
      c_2&=0 
      \end{split}
    \label{c1nores}
  \end{equation}

     If there are several point masses,  $\beta=1,2,\ldots$,
  orbiting the extended body, each one  in a precessing
   Keplerian orbit, with constant
   inclination, $\iota_{\beta}=$constant, and  with no spin-orbit resonance;
   as in the case where the Earth is the extended body and the Moon and the
   Sun represent
   the point masses; then 
   $c_2=0$ and (the index $p$ was omitted in several constants) 
 \begin{equation}
     c_1=\bigg(\sum_\beta s_\beta\bigg)\frac{1+3 \cos (2 \theta_g)}{4}\quad
     \text{where}\quad s_\beta=
     \frac{3{\cal G}m_\beta}{\omega^2 a_\beta^3} (1-e_\beta^2)^{-3/2}\,
     \frac{1+3 \cos (2 \iota_\beta)}{8}\,.
   \label{c1nores2}
 \end{equation}
In this case the average  rate of precession $\dot\psi_g$ of the spin axis of
the extended body is \cite{williams1994contributions}:
   \begin{equation}
     \frac{\dot \psi_g}{\omega}=-\big(\sum_\beta s_\beta\big)
     \frac{\ov I_3-\ov I_e}{\ov I_3}
     \cos \theta_g\approx -\big(\sum_\beta s_\beta\big)
    \ov \alpha_e\,
     \cos \theta_g\,;
     \label{dotpsi}
     \end{equation}
     and the motion of the slow frame is determined by $\mat R_s$ as given in equations (\ref{Rs}) and
(\ref{zeta}) with $\theta_g=$constant.

\section{A simple model:
  Considerations about the guiding frame.}\label{simple}

In this appendix we show by means of an example
 how the inertial forces that appear due to the   non-inertial character
 of the guiding frame are mostly cancelled out by external torques.
 This example will be used in Appendix \ref{offset}.

 The problem is to describe the axial precession of 
 an  extended rigid body of  mass $m$ and moment of inertia $\mat I$
  under the gravitational
force of a point
of  mass $m_p$ that moves in a circular orbit of radius $a_p$ and with 
angular frequency $n$.
We assume that the body is axisymmetric with  $\ov I_1=\ov I_2=\ov I_e<\ov I_3$. We define a body frame 
$\K$ such that  the $\mat e_3$-axis is the $\ov I_3$-principal axis
and an inertial frame  $\kappa$
such that the orbit of the point mass has components $(a_p\cos(nt),a_p\sin(nt),0)$.

Let $\omega>0$ be the  initial spin angular speed
    of the body that  is defined as the projection of the initial angular
    velocity vector $\boldsymbol{\omega}$ on the $I_3$-principal axis.
    We assume that the angular velocity is almost aligned with the  $I_3$-principal axis
     and that inertial forces prevail over  the gravitational torque, namely
\begin{equation}     \omega/\|\boldsymbol{\omega}\|\approx 1\quad\text{and}\quad
 \frac{3}{2} \frac{G m_p}{\omega^2 a^3}=s \ll 1
\label{simple0}
\end{equation}
In this case the torque-free inertial motion dominates and the body 
rotates almost steadily  about the  spin-axis. Note that $s$ is the quantity that we denoted as
$s_\beta$
 in equation (\ref{c1nores2}).

The tidal-force operator $\mat J=
\frac{3{\cal G}m_p}{r^5}\mathbf{r}\otimes\mathbf{r}$ can be decomposed into two terms
\begin{equation}
\mat J=
    \underbrace{s \, \omega^2\left(
\begin{array}{ccc}
 1 & 0 & 0 \\
 0 & 1 & 0 \\
 0 & 0 & 0 \\
\end{array}
\right)}_{= \mat J_0} +\underbrace{s\,\omega^2 \left(
\begin{array}{ccc}
 \cos (2 n t) & \sin (2 n t) & 0 \\
 \sin (2 n t) & -\cos (2 n t) & 0 \\
 0 & 0 & 0 \\
\end{array}
\right)}_{=\mat J_1}\,.\label{simpleJ}
\end{equation}
The axial precession is determined by the constant part $\mat J_0$, so in the following we consider
the problem of the motion of the extended body only under tidal-force operator $\mat J_0$, namely
\begin{equation}
  \begin{split}
    & \dot{\boldsymbol{\widehat \pi}}
    =[\mat I\, ,\mathbf{J}_0]
    \quad\text{where}\quad  \mat R:\K\to\kappa  \\
&\boldsymbol{\widehat \pi}=\Tr \big(\mat I\big) \boldsymbol{\widehat \omega}-
  \mat I \boldsymbol{\widehat \omega}-
    \boldsymbol{\widehat \omega}\mat I\,,\quad
    \boldsymbol{\widehat \omega}=
    \dot{\mat R}\mat R^{-1}\,.\\
    \end{split}\label{simple1}
    \end{equation}

A computation shows that $ \mat R=\mat R_3(\dot \psi \, t)\mat R_1(\theta)
       \mat R_3(-\dot \psi \, t\cos \theta)\mat R_3(\omega\, t)$
       is a solution to this equation, with $\dot\theta=\ddot \psi=0$,
       if
 \begin{equation}
  \left(s\frac{I_3-I_e}{I_3} -\frac{I_e}{I_3}
    \frac{\dot\psi^2}{\omega^2}\right) \cos \theta+\frac{\dot \psi}{\omega}=0
  \label{simple2.5}
\end{equation}
This equation has two solutions: one with $\dot\psi/\omega >0$ (prograde) and another
with  $\dot\psi/\omega <0$ (retrograde). Up to
leading
order in  the small parameter $s$
the prograde solution
\begin{equation}
\frac{\dot\psi}{\omega}=\frac{I_3}{I_e} \sec \theta\label{simple3}
\end{equation}
is the fast  torque-free precession and the retrograde solution
\begin{equation}
  \frac{\dot \psi}{\omega}= -s\frac{I_3-I_e}{I_3}
  \cos \theta\approx   -s\ov\alpha_e
  \cos \theta\label{simple4}
\end{equation}
is the slow precession that exists due  to the gravitational interaction.

We will use the retrograde solution to define the  guiding-frame:
\begin{equation}
  \mat R_g=\underbrace{\mat R_3(\dot \psi_g \, t)\mat R_1(\theta_g)
       \mat R_3(-\dot \psi_g \, t\cos \theta_g)}_{=\mat R_s}\mat R_3(\omega\, t)\,,
     \label{simple5}
   \end{equation}
   where $\dot\psi_g<0$ is the solution to  equation (\ref{simple4})
   with $\theta=\theta_g$. Notice that if $\theta_g=0$ then $\mat R_s=$Identity.

   The angular velocities of
   the guiding frame $\boldsymbol{\widehat \omega}_{g,g}=\mat R_g^{-1}
   \dot{\mat R}_g:\K_g\to\K_g$  and of the slow frame
   $\boldsymbol{\widehat \omega}_{s,s}=\mat R_s^{-1}
    \dot{\mat R}_s:\K_s\to\K_s$ are:
   \begin{equation}
    \boldsymbol{\omega}_{g,g}=\omega \mat e_3+\underbrace{\mat R_3^{-1}(\omega t)
       \boldsymbol{\omega}_{s,s}}_{\boldsymbol \omega_{s,g}}\,,\quad  \boldsymbol{\omega}_{s,s}=
     \dot \psi_g \sin \theta_g\left(
\begin{array}{c}
- \sin(t\dot\psi_g \cos\theta_g) \\
\ \  \cos(t\dot\psi_g \cos\theta_g)\\
 0 
\end{array}
\right)
\label{simple4.5}
\end{equation}

Since for the problem  considered in this section the guiding motion is a solution
to the equations of motion, the linearized equations about the guiding motion
are just the ordinary linearized equations  about a solution.
These equations can be obtained directly from  equation (\ref{lineq}), for the motion of the mantle,
with the following simplifications: $\boldsymbol\alpha_m\to\mat a$ ($\Id+\mat{\widehat a}:\K\to\K_g$),
$\mat I_m=\mat I$ and $\mat I_c=0$ (there is no fluid core), $\boldsymbol\delta \mat B_T=0$ (the body is rigid),
and $c_2=0\Longrightarrow \xi_1=\xi_2=1=c_1$ (there is no spin-orbit resonance) with
\begin{equation}
  c_1= s\frac{1+3 \cos (2 \theta_g)}{4} \quad\Big(\text{consequence of equation (\ref{c1nores2})}\Big)\,.
  \label{c1aux}
 \end{equation}

There are two forcing terms that appear in the right-hand side
of equation (\ref{lineq}):  the ``true-torque'' that comes from $\mat J_0$ and the
``fictitious-torque'' that comes from the non-inertial nature of the guiding frame.
The true-torque term is given by:
\begin{equation}
  [\ov{\mat I},\mat R_g^{-1}\mat J_0\mat R_g]^\vee=[\ov{\mat I},\boldsymbol \delta\mat J_{0,g}]^\vee=
  s\,\omega^2\,(\ov I_3-\ov I_e)\cos\theta_g\sin\theta_g
    \begin{bmatrix} -\cos \phi_g\\
     \ \ \, \sin \phi_g\\
      0
    \end{bmatrix}\,,
    \label{true-torque}
\end{equation}
where we used that: $[\ov {\mat I},\ov{\mat J}]=0$,
the check map $^\vee$ is the inverse of the hat map, and $\phi_g=t(\omega-\dot\psi_g\cos\theta_g)$
(see equations (\ref{Rs}) and (\ref{zeta})). The fictitious torque is
\begin{equation}
    \begin{bmatrix}
\ov I_{e}\dot{{\omega}}_{s,g1} +\omega\big(\ov I_3-\ov I_e\big){\omega}_{s,g2} \\
\ov I_{e}\dot{{\omega}}_{s,g2} -\omega\big(\ov I_3-\ov I_e\big){\omega}_{s,g1} \\
0   \end{bmatrix}=\sin\theta_g\Big(\ov I_e(\omega-\dot\psi_g\cos\theta_g)\dot\psi_g+\omega(\ov I_3-\ov I_e)
\dot\psi_g\Big)
 \begin{bmatrix} -\cos \phi_g\\
     \ \ \, \sin \phi_g\\
      0
    \end{bmatrix}\,,\label{fictitious-torque}
  \end{equation}
  where we used equation $\boldsymbol \omega_{s,g}$ as given in equation (\ref{simple4.5}).
If we use that $\dot\psi_g$ is a solution to equation (\ref{simple2.5}), then we obtain that
  the fictitious torque is equal to minus the true torque. So, the true torque cancels out the
  fictitious torque.

  In general the guiding motion is not a particular solution of the dynamical equations and so,
  we cannot expect the full cancellation of the fictitious torque. If  the guiding motion
  is a good approximation for the real motion, then the residue after
  the partial cancellation of the fictitious torque
  must be  of the order of the small terms in the equation.

  In conclusion, for equation (\ref{simple1}) the linearized equation about the guiding motion is
\begin{equation}
  \begin{split}
    &       \ov I_{e}\ddot{a}_{1}-\omega (2\ov I_{e}-\ov I_{3})\dot{a}_{2}
    +\omega^2\xi_1(\ov I_3-\ov I_e)a_{1}=0\\
 & \ov I_{e}\ddot{a}_{2}
  +\omega (2\ov I_e-\ov I_{3})\dot{a}_{1}
             +\omega^2\xi_2(\ov I_3-\ov I_e)a_{2}=0\\
             & \ov I_{3}\ddot{a}_{3}=0  \,.                            \end{split}
           \label{lineq2}                        
  \end{equation}

  \section{The classification of the roots of equation (\ref{charp}) into
    NDFW and FLL eigenvalues}
\label{class}

In this Appendix we classify the two roots of the equation (\ref{charp}), i.e.
$ x^2-x(f_0+1) (y+z )+(f_0+1)z\,y=0$.

  If  the discriminant $\Delta$ of equation (\ref{charp}) is different from zero, then the equation
  has two solutions. If the core is negligible $f_0=\frac{{\rm I}_{\circ c}}{{\rm I}_{\circ m}}=0$,
  then one of the root is $x_{dw}=y$ and is related to the Nearly Diurnal Free Wobble (NDFW)
  and the other  $x_{\ell a}=z$ is related to the Free Libration in Latitude (FLL).
In order to classify a given solution one must deform $f_0>0$ from
its current value to  $f_0=0$ keeping $(y,x)$ constant. If the function $f_0\to\Delta$  remains different
from zero 
during the deformation, then the root will move in the complex plane as a smooth function of $f_0$ and it
will eventually become either $y$, and the root will be classified as $x_{dw}$, or $z$, and
the root will be classified as $x_{\ell a}$.

\begin{figure}[hptb!]
\centering
\begin{minipage}{0.5\textwidth}
\centering
\includegraphics[width=0.9\textwidth]{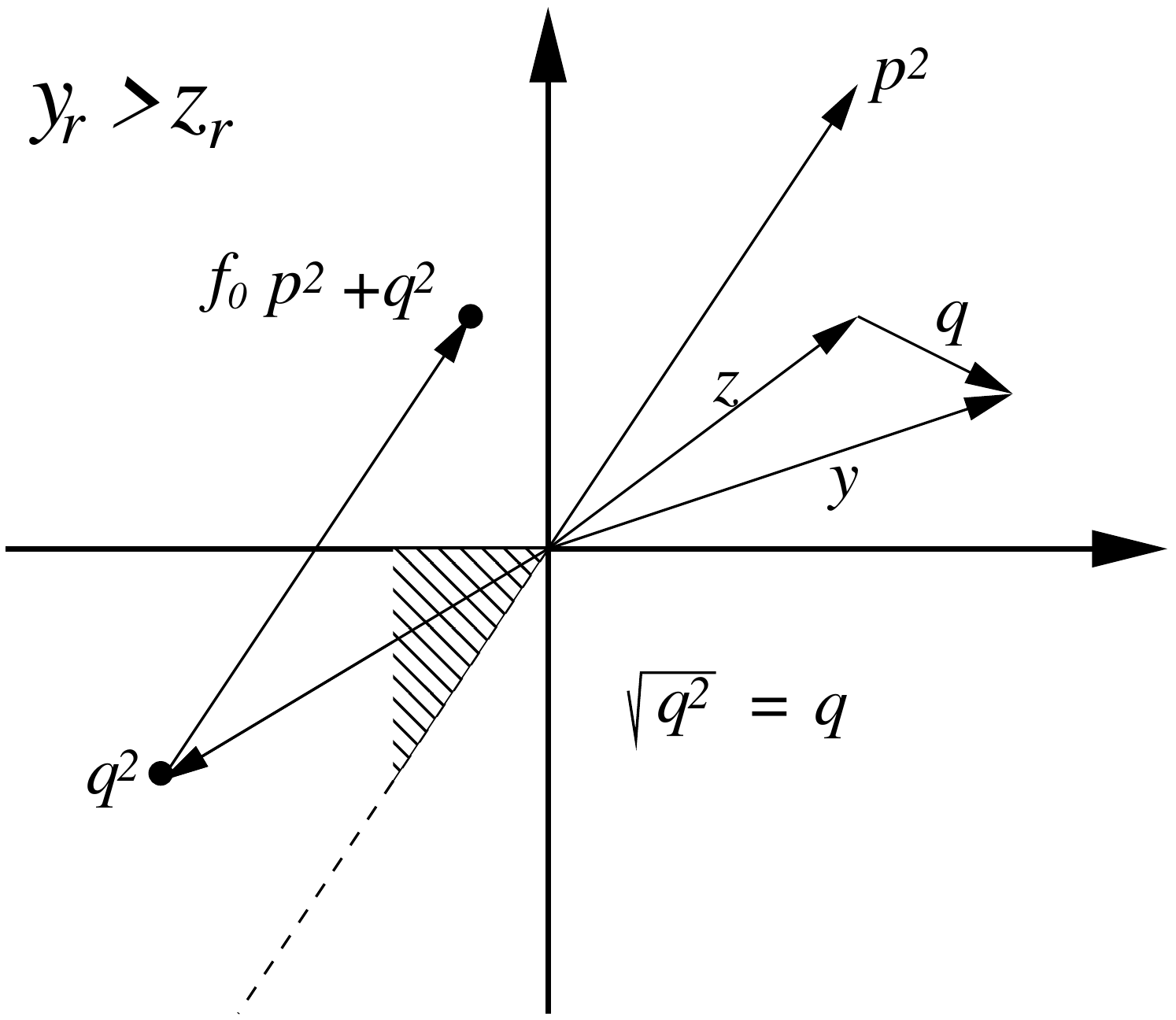}
\end{minipage}\hfill
\begin{minipage}{0.5\textwidth}
  \centering
  \includegraphics[width=0.9\textwidth]{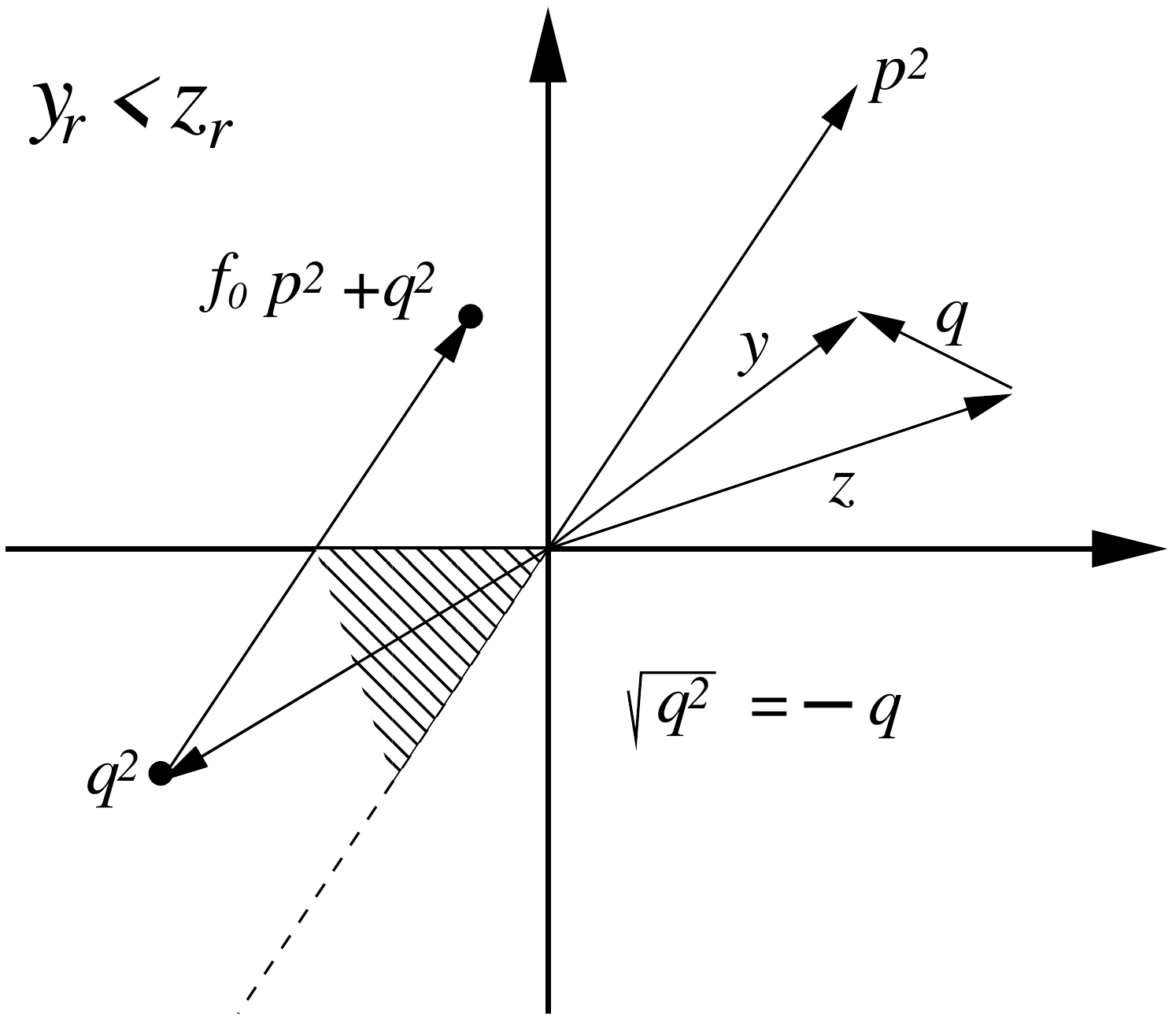}
\end{minipage}\hfill
\caption{Diagram  used in the classification of the eigenvalues of NDFW and FLL.
  Note that $y_r>z_r$ ( $y_r<z_r$)  implies: $q_r>0$
  ($q_r<0$),  $\sqrt{q^2}=q$ ($\sqrt{q^2}=-q$).
  \\  The inequality 
  $\{q_{2r}<0\,,q_{2i}<0\,,
  \frac{q_{2i}}{q_{2r}}<\frac{p_{2i}}{p_{2r}}\}$ is represented by the shaded areas in the figure and
    $f_0>-\frac{q_{2r}}{p_{2r}}$ implies  ${\rm Re}\,(f_0p^2+q^2)>0$.}
  \label{regions}
\end{figure}

The discriminant of equation (\ref{charp}) becomes simpler if we use the variables
  \begin{equation} \begin{split}
    & p:=\frac{1}{2}(y+ z)\,,\quad
q:=\frac{1}{2}(
  y-z)\Longrightarrow y=p+q\,,\quad z=p-q\,,\\ &\text{and}\quad
  \Delta=4 (f_0+1) \left(f_0 \,p^2+q^2\right)\,.
\end{split}
        \label{zpole2}
\end{equation}
The two solutions to equation (\ref{charp}) are 
$ (f_0+1)p\pm\sqrt{1+f_0}\sqrt{f_0 \,p^2+q^2}$, where $\sqrt{\ }$ denotes the principal value of the
square root, which is not continuous on the  non-positive real axis
and satisfies Re$\,\sqrt z\ge 0$ with $\sqrt{-1}=i$.
The lack of continuity of the principal value of the square root leads to some difficulties in the
classification of the roots because $f_0\to\Delta$ can cross the non-positive real
axis during the variation of $f_0$.  The  eigenvalues
are given by the following algorithm obtained essentially from Figure \ref{regions}.

Notation: (1) Given $(y,z)$ we define
  $q:=q_r+iq_i=(y_r-z_r)/2+i(y_i-z_i)/2$ and $p:=p_r+ip_i=(y_r+z_r)/2+i(y_i+z_i)/2$,
  as in equation (\ref{zpole2}), and $p^2:=p_{2r}+i p_{2i}=(p^2_r-p^2_i)+i2 p_rp_i $ and
  $=q^2:=q_{2r}+i q_{2i}=(q^2_r-q^2_i)+i2 q_rq_i $. (2) We define the set of inequalities
  \begin{equation}
      V=\bigg\{q_{2r}<0\,,q_{2i}<0\,,
      \frac{q_{2i}}{q_{2r}}<\frac{p_{2i}}{p_{2r}}\,, \ \text{and} \ f_0>-\frac{q_{2r}}{p_{2r}}\bigg\}\,.
      \label{V}
    \end{equation}
    All the inequalities in $V$ hold if $p^2$, $q^2$ and $f_0$  are arranged as illustrated in Figure
    \ref{regions}.
    
  Assumptions: (1) To  simplify the analysis we assume $p_i/p_r<1$, which implies that $p^2$ is in the
   positive quadrant of the complex plane, $p_{2r}>0$ and $p_{2i}>0$
   (the analysis could also be made without this hypothesis). (2) We also assume that 
   $|y|\ne|z|$ in order to avoid $\Delta=0$ during a variation of $f_0$.
\begin{itemize}
\item[$\bullet$]  If $ y_r>z_r$, then either all inequalities in $V$ 
  are true and  equation (\ref{xx2}) holds or at least one inequality in $V$ is not true
  and equation (\ref{xx1}) holds.
\item[$\bullet$]  If $ y_r<z_r$, then either all inequalities in $V$
    are true and  equation (\ref{xx1}) holds or at least one inequality in $V$ is not true
  and equation (\ref{xx2}) holds.
\end{itemize}
  \begin{eqnarray}
    &  & x_{dw}=(1+f_0)p+\frac{\sqrt\Delta}{2}\,,\quad
   x_{\ell a}=(1+f_0)p-\frac{\sqrt\Delta}{2}\,;\label{xx1}\\ & & \nonumber\\
& &  x_{dw}=(1+f_0)p-\frac{\sqrt\Delta}{2}\,,\quad
    x_{\ell a}=(1+f_0)p+\frac{\sqrt\Delta}{2}\,;\label{xx2}
      \end{eqnarray}
where $\Delta=4 (f_0+1) \left(f_0 \,p^2+q^2\right)$.      

The two eigenfrequencies coincide, $x_{dw}=x_{\ell a}$, when $\Delta=0$,
what  happens for $(y,z,f_0)$ in  the set
\begin{equation}
 \bigg\{ |z|=|y|\,, f_0=\tan^2(\psi/2)\bigg\}\,,
    \label{equal}
    \end{equation}
    where $\psi$  is the angle between $y$ and $z$, $y=\erm^{i\psi}z$.

    The NDFW and the FLL eigenfrequencies (this property does not extend to the eigenvectors)
    are dual to each other in the sense that if two bodies have
    the same $f_0$ but one has $(y,z)=(a_1,a_2)$ while the second  $(y,z)=(a_2,a_1)$, then
    the value of  $x_{dw}$ ($x_{\ell a}$) of the first is equal to the value of   $x_{\ell a}$
    ($x_{dw}$) of the second. This is a consequence of the symmetry of equation (\ref{charp})
    with respect to the permutation of $y$ and $z$.

 In the limit as the mantle becomes negligible, i.e.
 $f_0=\frac{{\rm I}_{\circ c}}{{\rm I}_{\circ m}}\to \infty$, we find two solutions
 to equation (\ref{charp}) 
       \begin{equation}
          x_c=\frac{yz}{y+z}\label{xc}
        \end{equation}
        and  $(1+f_0)(y+z)$. These solutions represent the limit of $x_{dw}(f_0)$
        and $x_{\ell a}(f_0)$ as $f_0\to\infty$. In order to decide whether  $x_{dw}$
        or      $x_{\ell a}$ is asymptotic to $x_c$, we solve  equation (\ref{charp}) 
       with $y=0$, the solutions being $x_{dw}(f_0)=0=x_c$ and   $x_{\ell a}(f_0)=(f_0+1)z$,
        and  with $z=0$, the solutions being $x_{\ell a}(f_0)=0=x_c$ and   $x_{dw}(f_0)=(f_0+1)y$.
        Since $\Delta\ne 0$, by continuity
        the same result  hold for $|y|$ small in the first case and $|z|$ small in the
        second. This fact,  the duality of  $x_{dw}$ and      $x_{\ell a}$ discussed in the
        previous paragraph,
        and the invariance of $x_c$ and  $(1+f_0)(y+z)$ with respect to the permutation
        $y\leftrightarrow z$,  lead us to:
        \begin{equation}\begin{split}
&          |y|<|z| \Longrightarrow  x_{dw}(f_0)\to x_c\ \ \text{and} \ \
x_{\ell a}(f_0)\to(f_0+1)z\ \ \text{as}\ \   f_0\to \infty\\
&          |z|<|y| \Longrightarrow  x_{dw}(f_0)\to(f_0+1)z \ \ \text{and} \ \
x_{\ell a}(f_0)\to x_c \ \ \text{as}\ \   f_0\to \infty\,.\end{split}\label{conj}
\end{equation}

 \section{The Poincar\'e-Hough flow.}

 \label{Poinc}

Following Poincar\'e (see  \cite{Lamb} paragraphs 146 and 384), let $\mat x_0\in\kappa$
be the position  at time $t=0$ of a fluid particle inside a triaxial ellipsoid
with semi-axes $A_1>A_2>A_3$. The mean radius of the ellipsoid is
 $R_c=(A_1A_2A_3)^{1/3}$. For any time the center
of the ellipsoid coincides with the origin of $\kappa$ and at $t=0$ the axes of the ellipsoid
are aligned with the axes of $\kappa$.
At $t=0$ the operator $\mat A:=$Diagonal$\{A_1,A_2,A_3\}$  maps
the ball of radius one onto the ellipsoid.  The ellipsoid  is fixed in the frame
of the mantle $\K_m$ and rotates  inside the inertial frame $\kappa$ according to $\mat R_m(t):\K_m\to\kappa$,
with $\mat R_m(0)=\Id$. The motion of the fluid particle initially at $\mat x_0$ is assumed to be:
\begin{equation}\label{Poinc1}
  \mat x(t)=\mat R_m(t)\mat A\mat R_f(t)\mat A^{-1}\mat x_0\,,
\end{equation}
where $t\to \mat R_m(t)$ is  known. The unknown rotation matrix $\mat R_f(t)$ must be determined from the
fluid dynamic equations.

The velocity field associated with the motion in equation (\ref{Poinc1})  is 
\begin{equation}\label{Poinc2}
  \mat v(\mat x, t)=\dot {\mat x}=\underbrace{\big(\boldsymbol{\widehat \omega}_m+
  \mat R_m\mat A\boldsymbol{\widehat \omega}_f\mat A^{-1} \mat R_m^{-1}\big)}_{:=\mat P}\mat x=\mat P \mat x
\end{equation}
It is convenient to split the operator $\mat A\boldsymbol{\widehat \omega}_f\mat A^{-1}$
into  symmetric and  anti-symmetric parts
\begin{equation}\label{Poinc3}
  \mat A\boldsymbol{\widehat \omega}_f\mat A^{-1}=\underbrace{\frac{1}{2}\Big(
    \mat A\boldsymbol{\widehat \omega}_f\mat A^{-1}+
    \big(\mat A\boldsymbol{\widehat \omega}_f\mat A^{-1}\big)^T\Big)}_{:=\mat S_A}+
  \underbrace{\frac{1}{2}\Big(
    \mat A\boldsymbol{\widehat \omega}_f\mat A^{-1}-
    \big(\mat A\boldsymbol{\widehat \omega}_f\mat A^{-1}\big)^T\Big)}_{:=\boldsymbol{\widehat \tau}}\,,
\end{equation}
such that
\begin{equation}
  \label{Poinc4}
  \mat P= \boldsymbol{\widehat \omega}_m+
  \mat R_m\boldsymbol{\widehat \tau}\mat R_m^{-1}+  \mat R_m\mat S_A \mat R_m^{-1}\,.
\end{equation}
The expressions for $\mat S_A$ and $\boldsymbol{\tau}$ are
\begin{equation}\label{Poinc4.1}
  \mat S_A=\frac{1}{2}\left(
\begin{array}{ccc}
  0 & \left(\frac{A_2}{A_1}-\frac{A_1}{A_2}\right)\omega_{f3} &
 \left(\frac{A_1}{A_3}-\frac{A_3}{A_1}\right)\omega_{f2} \\
   \left(\frac{A_2}{A_1}-\frac{A_1}{A_2}\right)\omega_{f3} & 0 & \left(\frac{A_3}{A_2}-\frac{A_2}{A_3}\right)\omega_{f1} \\  \left(\frac{A_1}{A_3}-\frac{A_3}{A_1}\right)\omega_{f2} 
    & \left(\frac{A_3}{A_2}-\frac{A_2}{A_3}\right)\omega_{f1}
      & 0 \\
\end{array}
\right)
\end{equation}
and
\begin{equation}\label{Poinc4.2}
  \boldsymbol{\widehat \tau}=\frac{1}{2}\left(
\begin{array}{ccc}
  0 & -\left(\frac{A_2}{A_1}+\frac{A_1}{A_2}\right)\omega_{f3} &
 \left(\frac{A_1}{A_3}+\frac{A_3}{A_1}\right)\omega_{f2} \\
   \left(\frac{A_2}{A_1}+\frac{A_1}{A_2}\right)\omega_{f3} & 0 &- \left(\frac{A_3}{A_2}+\frac{A_2}{A_3}\right)\omega_{f1} \\ - \left(\frac{A_1}{A_3}+\frac{A_3}{A_1}\right)\omega_{f2} 
    & \left(\frac{A_3}{A_2}+\frac{A_2}{A_3}\right)\omega_{f1}
      & 0 \\
\end{array}
\right)
\end{equation}
Using that 
$\mat v=\mat P x= \boldsymbol{ \omega}_m\times \mat x
+ (\mat R_m\boldsymbol{\tau})\times \mat x + \mat R_m\mat S_A \mat R_m^{-1}\mat x$ we obtain the vorticity 
field
\begin{equation}\label{Poinc5}
\mat w:=\mat{curl}\, \mat v= 2 \big(\boldsymbol{ \omega}_m+\mat R_m\boldsymbol{\tau}\big)\,,
\end{equation}
which implies
\begin{equation}
  \mat P= \frac{\mat{\widehat w}}{2}+  \mat R_m\mat S_A \mat R_m^{-1}\quad \text{and}\quad
  \mat v= \frac{\mat w}{2}\times \mat x + \mat R_m\mat S_A \mat R_m^{-1}\mat x\,.
  \label{Poinc5.1}
\end{equation}

The equations for the motion of an inviscid,  volume preserving, isentropic  fluid are (Euler):
\begin{equation}\begin{array}{lll}
    & \partial_t \mat v+\mat v\cdot\nabla\mat v=-\nabla h-\nabla\Phi\quad& (\text{dynamical equation})\,,\\
    &    \mat{div}\, \mat v=0\quad&(\text{incompressibility})\,,\\
                  &\partial_t \rho+  \mat{div}\,(\rho \mat v)=0\quad&(\text{conservation of mass})\,,\end{array}
                \label{Poinc6}
\end{equation}
where: $\rho$ is the density, $h$ is the enthalpy ($dh=\frac{dp}{\rho}$ where $p$ is the pressure),
and $\Phi$ is the external gravitational potential.

The divergence of $\mat v=\mat P x$ is zero because $\tr(\mat P)=0$. In order to fulfil the
equation of continuity we impose that,
\begin{equation}
  \rho(\mat x,t)=\rho_\circ(r)\,,\ \ \text{where}\ \ r=
  \|\mat A^{-1} \mat R_m^{-1}(t) \mat x\|\,, 0\le r\le R_c\label{Poinc7}
\end{equation}
If the fluid cavity would be  slowly  deformed to a round shape, then the density
of the fluid inside the round core would be the radial function $\rho_\circ(r)$.
Using that $\mat{div}\, \mat v=0$ the equation for conservation of mass can be written as
$\partial_t \rho+ \mat v\cdot\nabla \rho=\frac{d}{dt}\rho(\mat x(t),t)=0$. Equation
(\ref{Poinc1}) implies 
$\rho(\mat x(t),t)=\rho_\circ( \|\mat A^{-1} \mat R_m^{-1}(t) \mat x(t)\|)=
\rho_\circ( \|\mat A^{-1} \mat x_0\|)$, so $\frac{d}{dt}\rho(\mat x(t),t)=0$ and the equation for 
conservation of mass is verified. 

The last equation in (\ref{Poinc6}) to be verified is the dynamical one. If we take the curl of this
equation, then  we obtain $\partial_t \mat w+\mat v\cdot \nabla \mat w-\mat w\cdot \nabla\mat v=0$,
which is the dynamical equation for the vorticity. If the equation for the vorticity is verified,
then the enthalpy can be determined integrating the dynamical equation for $\mat v$. Since $\mat w$
does not depend on $\mat x$, $\mat v\cdot \nabla \mat w=0$ and, since $\mat v=\mat P x$ depends linearly
on  $\mat x$,   $\mat w\cdot \nabla\mat v=\mat P\mat w$, so the equation for the vorticity
reduces to
\begin{equation}
  \dot{\mat w}=\mat P \mat w\,.\label{Poinc8}
\end{equation}
This equation yields a simple  ordinary differential equation for the unknown function $\boldsymbol\omega_f$
from which we determine the Poincar\'e-Hough flow.

Equation (\ref{Poinc5.1}) implies that the
Tisserand angular velocity $\boldsymbol \omega_c$ associated with the Poincar\'e-Hough flow satisfies
\begin{equation}
 \boldsymbol \pi_c= \mat I_c \boldsymbol \omega_c=\int_{\mathcal B(t)}\rho(\mat x\times \mat v)d  x^3= \mat I_c \frac{\mat w}{2}
+  \int_{\mathcal B(t)}\rho \, \mat x\times\big(\mat R_m\mat S_A \mat R_m^{-1}\mat x)dx^3
  \label{Poinc9}
\end{equation}
where $\mathcal B(t)$ is the set of points inside the cavity at time $t$.
Since the density function is carried by the flow, 
$\mat I_c(t)=\mat R_m(t) \mat I_c(0)\mat R_m^{-1}(t)$.

In order to compute the moment of inertia
of the fluid at  $t=0$ we first compute the components of the density tensor $\mat M(0)$
\begin{equation}
  M_{ij}(0)=\int_{\|\mat A^{-1}\mat x\|\le 1}x_ix_j\rho_\circ(\|\mat A^{-1}\mat x\|)dx^3=
  \frac{ A_iA_j}{R_c^2}\int_{\|\mat y\|\le R_c}y_iy_j\rho_\circ(\|\mat y\|)dx^3\,,\label{Poinc10}
\end{equation}
where we did the change of variables $x_i=(A_i/R_c) y_i$, $i=1,2,3$.
Parity arguments imply that $M_{ij}(0)=0$ if $i\ne j$.
Now, consider a round cavity of radius $R_c$
filled with  a fluid with spherical density $\rho_\circ(r)$. The moment
of inertia about any axis, say $\mat e_3$, of the spherical mass of fluid is
\begin{equation}
  {\rm I}_{\circ c}=\int_{\|\mat y\|\le R_c}(y_1^2+y_2^2)\rho_\circ(\|\mat y\|)dx^3\label{Poinc11}
\end{equation}
So, by symmetry $\int_{\|\mat y\|\le R_c}y_i^2\rho_\circ(\|\mat y\|)dx^3= {\rm I}_{\circ c}/2$
for any $i=1,2,3$ and
\begin{equation}
  M_{ii}(0)=  \frac{ A_i^2}{R_c^2}\frac{{\rm I}_{\circ c}}{2}\,, \quad i=1,2,3\,.\label{Poinc12}
\end{equation}
Using that $\mat I_c=\tr\big(\mat M\big)\Id-\mat M$ we obtain
\begin{equation}
  \mat I_c(0)=\frac{{\rm I}_{\circ c}}{2 R_c^2}\left(\begin{array}{ccc} A_2^2+A_3^2&0&0\\0& A_1^2+A_3^2&0\\
                                                       0&0& A_1^2+A_2^2\end{array}\right)\label{Poinc13}
\end{equation}
  Note that $\tr \mat I_c(0)=    {\rm I}_{\circ c}\frac{A_1^2+A_2^2+A_3^2}{R_c^2}$, so, 
            ${\rm I}_{\circ c}$   may  not have the usual meaning $\frac{1}{3}\tr\mat I_c(0)$.                                 

  In order to compare the results obtained from the Poincar\'e-Hough flow with ours we have to assume that
  the fluid cavity is slightly aspherical, namely
  \begin{equation}
    A_1=R_c(1+\epsilon_1)\quad   A_2=R_c(1+\epsilon_2)\quad   A_3=R_c(1+\epsilon_3)\,,
  \end{equation}
  where $|\epsilon_1|,|\epsilon_2|,$ and $|\epsilon_3|$ are small. Since $A_1A_2A_3=R_c^3$,
  $\epsilon_1+\epsilon_2+\epsilon_3=0$ and 
equation (\ref{Poinc13}) implies
\begin{equation}
  \mat I_c(0)={\rm I}_{\circ c} \Id+
  {\rm I}_{\circ c}\left(\begin{array}{ccc} \epsilon_2+\epsilon_3&0&0\\0& \epsilon_1+\epsilon_3&0\\
                                                       0&0& \epsilon_1+\epsilon_2\end{array}\right)+\Oc(|\epsilon|^2)\,,\label{Poinc14.1}
   \end{equation}
   so, $\frac{1}{3}\tr \mat I_c(0)={\rm I}_{\circ c}$ holds up to second order in the ellipticity\footnote{
     Note that equation (\ref{Poinc14.1}) implies the relation
     relation 
     $\mat B=-$Diagonal$\{\epsilon_2+\epsilon_3,\epsilon_1+\epsilon_3, \epsilon_1+\epsilon_2\}$
     between the inertial deformation
     matrix  $\mat B$ associated with $ \mat I_c(0)$ and the geometric quantities
     $\epsilon_1,\epsilon_2,\epsilon_3$. This simple relation holds only because 
     the density of the fluid
     is constant over homothetic ellipsoids.}.
  Equation (\ref{Poinc4.1})
  implies
\begin{equation}\label{Poinc14}
  \mat S_A=\left(
\begin{array}{ccc}
  0 & \left(\epsilon_2-\epsilon_1\right)\omega_{f3} &
 \left(\epsilon_1-\epsilon_3\right)\omega_{f2} \\
  \left(\epsilon_2-\epsilon_1\right)\omega_{f3} & 0 & \left(\epsilon_3-\epsilon_2\right)\omega_{f1} \\
    \left(\epsilon_1-\epsilon_3\right)\omega_{f2}
    &  \left(\epsilon_3-\epsilon_2\right)\omega_{f1}
      & 0 \\
\end{array}
\right)+\Oc (|\epsilon|^3)\,,
\end{equation} and 
equation (\ref{Poinc4.2}) implies $  \boldsymbol{ \tau}=\boldsymbol \omega_f+\Oc(|\epsilon|^2)$.

A computation shows that the last term in the right-hand side of equation (\ref{Poinc9})
is of the order $\Oc(|\epsilon|^2)$ and, therefore
\begin{equation}
  \boldsymbol \pi_c= \mat I_c \boldsymbol \omega_c= \mat I_c \frac{\mat w}{2} +\Oc(|\epsilon|^2)
  \Longrightarrow  \boldsymbol \omega_c= \frac{\mat w}{2} +\Oc(|\epsilon|^2)=
  \boldsymbol\omega_m+\mat R_m  \boldsymbol\omega_f +\Oc(|\epsilon|^2)\label{Poinc15}
\end{equation}
The fact that $\boldsymbol{\omega}_c\approx \mat w/2$ for small ellipticities
has been noted in \cite{henrard2008rotation}.

For small ellipticities we can write $\boldsymbol{\omega}_c= \mat w/2$ and equation (\ref{Poinc8})
implies
\begin{equation}
  \dot{\boldsymbol{\omega}}_c=\mat P \boldsymbol{\omega}_c\,.\label{Poinc16}
\end{equation}
Therefore
\begin{equation}
  \dot {\boldsymbol \pi}_c=\dot{\mat I}_c\boldsymbol{\omega}_c+{\mat I}_c\dot{\boldsymbol{\omega}}_c=
  \boldsymbol \omega_m\times \mat I_c \boldsymbol{\omega}_c+ \mat I_c(-\boldsymbol{\widehat \omega}_m+\mat P)
  \boldsymbol{\omega}_c
  \end{equation}
Equation (\ref{Poinc2}) implies that the  last term in the right-hand side of this equation can be written as 
\[
  \mat I_c(-\boldsymbol{\widehat \omega}_m+\mat P)
  \boldsymbol{\omega}_c=\mat R_m \mat I_c(0)\mat A\boldsymbol{\widehat \omega}_f\mat A^{-1}
  \mat R_m^{-1} \boldsymbol{\omega}_c\,.
\]
A computation using equation (\ref{Poinc14.1})  shows that
$\mat I_c(0)\mat A\boldsymbol{\widehat \omega}_f\mat A^{-1}= \boldsymbol{\widehat \omega}_f \mat I_c(0)+
\Oc(|\epsilon|^2)$,  so up to second order in the ellipticities
\[
  \dot {\boldsymbol \pi}_c=
  \boldsymbol \omega_m\times \mat I_c \boldsymbol{\omega}_c+
  \mat R_m \boldsymbol{\widehat \omega}_f \mat I_c(0)
  \mat R_m^{-1} \boldsymbol{\omega}_c= \boldsymbol \omega_m\times \mat I_c \boldsymbol{\omega}_c+
  \mat R_m \boldsymbol{\widehat \omega}_f \mat R_m^{-1} \mat I_c \boldsymbol{\omega}_c
\]
and using equation (\ref{Poinc15}) we obtain
\begin{equation}
    \dot {\boldsymbol \pi}_c=
    \boldsymbol \omega_c\times \mat I_c \boldsymbol{\omega}_c
    \label{Poinc17}
  \end{equation}
  that is exactly the equation for the angular momentum of the core obtained in (\ref{eqrot3})
  with $k_c=0$.

  We remark that  the   hypotheses we used in this Appendix to obtain the general results
  in equation (\ref{eqrot3}) from the Poincar\'e-Hough flow, namely:  small
  asphericity, density stratification along concentric ellipsoidal shells and the volume preserving
  property of the fluid flow; are the same hypotheses we have  assumed since \cite{rr2017}.
  In this Appendix any rotational motion  $\mat R_m(t)$ of the ellipsoid is allowed and there is no reason
  to assume that   the principal axes of the mantle are aligned with   those of the fluid cavity
  (a hypothesis commonly assumed).
  Finally, it is crucial for the results in this Appendix that
  the centre of mass of the mantle coincides with that of the fluid core for all time.

 \section{
    The inertial   offset of the core
 rotation axis.}
 \label{offset}
 
 In this appendix we analyse the effect of the inertial torque studied in Section \ref{offsetsec}
 upon
 the fluid core.
  Two different  situations  will be considered:
 one without  spin-orbit resonance and another with spin-orbit resonance.

 \subsection{Precession under no spin-orbit resonance,
   the Tisserand angular velocity of the fluid core, 
    and the vorticity of
    Robert and Stewartson.}
  \label{Robert}

  In this Section we assume
  that both the mantle and the core are  axisymetric with $\ov \gamma=0$, 
  $\ov\alpha=\ov\beta=\ov\alpha_e$, and $I_{c1}=I_{c2}=I_{c3}(1-f_c)$, the mantle is rigid,
  and the precession rate
   is retrograde, $\dot\psi_g<0$.
  In this Section we further assume that $\eta_c/(\omega f_c)\ll 1$.
  The goal is to study  the angular velocity  of the core produced exclusively by the inertial torque.
  The axisymmetry of the problem implies that in the precessional frame (equation  (\ref{Kpr})) 
  both  $\boldsymbol \omega_{m,pr}$ and  $\boldsymbol \omega_{c,pr}$ are
  stationary, and according to equations 
  (\ref{alphamcdif}) and (\ref{ofdef})  are  given by
  \begin{equation}\begin{split}
        \boldsymbol \omega_{m,pr} &=
   \omega \mat e_3+\dot\psi_g\sin\theta_g\mat e_2+
   \omega\, \boldsymbol \delta_m\\     
  \boldsymbol \omega_{c,pr} &=
  \underbrace{ \omega \mat e_3+\dot\psi_g\sin\theta_g\mat e_2}_{\boldsymbol{\omega}_{g,pr}}+
   \omega\, \boldsymbol \delta_c\end{split}\label{omegamcpr}
  \end{equation}
  In order to write $ \boldsymbol \delta_c$ we follow the same steps we did to obtain
  $\boldsymbol \delta_m$ in equation (\ref{ofnon}). After doing this and using that the complex
  compliance is null (the mantle is rigid) and   $\eta_c/(\omega f_c)\ll 1$ ($\Rightarrow y\approx f_c$
  in equation  (\ref{ofnon}))
  we obtain
   \begin{equation}\label{xydif}   \begin{split}
      \boldsymbol\delta_m&= \frac{{\rm I_{\circ c}}}{\Io}\frac{ \dot\psi_g}{\omega f_c}
    \frac{2 \cos^2 \theta_g}{ \sin \theta_g}
    \left(
      \frac{ {\rm I}_{\circ m}}{\Io} \frac{\eta_c}{f_c\,\omega}\, \mat e_1-\mat e_2
    \right) \\     
    \boldsymbol\delta_c   &=\frac{\dot\psi_g}{\omega f_c} \sin \theta_g \left(1-
      2 \frac{{\rm I}_{\circ c}}{\Io}\cot^2\theta_g\right)
    \left(\mat e_2-
      \frac{ {\rm I}_{\circ m}}{\Io} \frac{\eta_c}{f_c\,\omega}\mat e_1\right)
    \\
    &=\boldsymbol\delta_m+ \frac{ \dot\psi_g\,\sin \theta_g}{\omega f_c}\left(\mat e_2-
     \frac{ {\rm I}_{\circ m}}{\Io} \frac{\eta_c}{f_c\,\omega}\mat e_1\right)
    \end{split}
  \end{equation}

  Roberts and Stewartson studied 
the motion  of  an incompressible fluid of constant density
inside  an ellipsoidal  shell of revolution  that precesses  in the same way as the guiding motion
used to obtain the angular velocities in equations (\ref{omegamcpr}) and (\ref{xydif}).
In  \cite{stewartson1963motion} and \cite{roberts1965motion}
the authors solved the Navier-Stokes equations by perturbation methods    (for a numerical study in the
 case of a spherical shell, see \cite{tilgner2001fluid}) assuming the
 two hypotheses
  \begin{equation}
    \frac{f_c \,\omega}{|\dot \psi_g|}\gg 1\quad\text{
      and}\quad  f_c \,\omega \frac{R_c^2}{\nu}=f_c\,\frac{\omega}{\eta_c}\gg 1\,,
    \label{robst}
  \end{equation}
  where: $R_c$ is the core
  mean radius,  $\nu$ is kinematic viscosity of the fluid, and
  $\eta_c= R_c^2/\nu$ is the viscosity coefficient defined
  in equation (\ref{etac}). These are the same hypotheses we  assumed to obtain equation
  (\ref{xydif}). 
  In the following we show that the difference
  $\boldsymbol \omega_{c,pr}-\boldsymbol \omega_{m,pr}=\omega(\boldsymbol\delta_c-\boldsymbol\delta_m)$
  is equal to  the average vorticity of the
 flow inside the cavity, computed by Robert and Stewartson,
minus  the vorticity of the flow induced by
a  rigid rotation. 

The velocity field of the fluid inside the  cavity with respect to the
precessional frame $\K_{pr}$ is \cite{stewartson1963motion}
\begin{equation}
  \mat u=\dot\phi_g\mat e_3\times \mat x+
  2\dot\psi_g\sin\theta_g \frac{a^2}{a^2-b^2}\left(x_3\mat e_1-\frac{b^2}{a^2} x_1\mat
    e_3\right)
  \label{rob1}
  \end{equation}
  where $b<a$ are the semi-axis of the cavity\footnote{\label{Stewartson} The coordinates
    in $\K_{pr}$ used in \cite{stewartson1963motion}
    are related to ours by the map 
    $(\mat e_1, \mat e_2)\to (-\mat e_2, \mat e_1)$ and their angle $\alpha$
    is equal to  $\pi-\theta_g$. In ibid.  the motion of
    the mantle is the motion of our guiding frame
    $\mat R_g=\mathbf{R_3}( \psi_g )\mathbf{R_1}(\theta_g)\mathbf{R_3}(\phi_g)
    :\K_m\to\kappa$,
    which implies that the motion of $\K_m$ with respect to $\K_{pr}$ is
    $\mathbf{R_3}(\phi_g):\K_m\to\K_{pr}$. Since the cavity is an ellipsoid
    of revolution it remains at rest in $\K_{pr}$ while its boundary rotate
    with angular velocity $\dot\phi_g\mat e_3$ (what we call $\dot\phi_g$ they call
    $\omega$).
    So, if the
    precessional velocity would be zero, then $\K_{pr}$ would be an inertial frame
    and viscosity would eventually bring the fluid to rotate, at least in the average,
    as a rigid body. This explains   the term $\dot\phi_g\mat e_3\times \mat x$ in the
    velocity field $\mat u$. The precessional angular velocity,
    which in $\K_{pr}$ is $\dot\psi_g(\sin\theta_g\mat e_2+\cos\theta_g\mat e_3)$,
    induces
    inertial forces upon the fluid that generate the additional non-rigid term
    to $\mat u$.}. The vorticity associated with this velocity field is
  \begin{equation}
    {\rm curl}\,\mat u=2\dot\phi\mat e_3+ 2\dot\psi_g\sin\theta_g
    \frac{a^2+b^2}{a^2-b^2}\mat e_2\label{rob2}
    \end{equation}
    If the fluid were moving as a rigid body attached to the mantle, then its
    vorticity would be  ${\rm curl} \dot\phi_g\mat e_3\times \mat x=
    2\dot\phi\mat e_3$. Recalling that vorticity is a measure of rotation
    of the vector field, as we have already mentioned in Appendix \ref{Poinc},
    \begin{equation}
     2\dot\psi_g\sin\theta_g
     \frac{a^2+b^2}{a^2-b^2}\mat e_2=
     \underbrace{{\rm curl}\, \mat u}_{rot. fluid}-
     \underbrace{2\dot\phi\mat e_3}_{rot.  mantle}\label{rob3}
   \end{equation}
   In order to relate equation (\ref{rob3}) to the difference
   $ \boldsymbol \omega_{c,pr}-\boldsymbol \omega_{m,pr}$ we use that the moments
   of inertia of an ellipsoid of revolution of constant density $\rho$  are
   $\ov I_{ce}=\frac{4\pi}{15}\rho (a^2+b^2)a^2b$  and
   $\ov I_{c3}=\frac{8\pi}{15}\rho a^4b$ that implies
   \begin{equation}
     f_c= \frac{\ov I_{c3}-\ov I_{ce}}{\ov I_{c3}}\approx
     \frac{\ov I_3-\ov I_{ce}}{\ov I_{ce}}=\frac{a^2-b^2}{a^2+b^2}\label{rob4}
   \end{equation}
Therefore, using the second hypothesis of Robert and Stewartson 
$\frac{\eta_c}{f_c\omega}\ll 1$ (equation (\ref{robst})) equations (\ref{omegamcpr}),
(\ref{xydif}), (\ref{rob3}), and (\ref{rob4}) imply the claimed result: 
\begin{equation}
  \boldsymbol \omega_{c,pr}-\boldsymbol \omega_{m,pr}\approx 
\frac{1}{2}\bigg({\rm curl}\, \mat u-
 2\dot\phi\mat e_3\bigg)= \frac{\dot\psi_g}{f_c}\sin\theta_g\mat e_2\,.
\end{equation}
We could also have computed the total angular momentum of the fluid with respect
to the precessional frame and arrived at the same result.

\subsection{The  resonant case and the Cassini states
  of G.  Bou\'e.}
\label{Boue}

The Cassini states of a body made of a rigid mantle and a fluid core with no
 viscous coupling  ($k_c=\eta_c=0$) were computed  in
 \cite{boue2020cassini}. In  ibid, the  difference
 between the  angular velocities of  mantle and core was not supposed  small and many
 possible Cassini states were found.  Our goal in the rest of this Appendix
 is to compare the inertial offsets $\boldsymbol \delta_m$ and $\boldsymbol \delta_c$
 with some of the  Cassini states found in \cite{boue2020cassini}.
 
 The expressions for $\boldsymbol \delta_m$ and $\boldsymbol \delta_c$  in equations
 (\ref{ofres})  simplify
 when we assume that:  the moment of inertia of the fluid core
 is not much smaller than that of the mantle,  the mantle is rigid,
 and $\eta_c=0$, 
\begin{equation}\begin{split}
  & \delta_m= \delta_{m2}= 
    - \frac{\frac{{\rm I}_{\circ c}}{{\rm I}_{\circ m}}
        \frac{\dot\psi_g}{\omega } \cos\theta_g} {\left(\frac{\dot\psi_g}{\omega} \cos\theta_g\right)^2
  +\frac{\dot\psi_g}{\omega} \cos\theta_g\frac{\Io}{{\rm I}_{\circ m}}(y+z)+\frac{\Io}{{\rm I}_{\circ m}}
  y z }
   \frac{\dot\psi_g}{\omega}\sin\theta_g\,,
   \\ & \delta_c=\delta_{c2}
 =
\frac{\frac{\Io}{{\rm I}_{\circ m}} z+ 
        \frac{\dot\psi_g}{\omega } \cos\theta_g} {\left(\frac{\dot\psi_g}{\omega} \cos\theta_g\right)^2
  +\frac{\dot\psi_g}{\omega} \cos\theta_g\frac{\Io}{{\rm I}_{\circ m}}(y+z)+\frac{\Io}{{\rm I}_{\circ m}}
  y z }\frac{\dot\psi_g}{\omega}\sin\theta_g\,,
  \end{split}\label{ofresmc}
\end{equation}
where $z=c_1 \ov \alpha_e +\frac{c_2}{2}\ov \gamma$ and $y=f_c$.
       According to
             equation (\ref{ofdef}) all the three vectors: the normal to the invariable plane
              $\mat e_3\in\kappa$, 
              the angular velocity of the mantle  $\langle \boldsymbol \omega_{m}\rangle_{pr}\in\kappa$,
              and
               the angular velocity of the core  $\langle \boldsymbol \omega_{c}\rangle_{pr}\in\kappa$,
                 both  averaged in the precessional frame; are contained in the same plane,
                 which is denoted by $\Sigma$ in Figure \ref{delta} LEFT.

                 The Cassini states    in \cite{boue2020cassini} are determined by the following two
                 equations (Equations (34a) and (34b) in ibid.)\footnote{The
                   correspondence between the notation in \cite{boue2020cassini}
                   and ours is: $\omega_p\to \dot \phi_g$, $\alpha_c\to f_c$, $g\to\dot\psi_g$,
                   $\alpha\to\ov\alpha_e$, $\beta\to\ov\gamma$, $C_c\to\ov I_{c3}$,
                   $C_m\to\ov I_{m3}$,$C\to\ov I_3$, $\theta_m^\prime\to \chi-\delta_m$,
                   $\theta_c^\prime\to \chi-\delta_c$,
                   $\theta_m^\prime-\iota\to \theta_g-\delta_m$, and
                   $\theta_c^\prime-\iota\to \theta_g-\delta_c$.
                   These correspondences  hold for both Cassini states 1 and 2. 
             }
                   \begin{equation}\begin{split}
      &   f_c\dot\phi_g\cos(\delta_m-\delta_c)\sin(\delta_m-\delta_c)+
      \dot\psi_g\sin(\theta_g-\delta_c)=0
      \\ & 
      \ov I_{c3}\Big(\dot\psi_g\sin(\theta_g-\delta_m)+f_c\dot\phi_g
\cos(\delta_m-\delta_c)\sin(\delta_m-\delta_c)\Big)-\frac{\ov I_3\omega^2}{\dot\phi_g}P_e(\delta_m)=0
    \end{split}\label{Boueeq}
    \end{equation}
    where
       \begin{equation}\begin{split}
           P_e(\delta_m)=&
     \frac{3}{2}\frac{G m_p}{a_p^3\omega^2}\bigg(\ov\alpha_e
       X^{-3,0}_0\cos(\chi_p-\delta_m)\sin(\chi_p-\delta_m)\\& 
       +\frac{\ov\gamma}{4}X^{-3,2}_s
   \big(1+\cos(\chi_p-\delta_m)\big)\sin(\chi_p-\delta_m)\bigg)
 +\frac{\dot\psi_g\dot\phi_g}{\omega^2}\sin(\theta_g-\delta_m)
 \,.\end{split}\label{peale3}
  \end{equation}
The identity  $P_e(0)=0$ holds due to Peale's equation (\ref{peale}).

Since our  inertial offset was computed by means of a perturbation of a rigid-body motion,
for which $\delta_m=\delta_c=0$, it is natural to look for solutions $(\delta_m,\delta_c)\approx (0,0)$
to equations (\ref{Boueeq}). For  $(\delta_m,\delta_c)=(0,0)$,  the first equation in  (\ref{Boueeq})
implies  $\dot\psi_g\sin(\theta_g)=0$ and then the second equation implies
\begin{equation}
  P_e(0)=0\Longrightarrow 
 \bigg(\ov\alpha_e
       X^{-3,0}_0\cos(\chi_p) 
       +\frac{\ov\gamma}{4}X^{-3,2}_s
   \big(1+\cos(\chi_p)\big)\bigg)\sin(\chi_p)=0\,.
 \end{equation}
 This equation has several solutions $\chi_p=0$, $\chi_p=\pi$, and
 $\cos\chi_p= -\frac{\ov \gamma  X^{-3,2}_s}{4 \ov\alpha_e  X^{-3,0}_0+\ov \gamma X^{-3,2}_s}$,
 where  $\chi_p$ is the obliquity of the body spin to the orbital plane. Among these solutions
 only $\chi_p=0$   has obliquity smaller then $\pi/2$
 (except
 $\cos\chi_p= -\frac{\ov \gamma  X^{-3,2}_s}{4 \ov\alpha_e  X^{-3,0}_0+\ov \gamma X^{-3,2}_s}$
 with s=1, for which the obliquity is smaller but close to $\pi/2$). We will only consider
 solutions to equations (\ref{Boueeq}) with obliquities smaller than $\pi/2$
 that originate from $ \chi_p=0$.

 In order to  show that the small solutions  $(\delta_m,\delta_c)$ to equations (\ref{Boueeq}) are
 given by equation (\ref{ofresmc}), it is enough to expand the functions in the left-hand side
 of equations (\ref{Boueeq}) up to first order in  $(\delta_m,\delta_c)$ and then to solve the
 linear system. This was done using the algebraic manipulator Mathematica.

 The Hessian determinant of equation  (\ref{Boueeq})  is proportional
 to $(x_{\ell a}+\frac{\dot\psi_g}{\omega } \cos\theta_g)
 (x_{dw}+\frac{\dot\psi_g}{\omega } \cos\theta_g)$,
  where  $\omega x_{dw}$ and $\omega x_{\ell a}$
  are the eigenfrequencies of the nearly diurnal free wobble (NDFW) and the
  free libration in latitude (FLL)
  in the inertial space. This shows that the critical
  flattening of the core $f_c\approx 0.005 \ov\alpha_c$ found for Mercury in Section 6.1 of
  \cite{boue2020cassini} (see also  Figure 3 in ibid.)
  is indeed a resonance  of the precession angular speed
  $-\dot \psi_d\cos\theta_g\approx$330 kyr
  with the NDFW eigenfrequency,  which  decreases to zero as $f_c\to 0$. The presence
  of dissipation attenuates the singularity, as illustrated in Figure  \ref{Mercury}.
 
\bibliographystyle{plainnat}
\bibliography{mybibliography}

\end{document}